\documentclass[12pt]{article}

\usepackage[utf8]{inputenc}
\usepackage{textcomp}
\usepackage{chetnew}
\usepackage{tikz}
\usepackage{amssymb,amsfonts,mathrsfs,dsfont,yfonts,bbm}\usetikzlibrary{calc}
\usepackage{xcolor}
\usepackage{braket}
\usepackage{outlines}
\usetikzlibrary{cd}

\newcommand{\be}{\begin{equation}}
\newcommand{\ee}{\end{equation}}
\newcommand{\bea}{\begin{eqnarray}}
\newcommand{\eea}{\end{eqnarray}}

\newcommand{\DZ}{\mathds{Z}}

\newcommand{\CA}{\mathcal{A}}

\newcommand{\CD}{\mathcal{D}}

\newcommand{\CQ}{\mathcal{Q}}
\newcommand{\CB}{\mathcal{B}}
\newcommand{\CC}{\mathcal{C}}

\newcommand{\CT}{\mathcal{T}}

\newcommand{\CS}{\mathcal{S}}
\newcommand{\CM}{\mathcal{M}}

\newcommand*{\boxcoloro}{orange}
\makeatletter
\newcommand{\boxedo}[1]{\textcolor{\boxcoloro}{%
\tikz[baseline={([yshift=-1ex]current bounding box.center)}] \node [rectangle, minimum width=1ex,rounded corners,draw] {\normalcolor\m@th$\displaystyle#1$};}}
 \makeatother
\newcommand*{\boxcolorr}{red}
\makeatletter
\newcommand{\boxedr}[1]{\textcolor{\boxcolorr}{%
\tikz[baseline={([yshift=-1ex]current bounding box.center)}] \node [rectangle, minimum width=1ex,rounded corners,draw] {\normalcolor\m@th$\displaystyle#1$};}}
 \makeatother
\newcommand*{\boxcolorb}{blue}
\makeatletter
\newcommand{\boxedb}[1]{\textcolor{\boxcolorb}{%
\tikz[baseline={([yshift=-1ex]current bounding box.center)}] \node [rectangle, minimum width=1ex,rounded corners,draw] {\normalcolor\m@th$\displaystyle#1$};}}
 \makeatother
\newcommand*{\boxcolorg}{green}
\makeatletter
\newcommand{\boxedg}[1]{\textcolor{\boxcolorg}{%
\tikz[baseline={([yshift=-1ex]current bounding box.center)}] \node [rectangle, minimum width=1ex,rounded corners,draw] {\normalcolor\m@th$\displaystyle#1$};}}
 \makeatother
 \newcommand*{\boxcolorp}{purple}
\makeatletter
\newcommand{\boxedp}[1]{\textcolor{\boxcolorp}{%
\tikz[baseline={([yshift=-1ex]current bounding box.center)}] \node [rectangle, minimum width=1ex,rounded corners,draw] {\normalcolor\m@th$\displaystyle#1$};}}
 \makeatother
  \newcommand*{\boxcolorc}{cyan}
\makeatletter
\newcommand{\boxedc}[1]{\textcolor{\boxcolorc}{%
\tikz[baseline={([yshift=-1ex]current bounding box.center)}] \node [rectangle, minimum width=1ex,rounded corners,draw] {\normalcolor\m@th$\displaystyle#1$};}}
 \makeatother
  \newcommand*{\boxcolory}{yellow}
\makeatletter
\newcommand{\boxedy}[1]{\textcolor{\boxcolory}{%
\tikz[baseline={([yshift=-1ex]current bounding box.center)}] \node [rectangle, minimum width=1ex,rounded corners,draw] {\normalcolor\m@th$\displaystyle#1$};}}
 \makeatother

\usepackage{amsmath}	% required for `\align' (yatex added)
\begin{document}

\title{Invertibility of Condensation Defects and \\[3mm] Symmetries of 2 + 1d QFTs}

\author{Matthew Buican$^{1}$ and Rajath Radhakrishnan$^{2}$}

\affiliation{\smallskip ${}^{1}$CTP and Department of Physics and Astronomy\\
Queen Mary University of London, London E1 4NS, UK\\
${}^{2}$International Centre for Theoretical Physics, Strada Costiera 11, Trieste 34151, Italy}

\abstract{We characterize discrete (anti-)unitary symmetries and their non-invertible generalizations in $2+1$d topological quantum field theories (TQFTs) through their actions on line operators and fusion spaces. We explain all possible sources of non-invertibility that can arise in this context. Our approach gives a simple $2+1$d proof that non-invertible generalizations of unitary symmetries exist if and only if a bosonic TQFT contains condensable bosonic line operators (i.e., these non-invertible symmetries are necessarily \lq\lq non-intrinsic"). Moving beyond unitary symmetries and their non-invertible cousins, we define a non-invertible generalization of time-reversal symmetries and derive various properties of TQFTs with such symmetries. Finally, using recent results on 2-categories, we extend our results to corresponding statements in $2+1$d quantum field theories that are not necessarily topological.}

\setcounter{tocdepth}{2}

\maketitle
\toc

\section{Introduction}
The modern perspective on symmetries in quantum field theory (QFT) is that they correspond to sectors of topological defects of various dimensions \cite{gaiotto2015generalized,cordova2022snowmass}. Fully understanding the symmetry of a given QFT then amounts to finding all topological operators in the theory and their correlations functions. Thinking in this way leads down a path beyond groups to the world of (non-)invertible symmetries, which are conjecturally characterized by higher-fusion categories \cite{Bhardwaj:2022yxj,Bhardwaj:2022kot,Bhardwaj:2022maz}. 

In spite of this understanding, even classifying the set of topological defects in a QFT can be difficult. However, some general structural properties of symmetries can be deduced from studying consistency conditions in higher-fusion categories.\footnote{For certain non-invertible symmetries, some of the associativity relations in the higher-fusion category have been worked out \cite{Copetti:2023mcq}.} For the purposes of this paper, we will be interested in the rich case of $2+1$d QFTs and their discrete symmetries. Mathematically, the relevant structures are fusion 2-categories, which capture the properties of topological surface and line operators \cite{kapustin2010surface,Fuchs_2013,Johnson_Freyd_2022}.

In this setting, it is interesting to understand how the different sub-sectors of fusion 2-categories interact with and constrain each other.\footnote{Understanding how more general sectors of QFTs constrain each other is of course also interesting, although typically much harder (e.g., see \cite{Buican:2021xhs} for a relation between non-topological local operators and topological surfaces in certain broad classes of theories).} To that end, the authors of \cite{JOHNSON_FREYD_2021} found an elegant and inspirational result: if a $2+1$d QFT does not contain any topological line operators, then all topological surfaces are invertible.

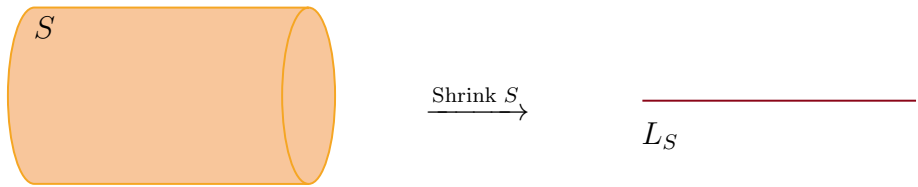
\begin{figure}[h!]
    \centering

\tikzset{every picture/.style={line width=0.75pt}} %set default line width to 0.75pt        

\begin{tikzpicture}[x=0.75pt,y=0.75pt,yscale=-1,xscale=1]

%Shape: Can [id:dp17309188464242953] 
\draw  [color={rgb, 255:red, 245; green, 166; blue, 35 }  ,draw opacity=1 ][fill={rgb, 255:red, 248; green, 198; blue, 155 }  ,fill opacity=1 ] (248.63,194.02) -- (110.33,193.98) .. controls (102.96,193.97) and (96.99,174.05) .. (97,149.47) .. controls (97.01,124.89) and (102.99,104.97) .. (110.37,104.98) -- (248.67,105.02) .. controls (256.04,105.03) and (262.01,124.95) .. (262,149.53) .. controls (261.99,174.11) and (256.01,194.03) .. (248.63,194.02) .. controls (241.26,194.02) and (235.29,174.1) .. (235.3,149.52) .. controls (235.31,124.94) and (241.29,105.02) .. (248.67,105.02) ;
%Straight Lines [id:da9030958192577389] 
\draw [color={rgb, 255:red, 139; green, 6; blue, 24 }  ,draw opacity=1 ]   (417,152) -- (555,152) ;

% Text Node
\draw (306,142.4) node [anchor=north west][inner sep=0.75pt]    {$\xrightarrow{\text{Shrink } S}$};
% Text Node
\draw (108.72,107.77) node [anchor=north west][inner sep=0.75pt]  [rotate=-0.93]  {$S$};
% Text Node
\draw (415,161.4) node [anchor=north west][inner sep=0.75pt]    {$L_{S}$};

\end{tikzpicture}
    \caption{Shrinking a topological surface operator, $S$, to get a topological line operator, $L_S$.}
    \label{fig:shrinkS}
\end{figure}

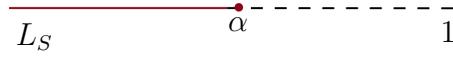
\begin{figure}[h!]
    \centering

\tikzset{every picture/.style={line width=0.75pt}} %set default line width to 0.75pt        

\begin{tikzpicture}[x=0.75pt,y=0.75pt,yscale=-1,xscale=1]

%Straight Lines [id:da682926502928498] 
\draw [color={rgb, 255:red, 139; green, 6; blue, 24 }  ,draw opacity=1 ]   (200.84,141.84) -- (310.84,141.84) ;
%Straight Lines [id:da8068039251009165] 
\draw  [dash pattern={on 4.5pt off 4.5pt}]  (310.84,141.84) -- (425,142) ;
%Shape: Circle [id:dp7591214709423143] 
\draw  [color={rgb, 255:red, 139; green, 6; blue, 24 }  ,draw opacity=1 ][fill={rgb, 255:red, 139; green, 6; blue, 24 }  ,fill opacity=1 ] (316.75,143.15) .. controls (315.84,143.15) and (315.1,142.41) .. (315.1,141.5) .. controls (315.1,140.59) and (315.84,139.85) .. (316.75,139.85) .. controls (317.66,139.85) and (318.4,140.59) .. (318.4,141.5) .. controls (318.4,142.41) and (317.66,143.15) .. (316.75,143.15) -- cycle ;

% Text Node
\draw (202.84,148.24) node [anchor=north west][inner sep=0.75pt]    {$L_{S}$};
% Text Node
\draw (417,147.4) node [anchor=north west][inner sep=0.75pt]    {$1$};
% Text Node
\draw (309.84,145.24) node [anchor=north west][inner sep=0.75pt]    {$\alpha $};

\end{tikzpicture}
    \caption{A junction of $L_S$ with the trivial line.}
    \label{fig:LSjunction}
\end{figure}

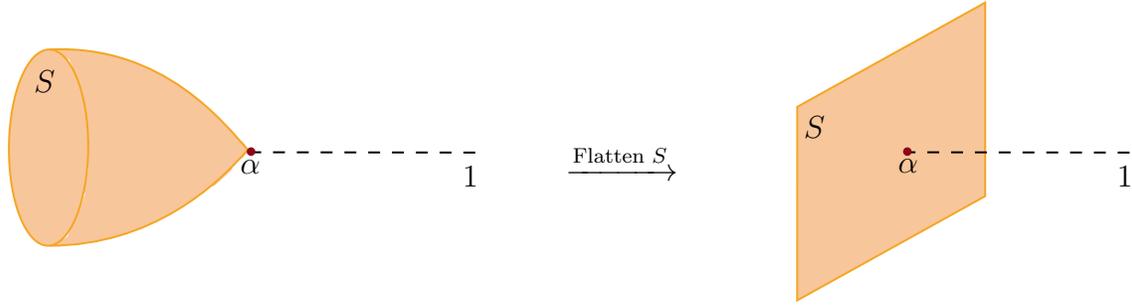
\begin{figure}[h!]
    \centering

\tikzset{every picture/.style={line width=0.75pt}} %set default line width to 0.75pt        

\begin{tikzpicture}[x=0.75pt,y=0.75pt,yscale=-1,xscale=1]

%Straight Lines [id:da8068039251009165] 
\draw  [dash pattern={on 4.5pt off 4.5pt}]  (169.84,140.84) -- (284,141) ;
%Shape: Free Drawing [id:dp4847252353474151] 
\draw  [color={rgb, 255:red, 248; green, 198; blue, 155 }  ,draw opacity=0 ][line width=6] [line join = round][line cap = round] (91.67,102.67) .. controls (94.56,104.11) and (97.44,105.56) .. (100.33,107) ;
%Shape: Free Drawing [id:dp5284603888042504] 
\draw  [color={rgb, 255:red, 248; green, 198; blue, 155 }  ,draw opacity=1 ][line width=6] [line join = round][line cap = round] (85.33,96) .. controls (85.89,96.89) and (86.44,97.78) .. (87,98.67) ;
%Shape: Free Drawing [id:dp7551231642930005] 
\draw  [color={rgb, 255:red, 248; green, 198; blue, 155 }  ,draw opacity=1 ][line width=6] [line join = round][line cap = round] (84.67,95.67) .. controls (84.67,95.67) and (84.67,95.67) .. (84.67,95.67) ;
%Shape: Free Drawing [id:dp26267675436228166] 
\draw  [color={rgb, 255:red, 248; green, 198; blue, 155 }  ,draw opacity=1 ][line width=6] [line join = round][line cap = round] (83.33,94) .. controls (86,97.67) and (88.67,101.33) .. (91.33,105) ;
%Shape: Free Drawing [id:dp291664790332813] 
\draw  [color={rgb, 255:red, 248; green, 198; blue, 155 }  ,draw opacity=1 ][line width=6] [line join = round][line cap = round] (89.67,98.67) .. controls (89.89,99.33) and (90.11,100) .. (90.33,100.67) ;
%Shape: Free Drawing [id:dp6685012187272841] 
\draw  [color={rgb, 255:red, 248; green, 198; blue, 155 }  ,draw opacity=1 ][line width=2.25] [line join = round][line cap = round] (86,100) .. controls (87.45,105.03) and (85.43,115) .. (90.67,115) ;
%Shape: Free Drawing [id:dp19319573740566387] 
\draw  [color={rgb, 255:red, 248; green, 198; blue, 155 }  ,draw opacity=1 ][line width=2.25] [line join = round][line cap = round] (86.67,103) .. controls (89.33,106) and (92,109) .. (94.67,112) ;
%Shape: Free Drawing [id:dp35956606857372364] 
\draw  [color={rgb, 255:red, 248; green, 198; blue, 155 }  ,draw opacity=1 ][line width=2.25] [line join = round][line cap = round] (93.67,102.33) .. controls (94.57,106.48) and (95.28,110.76) .. (95,115) ;
%Shape: Free Drawing [id:dp020879279713432952] 
\draw  [color={rgb, 255:red, 248; green, 198; blue, 155 }  ,draw opacity=1 ][line width=2.25] [line join = round][line cap = round] (90.67,107) .. controls (91.44,108.33) and (92.22,109.67) .. (93,111) ;
%Shape: Free Drawing [id:dp5693077215588687] 
\draw  [color={rgb, 255:red, 248; green, 198; blue, 155 }  ,draw opacity=1 ][line width=2.25] [line join = round][line cap = round] (88.33,107.67) .. controls (89.11,109.22) and (89.89,110.78) .. (90.67,112.33) ;
%Shape: Free Drawing [id:dp953762903631483] 
\draw  [color={rgb, 255:red, 248; green, 198; blue, 155 }  ,draw opacity=1 ][line width=2.25] [line join = round][line cap = round] (90.33,107) .. controls (91,108) and (91.67,109) .. (92.33,110) ;
%Shape: Free Drawing [id:dp7576982653891762] 
\draw  [color={rgb, 255:red, 248; green, 198; blue, 155 }  ,draw opacity=1 ][line width=2.25] [line join = round][line cap = round] (85,100.67) .. controls (84.67,100.33) and (88.46,103.62) .. (90,109) ;
%Shape: Free Drawing [id:dp9883133489648468] 
\draw  [color={rgb, 255:red, 248; green, 198; blue, 155 }  ,draw opacity=1 ][line width=2.25] [line join = round][line cap = round] (84.33,100) .. controls (86.7,101.57) and (87.64,104.69) .. (88.67,107.33) ;
%Shape: Free Drawing [id:dp9103152479208199] 
\draw  [color={rgb, 255:red, 248; green, 198; blue, 155 }  ,draw opacity=1 ][line width=2.25] [line join = round][line cap = round] (101.33,108.67) .. controls (104.22,113.11) and (107.26,117.46) .. (110,122) .. controls (115.29,130.76) and (88.31,100.72) .. (88,99.67) .. controls (87.02,96.29) and (94.05,103.33) .. (96.67,105.67) .. controls (100.95,109.5) and (104.46,114.57) .. (109.67,117) .. controls (110.09,117.2) and (109.08,116.23) .. (108.67,116) .. controls (107.28,115.22) and (105.62,114.94) .. (104.33,114) .. controls (100.14,110.92) and (96.13,107.56) .. (92.33,104) .. controls (90.31,102.1) and (96.65,107.52) .. (99,109) .. controls (101.78,110.75) and (104.64,112.39) .. (107.67,113.67) .. controls (108.31,113.94) and (107.51,112.15) .. (107,111.67) .. controls (105.62,110.38) and (101.09,109.05) .. (102.67,108) .. controls (104.36,106.87) and (106.22,110) .. (108,111) ;
%Shape: Free Drawing [id:dp384484064713041] 
\draw  [color={rgb, 255:red, 248; green, 198; blue, 155 }  ,draw opacity=1 ][line width=6] [line join = round][line cap = round] (98,114.67) .. controls (99.96,115.23) and (102.28,119.42) .. (103.33,117.67) .. controls (105.57,113.95) and (86.65,105.05) .. (94.33,104.33) .. controls (99.68,103.83) and (117.1,116.31) .. (123,119.33) .. controls (128.76,122.28) and (135,124.23) .. (140.67,127.33) .. controls (147.8,131.24) and (154.29,136.25) .. (161.33,140.33) .. controls (162.01,140.73) and (164.39,140.35) .. (163.67,140.67) .. controls (160.26,142.17) and (156.42,142.51) .. (153,144) .. controls (148.71,145.86) and (136.69,148.22) .. (140.67,150.67) .. controls (162.34,164) and (157,133.33) .. (161.33,137.67) ;
%Shape: Free Drawing [id:dp03152109070807674] 
\draw  [color={rgb, 255:red, 248; green, 198; blue, 155 }  ,draw opacity=1 ][line width=6] [line join = round][line cap = round] (157,147.67) .. controls (151.93,148.97) and (146.36,149.43) .. (142,152.33) ;
%Shape: Free Drawing [id:dp34147068020232396] 
\draw  [color={rgb, 255:red, 248; green, 198; blue, 155 }  ,draw opacity=1 ][line width=6] [line join = round][line cap = round] (150.33,141.67) .. controls (144.19,145.25) and (142.06,148.25) .. (135.33,153.33) .. controls (131.36,156.34) and (124.35,158.21) .. (124.67,158) .. controls (134.55,151.36) and (145.36,146.18) .. (155.33,139.67) .. controls (159.87,136.7) and (146.17,145.71) .. (141,147.33) .. controls (139.47,147.81) and (143.06,144.82) .. (143.67,143.33) .. controls (145.36,139.19) and (143.97,139.49) .. (141.33,141) .. controls (139.03,142.31) and (136.97,144.02) .. (134.67,145.33) .. controls (133.07,146.25) and (131.37,146.98) .. (129.67,147.67) .. controls (129.14,147.88) and (127.64,148.43) .. (128,148) .. controls (130.63,144.87) and (133.87,142.31) .. (136.67,139.33) .. controls (137.42,138.53) and (139.23,135.92) .. (139,137) .. controls (137.96,141.96) and (134.26,146.65) .. (135,151.67) .. controls (135.79,157.01) and (147.55,139.51) .. (147.67,139.33) .. controls (147.97,138.87) and (149.44,138.67) .. (149,138.33) .. controls (143.14,133.94) and (142.25,138.43) .. (135,136.33) .. controls (131.75,135.39) and (131.26,130.36) .. (128.33,128.67) .. controls (123.17,125.68) and (112,121.33) .. (112,121.33) .. controls (112,121.33) and (119.83,125.53) .. (124,127) .. controls (131.23,129.55) and (139.24,130.05) .. (146,133.67) .. controls (149.23,135.4) and (138.35,135.5) .. (135,134) .. controls (129.56,131.56) and (124.06,129.15) .. (119,126) .. controls (115.45,123.79) and (112.1,121.14) .. (109.33,118) .. controls (108.59,117.16) and (110.12,114.62) .. (109,114.67) .. controls (106.37,114.79) and (104.57,117.74) .. (102,118.33) .. controls (96.8,119.53) and (106.38,102.61) .. (96.33,110.33) .. controls (94.42,111.8) and (96.71,125.96) .. (97.33,128) .. controls (99.08,133.75) and (105.23,156.28) .. (112.33,164) .. controls (112.87,164.58) and (113.67,163.04) .. (114,162.33) .. controls (115.36,159.44) and (116.42,156.4) .. (117.33,153.33) .. controls (119.46,146.19) and (119.46,136.22) .. (118.67,129.67) .. controls (118.42,127.63) and (117.94,133.71) .. (117.33,135.67) .. controls (115.98,140.02) and (111.03,145.03) .. (113.67,150) .. controls (114.77,152.08) and (117.19,146.84) .. (118.67,145) .. controls (121.21,141.82) and (123.46,138.42) .. (125.67,135) .. controls (127.13,132.74) and (127.67,126.2) .. (129.67,128) .. controls (132.15,130.23) and (129.5,134.7) .. (129,138) .. controls (128.82,139.18) and (126.93,142.28) .. (127.67,141.33) .. controls (145.38,118.56) and (123.74,145.56) .. (123,147) .. controls (122.25,148.46) and (125.94,145.53) .. (127.33,144.67) .. controls (130.06,142.97) and (132.63,137.61) .. (135.33,139.33) .. controls (137.81,140.91) and (132.85,144.72) .. (131,147) .. controls (127.59,151.21) and (116.88,154.02) .. (119.67,158.67) .. controls (122.36,163.16) and (128.16,152.34) .. (133,150.33) .. controls (136.34,148.95) and (121.96,159.31) .. (125.33,158) .. controls (127.08,157.32) and (137.12,149.39) .. (141.67,151.67) .. controls (142.69,152.18) and (140.84,153.89) .. (140,154.67) .. controls (137.45,157.05) and (134.88,159.64) .. (131.67,161) .. controls (129.85,161.77) and (106.13,169.01) .. (101.67,167.33) .. controls (100.17,166.77) and (103.31,164.57) .. (104.33,163.33) .. controls (106.54,160.67) and (109.16,158.36) .. (111.33,155.67) .. controls (114.65,151.54) and (125.54,136.65) .. (122.67,132.33) .. controls (121.88,131.15) and (120.57,134.36) .. (120,135.67) .. controls (117.8,140.7) and (119.35,152.23) .. (114,151) .. controls (108.67,149.77) and (115.35,140.13) .. (115.67,134.67) .. controls (115.79,132.56) and (116.57,126.62) .. (115.33,128.33) .. controls (112.27,132.59) and (111.81,138.22) .. (109.67,143) .. controls (108.96,144.57) and (110.8,139.71) .. (111,138) .. controls (111.58,133.13) and (111.78,128.23) .. (112,123.33) .. controls (112.27,117.46) and (111,144.92) .. (107.33,140.33) .. controls (101.27,132.76) and (106.04,120.63) .. (102.33,111.67) .. controls (101.08,108.65) and (100.19,117.85) .. (99.33,121) .. controls (99.07,121.97) and (99.03,125.01) .. (99,124) .. controls (98.86,118.89) and (97.3,113.49) .. (99,108.67) .. controls (100.1,105.56) and (100.22,115.14) .. (101,118.33) .. controls (102.82,125.76) and (101.59,140.53) .. (103,144) .. controls (103.74,145.82) and (105.39,148.29) .. (107.33,148) .. controls (109.67,147.65) and (110.52,144.51) .. (112,142.67) .. controls (112.75,141.73) and (114.38,138.53) .. (114,139.67) .. controls (111.74,146.44) and (108.89,153) .. (106.33,159.67) ;
%Shape: Free Drawing [id:dp8730344396696194] 
\draw  [color={rgb, 255:red, 248; green, 198; blue, 155 }  ,draw opacity=1 ][line width=6] [line join = round][line cap = round] (108.33,140.67) .. controls (107.06,147.04) and (104.01,152.94) .. (101.67,159) .. controls (100.74,161.39) and (100.25,163.94) .. (99.33,166.33) .. controls (98.8,167.73) and (95.89,169.97) .. (97.33,170.33) .. controls (101.22,171.3) and (108.26,160.2) .. (110.67,162) .. controls (112.02,163.02) and (109.39,165.36) .. (108,166.33) .. controls (100.42,171.62) and (93.05,177.44) .. (84.67,181.33) .. controls (82.29,182.44) and (88.48,177.48) .. (89.33,175) .. controls (92.09,167) and (94.04,158.7) .. (95.33,150.33) .. controls (97.67,135.24) and (96.03,130.73) .. (92.67,116.67) .. controls (92.2,114.73) and (95.86,135.07) .. (96.33,137) .. controls (97.92,143.52) and (100.67,149.75) .. (102,156.33) .. controls (102.86,160.56) and (102.96,171.33) .. (99.33,169) .. controls (94.1,165.63) and (97.55,148.68) .. (97,144) .. controls (96.11,136.49) and (94.2,129.13) .. (93,121.67) .. controls (92.8,120.42) and (93.88,124.07) .. (94,125.33) .. controls (94.55,131.21) and (95.54,137.13) .. (95,143) .. controls (93.69,157.14) and (88.79,171.02) .. (82.33,183.67) ;
%Shape: Free Drawing [id:dp6699970926146334] 
\draw  [color={rgb, 255:red, 248; green, 198; blue, 155 }  ,draw opacity=1 ][line width=6] [line join = round][line cap = round] (90.33,159.33) .. controls (97.35,145.3) and (86.82,108.96) .. (84.33,99) ;
%Shape: Free Drawing [id:dp9548217182352592] 
\draw  [color={rgb, 255:red, 248; green, 198; blue, 155 }  ,draw opacity=1 ][line width=6] [line join = round][line cap = round] (147.67,144.33) .. controls (149.52,143.29) and (164.33,134.06) .. (164.33,140) ;
%Shape: Free Drawing [id:dp5373617437343092] 
\draw  [color={rgb, 255:red, 248; green, 198; blue, 155 }  ,draw opacity=1 ][line width=0.75] [line join = round][line cap = round] (166.33,141) .. controls (168.16,141) and (170.16,141) .. (168.33,141) ;
%Shape: Free Drawing [id:dp14018865309054773] 
\draw  [color={rgb, 255:red, 248; green, 198; blue, 155 }  ,draw opacity=1 ][line width=0.75] [line join = round][line cap = round] (165.67,140.33) .. controls (166.77,140.46) and (168.21,141.12) .. (169,140.33) ;
%Shape: Free Drawing [id:dp9074590754554999] 
\draw  [color={rgb, 255:red, 248; green, 198; blue, 155 }  ,draw opacity=1 ][line width=0.75] [line join = round][line cap = round] (166.33,139.33) .. controls (167,139.56) and (167.67,139.78) .. (168.33,140) ;
%Curve Lines [id:da32402854905225253] 
\draw [color={rgb, 255:red, 245; green, 166; blue, 35 }  ,draw opacity=1 ][fill={rgb, 255:red, 248; green, 198; blue, 155 }  ,fill opacity=1 ]   (69,89) .. controls (127,86) and (158.1,127.5) .. (169.1,139.5) ;
%Curve Lines [id:da5777106084262225] 
\draw [color={rgb, 255:red, 245; green, 166; blue, 35 }  ,draw opacity=1 ][fill={rgb, 255:red, 248; green, 198; blue, 155 }  ,fill opacity=1 ]   (69,188) .. controls (121,188) and (154,157) .. (168.84,139.84) ;
%Shape: Ellipse [id:dp7939336738432786] 
\draw  [color={rgb, 255:red, 245; green, 166; blue, 35 }  ,draw opacity=1 ][fill={rgb, 255:red, 248; green, 198; blue, 155 }  ,fill opacity=1 ] (68.26,187.99) .. controls (57.21,187.91) and (48.42,165.68) .. (48.63,138.35) .. controls (48.83,111.01) and (57.95,88.92) .. (69,89) .. controls (80.05,89.08) and (88.83,111.31) .. (88.63,138.65) .. controls (88.42,165.98) and (79.3,188.08) .. (68.26,187.99) -- cycle ;
%Shape: Free Drawing [id:dp514478035345145] 
\draw  [color={rgb, 255:red, 255; green, 255; blue, 255 }  ,draw opacity=1 ][line width=0.75] [line join = round][line cap = round] (161.33,151.33) .. controls (159.33,153.78) and (157.33,156.22) .. (155.33,158.67) ;
%Shape: Free Drawing [id:dp5036989569288491] 
\draw  [color={rgb, 255:red, 255; green, 255; blue, 255 }  ,draw opacity=1 ][line width=0.75] [line join = round][line cap = round] (161.33,151) .. controls (159,153.33) and (156.67,155.67) .. (154.33,158) ;
%Shape: Free Drawing [id:dp229510590365815] 
\draw  [color={rgb, 255:red, 255; green, 255; blue, 255 }  ,draw opacity=1 ][line width=0.75] [line join = round][line cap = round] (162,150) .. controls (159.11,152.78) and (156.22,155.56) .. (153.33,158.33) ;
%Shape: Free Drawing [id:dp7262107353776467] 
\draw  [color={rgb, 255:red, 255; green, 255; blue, 255 }  ,draw opacity=1 ][line width=0.75] [line join = round][line cap = round] (162.67,149) .. controls (159,152.33) and (155.33,155.67) .. (151.67,159) ;
%Shape: Parallelogram [id:dp8197481888863074] 
\draw  [color={rgb, 255:red, 245; green, 166; blue, 35 }  ,draw opacity=1 ][fill={rgb, 255:red, 248; green, 198; blue, 155 }  ,fill opacity=1 ] (446.2,215.61) -- (446.22,117.92) -- (541.02,65.31) -- (541,163) -- cycle ;
%Straight Lines [id:da9528777924274862] 
\draw  [dash pattern={on 4.5pt off 4.5pt}]  (499.84,140.84) -- (614,141) ;
%Shape: Circle [id:dp09435564196250168] 
\draw  [color={rgb, 255:red, 139; green, 6; blue, 24 }  ,draw opacity=1 ][fill={rgb, 255:red, 139; green, 6; blue, 24 }  ,fill opacity=1 ] (501.75,142.15) .. controls (500.84,142.15) and (500.1,141.41) .. (500.1,140.5) .. controls (500.1,139.59) and (500.84,138.85) .. (501.75,138.85) .. controls (502.66,138.85) and (503.4,139.59) .. (503.4,140.5) .. controls (503.4,141.41) and (502.66,142.15) .. (501.75,142.15) -- cycle ;
%Shape: Circle [id:dp7591214709423143] 
\draw  [color={rgb, 255:red, 139; green, 6; blue, 24 }  ,draw opacity=1 ][fill={rgb, 255:red, 139; green, 6; blue, 24 }  ,fill opacity=1 ] (170.75,142.15) .. controls (169.84,142.15) and (169.1,141.41) .. (169.1,140.5) .. controls (169.1,139.59) and (169.84,138.85) .. (170.75,138.85) .. controls (171.66,138.85) and (172.4,139.59) .. (172.4,140.5) .. controls (172.4,141.41) and (171.66,142.15) .. (170.75,142.15) -- cycle ;

% Text Node
\draw (59.84,98.24) node [anchor=north west][inner sep=0.75pt]    {$S$};
% Text Node
\draw (276,146.4) node [anchor=north west][inner sep=0.75pt]    {$1$};
% Text Node
\draw (448.22,121.32) node [anchor=north west][inner sep=0.75pt]    {$S$};
% Text Node
\draw (606,146.4) node [anchor=north west][inner sep=0.75pt]    {$1$};
% Text Node
\draw (163.84,143.24) node [anchor=north west][inner sep=0.75pt]    {$\alpha $};
% Text Node
\draw (495.61,142.86) node [anchor=north west][inner sep=0.75pt]    {$\alpha $};
% Text Node
\draw (328,135.4) node [anchor=north west][inner sep=0.75pt]    {$\xrightarrow{\text{Flatten } S}$};

\end{tikzpicture}
    \caption{The distinct junctions of $L_S$ with the trivial line operator imply distinct local operators on $S$.}
    \label{fig:flattenS}
\end{figure}

The argument is simple and ingenious: consider a topological surface operator, $S$, on a cylinder and shrink the cylinder to get a line operators, $L_S$ (see Fig. \ref{fig:shrinkS}). 
In general, $L_S$ decomposes into a sum of simple lines.\footnote{A simple line operator is a line that hosts only a single local topological operator on it.} Now, consider the junction of $L_S$ with the trivial line operator (see Fig. \ref{fig:LSjunction}).
The junction is labelled by a point operator, $\alpha \in \text{Hom}(L_S,1)$, where the dimension of $\text{Hom}(L_S,1)$ determines the number of copies of the trivial line in $L_S$. Replacing $L_S$ with the surface operator, $S$, we get Fig. \ref{fig:flattenS}.
Upon flattening $S$, we find that each $\alpha$ at the junction of $L_S$ with the trivial line implies a distinct local operator on $S$. This argument shows that if we shrink a simple surface operator,\footnote{A simple surface operator is a surface which hosts only a single local topological operator on it.} then $L_S$ contains only one copy of the trivial line operator.

Now, suppose a 2+1D QFT has no non-trivial topological line operators. Shrinking a simple surface operator, $S$, of this QFT, we get the trivial line. This implies that $S$ is invertible. Indeed, suppose 
\be
S \times \bar S= S_1 + S_2 + \cdots~,
\ee
for some surface operators, $S_1$, $S_2$, $\cdots$. Shrinking each of these surfaces, we get the fusion rule
\be
L_S \times L_{\bar S}= L_{S_1} + L_{S_2} + \cdots~.
\ee
Since the QFT only has the trivial line operator, we get
\be
1 \times 1 = 1 + 1 + \cdots~,
\ee
which is inconsistent. 

What about the converse of the above theorem? Does the existence of topological line operators guarantee that there are non-invertible surface operators? No. For example, $SU(2)_1$ Chern-Simons theory has a non-trivial line operator. However, this TQFT does not have any invertible or non-invertible surface operators. This discussion motivates us to study the following question in $2+1$d QFT:

\medskip
\noindent
{\bf Question:} {\it What properties of a theory's line operators determine whether there are non-invertible surface operators?} 
\medskip

Except for a discussion of non-genuine lines in Appendix \ref{sec:twisted}, we will only consider genuine line operators of a 2+1d QFT and symmetries implemented by topological surfaces acting on them. In particular, {\it we will only consider surface operators which act faithfully on the line operators and their fusion spaces.} In other words, we assume that a surface operator which acts trivially on the topological line operators and its fusion spaces is the trivial surface operator.\footnote{The non-invertible symmetries of 2+1D TQFTs can all be implemented by condensation defects \cite{Fuchs_2013,Roumpedakis:2022aik}. As we will discuss in section \ref{sec:faithfullness}, condensation defects satisfy this assumption.} 

Another important aspect of symmetries in QFT is their use in constraining the space of QFTs. Important examples of these ideas include $c$-theorems in different dimensions and their constraints on renormalization group flows, studies of metrics on conformal manifolds, and the constraints that topological modular forms place on more general deformations of 2d QFTs.

In many of these cases, the resulting relations between different theories can be understood in terms of interfaces of different kinds: RG interfaces (e.g., see \cite{Brunner:2007qu,Gaiotto:2012np,Konechny:2023xvo}), Janus interfaces (e.g., see \cite{Bachas:2001vj,Bachas:2013nxa}), and more general domain walls. Since we will be particularly interested in theories that can be related to each other through various topological manipulations like discrete gauging and condensation, the interfaces that are relevant to us are gapped domain walls.

Such gapped domain walls give, in a more mathematical language, theories with symmetries characterized by Morita equivalent (higher-)fusion categories. In the case of fusion 2-categories, D\'ecoppet showed that, up to Morita equivalence, any such category can be written in the following factorized form \cite{thibault2022drinfeld}
\begin{equation}\label{introMorita}
\text{2Vec}_{G}^{\omega} ~ \boxtimes \text{ Mod}(\CC)~,
\end{equation}
where the first factor denotes a set of invertible surface operators with fusion group $G$ and anomaly $\omega\in H^4(G,U(1))$, and the second factor denotes the line operators of a modular tensor category (MTC) \cite{Moore:1988qv,etingof2016tensor}, $\CC$, and the surfaces formed by all possible higher gaugings of these lines. In other words, $\text{ Mod}(\CC)$ denotes a (non-spin) $2+1$d topological quantum field theory (TQFT). Therefore, to answer our above question, it suffices to answer it in the context of $2+1$d TQFTs.

As a result, we embark on a study of the action of surfaces on lines and fusion spaces (i.e., trivalent junctions of lines) in different classes of $2+1$d TQFTs.\footnote{This approach can be thought of as a more \lq\lq active" version of the more \lq\lq passive" approach of understanding which lines can end on the boundary of a folded theory (or, in a more mathematical language, of understanding Lagrangian algebras of the folded theory) \cite{Kapustin:2010if,davydov2011witt,Fuchs_2013}.} As a simple starting point, we analyze $2+1$d non-spin Abelian TQFTs, where the action of surfaces on lines alone leads to a theorem on invertibility of surface operators. These are theories in which all line operators are invertible (i.e., their fusion is described by an Abelian group, $A$). As is well known, Abelian TQFTs are classified by the topological spins of their lines and $A$. In other words, these theories are described by their modular data. As we will see, modularity is then sufficiently powerful to lead to the following result:

\bigskip
\noindent
{\bf Theorem:} {\it In a $2+1$d Abelian TQFT (see table \ref{tb:terminology} for details), non-invertible surfaces exist if and only if there is a non-anomalous 1-form symmetry.}
\bigskip

We then turn our attention to non-Abelian TQFTs. In this case, the modular data does not determine the theory. Moreover, as we will see, the action of surfaces on lines does not suffice to give a generalization of Theorem 1. To make progress on such a generalization, we need to understand more precisely how a non-invertible symmetry acts on fusion spaces. To that end, we prove a lemma showing that fusion spaces are not new sources of non-invertibility: if a symmetry has an invertible action on lines, then it has an invertible action on fusion spaces. With this lemma in hand we prove that:

\bigskip
\noindent
{\bf Theorem:} {\it  $2+1$d TQFTs (see table \ref{tb:terminology} for details) without bosons only have invertible symmetries.}

\bigskip
\noindent
The above theorem generalizes one direction of our result in the Abelian case. To find a partial generalization of the converse, it turns out we need to delve deeper into the structure of the \lq\lq higher gauging" \cite{gaiotto2019condensations,Roumpedakis:2022aik,Choi:2022zal} that leads to the surfaces in the ${\rm Mod}(\CC)$ factor of the fusion 2-category in \eqref{introMorita}.\footnote{In studying condensation surface operators, there are two extreme cases where the classification becomes easier. One is when the the category of lines, $\CC$, is symmetric. In other words, the $\CS$-matrix of braidings of the lines is trivial. In particular, when the line operators are all bosonic, $\CC$ is equivalent to the representation category of a finite group, $\CC \simeq \text{Rep}(G)$ \cite{deligne2002categories}. The condensation surfaces obtained in this case were classified in terms of higher representation theory in \cite{Bhardwaj:2022lsg,bartsch2023noninvertible,greenough2010monoidal}. Various properties, like the fusion rules of these surfaces, can be determined explicitly. In particular, condensation surface operators obtained from higher-gauging line operators described by a Rep$(G)$ category always include non-invertible surface operators for any non-trivial group $G$. In a gauge theory, the fusion rules of these surface operators contain crucial information about the gauge group \cite{Radhakrishnan:2023zcq}. In deriving our results on TQFTs, we study the opposite case where the $\CS$-matrix of the line operators is non-degenerate.} In so doing we find the following general constraint:

\bigskip
\noindent
{\bf Theorem:} {\it A $2+1$d TQFT (see table \ref{tb:terminology} for details) has a non-invertible symmetry if and only if it contains condensable bosonic line operators.}

\bigskip
The above theorems follow from the existence of certain \lq\lq universal" non-invertible surfaces (i.e., surfaces that exist whenever there is a non-invertible symmetry).\footnote{These surface operators are a generalization of the universal surface operators found in \cite{Bhardwaj:2022lsg,bartsch2023noninvertible}.} However, given theories typically have a much more complicated zoology of non-invertible surfaces that includes these universal surfaces as a subsector. We give an abstract characterization of this zoology in terms of special Frobenius Algebras. Moreover, in the case of a condensation defect arising from gauging Abelian lines (even in a non-Abelian TQFT), we give a precise characterization of invertibility of the surface operator in terms of the data of the Abelian lines and their embedding in the larger theory.

Using the above results, combined with a simple pictorial derivation of the result of \cite{thibault2022drinfeld} that leads to the factorization in \eqref{introMorita}, we arrive at the following theorem:

\bigskip
\noindent
{\bf Theorem:} {\it A $2+1$d QFT (see table \ref{tb:terminology} for details) has a non-invertible symmetry if and only if it contains condensable bosonic line operators.}

\bigskip
\noindent
We conclude our discussion of non-invertible symmetries by rephrasing our results in terms of \lq\lq non-intrinsic" non-invertible symmetries \cite{Kaidi:2022cpf,Kaidi:2023maf}.

\begin{table}[h!]
\centering
 \begin{tabular}{||c | c||} 
 \hline
 Terminology & Meaning \\ [0.5ex] 
 \hline\hline
 Abelian TQFT  & non-spin Abelian TQFT\\
TQFT & non-spin TQFT\\
QFT & QFT without a transparent fermionic line \\
unitary symmetry & Invertible symmetry that preserves spin \\
anti-unitary / time-reversal symmetry & Invertible symmetry that conjugates spin \\
non-invertible symmetry & Non-invertible generalization of a unitary symmetry \\
non-invertible time-reversal symmetry & Non-invertible generalization of a time-reversal symmetry\\
line operator & genuine line operator \\
MTC & unitary MTC
\\ [1ex] 
 \hline
 \end{tabular}
 \caption{The terminology employed in this paper. We adopt a shorthand so as not to overload definitions, theorems, and discussions with too many adjectives. In particular, note that {\it all TQFTs we study in this paper are non-spin (i.e., they do not require a spin structure to be well-defined)}.}
 \label{tb:terminology}
\end{table}

The final topic we tackle in this paper is to give a definition of a non-invertible generalization of anti-unitary or \lq\lq time-reversal" symmetries (e.g., see  \cite{Choi:2022rfe} for discussion in the context of Maxwell theory and QCD). We point out various peculiar features of these symmetries, including the fact that it is not always possible to distinguish them from standard non-invertible symmetries (see table \ref{tb:terminology} for an explanation of our terminology) via their actions on lines, even when the theory contains lines that are genuinely anyonic (i.e., not bosonic or fermionic).

The structure of the paper is as follows. In the next section, we start with a review of what is meant by a (non-)invertible surface. After that, we proceed to a discussion of symmetries of abelian TQFTs in Sec. \ref{sec: Non-inv sym abelian TQFTs}. We define a non-invertible symmetry of an abelian TQFT, derive various consequences, and prove a theorem on the existence of non-invertible symmetries in such theories. We move on to define non-invertible symmetries of non-Abelian TQFTs in Sec. \ref{sec:noninv sym of non-abelian TQFTs}, and we find a partial generalization of our theorem in Abelian theories. We then discuss higher-gauging and condensation surfaces in Sec. \ref{resolution}. We use this machinery to prove a general theorem on the existence of non-invertible symmetries in $2+1$d TQFT. We then discuss how our results fit in to a general theorem on $2+1$d QFT in Sec. \ref{sec:QFTembedd}. In Sec. \ref{non-inv-TR} we move on to non-invertible time-reversal symmetries and describe certain peculiar features of these symmetries. After concluding with some open questions, we include various appendices that show how our \lq\lq active" definitions of symmetries relate to more standard \lq\lq passive" definitions in terms of Lagrangian algebras in folded theories and also explain some basic aspects of how our work extends to non-genuine lines.

\section{Basics of surface fusion and invertibility}

\label{sec:invertibility_basic}
In this section we give a basic idea of what it means for a topological surface operator to be invertible, and we sketch out some facts about the action of (non-)invertible surface operators on lines. In subsequent sections, we give more precise definitions of surfaces and derive various consequences.

A surface operator, $S$, is said to be invertible if there exists another surface operator, $\bar S$, such that
\be
\label{eq:invfusion}
S\times \bar S=\bar S\times S= \mathds{1}~,
\ee
where $\mathds{1}$ is the trivial surface operator. The invertible surface operators form a group under fusion.

To prove this statement, note that, if $S_1$ and $S_2$ are invertible surface operators, then $S_1 \times S_2\ne0$. Indeed, suppose this fusion was zero, then multiplying both sides on the right by $\bar S_2$ implies that $S_1$ is the zero operator, which is not consistent with invertibility of $S_1$. Given that $S_1 \times S_2\ne0$, this product is itself an invertible surface operator with inverse $\bar S_2 \times \bar S_1$. Therefore, the invertible surface operators are closed under fusion. Assuming that the fusion is associative, we have an associative multiplication on the surface operators with $\mathds{1}$ being the identity. This means that the invertible surface operators form a monoid. The inverse of $S$ is unique because otherwise
\be
S \times \bar S = \mathds{1}= S \times \bar S'~,
\ee
implies that $\bar S=\bar S'$. Therefore, the invertible surface operators indeed form a group. 

Since $\mathds{1}$ acts trivially on all line operators, it is clear that the action of $S$ and $\bar S$ on any line operator cannot be zero. In fact, the action of an invertible surface operator on a simple line operator should result in a single outcome. To understand this, suppose we have instead
\be
S \cdot p= p_1 + p_2~,
\ee
for some simple lines, $p_1$ and $p_2$. Then, we find
\be
\bar S \times S \cdot p= \bar S \cdot p_1 + \bar S \cdot p_2~.
\ee
Since the action of $\bar S$ on a line operator cannot be trivial, the R.H.S of this expression has at least two line operators. This contradicts the trivial action of $\mathds{1}$ on $p$. Therefore, the action of an invertible surface operator on a simple line operator should result in a single outcome. Equation \eqref{eq:invfusion} implies that the action of $S$ and $\bar S$  should be invertible functions on the set of simple line operators: permutations. 

Next let us suppose $S$ is a non-invertible surface, and let us further assume that the action of surface operators on line operators is faithful. It is easy to then show that the action of $S$ on the simple lines cannot be a permutation. To the contrary, assume that the action of $S$  on the simple lines is a permutation. Let $N$ be the order of this permutation. Then, the action of $S^N$ on the line operators is trivial. If the action of the surface operators on the lines is faithful, then 
\be
S^N= \mathds{1}~,
\ee
and we have
\be
S \times \bar S= \mathds{1}~,
\ee
where $\bar S:=S^{N-1}$. This result contradicts the assumption that $S$ is non-invertible. As we will see in section \ref{sec:noninv sym of non-abelian TQFTs}, surfaces in non-Abelian TQFTs can act trivially on lines but still act non-trivially on fusion spaces. However, we will show that such surfaces (including the hypothetical $S^N$ above) are necessarily invertible and so our above discussion can be appropriately modified to show that non-invertible surfaces cannot act via permutations.\footnote{Since we are dealing with surface operators which from a fusion 2-category, there must be a finite $N$ such that $\mathds{1}\in S^N$. In particular, a surface which acts trivially on the lines and non-trivially on the fusion spaces also has a finite order, and therefore the above argument holds for such surface operators.}

In the next section we give a precise definition of a symmetry in an Abelian TQFT and derive various consequences.

\section{(Non-)invertible symmetries of Abelian TQFTs}

\label{sec: Non-inv sym abelian TQFTs}
In this section we discuss basic properties of Abelian TQFTs (recall from table \ref{tb:terminology} that we will only consider non-spin theories) before moving on to a definition of topological surfaces in these theories. We then derive some consequences of our definition, culminating in theorem 1.

The line operators of an abelian TQFT, $\CT$, have fusion rules given by an abelian group, $A$. The correlation functions of the line operators are then fixed by the choice of a non-degenerate quadratic form on $A$ called the topological spin \cite{Belov:2005ze,stirling2008abelian,Kapustin_2011}
\be
\theta: A \to U(1)~.
\ee
The non-degeneracy of the quadratic form is, by definition, the invertiblity of the matrix
\be
\label{eq:Smatdef}
\CS_{a,b}:=\frac{1}{\sqrt{|A|}} \frac{\theta_{a\times b}}{\theta_a\theta_b}~,
\ee
where $|A|$ is the order of $A$. The group, $A$, along with the topological spin, $\theta$, define a pointed MTC, $\CC$ \cite{etingof2016tensor}. In fact, since we can reconstruct $A$ from $\CS$ via the Verlinde formula, we can equivalently think of the MTC as being defined by the modular data, $(\CS,\theta)$. We will use the notation $\CT$ and $\CC$ interchangeably to describe a TQFT.

In the next subsection, we review the definition of a unitary invertible symmetry of an abelian TQFT and interpret it in terms of the action of a surface operator on the line operators. Then we will define non-invertible generalizations of these symmetries (which we will simply call \lq\lq non-invertible symmetries") and study their properties.

\subsection{Characterizing unitary symmetries of Abelian TQFTs and their non-invertible generalizations}

\label{sec:char sym of abelian TQFTs}

A unitary 0-form symmetry of an abelian TQFT must preserve the fusion rules of the lines. Therefore, it is a permutation, $\sigma$, of the line operators that is an automorphism of the fusion group, $A$. This permutation should also preserve the correlation functions of line operators. Since the correlation functions are completely determined by the topological spins, we require
\be
\theta_{\sigma(a)}=\theta_a~.
\ee
Using \eqref{eq:Smatdef}, it follows that $\CS$ is also preserved under the symmetry action
\be
\label{eq:SmatScommut}
\CS_{\sigma(a)\sigma(b)}=\CS_{ab}~.
\ee
Therefore, $\sigma$, viewed as an $|A| \times |A|$ permutation matrix, commutes with $\CS$ and $\CT$, where
\be
\CT_{ab}:=\delta_{a,b}\theta_b~.
\ee
The group of unitary invertible symmetries of abelian TQFTs have been classified for many types of abelian TQFTs \cite{Delmastro:2019vnj,Wang_2020}.

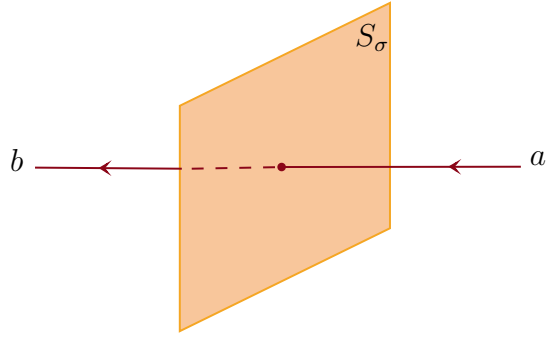
\begin{figure}[h!]
    \centering

\tikzset{every picture/.style={line width=0.75pt}} %set default line width to 0.75pt        

\begin{tikzpicture}[x=0.75pt,y=0.75pt,yscale=-1,xscale=1]

%Shape: Parallelogram [id:dp6562840942953989] 
\draw  [color={rgb, 255:red, 245; green, 166; blue, 35 }  ,draw opacity=1 ][fill={rgb, 255:red, 248; green, 198; blue, 155 }  ,fill opacity=1 ] (263.98,219.76) -- (264,106) -- (370,54.06) -- (369.98,167.82) -- cycle ;
%Straight Lines [id:da007882411965093872] 
\draw [color={rgb, 255:red, 139; green, 6; blue, 24 }  ,draw opacity=1 ]   (316.99,136.91) -- (436,136.76) ;
%Straight Lines [id:da550732043590615] 
\draw [color={rgb, 255:red, 139; green, 6; blue, 24 }  ,draw opacity=1 ] [dash pattern={on 4.5pt off 4.5pt}]  (270,137.76) -- (316.99,136.91) ;
%Straight Lines [id:da3364633224260798] 
\draw [color={rgb, 255:red, 139; green, 6; blue, 24 }  ,draw opacity=1 ]   (190.97,137.21) -- (265,137.76) ;
\draw  [color={rgb, 255:red, 139; green, 6; blue, 24 }  ,draw opacity=1 ][fill={rgb, 255:red, 139; green, 6; blue, 24 }  ,fill opacity=1 ] (227.33,139.5) -- (223.98,137.39) -- (227.4,135.38) -- (225.67,137.42) -- cycle ;
%Shape: Circle [id:dp6782029889732475] 
\draw  [color={rgb, 255:red, 139; green, 6; blue, 24 }  ,draw opacity=1 ][fill={rgb, 255:red, 139; green, 6; blue, 24 }  ,fill opacity=1 ] (313.69,136.91) .. controls (313.69,136) and (314.43,135.26) .. (315.34,135.26) .. controls (316.25,135.26) and (316.99,136) .. (316.99,136.91) .. controls (316.99,137.82) and (316.25,138.56) .. (315.34,138.56) .. controls (314.43,138.56) and (313.69,137.82) .. (313.69,136.91) -- cycle ;
\draw  [color={rgb, 255:red, 139; green, 6; blue, 24 }  ,draw opacity=1 ][fill={rgb, 255:red, 139; green, 6; blue, 24 }  ,fill opacity=1 ] (403.67,138.83) -- (400.31,136.72) -- (403.73,134.72) -- (402,136.75) -- cycle ;

% Text Node
\draw (351,63.4) node [anchor=north west][inner sep=0.75pt]    {$S_{\sigma }$};
% Text Node
\draw (439,127.4) node [anchor=north west][inner sep=0.75pt]    {$a$};
% Text Node
\draw (177,126.4) node [anchor=north west][inner sep=0.75pt]    {$b$};

\end{tikzpicture}
\caption{The symmetry action of a surface on a line operator.}
\label{action}
\end{figure}

\begin{figure}
    \centering

\tikzset{every picture/.style={line width=0.75pt}} %set default line width to 0.75pt        

\begin{tikzpicture}[x=0.75pt,y=0.75pt,yscale=-1,xscale=1]

%Shape: Parallelogram [id:dp10998265343804803] 
\draw  [color={rgb, 255:red, 245; green, 166; blue, 35 }  ,draw opacity=1 ][fill={rgb, 255:red, 248; green, 198; blue, 155 }  ,fill opacity=1 ] (90.98,230.76) -- (91,117) -- (197,65.06) -- (196.98,178.82) -- cycle ;
%Straight Lines [id:da958904988416184] 
\draw [color={rgb, 255:red, 139; green, 6; blue, 24 }  ,draw opacity=1 ]   (152.99,137.91) -- (248,137.76) ;
%Straight Lines [id:da5224461561005044] 
\draw [color={rgb, 255:red, 139; green, 6; blue, 24 }  ,draw opacity=1 ] [dash pattern={on 4.5pt off 4.5pt}]  (96,138.76) -- (154.99,137.91) ;
%Straight Lines [id:da43831028924332205] 
\draw [color={rgb, 255:red, 139; green, 6; blue, 24 }  ,draw opacity=1 ]   (28.97,138.21) -- (91,138.76) ;
\draw  [color={rgb, 255:red, 139; green, 6; blue, 24 }  ,draw opacity=1 ][fill={rgb, 255:red, 139; green, 6; blue, 24 }  ,fill opacity=1 ] (59.33,140.5) -- (55.98,138.39) -- (59.4,136.38) -- (57.67,138.42) -- cycle ;
%Shape: Circle [id:dp6562124981161794] 
\draw  [color={rgb, 255:red, 139; green, 6; blue, 24 }  ,draw opacity=1 ][fill={rgb, 255:red, 139; green, 6; blue, 24 }  ,fill opacity=1 ] (151.69,137.91) .. controls (151.69,137) and (152.43,136.26) .. (153.34,136.26) .. controls (154.25,136.26) and (154.99,137) .. (154.99,137.91) .. controls (154.99,138.82) and (154.25,139.56) .. (153.34,139.56) .. controls (152.43,139.56) and (151.69,138.82) .. (151.69,137.91) -- cycle ;
\draw  [color={rgb, 255:red, 139; green, 6; blue, 24 }  ,draw opacity=1 ][fill={rgb, 255:red, 139; green, 6; blue, 24 }  ,fill opacity=1 ] (222.67,139.83) -- (219.31,137.72) -- (222.73,135.72) -- (221,137.75) -- cycle ;
%Straight Lines [id:da6916224948050725] 
\draw [color={rgb, 255:red, 139; green, 6; blue, 24 }  ,draw opacity=1 ]   (126.99,166.91) -- (245,166.76) ;
%Straight Lines [id:da27680946277590823] 
\draw [color={rgb, 255:red, 139; green, 6; blue, 24 }  ,draw opacity=1 ] [dash pattern={on 4.5pt off 4.5pt}]  (93,167.76) -- (125.34,167.26) ;
%Straight Lines [id:da8199878145279552] 
\draw [color={rgb, 255:red, 139; green, 6; blue, 24 }  ,draw opacity=1 ]   (27.97,167.21) -- (88,167.76) ;
\draw  [color={rgb, 255:red, 139; green, 6; blue, 24 }  ,draw opacity=1 ][fill={rgb, 255:red, 139; green, 6; blue, 24 }  ,fill opacity=1 ] (57.33,169.5) -- (53.98,167.39) -- (57.4,165.38) -- (55.67,167.42) -- cycle ;
%Shape: Circle [id:dp8422080219608383] 
\draw  [color={rgb, 255:red, 139; green, 6; blue, 24 }  ,draw opacity=1 ][fill={rgb, 255:red, 139; green, 6; blue, 24 }  ,fill opacity=1 ] (124.69,166.91) .. controls (124.69,166) and (125.43,165.26) .. (126.34,165.26) .. controls (127.25,165.26) and (127.99,166) .. (127.99,166.91) .. controls (127.99,167.82) and (127.25,168.56) .. (126.34,168.56) .. controls (125.43,168.56) and (124.69,167.82) .. (124.69,166.91) -- cycle ;
\draw  [color={rgb, 255:red, 139; green, 6; blue, 24 }  ,draw opacity=1 ][fill={rgb, 255:red, 139; green, 6; blue, 24 }  ,fill opacity=1 ] (221.67,168.83) -- (218.31,166.72) -- (221.73,164.72) -- (220,166.75) -- cycle ;
%Shape: Parallelogram [id:dp15296129489472066] 
\draw  [color={rgb, 255:red, 245; green, 166; blue, 35 }  ,draw opacity=1 ][fill={rgb, 255:red, 248; green, 198; blue, 155 }  ,fill opacity=1 ] (456.98,234.76) -- (457,121) -- (563,69.06) -- (562.98,182.82) -- cycle ;
%Straight Lines [id:da7219026194895927] 
\draw [color={rgb, 255:red, 139; green, 6; blue, 24 }  ,draw opacity=1 ]   (509.99,151.91) -- (607,151.76) ;
%Straight Lines [id:da8396749363775703] 
\draw [color={rgb, 255:red, 139; green, 6; blue, 24 }  ,draw opacity=1 ] [dash pattern={on 4.5pt off 4.5pt}]  (463,152.76) -- (509.99,151.91) ;
%Straight Lines [id:da6356895489227852] 
\draw [color={rgb, 255:red, 139; green, 6; blue, 24 }  ,draw opacity=1 ]   (394.97,152.21) -- (458,152.76) ;
\draw  [color={rgb, 255:red, 139; green, 6; blue, 24 }  ,draw opacity=1 ][fill={rgb, 255:red, 139; green, 6; blue, 24 }  ,fill opacity=1 ] (422.33,154.5) -- (418.98,152.39) -- (422.4,150.38) -- (420.67,152.42) -- cycle ;
%Shape: Circle [id:dp7128669118673362] 
\draw  [color={rgb, 255:red, 139; green, 6; blue, 24 }  ,draw opacity=1 ][fill={rgb, 255:red, 139; green, 6; blue, 24 }  ,fill opacity=1 ] (506.69,151.91) .. controls (506.69,151) and (507.43,150.26) .. (508.34,150.26) .. controls (509.25,150.26) and (509.99,151) .. (509.99,151.91) .. controls (509.99,152.82) and (509.25,153.56) .. (508.34,153.56) .. controls (507.43,153.56) and (506.69,152.82) .. (506.69,151.91) -- cycle ;
\draw  [color={rgb, 255:red, 139; green, 6; blue, 24 }  ,draw opacity=1 ][fill={rgb, 255:red, 139; green, 6; blue, 24 }  ,fill opacity=1 ] (587.67,153.83) -- (584.31,151.72) -- (587.73,149.72) -- (586,151.75) -- cycle ;

% Text Node
\draw (253,124.4) node [anchor=north west][inner sep=0.75pt]    {$c$};
% Text Node
\draw (14,127.4) node [anchor=north west][inner sep=0.75pt]    {$d$};
% Text Node
\draw (250,155.4) node [anchor=north west][inner sep=0.75pt]    {$a$};
% Text Node
\draw (13,156.4) node [anchor=north west][inner sep=0.75pt]    {$b$};
% Text Node
\draw (544,77.4) node [anchor=north west][inner sep=0.75pt]    {$S_{\sigma }$};
% Text Node
\draw (357,142.4) node [anchor=north west][inner sep=0.75pt]    {$b\times d$};
% Text Node
\draw (612,140.4) node [anchor=north west][inner sep=0.75pt]    {$a\ \times \ c$};
% Text Node
\draw (296,143.4) node [anchor=north west][inner sep=0.75pt]    {$\Longrightarrow $};
% Text Node
\draw (180,71.4) node [anchor=north west][inner sep=0.75pt]    {$S_{\sigma }$};

\end{tikzpicture}
\caption{A necessary condition for the $S_{\sigma}$ action to be a symmetry.}
\label{fig:Sactioncond}
\end{figure}
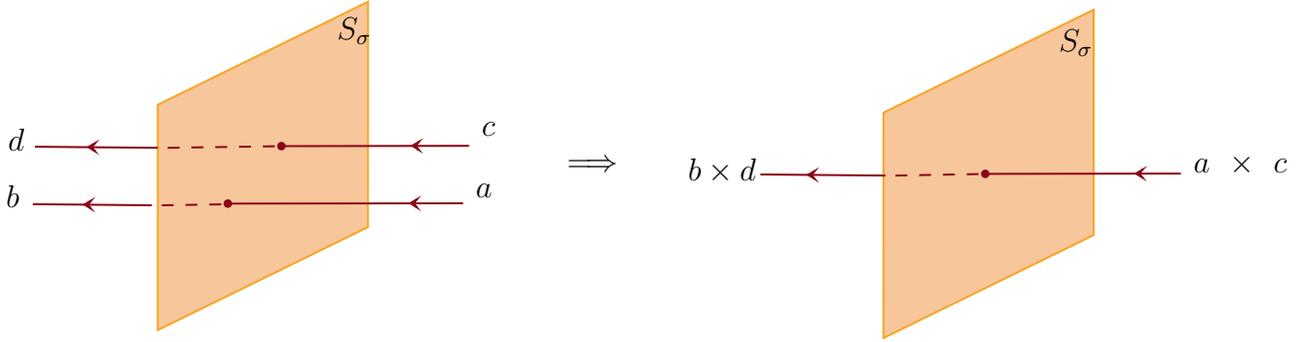

A unitary 0-form symmetry is implemented by an invertible surface operator, $S_{\sigma}$. Let us depict the conditions described above diagrammatically. If we have
\be\label{SurfaceLine}
S_{\sigma} \cdot a=b~,
\ee
then the line operators $a$ and $b$ can form a junction on the surface operator, $S_{\sigma}$, as in Fig. \ref{action}. 
$S_{\sigma}$ implements a symmetry if it preserves twists and if the condition in Fig. \ref{fig:Sactioncond} is satisfied. 
This condition is natural, since we can bring the two lines, $a$ and $c$, on the right arbitrarily close to each other and fuse them to form $a \times c$. Since the junction on the surface is topological, this maneuver brings the lines, $b$ and $d$, arbitrarily close to each other, which then fuse to give the outcome $b \times d$. 

To describe non-invertible symmetries, let us relax the assumption that $S$ should act as a permutation (see the discussion in Sec. \ref{sec:invertibility_basic}). In this case, $S$ can be a function on the set of line operators to itself which has a non-trivial kernel. In other words, defining
\begin{equation}
{\rm Ker}(S):=\left\{a\in\CT| S\cdot a=1\right\}~,
\end{equation}
we could have $1\ne a\in{\rm Ker}(S)$ (here $1$ is the trivial line). Moreover, we could have $S\cdot b=0$ for some line operator $b\ne1$. We will denote this \lq\lq null" set
\begin{equation}\label{nullS}
{\rm Null}(S):=\left\{a\in\CT|S\cdot a=0\right\}~.
\end{equation}
Lines in ${\rm Null}(S)$ cannot form a junction on the surface operator with any other line operator.\footnote{Note that lines in ${\rm Null}(S)$ may be mapped by $S$ to non-genuine / twisted sector lines (i.e., to lines that are attached to surfaces). Otherwise, moving $S$ through certain non-vanishing configurations of lines leads to discontinuities in the space of the corresponding correlation functions. Although it will be irrelevant for our discussion of genuine lines in the main text (we will briefly discuss certain non-genuine lines in App. \ref{sec:twisted}.), it would be interesting to explore these ideas further and understand if such discontinuities are allowed or forbidden.} Note that ${\rm Ker}(S)$ and ${\rm Null}(S)$ are distinct spaces and that
\begin{equation}
{\rm Ker}(S)\cap{\rm Null}(S)=\emptyset~.
\end{equation}

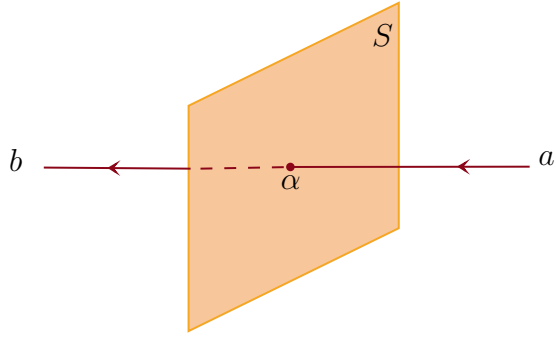
\begin{figure}[h!]
    \centering

\tikzset{every picture/.style={line width=0.75pt}} %set default line width to 0.75pt        

\begin{tikzpicture}[x=0.75pt,y=0.75pt,yscale=-1,xscale=1]

%Shape: Parallelogram [id:dp9547745150430316] 
\draw  [color={rgb, 255:red, 245; green, 166; blue, 35 }  ,draw opacity=1 ][fill={rgb, 255:red, 248; green, 198; blue, 155 }  ,fill opacity=1 ] (263.98,219.76) -- (264,106) -- (370,54.06) -- (369.98,167.82) -- cycle ;
%Straight Lines [id:da9423078429202264] 
\draw [color={rgb, 255:red, 139; green, 6; blue, 24 }  ,draw opacity=1 ]   (316.99,136.91) -- (436,136.76) ;
%Straight Lines [id:da8558727069000894] 
\draw [color={rgb, 255:red, 139; green, 6; blue, 24 }  ,draw opacity=1 ] [dash pattern={on 4.5pt off 4.5pt}]  (270,137.76) -- (316.99,136.91) ;
%Straight Lines [id:da8364306225426728] 
\draw [color={rgb, 255:red, 139; green, 6; blue, 24 }  ,draw opacity=1 ]   (190.97,137.21) -- (265,137.76) ;
\draw  [color={rgb, 255:red, 139; green, 6; blue, 24 }  ,draw opacity=1 ][fill={rgb, 255:red, 139; green, 6; blue, 24 }  ,fill opacity=1 ] (227.33,139.5) -- (223.98,137.39) -- (227.4,135.38) -- (225.67,137.42) -- cycle ;
%Shape: Circle [id:dp5656852943928684] 
\draw  [color={rgb, 255:red, 139; green, 6; blue, 24 }  ,draw opacity=1 ][fill={rgb, 255:red, 139; green, 6; blue, 24 }  ,fill opacity=1 ] (313.69,136.91) .. controls (313.69,136) and (314.43,135.26) .. (315.34,135.26) .. controls (316.25,135.26) and (316.99,136) .. (316.99,136.91) .. controls (316.99,137.82) and (316.25,138.56) .. (315.34,138.56) .. controls (314.43,138.56) and (313.69,137.82) .. (313.69,136.91) -- cycle ;
\draw  [color={rgb, 255:red, 139; green, 6; blue, 24 }  ,draw opacity=1 ][fill={rgb, 255:red, 139; green, 6; blue, 24 }  ,fill opacity=1 ] (403.67,138.83) -- (400.31,136.72) -- (403.73,134.72) -- (402,136.75) -- cycle ;

% Text Node
\draw (355,63.4) node [anchor=north west][inner sep=0.75pt]    {$S$};
% Text Node
\draw (439,127.4) node [anchor=north west][inner sep=0.75pt]    {$a$};
% Text Node
\draw (172,126.4) node [anchor=north west][inner sep=0.75pt]    {$b$};
% Text Node
\draw (309,139.4) node [anchor=north west][inner sep=0.75pt]    {$\alpha $};

\end{tikzpicture}
    \caption{If $b \in S \cdot a$, then $a$ and $b$ can form a junction on $S$. Here $\alpha$ is a local operator that lives at the junction and belongs to a vector space, $V_{S\cdot a}^{b}$, with dimension $n_{S\cdot a}^{b}$.}
    \label{fig:noninvsymact}
\end{figure}

More generally, $S \cdot a$ is of the form
\be
S \cdot a= \sum_{b} n_{S\cdot a}^b b~,\ \ \ n_{S\cdot a}^b\in\mathbb{N}~.
\ee
If $n_{S\cdot a}^b\neq 0$, we get Fig. \ref{fig:noninvsymact}.

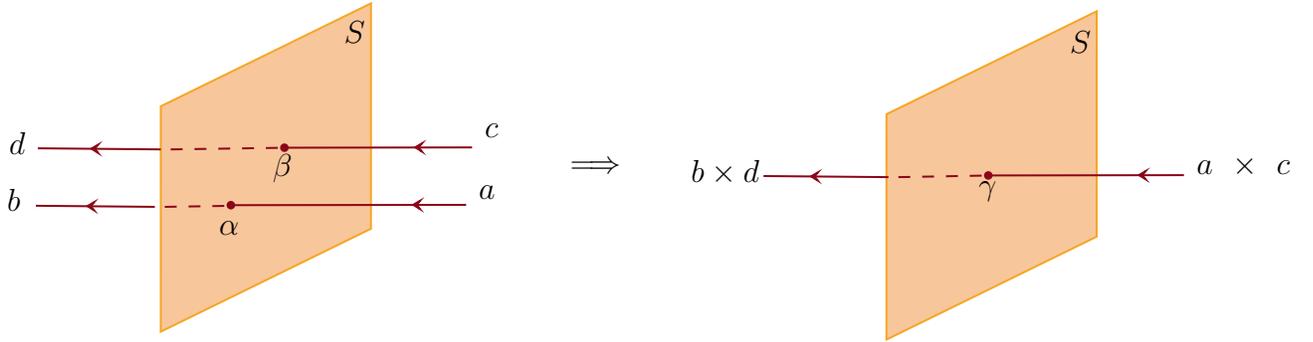
\begin{figure}[h!]
    \centering

\tikzset{every picture/.style={line width=0.75pt}} %set default line width to 0.75pt        

\begin{tikzpicture}[x=0.75pt,y=0.75pt,yscale=-1,xscale=1]

%Shape: Parallelogram [id:dp5816662440710975] 
\draw  [color={rgb, 255:red, 245; green, 166; blue, 35 }  ,draw opacity=1 ][fill={rgb, 255:red, 248; green, 198; blue, 155 }  ,fill opacity=1 ] (87.98,236.76) -- (88,123) -- (194,71.06) -- (193.98,184.82) -- cycle ;
%Straight Lines [id:da7109655491969266] 
\draw [color={rgb, 255:red, 139; green, 6; blue, 24 }  ,draw opacity=1 ]   (149.99,143.91) -- (245,143.76) ;
%Straight Lines [id:da016823770173435726] 
\draw [color={rgb, 255:red, 139; green, 6; blue, 24 }  ,draw opacity=1 ] [dash pattern={on 4.5pt off 4.5pt}]  (93,144.76) -- (151.99,143.91) ;
%Straight Lines [id:da63861834072555] 
\draw [color={rgb, 255:red, 139; green, 6; blue, 24 }  ,draw opacity=1 ]   (25.97,144.21) -- (88,144.76) ;
\draw  [color={rgb, 255:red, 139; green, 6; blue, 24 }  ,draw opacity=1 ][fill={rgb, 255:red, 139; green, 6; blue, 24 }  ,fill opacity=1 ] (56.33,146.5) -- (52.98,144.39) -- (56.4,142.38) -- (54.67,144.42) -- cycle ;
%Shape: Circle [id:dp8480787293657308] 
\draw  [color={rgb, 255:red, 139; green, 6; blue, 24 }  ,draw opacity=1 ][fill={rgb, 255:red, 139; green, 6; blue, 24 }  ,fill opacity=1 ] (148.69,143.91) .. controls (148.69,143) and (149.43,142.26) .. (150.34,142.26) .. controls (151.25,142.26) and (151.99,143) .. (151.99,143.91) .. controls (151.99,144.82) and (151.25,145.56) .. (150.34,145.56) .. controls (149.43,145.56) and (148.69,144.82) .. (148.69,143.91) -- cycle ;
\draw  [color={rgb, 255:red, 139; green, 6; blue, 24 }  ,draw opacity=1 ][fill={rgb, 255:red, 139; green, 6; blue, 24 }  ,fill opacity=1 ] (219.67,145.83) -- (216.31,143.72) -- (219.73,141.72) -- (218,143.75) -- cycle ;
%Straight Lines [id:da2071164666425508] 
\draw [color={rgb, 255:red, 139; green, 6; blue, 24 }  ,draw opacity=1 ]   (123.99,172.91) -- (242,172.76) ;
%Straight Lines [id:da699777348198614] 
\draw [color={rgb, 255:red, 139; green, 6; blue, 24 }  ,draw opacity=1 ] [dash pattern={on 4.5pt off 4.5pt}]  (90,173.76) -- (122.34,173.26) ;
%Straight Lines [id:da710506835529367] 
\draw [color={rgb, 255:red, 139; green, 6; blue, 24 }  ,draw opacity=1 ]   (24.97,173.21) -- (85,173.76) ;
\draw  [color={rgb, 255:red, 139; green, 6; blue, 24 }  ,draw opacity=1 ][fill={rgb, 255:red, 139; green, 6; blue, 24 }  ,fill opacity=1 ] (54.33,175.5) -- (50.98,173.39) -- (54.4,171.38) -- (52.67,173.42) -- cycle ;
%Shape: Circle [id:dp7915686939135019] 
\draw  [color={rgb, 255:red, 139; green, 6; blue, 24 }  ,draw opacity=1 ][fill={rgb, 255:red, 139; green, 6; blue, 24 }  ,fill opacity=1 ] (121.69,172.91) .. controls (121.69,172) and (122.43,171.26) .. (123.34,171.26) .. controls (124.25,171.26) and (124.99,172) .. (124.99,172.91) .. controls (124.99,173.82) and (124.25,174.56) .. (123.34,174.56) .. controls (122.43,174.56) and (121.69,173.82) .. (121.69,172.91) -- cycle ;
\draw  [color={rgb, 255:red, 139; green, 6; blue, 24 }  ,draw opacity=1 ][fill={rgb, 255:red, 139; green, 6; blue, 24 }  ,fill opacity=1 ] (218.67,174.83) -- (215.31,172.72) -- (218.73,170.72) -- (217,172.75) -- cycle ;
%Shape: Parallelogram [id:dp3343700084822019] 
\draw  [color={rgb, 255:red, 245; green, 166; blue, 35 }  ,draw opacity=1 ][fill={rgb, 255:red, 248; green, 198; blue, 155 }  ,fill opacity=1 ] (453.98,240.76) -- (454,127) -- (560,75.06) -- (559.98,188.82) -- cycle ;
%Straight Lines [id:da8545060338914964] 
\draw [color={rgb, 255:red, 139; green, 6; blue, 24 }  ,draw opacity=1 ]   (506.99,157.91) -- (604,157.76) ;
%Straight Lines [id:da7151288174172565] 
\draw [color={rgb, 255:red, 139; green, 6; blue, 24 }  ,draw opacity=1 ] [dash pattern={on 4.5pt off 4.5pt}]  (460,158.76) -- (506.99,157.91) ;
%Straight Lines [id:da7755672226340434] 
\draw [color={rgb, 255:red, 139; green, 6; blue, 24 }  ,draw opacity=1 ]   (391.97,158.21) -- (455,158.76) ;
\draw  [color={rgb, 255:red, 139; green, 6; blue, 24 }  ,draw opacity=1 ][fill={rgb, 255:red, 139; green, 6; blue, 24 }  ,fill opacity=1 ] (419.33,160.5) -- (415.98,158.39) -- (419.4,156.38) -- (417.67,158.42) -- cycle ;
%Shape: Circle [id:dp3613317453965048] 
\draw  [color={rgb, 255:red, 139; green, 6; blue, 24 }  ,draw opacity=1 ][fill={rgb, 255:red, 139; green, 6; blue, 24 }  ,fill opacity=1 ] (503.69,157.91) .. controls (503.69,157) and (504.43,156.26) .. (505.34,156.26) .. controls (506.25,156.26) and (506.99,157) .. (506.99,157.91) .. controls (506.99,158.82) and (506.25,159.56) .. (505.34,159.56) .. controls (504.43,159.56) and (503.69,158.82) .. (503.69,157.91) -- cycle ;
\draw  [color={rgb, 255:red, 139; green, 6; blue, 24 }  ,draw opacity=1 ][fill={rgb, 255:red, 139; green, 6; blue, 24 }  ,fill opacity=1 ] (584.67,159.83) -- (581.31,157.72) -- (584.73,155.72) -- (583,157.75) -- cycle ;

% Text Node
\draw (250,130.4) node [anchor=north west][inner sep=0.75pt]    {$c$};
% Text Node
\draw (247,161.4) node [anchor=north west][inner sep=0.75pt]    {$a$};
% Text Node
\draw (545,83.4) node [anchor=north west][inner sep=0.75pt]    {$S$};
% Text Node
\draw (354,148.4) node [anchor=north west][inner sep=0.75pt]    {$b\times d$};
% Text Node
\draw (609,146.4) node [anchor=north west][inner sep=0.75pt]    {$a\ \times \ c$};
% Text Node
\draw (293,149.4) node [anchor=north west][inner sep=0.75pt]    {$\Longrightarrow $};
% Text Node
\draw (143,145.4) node [anchor=north west][inner sep=0.75pt]    {$\beta $};
% Text Node
\draw (116,179.4) node [anchor=north west][inner sep=0.75pt]    {$\alpha $};
% Text Node
\draw (499,159.4) node [anchor=north west][inner sep=0.75pt]    {$\gamma $};
% Text Node
\draw (10,134.4) node [anchor=north west][inner sep=0.75pt]    {$d$};
% Text Node
\draw (9,163.4) node [anchor=north west][inner sep=0.75pt]    {$b$};
% Text Node
\draw (179,78.4) node [anchor=north west][inner sep=0.75pt]    {$S$};

\end{tikzpicture}
    \caption{A necessary condition for the $S$ action to be a non-invertible symmetry.}
    \label{fig:noninvSactioncond}
\end{figure}

Let us repeat the analysis in Fig. \ref{fig:Sactioncond} but now keeping track of the local operators, $\alpha$, at the junctions. We get Fig. \ref{fig:noninvSactioncond}.
For every choice of $\alpha \in V_{S\cdot a}^b$ and $\beta \in V_{S\cdot c}^d$, there should be a $\gamma \in V_{S\cdot(a\times b)}^{b\times d}$. This logic implies that the following inequality should be satisfied
\be
n_{S\cdot a}^bn_{S\cdot c}^d\leq n_{S\cdot (a\times c)}^{b\times d}~.
\ee
From this definition, it is possible to show that, for any unitary symmetry (including the trivial one), we have $|A|^2(|A|-1)$ strict inequalities and $|A|^4-|A|^3+|A|^2$ equalities. Interestingly, this statement still holds for the non-invertible generalizations we will discuss in this section. Also, we must have $n_{S\cdot1}^1\neq 0$, since the identity line forms a junction with itself on the surface. We will only consider simple surface operators (i.e., surfaces that do not have any non-trivial local operators on them). A generic non-simple surface operator is a direct sum of simple surfaces. Therefore, we have $n_{S\cdot1}^1=1$. 

To find non-invertible generalizations of unitary symmetries (hereafter referred to as \lq\lq non-invertible symmetries"), we require that $\theta_a=\theta_b$ if $n_{S\cdot a}^b\neq 0$. This demand is equivalent to requiring that $n_{S\cdot a}^b$ commutes with $\CT$ as a matrix. For unitary symmetries, this implied that components of the $\CS$-matrix related by the symmetry had to be equal.

On the other hand, non-invertible symmetries can (and as we will see below do) act by mapping some simple line operators to direct sums of simple lines. Moreover, since the modular data defines an Abelian TQFT, the $n_S$ matrix should also commute with $\CS$,\footnote{This fact also follows from the sandwich picture of Fig. \ref{fig:3d2d}, where the TQFT sits in a slab bounded by surfaces supporting 2d chiral and anti-chiral CFT modes. Our statements then amount to the modular invariance of the CFT torus partition function.} and the unitary condition generalizes to
\be
\label{eq:Scommutnoninv}
\sum_in_{S\cdot a}^i\CS_{ib}=\sum_j\CS_{aj}n_{S\cdot j}^b\ \Rightarrow\ [n_S,\CS]=0~,
\ee
or equivalently
\be
n_{S\cdot a}^b=\sum_{i,j} \CS_{ai}n_{S\cdot i}^j\CS_{jb}^{-1}~.
\ee
In particular, choosing $a$ and $b$ in the above equation to be the trivial line operator, we get
\be
\sum_{i,j} n_{S\cdot i}^j =|A|~.
\ee

As a consequence of this discussion, we define a (non-invertible) symmetry of an abelian TQFT as follows:
\vspace{0.2cm}

\noindent \textbf{Definition:} A (non-invertible) symmetry of an abelian TQFT is a map, $S$, on the ring of line operators such that 
\begin{enumerate}
    \item \label{def:noninvsym1} If $n_{S\cdot a}^b \neq 0$, then $\theta_a=\theta_b$.
    \item \label{def:noninvsym2} $n_{S\cdot a}^bn_{S\cdot c}^d\leq n_{S\cdot(a\times c)}^{b\times d}$ and $n_{S\cdot 1}^1=1$. 
    \item \label{def:noninvsym3} $\sum_{a,b\in A} n_{S\cdot a}^b=|A|$.
\end{enumerate}
Condition \ref{def:noninvsym1} in the definition above ensures that $n_{S\cdot a}^b$ commutes with $\CT$, and condition \ref{def:noninvsym3} ensures that $n_{S\cdot a}^b$ commutes with $\CS$ (for example, see \cite[Theorem 3.4]{Kong_2009}). In appendix \ref{AppA}, we show that this definition is equivalent to the definition of boundary conditions in terms of Lagrangian subgroups in the folded theory.

Let us derive some consequences of this definition. To that end:

\medskip
\noindent
{\it If $S$ implements a (non-invertible) symmetry of an abelian TQFT, $\CT$, then the integers $n_{S\cdot a}^b\in\left\{0,1\right\}$.}

\medskip
\noindent
To understand this statement, suppose $n_{S\cdot a}^{b}$ is non-zero. Using condition \eqref{def:noninvsym2}, we get
\be
n_{S\cdot a}^bn_{S\cdot\bar a}^{\bar b}\leq n_{S\cdot1}^{1}=1~,
\ee
where $\bar a$ is the inverse of $a\in A$. We have the following possibilities: $n_{S\cdot\bar a}^{\bar b}=0$ or $n_{S\cdot a}^b=n_{S\cdot\bar a}^{\bar b}=1$. Let us show that $n_{S\cdot\bar a}^{\bar b}$ cannot be zero. Since $n_{S\cdot a}^b$ is non-zero, we have
\be
n_{S\cdot a}^bn_{S\cdot a}^b\leq n_{S\cdot a^2}^{b^2} \implies n_{S\cdot a^2}^{b^2}\neq 0~.
\ee
Similarly, we can show that $n_{S\cdot a^k}^{b^k}\neq 0$ for any $k$. Now, let $k'=\text{lcm}(O_a,O_b)-1$, where $O_a$ is the order of $a\in A$. Then $n_{S\cdot  a^{k'}}^{b^{k'}}=n_{S\cdot \bar a}^{\bar b} \neq 0$. $\square$ 

Let us derive some more properties of non-invertible symmetries of abelian TQFTs. First, we claim:

\medskip\noindent
{\it Under the action of a non-invertible symmetry on an abelian TQFT, there is at least one line operator which is mapped to a non-simple line.}

\medskip
\noindent
To prove this assertion, suppose $S$ acts on the line operators to give a unique non-zero outcome so that $n_{S\cdot a}^b=n_{S\cdot c}^d=1$. Condition \ref{def:noninvsym2} simplifies to
\be\label{endo}
n_{S\cdot a}^bn_{S\cdot c}^d= n_{S\cdot(a\times c)}^{b\times d}~,
\ee
where the integers $n_{S\cdot a}^b\in\{0,1\}$. In other words,
\be
\label{eq:propnoninv}
S\cdot (a\times b)= S\cdot a \times S \cdot b~.
\ee
That is, $S$ is an endomorphism of the fusion group, $A$. Now, for $S$ to be non-invertible, it has to map some non-trivial line operator to the trivial line (it has to be a non-injective endomorphism on the group of simple lines $A$). In other words, we require: 

\medskip
\noindent
{\it For a non-invertible symmetry, $S$, ${\rm Ker}(S)$ is non-trivial. In other words, there exists a non-trivial $\ell\in{\rm Ker}(S)$.}

\medskip

Now, from condition \ref{def:noninvsym1}, we know that all lines in Ker$(S)$ must be bosons. We have (using \eqref{eq:propnoninv})
\be
S\cdot a= S \cdot (e \times a)~,\ \forall a\in\CT\ {\rm and}\ \forall e\in{\rm Ker}(S)~.
\ee
Condition \ref{def:noninvsym1} implies
\be
\theta_{a} =\theta_{e \times a} \implies S_{ea}= \frac{\theta_{e\times a}}{\theta_e\theta_a}=1~,\ \forall a\in\CT~,
\ee
which is inconsistent with modularity of the TQFT. Therefore, a non-invertible symmetry cannot act on line operators to give a unique outcome. There is at least one line operator on which $S$ acts to give a sum of simple lines as claimed. $\square$

This result in turn implies that, to satisfy condition \ref{def:noninvsym3}, the surface operator must send at least one line operator to $0$:

\medskip
\noindent
{\it For a non-invertible symmetry, $S$, ${\rm Null}(S)$ is non-trivial. In other words, there exists non-zero $n\in{\rm Null}(S)$.}

\medskip
\noindent
In other words, there is at least one line operator which cannot form a junction on the surface with any line operator. This should be contrasted with a non-invertible symmetry action of line operators on other lines operators through linking in $2+1$d or line operators on local operators in $1+1$d, where the non-invertibility of the lines does not necessarily imply that the result of the symmetry action is zero (for example, consider the Fibonacci category).\footnote{Note that in $2+1$d TQFTs described by weakly integral modular categories, the non-invertible symmetries implemented by line operators always set some line to zero. This statement follows from the existence of zeroes in the modular $S$ matrix \cite{gelaki2009centers}.}

Given the universality of the trivial line (i.e., it is always present), we next argue for the following useful statement:

\medskip
\noindent
{\it The number of simple line operators that can appear on the R.H.S. of the non-invertible symmetry action is completely determined by the symmetry action on the trivial line.} 

\medskip
\noindent
To understand this claim, consider the action of $S$ on the trivial line. The trivial line can form a junction with itself on the surface operator. Therefore, we have
\be
S\cdot 1= 1 + \sum_{\ell} n_{S\cdot1}^{\ell}  ~ \ell~.
\ee
If $n_{S\cdot1}^{\ell}\ne0$, then, from condition \ref{def:noninvsym1}, $\ell$ is a boson. Let us show that the integers $n_{S\cdot1}^{\ell}$ cannot all be zero. From our discussion above, we know that a surface operator, $S$, implementing a non-invertible symmetry acts on some line operator, $a$, to give more than one outcome. Suppose we have
\be\label{aact}
S \cdot a= \sum_{i=1}^N b_i~, \ N>1~.
\ee
In particular, $n_{S\cdot a}^{b_i}=n_{S\cdot a}^{b_j}\neq 0$ with $i\ne j$. Then, using \eqref{endo}, we have 
\be
n_{S\cdot a}^{b_i}n_{S\cdot\bar a}^{\bar b_j}=n_{S\cdot1}^{b_i \times \bar b_j}\neq 0~.
\ee
Therefore, the action of $S$ on the trivial line contains the outcome $b_i \times \bar b_j$. If $i \neq j$, then $b_i\times \bar b_j$ is a non-trivial line operator. This shows that $b_i\times \bar b_j$ for all $j\neq i \in \{1,...,N\}$ should be one of the outcomes in the fusion $S \cdot 1$. Therefore:

\medskip
\noindent
{\it A symmetry of an abelian TQFT is non-invertible if and only if its action on the trivial line operator results in more than one outcome.}

\medskip
In particular, if $S \cdot 1$ is a sum of $N$ line operators, then the action of $S$ on any other line operator (if non-trivial) also results in $N$ outcomes. If $b \in S\cdot a$, then $b \times l_i \in S\cdot a$ where 
\be\label{1act}
S\cdot 1 = 1 + \sum_{i=1}^{N-1} l_i~.
\ee
This proves our assertion. $\square$

Moreover, our discussion implies the following:

\medskip
\noindent
{\it If an abelian TQFT does not have any non-trivial bosons, then it does not have any non-invertible symmetries.}

\medskip
Let us now find a more precise characterization of $N$. To that end, it is easy to see that the larger the number of simple lines, $N$, appearing on the R.H.S. of \eqref{1act}, the larger $|{\rm Null}(S)|$ is (i.e., the larger the number of simple lines that cannot form a junction on $S$). To derive this statement, let $S \cdot a =0$, and take $n:=|A|-|{\rm Null}(A)|$ be the number of simple lines on which $S$ acts non-trivially. Clearly $n\geq 1$, since $S$ acts non-trivially on $1$. In fact $n\geq 2$, because, if $n=1$, then, to satisfy condition \ref{def:noninvsym3}, $S\cdot 1$ should be the sum of all anyons in the TQFT. This scenario requires all anyons in the TQFT to be bosons, which is inconsistent with modularity. Moreover, we know that on each of the $n$ simple lines, $S$ acts to give $N$ outcomes. From condition \ref{def:noninvsym3}, we get
\be
\label{eq:noninvsymcond}
Nn=N(|A|-|{\rm Null}(A)|)=|A|~.
\ee
Then, the action of $S$ on $|A|-n=|{\rm Null}(A)|$ simple line operators is zero.  If $N=1$, then $n=|A|$ and we have an invertible symmetry (clearly $|{\rm Null}(A)|=0$ in the invertible case).  

If there are non-trivial bosons in the TQFT, is there always a non-invertible symmetry? Yes! Let $B$ be a non-anomalous bosonic 1-form symmetry of $\CT$.\footnote{Given a boson, $b$, in an Abelian TQFT, $b^n$ is also a boson for all $n\in\left\{0,1,\cdots O_b\right\}$ (where $O_b$ is the order of $b\in A$), since $\theta$ is a quadratic form. This $\mathbb{Z}_{O_b}$ 1-form symmetry is then non-anomalous since $\CS$ restricted to this subspace is completely degenerate.} Then we can define the non-invertible symmetry implemented by a surface, $S_B$, which acts on the anyons as follows
\be
\label{eq:SBsym}
S_B \cdot a=\begin{cases} 
      B\text{-orbit of } a & \text{ if a braids trivially with }B~, \\
       0 & \text{ if a braids non-trivially with }B~. 
   \end{cases}
\ee
As we will see in Section \ref{resolution}, a typical theory has a much larger set of non-invertible symmetries. However, the symmetries described in \eqref{eq:SBsym} are {\it universal} in the sense that they are present in all theories with non-invertible symmetries.\footnote{Let us verify that \eqref{eq:SBsym} indeed gives a symmetry according to our definition. To that end, the fact that $n_{S\cdot a}^c\ne0$ if and only if $a$ braids trivially with $B$, and $c=\beta\times a$ for some (bosonic) $\beta\in B$ means that $\theta_a=\theta_c$ (i.e., condition \ref{def:noninvsym1} of our definition is satisfied). Now, if $a$ and $c$ braid trivially with $B$, then so too does $a\times c$. Moreover, if $b$ and $d$ are in the $B$-orbits of $a$ and $c$ respectively, then $b\times d$ is in the $B$-orbit of $a\times c$. As a result, condition \ref{def:noninvsym2} is also satisfied. A result of M\"uger \cite{muger2003structure} implies that $|B|\cdot|\CC_B|=|A|$, where $\CC_B$ consists of the lines that braid trivially with $B$. As a result, condition \ref{def:noninvsym3} is also satisfied.}

We have therefore arrived at the following result:

 \bigskip
 \noindent
{\bf Theorem 1:} \textit{An abelian TQFT has non-invertible symmetries if and only if it has a non-trivial non-anomalous bosonic 1-form symmetry.}\footnote{This is a stricter version of the result in \cite{JOHNSON_FREYD_2021} described in the introduction. We have shown that the fusion 2-category of modules over a pointed modular category $\CC$, Mod$(\CC)$, is group-like if and only if $\CC$ has no non-trivial objects on which the non-degenerate quadratic form, $\theta$, vanishes.}

\bigskip
Abelian TQFTs without non-invertible symmetries are precisely those which do not contain any bosonic line operators. These correspond to anisotropic pointed categories. Anisotropic pointed categories have been completely classified and are \cite{braidedcat1}: 
\begin{enumerate}
\item $A_p$ TQFT $\cong SU(p)_1$ Chern-Simons theory or its Galois conjugate depending on the odd prime $p$.
\item $B_p$ TQFT $\cong$ Galois conjugate of $A_p$.
\item $A_p \boxtimes A_p= B_p \boxtimes B_p$ TQFT.
\item Semion  and $\overline{\text{Semion}}$.
\item Semion $\boxtimes$ Semion and $\overline{\text{Semion}} \boxtimes \overline{\text{Semion}}$.
\item 3-Fermion Model $\cong$ Spin$(8)_1$ Chern-Simons theory.
\item $SU(4)_1$ Chern-Simons theory  and Galois conjugates.
\item $SU(4)_1$ Chern-Simons theory $~ \boxtimes$ Semion, $SU(4)_1$ Chern-Simons theory $\boxtimes ~\overline{\text{Semion}}$ and Galois conjugates.  
\end{enumerate}
Here, Semion is $U(1)_2$ Chern-Simons theory, and $p$ is an odd prime.\footnote{A subset of these abelian TQFTs also play a crucial role in the classification of TQFTs invariant under Galois conjugations \cite{Buican:2021axn}.}
All abelian TQFTs with non-invertible symmetry come from gauging a 0-form symmetry of an anisotropic abelian TQFT. Note that this 0-form symmetry must act trivially on the line operators. If not, the theory obtained after gauging will be a non-abelian TQFT.  

Another way to present the above results is to say that non-invertible symmetries of Abelian TQFTs are emergent or \lq\lq non-intrinsic" in the language of \cite{Kaidi:2022cpf,Kaidi:2023maf}. This characterization follows because non-invertible symmetries arise from performing a topological operation (0-form gauging) on an input theory that lacks non-invertible symmetries (in this case, an anisotropic Abelian TQFT). 

Finally, we conclude by noting that the presence of a non-anomalous $2+1$d 1-form symmetry is the condition for an orbifold to produce a 2d \lq\lq code" CFT in the construction of \cite{Buican:2021uyp} (see other related constructions in \cite{Kawabata:2023rlt,Kawabata:2023iss,Kawabata:2023usr,Alam:2023qac,Furuta:2023xwl,Kawabata:2023nlt,Dymarsky:2020qom}). This is a CFT living on the boundary of a slab of non-spin Abelian TQFT (as in Fig. \ref{fig:3d2d}) related to a quantum code by the quantum stabilizer code / CFT map of \cite{Buican:2021uyp}. Therefore, from Theorem 1, we see that theories with non-invertible symmetries give rise to orbifold 2d code CFTs in the language of \cite{Buican:2021uyp} (see also \cite{Kibe:2023nnj} for an appearance of non-invertible symmetries in the context of stabilizer codes).\footnote{The discussion in \cite{Buican:2021uyp} applies to $\mathbb{Z}_2^k$ 1-form symmetry, but this discussion can be generalized.} It would be interesting to explore and generalize this connection further.

In the next section, we will shift our attention to non-Abelian TQFTs. However, we will revisit Abelian TQFTs and their topological surfaces from the perspective of higher-gauging in Sec. \ref{resolution}. This perspective will allow us to better understand the action of more general topological surfaces on the line operators.

\section{(Non-)invertible symmetries of non-Abelian TQFTs}

\label{sec:noninv sym of non-abelian TQFTs}
\begin{figure}[h!]
    \centering

\tikzset{every picture/.style={line width=0.75pt}} %set default line width to 0.75pt        

\begin{tikzpicture}[x=0.75pt,y=0.75pt,yscale=-1,xscale=1]

%Shape: Parallelogram [id:dp4548180053301537] 
\draw  [color={rgb, 255:red, 245; green, 166; blue, 35 }  ,draw opacity=1 ][fill={rgb, 255:red, 248; green, 198; blue, 155 }  ,fill opacity=1 ] (87.98,236.76) -- (88,123) -- (194,71.06) -- (193.98,184.82) -- cycle ;
%Straight Lines [id:da11633765824580555] 
\draw [color={rgb, 255:red, 139; green, 6; blue, 24 }  ,draw opacity=1 ]   (149.99,143.91) -- (245,143.76) ;
%Straight Lines [id:da7168068786631839] 
\draw [color={rgb, 255:red, 139; green, 6; blue, 24 }  ,draw opacity=1 ] [dash pattern={on 4.5pt off 4.5pt}]  (93,144.76) -- (151.99,143.91) ;
%Straight Lines [id:da6478115449463211] 
\draw [color={rgb, 255:red, 139; green, 6; blue, 24 }  ,draw opacity=1 ]   (25.97,144.21) -- (88,144.76) ;
\draw  [color={rgb, 255:red, 139; green, 6; blue, 24 }  ,draw opacity=1 ][fill={rgb, 255:red, 139; green, 6; blue, 24 }  ,fill opacity=1 ] (56.33,146.5) -- (52.98,144.39) -- (56.4,142.38) -- (54.67,144.42) -- cycle ;
%Shape: Circle [id:dp0371007584477937] 
\draw  [color={rgb, 255:red, 139; green, 6; blue, 24 }  ,draw opacity=1 ][fill={rgb, 255:red, 139; green, 6; blue, 24 }  ,fill opacity=1 ] (148.69,143.91) .. controls (148.69,143) and (149.43,142.26) .. (150.34,142.26) .. controls (151.25,142.26) and (151.99,143) .. (151.99,143.91) .. controls (151.99,144.82) and (151.25,145.56) .. (150.34,145.56) .. controls (149.43,145.56) and (148.69,144.82) .. (148.69,143.91) -- cycle ;
\draw  [color={rgb, 255:red, 139; green, 6; blue, 24 }  ,draw opacity=1 ][fill={rgb, 255:red, 139; green, 6; blue, 24 }  ,fill opacity=1 ] (219.67,145.83) -- (216.31,143.72) -- (219.73,141.72) -- (218,143.75) -- cycle ;
%Straight Lines [id:da34571177189764013] 
\draw [color={rgb, 255:red, 139; green, 6; blue, 24 }  ,draw opacity=1 ]   (123.99,172.91) -- (242,172.76) ;
%Straight Lines [id:da3540917782373827] 
\draw [color={rgb, 255:red, 139; green, 6; blue, 24 }  ,draw opacity=1 ] [dash pattern={on 4.5pt off 4.5pt}]  (90,173.76) -- (122.34,173.26) ;
%Straight Lines [id:da4687584988372703] 
\draw [color={rgb, 255:red, 139; green, 6; blue, 24 }  ,draw opacity=1 ]   (24.97,173.21) -- (85,173.76) ;
\draw  [color={rgb, 255:red, 139; green, 6; blue, 24 }  ,draw opacity=1 ][fill={rgb, 255:red, 139; green, 6; blue, 24 }  ,fill opacity=1 ] (54.33,175.5) -- (50.98,173.39) -- (54.4,171.38) -- (52.67,173.42) -- cycle ;
%Shape: Circle [id:dp9109189507562275] 
\draw  [color={rgb, 255:red, 139; green, 6; blue, 24 }  ,draw opacity=1 ][fill={rgb, 255:red, 139; green, 6; blue, 24 }  ,fill opacity=1 ] (121.69,172.91) .. controls (121.69,172) and (122.43,171.26) .. (123.34,171.26) .. controls (124.25,171.26) and (124.99,172) .. (124.99,172.91) .. controls (124.99,173.82) and (124.25,174.56) .. (123.34,174.56) .. controls (122.43,174.56) and (121.69,173.82) .. (121.69,172.91) -- cycle ;
\draw  [color={rgb, 255:red, 139; green, 6; blue, 24 }  ,draw opacity=1 ][fill={rgb, 255:red, 139; green, 6; blue, 24 }  ,fill opacity=1 ] (218.67,174.83) -- (215.31,172.72) -- (218.73,170.72) -- (217,172.75) -- cycle ;
%Shape: Parallelogram [id:dp7614777382120138] 
\draw  [color={rgb, 255:red, 245; green, 166; blue, 35 }  ,draw opacity=1 ][fill={rgb, 255:red, 248; green, 198; blue, 155 }  ,fill opacity=1 ] (453.98,240.76) -- (454,127) -- (560,75.06) -- (559.98,188.82) -- cycle ;
%Straight Lines [id:da5544265540675392] 
\draw [color={rgb, 255:red, 139; green, 6; blue, 24 }  ,draw opacity=1 ]   (506.99,157.91) -- (604,157.76) ;
%Straight Lines [id:da11241939975796655] 
\draw [color={rgb, 255:red, 139; green, 6; blue, 24 }  ,draw opacity=1 ] [dash pattern={on 4.5pt off 4.5pt}]  (460,158.76) -- (506.99,157.91) ;
%Straight Lines [id:da9830500469004758] 
\draw [color={rgb, 255:red, 139; green, 6; blue, 24 }  ,draw opacity=1 ]   (391.97,158.21) -- (455,158.76) ;
\draw  [color={rgb, 255:red, 139; green, 6; blue, 24 }  ,draw opacity=1 ][fill={rgb, 255:red, 139; green, 6; blue, 24 }  ,fill opacity=1 ] (419.33,160.5) -- (415.98,158.39) -- (419.4,156.38) -- (417.67,158.42) -- cycle ;
%Shape: Circle [id:dp9278935458291747] 
\draw  [color={rgb, 255:red, 139; green, 6; blue, 24 }  ,draw opacity=1 ][fill={rgb, 255:red, 139; green, 6; blue, 24 }  ,fill opacity=1 ] (503.69,157.91) .. controls (503.69,157) and (504.43,156.26) .. (505.34,156.26) .. controls (506.25,156.26) and (506.99,157) .. (506.99,157.91) .. controls (506.99,158.82) and (506.25,159.56) .. (505.34,159.56) .. controls (504.43,159.56) and (503.69,158.82) .. (503.69,157.91) -- cycle ;
\draw  [color={rgb, 255:red, 139; green, 6; blue, 24 }  ,draw opacity=1 ][fill={rgb, 255:red, 139; green, 6; blue, 24 }  ,fill opacity=1 ] (584.67,159.83) -- (581.31,157.72) -- (584.73,155.72) -- (583,157.75) -- cycle ;

% Text Node
\draw (250,130.4) node [anchor=north west][inner sep=0.75pt]    {$c$};
% Text Node
\draw (247,161.4) node [anchor=north west][inner sep=0.75pt]    {$a$};
% Text Node
\draw (546,82.4) node [anchor=north west][inner sep=0.75pt]    {$S$};
% Text Node
\draw (378,148.4) node [anchor=north west][inner sep=0.75pt]    {$e$};
% Text Node
\draw (612,150.4) node [anchor=north west][inner sep=0.75pt]    {$f$};
% Text Node
\draw (293,149.4) node [anchor=north west][inner sep=0.75pt]    {$\Longrightarrow $};
% Text Node
\draw (143,145.4) node [anchor=north west][inner sep=0.75pt]    {$\beta $};
% Text Node
\draw (116,179.4) node [anchor=north west][inner sep=0.75pt]    {$\alpha $};
% Text Node
\draw (499,159.4) node [anchor=north west][inner sep=0.75pt]    {$\gamma $};
% Text Node
\draw (10,134.4) node [anchor=north west][inner sep=0.75pt]    {$d$};
% Text Node
\draw (9,163.4) node [anchor=north west][inner sep=0.75pt]    {$b$};
% Text Node
\draw (179,79.4) node [anchor=north west][inner sep=0.75pt]    {$S$};

\end{tikzpicture}
    \caption{A necessary condition for $S$ to generate a non-invertible symmetry.}
    \label{fig:nonabnoninvcond}
\end{figure}
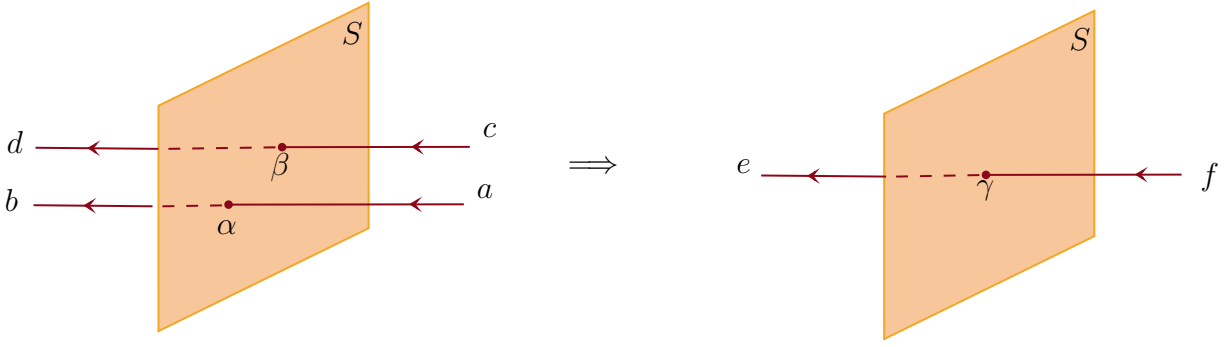

\begin{figure}[h!]
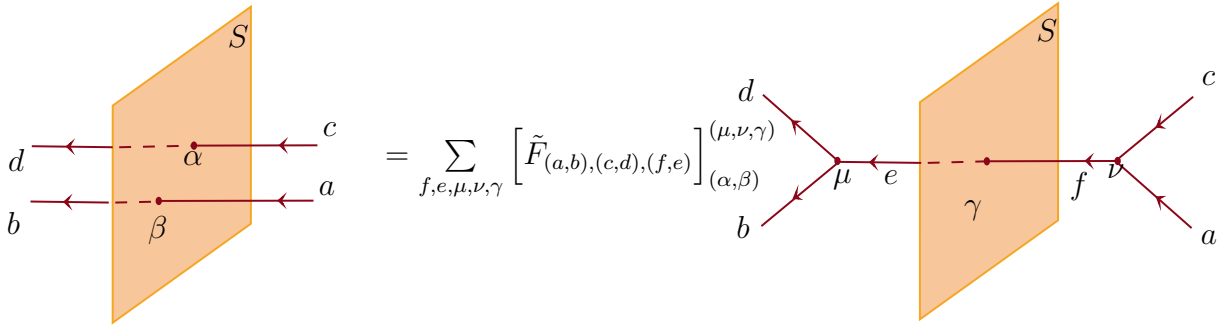

    \centering

\tikzset{every picture/.style={line width=0.75pt}} %set default line width to 0.75pt        


    \caption{A morphism between junction vector spaces.}
    \label{fig:SactFdef}
\end{figure}

\begin{figure}[h!]
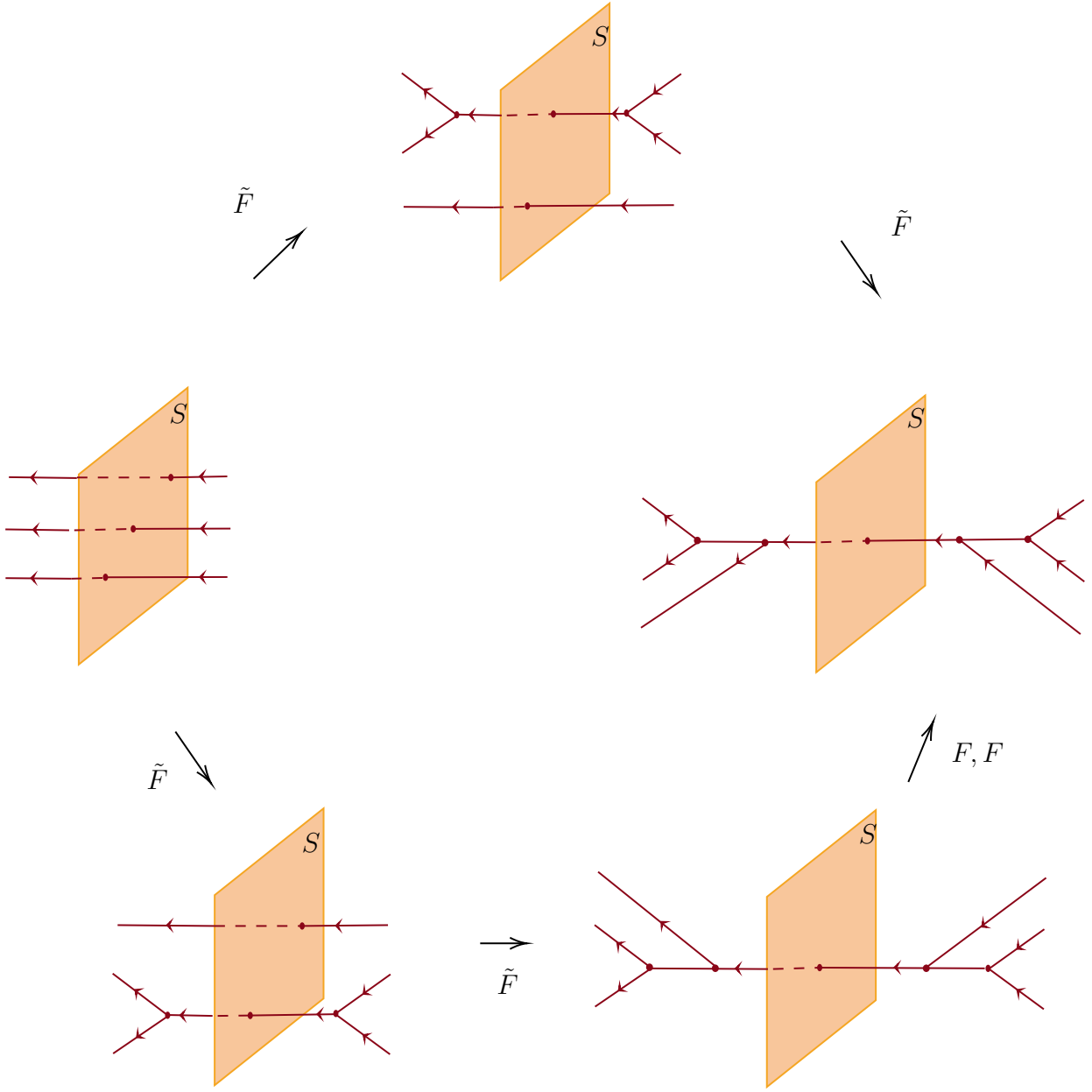

    \centering
\tikzset{every picture/.style={line width=0.75pt}} %set default line width to 0.75pt        

%
    \caption{Constraint on $\tilde F$ matrices from associativity of fusion. On the right bottom of the diagram, the $F,F$ action indicates the $F$-move on either side of the surface operator.}
    \label{fig:tildeFconst}
\end{figure}

In this section, our goal is to begin generalizing the results of the previous section to non-Abelian TQFTs (again, recall from table \ref{tb:terminology} that we only study non-spin theories). To that end, consider a non-abelian TQFT, $\CT$, whose lines are described by some unitary MTC, $\CC$. To characterise non-invertible symmetries of $\CT$, we generalize Fig. \ref{fig:noninvSactioncond} to non-abelian TQFTs as in Fig. \ref{fig:nonabnoninvcond}.
In this latter figure, $e \in b\times d$ and $f \in a \times c$. Here $\alpha$ and $\beta$ belong to the vector spaces $V_{S\cdot a}^b$ and $V_{S\cdot c}^d$ respectively. There are a total of $n_{S\cdot a}^bn_{S\cdot c}^d$ possible such junctions. For each of these junctions, there should be some $e \in b\times d$, $f \in a \times c$, and $\gamma \in V_{Sf}^e$ such that the diagram on the R.H.S. is allowed. Therefore, we get the constraint
\be
n_{S\cdot a}^bn_{S\cdot c}^d \leq  \sum_{e,f} N_{ac}^fN_{bd}^e n_{S\cdot f}^e~.
\ee
Moreover, if dim$(V_{S \cdot f}^e)\neq 0$, there must be a morphism between junction vector spaces $V_{S\cdot a}^b \otimes V_{S\cdot c}^d$ and $\bigoplus_{f,e} n_{S \cdot f}^{e} ~ V_{S\cdot f}^{e}$. In an explicit choice of basis, we get the constraint in Fig. \ref{fig:SactFdef}.
Considering the action of a surface operator on three line operators and using the associativity of the fusion rules, we get a constraint on the $\tilde F$ matrices given by Fig. \ref{fig:tildeFconst}.\footnote{See \cite{ostrik2001module,Fuchs:2002cm} for a discussion of these constraints in the context of module categories.}

\begin{figure}[h!]
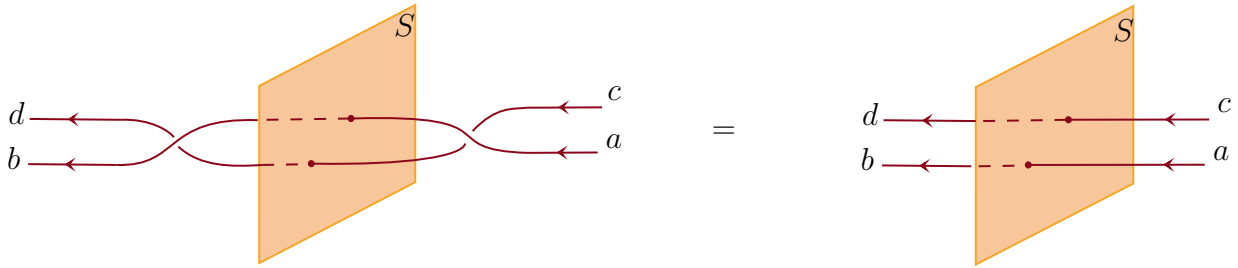

    \centering

\tikzset{every picture/.style={line width=0.75pt}} %set default line width to 0.75pt        

%
    \caption{Compatibility of $S$ with braiding.}
    \label{fig:constfrombraiding}
\end{figure}

In addition, the action of the surface operator must be compatible with the braiding of line operators. This reasoning leads to the constraint in Fig. \ref{fig:constfrombraiding}, which leads to the equation
\be\label{RFtRI}
R_{bd}^{e} \tilde F_{(c,d),(a,b),(f,e)} (R_{ac}^f)^{-1}= \tilde F_{(a,b),(c,d),(f,e)}~.
\ee
Following this discussion, we can define a (non-invertible) symmetry of a non-abelian TQFT as follows:
\vspace{0.2cm}

\noindent \textbf{Definition:} A (non-invertible) symmetry of a non-abelian TQFT, $\CT$, with MTC, $\CC$, is a map, $S$, on the ring of line operators such that:
\begin{enumerate}
    \item \label{def:nonabnoninvsym3} The $\tilde F$ matrices defined in Fig. \ref{fig:SactFdef} satisfying the conditions in Figs. \ref{fig:tildeFconst} and \ref{fig:constfrombraiding} exist.
    \item \label{def:nonabnoninvsym1} $n_{S\cdot a}^bn_{S\cdot c}^d \leq  \sum_{e,f} N_{ac}^fN_{bd}^e n_{S\cdot f}^e$ and $n_{S\cdot1}^1=1$. 
    \item \label{def:nonabnoninvsym2} $\sum_{a,b\in \CC} n_{S\cdot a}^b d_a d_b =\text{dim}(\CC)$.
\end{enumerate}
In appendix \ref{AppA}, we show that this definition is equivalent to the definition of boundary conditions in terms of Lagrangian algebras in $\CT\boxtimes\bar\CT$.

Note that the last two conditions above involve gauge-invariant quantities (i.e., quantities independent of fusion spaces), while the first condition involves $\tilde F$ and $R$, which are gauge-dependent. The final condition can be derived, as in the case of the corresponding Abelian condition \ref{def:noninvsym3}, from commutation of $n_S$ with the $\CS$ matrix (see the discussion around \eqref{eq:Scommutnoninv}). 

Furthermore, note that the last two conditions are necessary but not sufficient to have a non-invertible symmetry \cite{davydov2015unphysical,Kawahigashi_2015}. For example, in \cite{davydov2015unphysical} Davydov shows that, while the choice $n_{S\cdot a}^b=\delta_{a, b}$ is always physical/consistent (since it holds for the identity interface), the choice $n_{S\cdot a}^b=\delta_{a,\bar a}$ is not always consistent. In other words, charge conjugation is not always a symmetry of a TQFT, even though it is a symmetry of the modular data and hence of the fusion rules (see also \cite{Buican:2020who}). It is therefore not a surprise that $n_{S\cdot a}^b=\delta_{a,\bar a}$ satisfies the last two conditions in our definition.\footnote{It would be very interesting to obtain a set of constraints using only gauge-invariant quantities to characterise a non-invertible symmetry.}

In addition, note that, for a given choice of integers, $n_{S\cdot a}^{b}$, satisfying conditions \ref{def:nonabnoninvsym1} and \ref{def:nonabnoninvsym2}, we may have distinct $\tilde F$ solutions satisfying \ref{def:nonabnoninvsym3} \cite{Davydov_2014}\cite[Remark 3.9]{Cong_2017}. Such surface operators have identical action on the line operators and correspond to distinct isospectral boundary CFTs in the configuration in Fig. \ref{fig:3d2d}. For example, in \cite{Davydov_2014} Davydov finds two isospectral CFTs with the charge-conjugation modular invariant. This result implies that, in the corresponding bulk TQFT, there is an invertible surface operator which acts trivially on all line operators, but acts non-trivially on the fusion spaces. 

\begin{figure}[h!]
    \centering

\tikzset{every picture/.style={line width=0.75pt}} %set default line width to 0.75pt        

\begin{tikzpicture}[x=0.75pt,y=0.75pt,yscale=-1,xscale=1]

%Shape: Parallelogram [id:dp5722386809983322] 
\draw  [color={rgb, 255:red, 245; green, 166; blue, 35 }  ,draw opacity=1 ][fill={rgb, 255:red, 248; green, 198; blue, 155 }  ,fill opacity=1 ] (108.37,220.27) -- (108.39,110.69) -- (178.09,60.72) -- (178.07,170.3) -- cycle ;
%Straight Lines [id:da020842549368530472] 
\draw [color={rgb, 255:red, 139; green, 6; blue, 24 }  ,draw opacity=1 ]   (143.23,140.49) -- (207.03,140.35) ;
%Straight Lines [id:da23381645043365973] 
\draw [color={rgb, 255:red, 139; green, 6; blue, 24 }  ,draw opacity=1 ] [dash pattern={on 4.5pt off 4.5pt}]  (112.33,141.31) -- (143.23,140.49) ;
%Straight Lines [id:da203893727685442] 
\draw [color={rgb, 255:red, 139; green, 6; blue, 24 }  ,draw opacity=1 ]   (67.59,140.79) -- (109.04,141.31) ;
\draw  [color={rgb, 255:red, 139; green, 6; blue, 24 }  ,draw opacity=1 ][fill={rgb, 255:red, 139; green, 6; blue, 24 }  ,fill opacity=1 ] (85.59,142.99) -- (83.38,140.96) -- (85.63,139.03) -- (84.49,140.98) -- cycle ;
%Shape: Ellipse [id:dp5924824716147594] 
\draw  [color={rgb, 255:red, 139; green, 6; blue, 24 }  ,draw opacity=1 ][fill={rgb, 255:red, 139; green, 6; blue, 24 }  ,fill opacity=1 ] (141.06,140.49) .. controls (141.06,139.62) and (141.55,138.91) .. (142.15,138.91) .. controls (142.75,138.91) and (143.23,139.62) .. (143.23,140.49) .. controls (143.23,141.37) and (142.75,142.08) .. (142.15,142.08) .. controls (141.55,142.08) and (141.06,141.37) .. (141.06,140.49) -- cycle ;
\draw  [color={rgb, 255:red, 139; green, 6; blue, 24 }  ,draw opacity=1 ][fill={rgb, 255:red, 139; green, 6; blue, 24 }  ,fill opacity=1 ] (194.32,142.35) -- (192.11,140.31) -- (194.36,138.38) -- (193.22,140.34) -- cycle ;
%Straight Lines [id:da40908812388660176] 
\draw [color={rgb, 255:red, 139; green, 6; blue, 24 }  ,draw opacity=1 ]   (207.03,140.35) -- (246.2,107.92) ;
\draw  [color={rgb, 255:red, 139; green, 6; blue, 24 }  ,draw opacity=1 ][fill={rgb, 255:red, 139; green, 6; blue, 24 }  ,fill opacity=1 ] (230.52,123.38) -- (227.9,123.09) -- (229.22,119.87) -- (228.89,122.36) -- cycle ;
%Straight Lines [id:da9316886647264938] 
\draw [color={rgb, 255:red, 139; green, 6; blue, 24 }  ,draw opacity=1 ]   (207.03,140.35) -- (245.41,174.21) ;
\draw  [color={rgb, 255:red, 139; green, 6; blue, 24 }  ,draw opacity=1 ][fill={rgb, 255:red, 139; green, 6; blue, 24 }  ,fill opacity=1 ] (229.13,161.95) -- (227.93,158.63) -- (230.56,158.54) -- (228.89,159.44) -- cycle ;
%Shape: Ellipse [id:dp4820643679572033] 
\draw  [color={rgb, 255:red, 139; green, 6; blue, 24 }  ,draw opacity=1 ][fill={rgb, 255:red, 139; green, 6; blue, 24 }  ,fill opacity=1 ] (207.03,140.35) .. controls (207.03,139.47) and (207.51,138.76) .. (208.11,138.76) .. controls (208.71,138.76) and (209.2,139.47) .. (209.2,140.35) .. controls (209.2,141.22) and (208.71,141.93) .. (208.11,141.93) .. controls (207.51,141.93) and (207.03,141.22) .. (207.03,140.35) -- cycle ;
%Shape: Parallelogram [id:dp2785310500062418] 
\draw  [color={rgb, 255:red, 245; green, 166; blue, 35 }  ,draw opacity=1 ][fill={rgb, 255:red, 248; green, 198; blue, 155 }  ,fill opacity=1 ] (523.85,213.77) -- (523.87,104.19) -- (593.56,54.22) -- (593.54,163.8) -- cycle ;
%Straight Lines [id:da9321071574235893] 
\draw [color={rgb, 255:red, 139; green, 6; blue, 24 }  ,draw opacity=1 ]   (473.29,132.5) -- (528.29,133) ;
%Shape: Ellipse [id:dp3909668504491136] 
\draw  [color={rgb, 255:red, 139; green, 6; blue, 24 }  ,draw opacity=1 ][fill={rgb, 255:red, 139; green, 6; blue, 24 }  ,fill opacity=1 ] (577.35,147.22) .. controls (577.35,146.35) and (577.84,145.64) .. (578.44,145.64) .. controls (579.04,145.64) and (579.52,146.35) .. (579.52,147.22) .. controls (579.52,148.1) and (579.04,148.81) .. (578.44,148.81) .. controls (577.84,148.81) and (577.35,148.1) .. (577.35,147.22) -- cycle ;
\draw  [color={rgb, 255:red, 139; green, 6; blue, 24 }  ,draw opacity=1 ][fill={rgb, 255:red, 139; green, 6; blue, 24 }  ,fill opacity=1 ] (502.32,134.85) -- (500.11,132.81) -- (502.36,130.88) -- (501.22,132.84) -- cycle ;
%Straight Lines [id:da9323998932557129] 
\draw [color={rgb, 255:red, 139; green, 6; blue, 24 }  ,draw opacity=1 ]   (578.79,119.25) -- (601.29,99.33) ;
\draw  [color={rgb, 255:red, 139; green, 6; blue, 24 }  ,draw opacity=1 ][fill={rgb, 255:red, 139; green, 6; blue, 24 }  ,fill opacity=1 ] (594.52,107.38) -- (591.9,107.09) -- (593.22,103.87) -- (592.89,106.36) -- cycle ;
%Straight Lines [id:da2993889469087605] 
\draw [color={rgb, 255:red, 139; green, 6; blue, 24 }  ,draw opacity=1 ]   (578.79,147.75) -- (599.41,167.21) ;
\draw  [color={rgb, 255:red, 139; green, 6; blue, 24 }  ,draw opacity=1 ][fill={rgb, 255:red, 139; green, 6; blue, 24 }  ,fill opacity=1 ] (588.62,159.95) -- (588.11,156.46) -- (590.7,156.9) -- (588.89,157.44) -- cycle ;
%Shape: Ellipse [id:dp6119767286734226] 
\draw  [color={rgb, 255:red, 139; green, 6; blue, 24 }  ,draw opacity=1 ][fill={rgb, 255:red, 139; green, 6; blue, 24 }  ,fill opacity=1 ] (577.7,119.66) .. controls (577.7,118.78) and (578.19,118.07) .. (578.79,118.07) .. controls (579.39,118.07) and (579.87,118.78) .. (579.87,119.66) .. controls (579.87,120.54) and (579.39,121.25) .. (578.79,121.25) .. controls (578.19,121.25) and (577.7,120.54) .. (577.7,119.66) -- cycle ;
%Straight Lines [id:da9878740718095115] 
\draw [color={rgb, 255:red, 139; green, 6; blue, 24 }  ,draw opacity=1 ] [dash pattern={on 4.5pt off 4.5pt}]  (561.03,133.35) -- (578.79,119.25) ;
%Straight Lines [id:da493198698244399] 
\draw [color={rgb, 255:red, 139; green, 6; blue, 24 }  ,draw opacity=1 ] [dash pattern={on 4.5pt off 4.5pt}]  (578.99,147.67) -- (578.81,147.52) -- (578.44,147.22) -- (577.76,146.67) -- (561.23,133.27) ;
%Straight Lines [id:da08931882315716533] 
\draw [color={rgb, 255:red, 139; green, 6; blue, 24 }  ,draw opacity=1 ] [dash pattern={on 4.5pt off 4.5pt}]  (528.29,133) -- (561.23,133.27) ;
%Shape: Ellipse [id:dp7055888301958663] 
\draw  [color={rgb, 255:red, 139; green, 6; blue, 24 }  ,draw opacity=1 ][fill={rgb, 255:red, 139; green, 6; blue, 24 }  ,fill opacity=1 ] (559.06,133.27) .. controls (559.06,132.39) and (559.55,131.68) .. (560.15,131.68) .. controls (560.75,131.68) and (561.23,132.39) .. (561.23,133.27) .. controls (561.23,134.15) and (560.75,134.86) .. (560.15,134.86) .. controls (559.55,134.86) and (559.06,134.15) .. (559.06,133.27) -- cycle ;

% Text Node
\draw (165.64,67.46) node [anchor=north west][inner sep=0.75pt]    {$S$};
% Text Node
\draw (88.66,143.56) node [anchor=north west][inner sep=0.75pt]    {$f$};
% Text Node
\draw (184.5,143.22) node [anchor=north west][inner sep=0.75pt]    {$c$};
% Text Node
\draw (135.19,144.5) node [anchor=north west][inner sep=0.75pt]    {$\gamma $};
% Text Node
\draw (248.56,173.67) node [anchor=north west][inner sep=0.75pt]    {$a$};
% Text Node
\draw (249,93.67) node [anchor=north west][inner sep=0.75pt]    {$b$};
% Text Node
\draw (201.37,139.77) node [anchor=north west][inner sep=0.75pt]    {$\nu $};
% Text Node
\draw (580.64,62.46) node [anchor=north west][inner sep=0.75pt]    {$S$};
% Text Node
\draw (497.09,135.56) node [anchor=north west][inner sep=0.75pt]    {$f$};
% Text Node
\draw (602.56,166.67) node [anchor=north west][inner sep=0.75pt]    {$a$};
% Text Node
\draw (603,86.67) node [anchor=north west][inner sep=0.75pt]    {$b$};
% Text Node
\draw (539.76,134.53) node [anchor=north west][inner sep=0.75pt]    {$\nu '$};
% Text Node
\draw (558.56,140.67) node [anchor=north west][inner sep=0.75pt]    {$d$};
% Text Node
\draw (562,108.67) node [anchor=north west][inner sep=0.75pt]    {$e$};
% Text Node
\draw (276,117.4) node [anchor=north west][inner sep=0.75pt]    {$=\ \sum\limits_{e,d,\nu '}[ U_{S}( a,b,c,f)_{e,d}]_{\nu \nu '}$};

\end{tikzpicture}
    \caption{Action of $S$ on fusion spaces.}
    \label{fig:S act fusion space}
\end{figure}
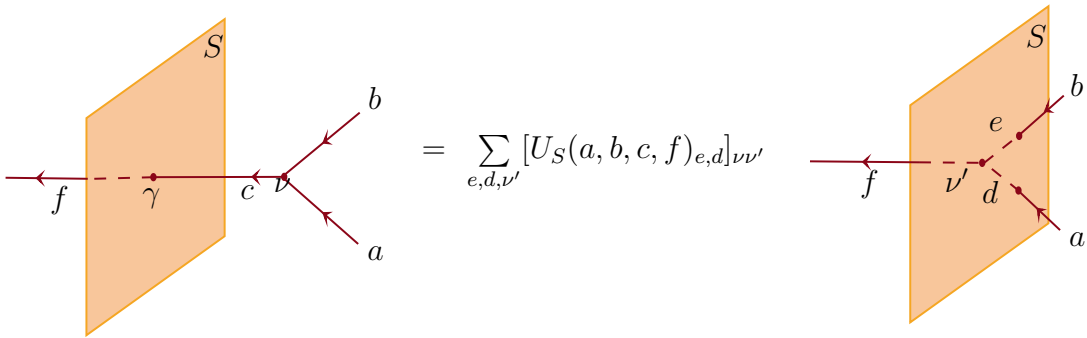

This discussion motivates us to more explicitly consider the action of symmetries on fusion spaces. To that end, take a junction of three line operators, $a$, $b$, and $c$. We can move a (generically non-invertible) surface operator operator across the junction as in Fig. \ref{fig:S act fusion space} to get a matrix, $U_{S}( a,b,c,f)_{e,d}$, acting on the fusion space, $V_{ed}^f$. 
At the level of fusion spaces, this action can be written as
\be
\label{eq:Sact fusion space}
U_S(a,b,c): \bigoplus_{d,e,f} N_{de}^f ~ n_{Sa}^{d} ~ n_{Sb}^{e} ~ n_{Sc}^{f} ~ V_{de}^{f} \rightarrow V_{ab}^c~, \ee
where $U_S(a,b,c)$ is a block-diagonal matrix, with the block $U_S(a,b,c,f)_{d,e}$ acting on $V_{de}^f$. Note that in general $U_S$ is a rectangular matrix, and is therefore non-invertible.

The collection of matrices $\{U_S(a,b,c)\}$ cannot be chosen arbitrarily. They must satisfy consistency conditions with the $\tilde F$ matrices. Writing these full consistency conditions and deriving their consequences are beyond the scope of this paper (this route would amount to a non-invertible generalization of \cite{barkeshli2019symmetry}).\footnote{In principle, the full set of consistency conditions is already encoded in the definition of the 2-category of module categories over a modular tensor category. However, it would be illuminating to reduce the consistency conditions to a set of polynomial equations to enable explicit calculations.} However, the discussion in \cite{Davydov_2014} motivates us to ask the following simpler question: {\it if a surface acts invertibly on lines, can it act non-invertibly on fusion spaces?}

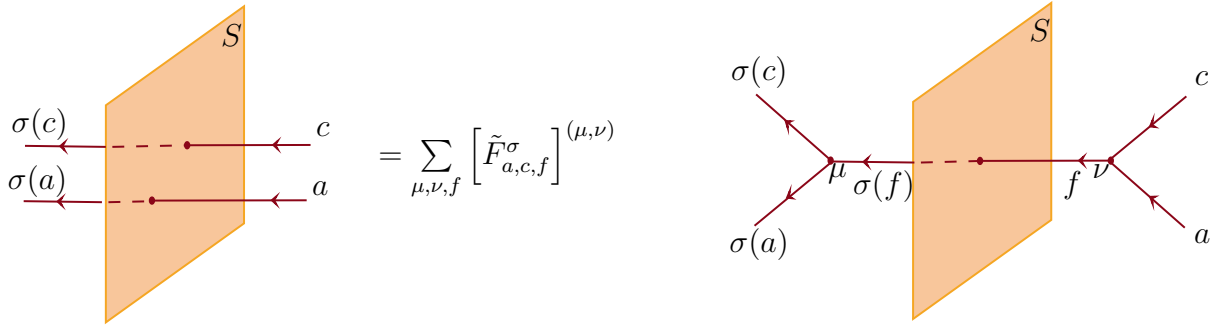
\begin{figure}
    \centering

\tikzset{every picture/.style={line width=0.75pt}} %set default line width to 0.75pt        

\begin{tikzpicture}[x=0.75pt,y=0.75pt,yscale=-1,xscale=1]

%Shape: Parallelogram [id:dp4548180053301537] 
\draw  [color={rgb, 255:red, 245; green, 166; blue, 35 }  ,draw opacity=1 ][fill={rgb, 255:red, 248; green, 198; blue, 155 }  ,fill opacity=1 ] (75.97,233) -- (75.99,123.41) -- (145.68,73.45) -- (145.66,183.03) -- cycle ;
%Straight Lines [id:da11633765824580555] 
\draw [color={rgb, 255:red, 139; green, 6; blue, 24 }  ,draw opacity=1 ]   (116.74,143.59) -- (179.22,143.44) ;
%Straight Lines [id:da7168068786631839] 
\draw [color={rgb, 255:red, 139; green, 6; blue, 24 }  ,draw opacity=1 ] [dash pattern={on 4.5pt off 4.5pt}]  (79.26,144.41) -- (118.06,143.59) ;
%Straight Lines [id:da6478115449463211] 
\draw [color={rgb, 255:red, 139; green, 6; blue, 24 }  ,draw opacity=1 ]   (35.19,143.89) -- (75.97,144.41) ;
\draw  [color={rgb, 255:red, 139; green, 6; blue, 24 }  ,draw opacity=1 ][fill={rgb, 255:red, 139; green, 6; blue, 24 }  ,fill opacity=1 ] (55.15,146.09) -- (52.95,144.05) -- (55.19,142.12) -- (54.06,144.08) -- cycle ;
%Shape: Ellipse [id:dp0371007584477937] 
\draw  [color={rgb, 255:red, 139; green, 6; blue, 24 }  ,draw opacity=1 ][fill={rgb, 255:red, 139; green, 6; blue, 24 }  ,fill opacity=1 ] (115.89,143.59) .. controls (115.89,142.71) and (116.38,142) .. (116.97,142) .. controls (117.57,142) and (118.06,142.71) .. (118.06,143.59) .. controls (118.06,144.47) and (117.57,145.18) .. (116.97,145.18) .. controls (116.38,145.18) and (115.89,144.47) .. (115.89,143.59) -- cycle ;
\draw  [color={rgb, 255:red, 139; green, 6; blue, 24 }  ,draw opacity=1 ][fill={rgb, 255:red, 139; green, 6; blue, 24 }  ,fill opacity=1 ] (162.57,145.44) -- (160.36,143.41) -- (162.61,141.48) -- (161.47,143.44) -- cycle ;
%Straight Lines [id:da34571177189764013] 
\draw [color={rgb, 255:red, 139; green, 6; blue, 24 }  ,draw opacity=1 ]   (99.65,171.52) -- (177.25,171.37) ;
%Straight Lines [id:da3540917782373827] 
\draw [color={rgb, 255:red, 139; green, 6; blue, 24 }  ,draw opacity=1 ] [dash pattern={on 4.5pt off 4.5pt}]  (77.29,172.33) -- (98.56,171.86) ;
%Straight Lines [id:da4687584988372703] 
\draw [color={rgb, 255:red, 139; green, 6; blue, 24 }  ,draw opacity=1 ]   (34.53,171.81) -- (74,172.33) ;
\draw  [color={rgb, 255:red, 139; green, 6; blue, 24 }  ,draw opacity=1 ][fill={rgb, 255:red, 139; green, 6; blue, 24 }  ,fill opacity=1 ] (53.84,174.01) -- (51.63,171.98) -- (53.88,170.05) -- (52.74,172.01) -- cycle ;
%Shape: Ellipse [id:dp9109189507562275] 
\draw  [color={rgb, 255:red, 139; green, 6; blue, 24 }  ,draw opacity=1 ][fill={rgb, 255:red, 139; green, 6; blue, 24 }  ,fill opacity=1 ] (98.13,171.52) .. controls (98.13,170.64) and (98.62,169.93) .. (99.22,169.93) .. controls (99.82,169.93) and (100.3,170.64) .. (100.3,171.52) .. controls (100.3,172.4) and (99.82,173.11) .. (99.22,173.11) .. controls (98.62,173.11) and (98.13,172.4) .. (98.13,171.52) -- cycle ;
\draw  [color={rgb, 255:red, 139; green, 6; blue, 24 }  ,draw opacity=1 ][fill={rgb, 255:red, 139; green, 6; blue, 24 }  ,fill opacity=1 ] (161.91,173.37) -- (159.7,171.34) -- (161.95,169.41) -- (160.82,171.36) -- cycle ;
%Shape: Parallelogram [id:dp7614777382120138] 
\draw  [color={rgb, 255:red, 245; green, 166; blue, 35 }  ,draw opacity=1 ][fill={rgb, 255:red, 248; green, 198; blue, 155 }  ,fill opacity=1 ] (483.09,231.27) -- (483.11,121.69) -- (552.8,71.72) -- (552.78,181.3) -- cycle ;
%Straight Lines [id:da5544265540675392] 
\draw [color={rgb, 255:red, 139; green, 6; blue, 24 }  ,draw opacity=1 ]   (517.95,151.49) -- (581.74,151.35) ;
%Straight Lines [id:da11241939975796655] 
\draw [color={rgb, 255:red, 139; green, 6; blue, 24 }  ,draw opacity=1 ] [dash pattern={on 4.5pt off 4.5pt}]  (487.04,152.31) -- (517.95,151.49) ;
%Straight Lines [id:da9830500469004758] 
\draw [color={rgb, 255:red, 139; green, 6; blue, 24 }  ,draw opacity=1 ]   (442.31,151.79) -- (483.75,152.31) ;
\draw  [color={rgb, 255:red, 139; green, 6; blue, 24 }  ,draw opacity=1 ][fill={rgb, 255:red, 139; green, 6; blue, 24 }  ,fill opacity=1 ] (460.3,153.99) -- (458.09,151.96) -- (460.34,150.03) -- (459.21,151.98) -- cycle ;
%Shape: Ellipse [id:dp9278935458291747] 
\draw  [color={rgb, 255:red, 139; green, 6; blue, 24 }  ,draw opacity=1 ][fill={rgb, 255:red, 139; green, 6; blue, 24 }  ,fill opacity=1 ] (515.78,151.49) .. controls (515.78,150.62) and (516.26,149.91) .. (516.86,149.91) .. controls (517.46,149.91) and (517.95,150.62) .. (517.95,151.49) .. controls (517.95,152.37) and (517.46,153.08) .. (516.86,153.08) .. controls (516.26,153.08) and (515.78,152.37) .. (515.78,151.49) -- cycle ;
\draw  [color={rgb, 255:red, 139; green, 6; blue, 24 }  ,draw opacity=1 ][fill={rgb, 255:red, 139; green, 6; blue, 24 }  ,fill opacity=1 ] (569.03,153.35) -- (566.82,151.31) -- (569.07,149.38) -- (567.94,151.34) -- cycle ;
%Straight Lines [id:da35043928515103373] 
\draw [color={rgb, 255:red, 139; green, 6; blue, 24 }  ,draw opacity=1 ]   (581.74,151.35) -- (620.92,118.92) ;
\draw  [color={rgb, 255:red, 139; green, 6; blue, 24 }  ,draw opacity=1 ][fill={rgb, 255:red, 139; green, 6; blue, 24 }  ,fill opacity=1 ] (605.24,134.38) -- (602.61,134.09) -- (603.94,130.87) -- (603.6,133.36) -- cycle ;
%Straight Lines [id:da837740674165147] 
\draw [color={rgb, 255:red, 139; green, 6; blue, 24 }  ,draw opacity=1 ]   (581.74,151.35) -- (620.12,185.21) ;
%Straight Lines [id:da7371922708896648] 
\draw [color={rgb, 255:red, 139; green, 6; blue, 24 }  ,draw opacity=1 ]   (403.13,184.22) -- (442.31,151.79) ;
%Straight Lines [id:da8672294699494015] 
\draw [color={rgb, 255:red, 139; green, 6; blue, 24 }  ,draw opacity=1 ]   (403.93,117.92) -- (442.31,151.79) ;
\draw  [color={rgb, 255:red, 139; green, 6; blue, 24 }  ,draw opacity=1 ][fill={rgb, 255:red, 139; green, 6; blue, 24 }  ,fill opacity=1 ] (422.17,170.34) -- (419.54,170.05) -- (420.87,166.83) -- (420.53,169.32) -- cycle ;
\draw  [color={rgb, 255:red, 139; green, 6; blue, 24 }  ,draw opacity=1 ][fill={rgb, 255:red, 139; green, 6; blue, 24 }  ,fill opacity=1 ] (603.85,172.95) -- (602.64,169.63) -- (605.27,169.54) -- (603.6,170.44) -- cycle ;
\draw  [color={rgb, 255:red, 139; green, 6; blue, 24 }  ,draw opacity=1 ][fill={rgb, 255:red, 139; green, 6; blue, 24 }  ,fill opacity=1 ] (419.86,134.77) -- (418.82,131.34) -- (421.45,131.51) -- (419.74,132.24) -- cycle ;
%Shape: Ellipse [id:dp6440090753959639] 
\draw  [color={rgb, 255:red, 139; green, 6; blue, 24 }  ,draw opacity=1 ][fill={rgb, 255:red, 139; green, 6; blue, 24 }  ,fill opacity=1 ] (581.74,151.35) .. controls (581.74,150.47) and (582.23,149.76) .. (582.83,149.76) .. controls (583.42,149.76) and (583.91,150.47) .. (583.91,151.35) .. controls (583.91,152.22) and (583.42,152.93) .. (582.83,152.93) .. controls (582.23,152.93) and (581.74,152.22) .. (581.74,151.35) -- cycle ;
%Shape: Ellipse [id:dp3944982166605865] 
\draw  [color={rgb, 255:red, 139; green, 6; blue, 24 }  ,draw opacity=1 ][fill={rgb, 255:red, 139; green, 6; blue, 24 }  ,fill opacity=1 ] (440.43,151.32) .. controls (440.43,150.45) and (440.92,149.74) .. (441.51,149.74) .. controls (442.11,149.74) and (442.6,150.45) .. (442.6,151.32) .. controls (442.6,152.2) and (442.11,152.91) .. (441.51,152.91) .. controls (440.92,152.91) and (440.43,152.2) .. (440.43,151.32) -- cycle ;

% Text Node
\draw (180.46,130.3) node [anchor=north west][inner sep=0.75pt]    {$c$};
% Text Node
\draw (178.49,160.15) node [anchor=north west][inner sep=0.75pt]    {$a$};
% Text Node
\draw (540.35,78.46) node [anchor=north west][inner sep=0.75pt]    {$S$};
% Text Node
\draw (451.37,154.56) node [anchor=north west][inner sep=0.75pt]    {$\sigma(f)$};
% Text Node
\draw (557.21,155.22) node [anchor=north west][inner sep=0.75pt]    {$f$};
% Text Node
\draw (26.84,123.58) node [anchor=north west][inner sep=0.75pt]    {$\sigma(c)$};
% Text Node
\draw (25.1,152.36) node [anchor=north west][inner sep=0.75pt]    {$\sigma(a)$};
% Text Node
\draw (132.57,81.15) node [anchor=north west][inner sep=0.75pt]    {$S$};
% Text Node
\draw (623.27,184.67) node [anchor=north west][inner sep=0.75pt]    {$a$};
% Text Node
\draw (623.71,104.67) node [anchor=north west][inner sep=0.75pt]    {$c$};
% Text Node
\draw (389.84,97.35) node [anchor=north west][inner sep=0.75pt]    {$\sigma(c)$};
% Text Node
\draw (389.05,183.85) node [anchor=north west][inner sep=0.75pt]    {$\sigma(a)$};
% Text Node
\draw (438.79,151.02) node [anchor=north west][inner sep=0.75pt]    {$\mu $};
% Text Node
\draw (572.08,151.77) node [anchor=north west][inner sep=0.75pt]    {$\nu $};
% Text Node
\draw (211.95,129.18) node [anchor=north west][inner sep=0.75pt]    {$=\sum\limits_{\mu ,\nu,f}\left[\tilde{F}^{\sigma}_{a,c,f}\right]^{( \mu ,\nu )} $};

\end{tikzpicture}
    \caption{A surface, $S$, with a permutation action on lines.}
    \label{fig:trivactconst1}
\end{figure}

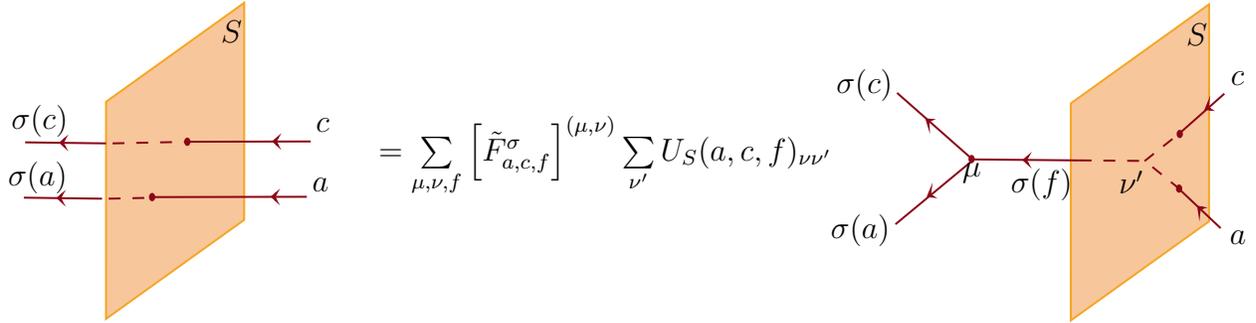
\begin{figure}
    \centering

\tikzset{every picture/.style={line width=0.75pt}} %set default line width to 0.75pt        

\begin{tikzpicture}[x=0.75pt,y=0.75pt,yscale=-1,xscale=1]
%uncomment if require: \path (0,523); %set diagram left start at 0, and has height of 523

%Shape: Parallelogram [id:dp4548180053301537] 
\draw  [color={rgb, 255:red, 245; green, 166; blue, 35 }  ,draw opacity=1 ][fill={rgb, 255:red, 248; green, 198; blue, 155 }  ,fill opacity=1 ] (75.97,233) -- (75.99,123.41) -- (145.68,73.45) -- (145.66,183.03) -- cycle ;
%Straight Lines [id:da11633765824580555] 
\draw [color={rgb, 255:red, 139; green, 6; blue, 24 }  ,draw opacity=1 ]   (116.74,143.59) -- (179.22,143.44) ;
%Straight Lines [id:da7168068786631839] 
\draw [color={rgb, 255:red, 139; green, 6; blue, 24 }  ,draw opacity=1 ] [dash pattern={on 4.5pt off 4.5pt}]  (79.26,144.41) -- (118.06,143.59) ;
%Straight Lines [id:da6478115449463211] 
\draw [color={rgb, 255:red, 139; green, 6; blue, 24 }  ,draw opacity=1 ]   (35.19,143.89) -- (75.97,144.41) ;
\draw  [color={rgb, 255:red, 139; green, 6; blue, 24 }  ,draw opacity=1 ][fill={rgb, 255:red, 139; green, 6; blue, 24 }  ,fill opacity=1 ] (55.15,146.09) -- (52.95,144.05) -- (55.19,142.12) -- (54.06,144.08) -- cycle ;
%Shape: Ellipse [id:dp0371007584477937] 
\draw  [color={rgb, 255:red, 139; green, 6; blue, 24 }  ,draw opacity=1 ][fill={rgb, 255:red, 139; green, 6; blue, 24 }  ,fill opacity=1 ] (115.89,143.59) .. controls (115.89,142.71) and (116.38,142) .. (116.97,142) .. controls (117.57,142) and (118.06,142.71) .. (118.06,143.59) .. controls (118.06,144.47) and (117.57,145.18) .. (116.97,145.18) .. controls (116.38,145.18) and (115.89,144.47) .. (115.89,143.59) -- cycle ;
\draw  [color={rgb, 255:red, 139; green, 6; blue, 24 }  ,draw opacity=1 ][fill={rgb, 255:red, 139; green, 6; blue, 24 }  ,fill opacity=1 ] (162.57,145.44) -- (160.36,143.41) -- (162.61,141.48) -- (161.47,143.44) -- cycle ;
%Straight Lines [id:da34571177189764013] 
\draw [color={rgb, 255:red, 139; green, 6; blue, 24 }  ,draw opacity=1 ]   (99.65,171.52) -- (177.25,171.37) ;
%Straight Lines [id:da3540917782373827] 
\draw [color={rgb, 255:red, 139; green, 6; blue, 24 }  ,draw opacity=1 ] [dash pattern={on 4.5pt off 4.5pt}]  (77.29,172.33) -- (98.56,171.86) ;
%Straight Lines [id:da4687584988372703] 
\draw [color={rgb, 255:red, 139; green, 6; blue, 24 }  ,draw opacity=1 ]   (34.53,171.81) -- (74,172.33) ;
\draw  [color={rgb, 255:red, 139; green, 6; blue, 24 }  ,draw opacity=1 ][fill={rgb, 255:red, 139; green, 6; blue, 24 }  ,fill opacity=1 ] (53.84,174.01) -- (51.63,171.98) -- (53.88,170.05) -- (52.74,172.01) -- cycle ;
%Shape: Ellipse [id:dp9109189507562275] 
\draw  [color={rgb, 255:red, 139; green, 6; blue, 24 }  ,draw opacity=1 ][fill={rgb, 255:red, 139; green, 6; blue, 24 }  ,fill opacity=1 ] (98.13,171.52) .. controls (98.13,170.64) and (98.62,169.93) .. (99.22,169.93) .. controls (99.82,169.93) and (100.3,170.64) .. (100.3,171.52) .. controls (100.3,172.4) and (99.82,173.11) .. (99.22,173.11) .. controls (98.62,173.11) and (98.13,172.4) .. (98.13,171.52) -- cycle ;
\draw  [color={rgb, 255:red, 139; green, 6; blue, 24 }  ,draw opacity=1 ][fill={rgb, 255:red, 139; green, 6; blue, 24 }  ,fill opacity=1 ] (161.91,173.37) -- (159.7,171.34) -- (161.95,169.41) -- (160.82,171.36) -- cycle ;
%Shape: Parallelogram [id:dp7614777382120138] 
\draw  [color={rgb, 255:red, 245; green, 166; blue, 35 }  ,draw opacity=1 ][fill={rgb, 255:red, 248; green, 198; blue, 155 }  ,fill opacity=1 ] (562.56,233.77) -- (562.58,124.19) -- (632.28,74.22) -- (632.26,183.8) -- cycle ;
%Straight Lines [id:da5544265540675392] 
\draw [color={rgb, 255:red, 139; green, 6; blue, 24 }  ,draw opacity=1 ]   (512,152.5) -- (567,153) ;
%Shape: Ellipse [id:dp9278935458291747] 
\draw  [color={rgb, 255:red, 139; green, 6; blue, 24 }  ,draw opacity=1 ][fill={rgb, 255:red, 139; green, 6; blue, 24 }  ,fill opacity=1 ] (616.07,167.22) .. controls (616.07,166.35) and (616.55,165.64) .. (617.15,165.64) .. controls (617.75,165.64) and (618.24,166.35) .. (618.24,167.22) .. controls (618.24,168.1) and (617.75,168.81) .. (617.15,168.81) .. controls (616.55,168.81) and (616.07,168.1) .. (616.07,167.22) -- cycle ;
\draw  [color={rgb, 255:red, 139; green, 6; blue, 24 }  ,draw opacity=1 ][fill={rgb, 255:red, 139; green, 6; blue, 24 }  ,fill opacity=1 ] (541.03,154.85) -- (538.82,152.81) -- (541.07,150.88) -- (539.94,152.84) -- cycle ;
%Straight Lines [id:da35043928515103373] 
\draw [color={rgb, 255:red, 139; green, 6; blue, 24 }  ,draw opacity=1 ]   (617.5,139.25) -- (640,119.33) ;
\draw  [color={rgb, 255:red, 139; green, 6; blue, 24 }  ,draw opacity=1 ][fill={rgb, 255:red, 139; green, 6; blue, 24 }  ,fill opacity=1 ] (633.24,127.38) -- (630.61,127.09) -- (631.94,123.87) -- (631.6,126.36) -- cycle ;
%Straight Lines [id:da837740674165147] 
\draw [color={rgb, 255:red, 139; green, 6; blue, 24 }  ,draw opacity=1 ]   (617.5,167.75) -- (638.12,187.21) ;
%Straight Lines [id:da7371922708896648] 
\draw [color={rgb, 255:red, 139; green, 6; blue, 24 }  ,draw opacity=1 ]   (474.13,185.22) -- (513.31,152.79) ;
%Straight Lines [id:da8672294699494015] 
\draw [color={rgb, 255:red, 139; green, 6; blue, 24 }  ,draw opacity=1 ]   (474.93,118.92) -- (513.31,152.79) ;
\draw  [color={rgb, 255:red, 139; green, 6; blue, 24 }  ,draw opacity=1 ][fill={rgb, 255:red, 139; green, 6; blue, 24 }  ,fill opacity=1 ] (493.17,171.34) -- (490.54,171.05) -- (491.87,167.83) -- (491.53,170.32) -- cycle ;
\draw  [color={rgb, 255:red, 139; green, 6; blue, 24 }  ,draw opacity=1 ][fill={rgb, 255:red, 139; green, 6; blue, 24 }  ,fill opacity=1 ] (627.34,179.95) -- (626.82,176.46) -- (629.42,176.9) -- (627.6,177.44) -- cycle ;
\draw  [color={rgb, 255:red, 139; green, 6; blue, 24 }  ,draw opacity=1 ][fill={rgb, 255:red, 139; green, 6; blue, 24 }  ,fill opacity=1 ] (490.86,135.77) -- (489.82,132.34) -- (492.45,132.51) -- (490.74,133.24) -- cycle ;
%Shape: Ellipse [id:dp6440090753959639] 
\draw  [color={rgb, 255:red, 139; green, 6; blue, 24 }  ,draw opacity=1 ][fill={rgb, 255:red, 139; green, 6; blue, 24 }  ,fill opacity=1 ] (616.42,139.66) .. controls (616.42,138.78) and (616.9,138.07) .. (617.5,138.07) .. controls (618.1,138.07) and (618.58,138.78) .. (618.58,139.66) .. controls (618.58,140.54) and (618.1,141.25) .. (617.5,141.25) .. controls (616.9,141.25) and (616.42,140.54) .. (616.42,139.66) -- cycle ;
%Shape: Ellipse [id:dp3944982166605865] 
\draw  [color={rgb, 255:red, 139; green, 6; blue, 24 }  ,draw opacity=1 ][fill={rgb, 255:red, 139; green, 6; blue, 24 }  ,fill opacity=1 ] (511.43,152.32) .. controls (511.43,151.45) and (511.92,150.74) .. (512.51,150.74) .. controls (513.11,150.74) and (513.6,151.45) .. (513.6,152.32) .. controls (513.6,153.2) and (513.11,153.91) .. (512.51,153.91) .. controls (511.92,153.91) and (511.43,153.2) .. (511.43,152.32) -- cycle ;
%Straight Lines [id:da12007635074139289] 
\draw [color={rgb, 255:red, 139; green, 6; blue, 24 }  ,draw opacity=1 ] [dash pattern={on 4.5pt off 4.5pt}]  (599.74,153.35) -- (617.5,139.25) ;
%Straight Lines [id:da5619431705235536] 
\draw [color={rgb, 255:red, 139; green, 6; blue, 24 }  ,draw opacity=1 ] [dash pattern={on 4.5pt off 4.5pt}]  (617.71,167.67) -- (617.52,167.52) -- (617.15,167.22) -- (616.47,166.67) -- (599.95,153.27) ;
%Straight Lines [id:da08930283011741758] 
\draw [color={rgb, 255:red, 139; green, 6; blue, 24 }  ,draw opacity=1 ] [dash pattern={on 4.5pt off 4.5pt}]  (567,153) -- (599.95,153.27) ;

% Text Node
\draw (180.46,130.3) node [anchor=north west][inner sep=0.75pt]    {$c$};
% Text Node
\draw (178.49,160.15) node [anchor=north west][inner sep=0.75pt]    {$a$};
% Text Node
\draw (619.35,82.46) node [anchor=north west][inner sep=0.75pt]    {$S$};
% Text Node
\draw (530.8,155.56) node [anchor=north west][inner sep=0.75pt]    {$\sigma(f)$};
% Text Node
\draw (26.84,123.58) node [anchor=north west][inner sep=0.75pt]    {$\sigma(c)$};
% Text Node
\draw (25.1,152.36) node [anchor=north west][inner sep=0.75pt]    {$\sigma(a)$};
% Text Node
\draw (132.57,81.15) node [anchor=north west][inner sep=0.75pt]    {$S$};
% Text Node
\draw (641.27,186.67) node [anchor=north west][inner sep=0.75pt]    {$a$};
% Text Node
\draw (641.71,106.67) node [anchor=north west][inner sep=0.75pt]    {$c$};
% Text Node
\draw (442.84,105.35) node [anchor=north west][inner sep=0.75pt]    {$\sigma(c)$};
% Text Node
\draw (440.05,178.85) node [anchor=north west][inner sep=0.75pt]    {$\sigma (a)$};
% Text Node
\draw (506.43,152.72) node [anchor=north west][inner sep=0.75pt]    {$\mu $};
% Text Node
\draw (585.47,156.53) node [anchor=north west][inner sep=0.75pt]    {$\nu '$};
% Text Node
\draw (211.95,129.18) node [anchor=north west][inner sep=0.75pt]    {$=\sum\limits_{\mu ,\nu,f}\left[\tilde{F}^{\sigma}_{a,c,f}\right]^{( \mu ,\nu )}\sum\limits_{\nu '} U_{S}( a,c,f)_{\nu \nu '}$};

\end{tikzpicture}
    \caption{The action of $S$ on fusion spaces.}
    \label{fig:trivactconst2}
\end{figure}

To answer this question, we first return to Fig. \ref{fig:SactFdef} and specialize to the case that the surface acts invertibly (i.e., as a permutation) on the lines. We get Fig. \ref{fig:trivactconst1}, where we have simplified the notation for $\tilde F$. 
Let the action on $S$ on the fusion space, $V_{ab}^c$, be given by the matrix 
\be
U_{S}(a,b,c)_{\mu,\nu}~,
\ee
where $\mu,\nu$ runs over an orthonormal basis for $V_{ab}^c$. Then, moving the surface on the R.H.S. of Fig. \ref{fig:trivactconst1} over the fusion space, we get Fig. \ref{fig:trivactconst2}.
Using the relation in Fig. \ref{RHSrel} on the L.H.S. of Fig. \ref{fig:trivactconst2} we get Fig. \ref{fig:trivactconst3}. Since the sum in the L.H.S and R.H.S of this diagram is over the same basis, we can equate the components to get the equation
\be\label{invRel}
\sqrt{\frac{d_f}{d_ad_c}} \delta_{\nu',\mu^{\dagger}}= \sum_{\nu'}  [\tilde F^{\sigma}_{a,c,f}]^{\mu,\nu} ~ U_{S}(a,c,f)_{\nu\nu'}~.
\ee

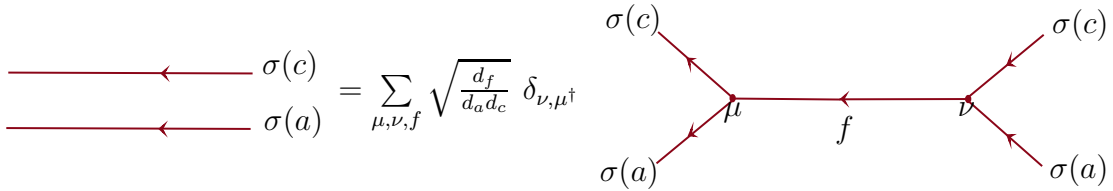
\begin{figure}[h!]
    \centering

\tikzset{every picture/.style={line width=0.75pt}} %set default line width to 0.75pt        

\begin{tikzpicture}[x=0.75pt,y=0.75pt,yscale=-1,xscale=1]
%uncomment if require: \path (0,523); %set diagram left start at 0, and has height of 523

%Straight Lines [id:da8986001156625869] 
\draw [color={rgb, 255:red, 139; green, 6; blue, 24 }  ,draw opacity=1 ]   (56,143) -- (179.22,143.44) ;
\draw  [color={rgb, 255:red, 139; green, 6; blue, 24 }  ,draw opacity=1 ][fill={rgb, 255:red, 139; green, 6; blue, 24 }  ,fill opacity=1 ] (135.57,145.44) -- (133.36,143.41) -- (135.61,141.48) -- (134.47,143.44) -- cycle ;
%Straight Lines [id:da6313377151666129] 
\draw [color={rgb, 255:red, 139; green, 6; blue, 24 }  ,draw opacity=1 ]   (55,171) -- (178.22,171.44) ;
\draw  [color={rgb, 255:red, 139; green, 6; blue, 24 }  ,draw opacity=1 ][fill={rgb, 255:red, 139; green, 6; blue, 24 }  ,fill opacity=1 ] (134.57,173.44) -- (132.36,171.41) -- (134.61,169.48) -- (133.47,171.44) -- cycle ;
%Straight Lines [id:da6106243776860283] 
\draw [color={rgb, 255:red, 139; green, 6; blue, 24 }  ,draw opacity=1 ]   (475.23,156.49) -- (539.03,156.35) ;
\draw  [color={rgb, 255:red, 139; green, 6; blue, 24 }  ,draw opacity=1 ][fill={rgb, 255:red, 139; green, 6; blue, 24 }  ,fill opacity=1 ] (478.32,158.35) -- (476.11,156.31) -- (478.36,154.38) -- (477.22,156.34) -- cycle ;
%Straight Lines [id:da6550924594683141] 
\draw [color={rgb, 255:red, 139; green, 6; blue, 24 }  ,draw opacity=1 ]   (539.03,156.35) -- (578.2,123.92) ;
\draw  [color={rgb, 255:red, 139; green, 6; blue, 24 }  ,draw opacity=1 ][fill={rgb, 255:red, 139; green, 6; blue, 24 }  ,fill opacity=1 ] (562.52,139.38) -- (559.9,139.09) -- (561.22,135.87) -- (560.89,138.36) -- cycle ;
%Straight Lines [id:da7660814008989012] 
\draw [color={rgb, 255:red, 139; green, 6; blue, 24 }  ,draw opacity=1 ]   (539.03,156.35) -- (577.41,190.21) ;
\draw  [color={rgb, 255:red, 139; green, 6; blue, 24 }  ,draw opacity=1 ][fill={rgb, 255:red, 139; green, 6; blue, 24 }  ,fill opacity=1 ] (561.13,177.95) -- (559.93,174.63) -- (562.56,174.54) -- (560.89,175.44) -- cycle ;
%Shape: Ellipse [id:dp43009066398137696] 
\draw  [color={rgb, 255:red, 139; green, 6; blue, 24 }  ,draw opacity=1 ][fill={rgb, 255:red, 139; green, 6; blue, 24 }  ,fill opacity=1 ] (539.03,156.35) .. controls (539.03,155.47) and (539.51,154.76) .. (540.11,154.76) .. controls (540.71,154.76) and (541.2,155.47) .. (541.2,156.35) .. controls (541.2,157.22) and (540.71,157.93) .. (540.11,157.93) .. controls (539.51,157.93) and (539.03,157.22) .. (539.03,156.35) -- cycle ;
%Straight Lines [id:da6617957667123842] 
\draw [color={rgb, 255:red, 139; green, 6; blue, 24 }  ,draw opacity=1 ]   (420.8,156.1) -- (475.8,156.6) ;
%Straight Lines [id:da12895020696997994] 
\draw [color={rgb, 255:red, 139; green, 6; blue, 24 }  ,draw opacity=1 ]   (382.93,188.82) -- (422.11,156.39) ;
%Straight Lines [id:da7010717480231173] 
\draw [color={rgb, 255:red, 139; green, 6; blue, 24 }  ,draw opacity=1 ]   (383.73,122.52) -- (422.11,156.39) ;
\draw  [color={rgb, 255:red, 139; green, 6; blue, 24 }  ,draw opacity=1 ][fill={rgb, 255:red, 139; green, 6; blue, 24 }  ,fill opacity=1 ] (401.97,174.94) -- (399.34,174.65) -- (400.67,171.43) -- (400.33,173.92) -- cycle ;
\draw  [color={rgb, 255:red, 139; green, 6; blue, 24 }  ,draw opacity=1 ][fill={rgb, 255:red, 139; green, 6; blue, 24 }  ,fill opacity=1 ] (399.66,139.37) -- (398.62,135.94) -- (401.25,136.11) -- (399.54,136.84) -- cycle ;
%Shape: Ellipse [id:dp2074355275095363] 
\draw  [color={rgb, 255:red, 139; green, 6; blue, 24 }  ,draw opacity=1 ][fill={rgb, 255:red, 139; green, 6; blue, 24 }  ,fill opacity=1 ] (420.23,155.92) .. controls (420.23,155.05) and (420.72,154.34) .. (421.31,154.34) .. controls (421.91,154.34) and (422.4,155.05) .. (422.4,155.92) .. controls (422.4,156.8) and (421.91,157.51) .. (421.31,157.51) .. controls (420.72,157.51) and (420.23,156.8) .. (420.23,155.92) -- cycle ;

% Text Node
\draw (182.46,130.3) node [anchor=north west][inner sep=0.75pt]    {$\sigma(c)$};
% Text Node
\draw (183.46,158.3) node [anchor=north west][inner sep=0.75pt]    {$\sigma(a)$};
% Text Node
\draw (471.5,163.22) node [anchor=north west][inner sep=0.75pt]    {$f$};
% Text Node
\draw (579.56,182.67) node [anchor=north west][inner sep=0.75pt]    {$\sigma(a)$};
% Text Node
\draw (581,109.67) node [anchor=north west][inner sep=0.75pt]    {$\sigma(c)$};
% Text Node
\draw (533.37,156.77) node [anchor=north west][inner sep=0.75pt]    {$\nu $};
% Text Node
\draw (355.64,106.95) node [anchor=north west][inner sep=0.75pt]    {$\sigma(c)$};
% Text Node
\draw (352.85,184.45) node [anchor=north west][inner sep=0.75pt]    {$\sigma(a)$};
% Text Node
\draw (415.23,156.32) node [anchor=north west][inner sep=0.75pt]    {$\mu $};
% Text Node
\draw (221,134.4) node [anchor=north west][inner sep=0.75pt]    {$=\sum \limits_{\mu,\nu,f}\sqrt{\frac{d_{f}}{d_{a} d_{c}}} \ \delta _{\nu ,\mu ^{\dagger }}$};
\end{tikzpicture}
\caption{The completeness relation for lines. Note that $d_{\sigma(\ell)}=d_{\ell}$ by commutativity with $\CS$.}
\label{RHSrel}
\end{figure}
\begin{figure}[h!]
    \centering

\tikzset{every picture/.style={line width=0.75pt}} %set default line width to 0.75pt        

\begin{tikzpicture}[x=0.75pt,y=0.75pt,yscale=-1,xscale=1]
%uncomment if require: \path (0,523); %set diagram left start at 0, and has height of 523

%Shape: Parallelogram [id:dp7614777382120138] 
\draw  [color={rgb, 255:red, 245; green, 166; blue, 35 }  ,draw opacity=1 ][fill={rgb, 255:red, 248; green, 198; blue, 155 }  ,fill opacity=1 ] (582.08,234) -- (582.1,150.91) -- (640.41,113) -- (640.39,196.09) -- cycle ;
%Straight Lines [id:da5544265540675392] 
\draw [color={rgb, 255:red, 139; green, 6; blue, 24 }  ,draw opacity=1 ]   (539.79,172.37) -- (585.8,172.75) ;
%Shape: Ellipse [id:dp9278935458291747] 
\draw  [color={rgb, 255:red, 139; green, 6; blue, 24 }  ,draw opacity=1 ][fill={rgb, 255:red, 139; green, 6; blue, 24 }  ,fill opacity=1 ] (626.85,183.53) .. controls (626.85,182.87) and (627.25,182.33) .. (627.75,182.33) .. controls (628.26,182.33) and (628.66,182.87) .. (628.66,183.53) .. controls (628.66,184.2) and (628.26,184.74) .. (627.75,184.74) .. controls (627.25,184.74) and (626.85,184.2) .. (626.85,183.53) -- cycle ;
\draw  [color={rgb, 255:red, 139; green, 6; blue, 24 }  ,draw opacity=1 ][fill={rgb, 255:red, 139; green, 6; blue, 24 }  ,fill opacity=1 ] (564.07,174.15) -- (562.23,172.6) -- (564.11,171.14) -- (563.16,172.62) -- cycle ;
%Straight Lines [id:da35043928515103373] 
\draw [color={rgb, 255:red, 139; green, 6; blue, 24 }  ,draw opacity=1 ]   (628.05,162.32) -- (646.87,147.21) ;
\draw  [color={rgb, 255:red, 139; green, 6; blue, 24 }  ,draw opacity=1 ][fill={rgb, 255:red, 139; green, 6; blue, 24 }  ,fill opacity=1 ] (641.21,153.32) -- (639.02,153.1) -- (640.12,150.66) -- (639.84,152.54) -- cycle ;
%Straight Lines [id:da837740674165147] 
\draw [color={rgb, 255:red, 139; green, 6; blue, 24 }  ,draw opacity=1 ]   (628.05,183.93) -- (645.3,198.69) ;
%Straight Lines [id:da7371922708896648] 
\draw [color={rgb, 255:red, 139; green, 6; blue, 24 }  ,draw opacity=1 ]   (508.11,197.18) -- (540.88,172.58) ;
%Straight Lines [id:da8672294699494015] 
\draw [color={rgb, 255:red, 139; green, 6; blue, 24 }  ,draw opacity=1 ]   (508.77,146.9) -- (540.88,172.58) ;
\draw  [color={rgb, 255:red, 139; green, 6; blue, 24 }  ,draw opacity=1 ][fill={rgb, 255:red, 139; green, 6; blue, 24 }  ,fill opacity=1 ] (524.03,186.65) -- (521.84,186.43) -- (522.94,183.99) -- (522.66,185.88) -- cycle ;
\draw  [color={rgb, 255:red, 139; green, 6; blue, 24 }  ,draw opacity=1 ][fill={rgb, 255:red, 139; green, 6; blue, 24 }  ,fill opacity=1 ] (636.27,193.18) -- (635.85,190.54) -- (638.01,190.87) -- (636.5,191.28) -- cycle ;
\draw  [color={rgb, 255:red, 139; green, 6; blue, 24 }  ,draw opacity=1 ][fill={rgb, 255:red, 139; green, 6; blue, 24 }  ,fill opacity=1 ] (522.1,159.68) -- (521.23,157.08) -- (523.43,157.2) -- (522,157.76) -- cycle ;
%Shape: Ellipse [id:dp6440090753959639] 
\draw  [color={rgb, 255:red, 139; green, 6; blue, 24 }  ,draw opacity=1 ][fill={rgb, 255:red, 139; green, 6; blue, 24 }  ,fill opacity=1 ] (627.14,162.63) .. controls (627.14,161.97) and (627.54,161.43) .. (628.05,161.43) .. controls (628.55,161.43) and (628.95,161.97) .. (628.95,162.63) .. controls (628.95,163.3) and (628.55,163.83) .. (628.05,163.83) .. controls (627.54,163.83) and (627.14,163.3) .. (627.14,162.63) -- cycle ;
%Shape: Ellipse [id:dp3944982166605865] 
\draw  [color={rgb, 255:red, 139; green, 6; blue, 24 }  ,draw opacity=1 ][fill={rgb, 255:red, 139; green, 6; blue, 24 }  ,fill opacity=1 ] (539.31,172.23) .. controls (539.31,171.57) and (539.72,171.03) .. (540.22,171.03) .. controls (540.72,171.03) and (541.13,171.57) .. (541.13,172.23) .. controls (541.13,172.9) and (540.72,173.44) .. (540.22,173.44) .. controls (539.72,173.44) and (539.31,172.9) .. (539.31,172.23) -- cycle ;
%Straight Lines [id:da12007635074139289] 
\draw [color={rgb, 255:red, 139; green, 6; blue, 24 }  ,draw opacity=1 ] [dash pattern={on 4.5pt off 4.5pt}]  (613.19,173.01) -- (628.05,162.32) ;
%Straight Lines [id:da5619431705235536] 
\draw [color={rgb, 255:red, 139; green, 6; blue, 24 }  ,draw opacity=1 ] [dash pattern={on 4.5pt off 4.5pt}]  (628.22,183.87) -- (628.06,183.76) -- (627.75,183.53) -- (627.19,183.11) -- (613.36,172.95) ;
%Straight Lines [id:da08930283011741758] 
\draw [color={rgb, 255:red, 139; green, 6; blue, 24 }  ,draw opacity=1 ] [dash pattern={on 4.5pt off 4.5pt}]  (585.8,172.75) -- (613.36,172.95) ;
%Shape: Parallelogram [id:dp9617638601663994] 
\draw  [color={rgb, 255:red, 245; green, 166; blue, 35 }  ,draw opacity=1 ][fill={rgb, 255:red, 248; green, 198; blue, 155 }  ,fill opacity=1 ] (202.25,235) -- (202.26,151.91) -- (260.57,114) -- (260.56,197.09) -- cycle ;
%Straight Lines [id:da6178023407533235] 
\draw [color={rgb, 255:red, 139; green, 6; blue, 24 }  ,draw opacity=1 ]   (159.95,173.37) -- (205.96,173.75) ;
%Shape: Ellipse [id:dp6427707763468771] 
\draw  [color={rgb, 255:red, 139; green, 6; blue, 24 }  ,draw opacity=1 ][fill={rgb, 255:red, 139; green, 6; blue, 24 }  ,fill opacity=1 ] (247.01,184.53) .. controls (247.01,183.87) and (247.42,183.33) .. (247.92,183.33) .. controls (248.42,183.33) and (248.83,183.87) .. (248.83,184.53) .. controls (248.83,185.2) and (248.42,185.74) .. (247.92,185.74) .. controls (247.42,185.74) and (247.01,185.2) .. (247.01,184.53) -- cycle ;
\draw  [color={rgb, 255:red, 139; green, 6; blue, 24 }  ,draw opacity=1 ][fill={rgb, 255:red, 139; green, 6; blue, 24 }  ,fill opacity=1 ] (184.24,175.15) -- (182.39,173.6) -- (184.27,172.14) -- (183.32,173.62) -- cycle ;
%Straight Lines [id:da5021996317671136] 
\draw [color={rgb, 255:red, 139; green, 6; blue, 24 }  ,draw opacity=1 ]   (248.21,163.32) -- (267.03,148.21) ;
\draw  [color={rgb, 255:red, 139; green, 6; blue, 24 }  ,draw opacity=1 ][fill={rgb, 255:red, 139; green, 6; blue, 24 }  ,fill opacity=1 ] (261.38,154.32) -- (259.18,154.1) -- (260.29,151.66) -- (260.01,153.54) -- cycle ;
%Straight Lines [id:da20339763223698148] 
\draw [color={rgb, 255:red, 139; green, 6; blue, 24 }  ,draw opacity=1 ]   (248.21,184.93) -- (265.46,199.69) ;
%Straight Lines [id:da40847651265023954] 
\draw [color={rgb, 255:red, 139; green, 6; blue, 24 }  ,draw opacity=1 ]   (128.28,198.18) -- (161.05,173.58) ;
%Straight Lines [id:da11137207704624841] 
\draw [color={rgb, 255:red, 139; green, 6; blue, 24 }  ,draw opacity=1 ]   (128.94,147.9) -- (161.05,173.58) ;
\draw  [color={rgb, 255:red, 139; green, 6; blue, 24 }  ,draw opacity=1 ][fill={rgb, 255:red, 139; green, 6; blue, 24 }  ,fill opacity=1 ] (144.2,187.65) -- (142,187.43) -- (143.11,184.99) -- (142.83,186.88) -- cycle ;
\draw  [color={rgb, 255:red, 139; green, 6; blue, 24 }  ,draw opacity=1 ][fill={rgb, 255:red, 139; green, 6; blue, 24 }  ,fill opacity=1 ] (256.44,194.18) -- (256.01,191.54) -- (258.18,191.87) -- (256.66,192.28) -- cycle ;
\draw  [color={rgb, 255:red, 139; green, 6; blue, 24 }  ,draw opacity=1 ][fill={rgb, 255:red, 139; green, 6; blue, 24 }  ,fill opacity=1 ] (142.27,160.68) -- (141.4,158.08) -- (143.6,158.2) -- (142.17,158.76) -- cycle ;
%Shape: Ellipse [id:dp1440659941415816] 
\draw  [color={rgb, 255:red, 139; green, 6; blue, 24 }  ,draw opacity=1 ][fill={rgb, 255:red, 139; green, 6; blue, 24 }  ,fill opacity=1 ] (247.3,163.63) .. controls (247.3,162.97) and (247.71,162.43) .. (248.21,162.43) .. controls (248.71,162.43) and (249.12,162.97) .. (249.12,163.63) .. controls (249.12,164.3) and (248.71,164.83) .. (248.21,164.83) .. controls (247.71,164.83) and (247.3,164.3) .. (247.3,163.63) -- cycle ;
%Shape: Ellipse [id:dp46742550606707856] 
\draw  [color={rgb, 255:red, 139; green, 6; blue, 24 }  ,draw opacity=1 ][fill={rgb, 255:red, 139; green, 6; blue, 24 }  ,fill opacity=1 ] (159.48,173.23) .. controls (159.48,172.57) and (159.88,172.03) .. (160.38,172.03) .. controls (160.89,172.03) and (161.29,172.57) .. (161.29,173.23) .. controls (161.29,173.9) and (160.89,174.44) .. (160.38,174.44) .. controls (159.88,174.44) and (159.48,173.9) .. (159.48,173.23) -- cycle ;
%Straight Lines [id:da41917400320313514] 
\draw [color={rgb, 255:red, 139; green, 6; blue, 24 }  ,draw opacity=1 ] [dash pattern={on 4.5pt off 4.5pt}]  (233.36,174.01) -- (248.21,163.32) ;
%Straight Lines [id:da1729565061850733] 
\draw [color={rgb, 255:red, 139; green, 6; blue, 24 }  ,draw opacity=1 ] [dash pattern={on 4.5pt off 4.5pt}]  (248.38,184.87) -- (248.23,184.76) -- (247.92,184.53) -- (247.35,184.11) -- (233.53,173.95) ;
%Straight Lines [id:da15365490860804665] 
\draw [color={rgb, 255:red, 139; green, 6; blue, 24 }  ,draw opacity=1 ] [dash pattern={on 4.5pt off 4.5pt}]  (205.96,173.75) -- (233.53,173.95) ;

% Text Node
\draw (628.04,117.17) node [anchor=north west][inner sep=0.75pt]    {$S$};
% Text Node
\draw (646.95,196.45) node [anchor=north west][inner sep=0.75pt]    {$a$};
% Text Node
\draw (647.32,135.77) node [anchor=north west][inner sep=0.75pt]    {$c$};
% Text Node
\draw (534.15,170.7) node [anchor=north west][inner sep=0.75pt]    {$\mu $};
% Text Node
\draw (600.03,173.59) node [anchor=north west][inner sep=0.75pt]    {$\nu '$};
% Text Node
\draw (277.49,151.5) node [anchor=north west][inner sep=0.75pt]  [font=\small]  {$=\sum\limits_{\mu ,\nu ,f}\left[\tilde{F}_{a,c,f}^{\sigma }\right]^{( \mu ,\nu )}\sum\limits_{\nu '} U_{S}( a,c,f)_{\nu \nu '}$};
% Text Node
\draw (248.21,118.17) node [anchor=north west][inner sep=0.75pt]    {$S$};
% Text Node
\draw (169.88,173.85) node [anchor=north west][inner sep=0.75pt]    {$\sigma ( f)$};
% Text Node
\draw (100.18,138.8) node [anchor=north west][inner sep=0.75pt]    {$\sigma ( c)$};
% Text Node
\draw (98.51,191.51) node [anchor=north west][inner sep=0.75pt]    {$\sigma ( a)$};
% Text Node
\draw (154.31,171.7) node [anchor=north west][inner sep=0.75pt]    {$\mu $};
% Text Node
\draw (220.19,174.59) node [anchor=north west][inner sep=0.75pt]    {$\nu '$};
% Text Node
\draw (6,156.4) node [anchor=north west][inner sep=0.75pt]  [font=\small]  {$\sum\limits_{\mu ,\nu 'f}\sqrt{\frac{d_{f}}{d_{a} d_{c}}} \ \delta _{\nu ',\mu ^{\dagger }}$};
% Text Node
\draw (271,136.77) node [anchor=north west][inner sep=0.75pt]    {$c$};
% Text Node
\draw (272,190.45) node [anchor=north west][inner sep=0.75pt]    {$a$};
% Text Node
\draw (482.18,134.8) node [anchor=north west][inner sep=0.75pt]    {$\sigma ( c)$};
% Text Node
\draw (478.51,188.51) node [anchor=north west][inner sep=0.75pt]    {$\sigma ( a)$};
% Text Node
\draw (549.88,175.85) node [anchor=north west][inner sep=0.75pt]    {$\sigma ( f)$};

\end{tikzpicture}
    \caption{Using completeness relation for lines on the L.H.S of Fig. \ref{fig:trivactconst2}.}
    \label{fig:trivactconst3}
\end{figure}
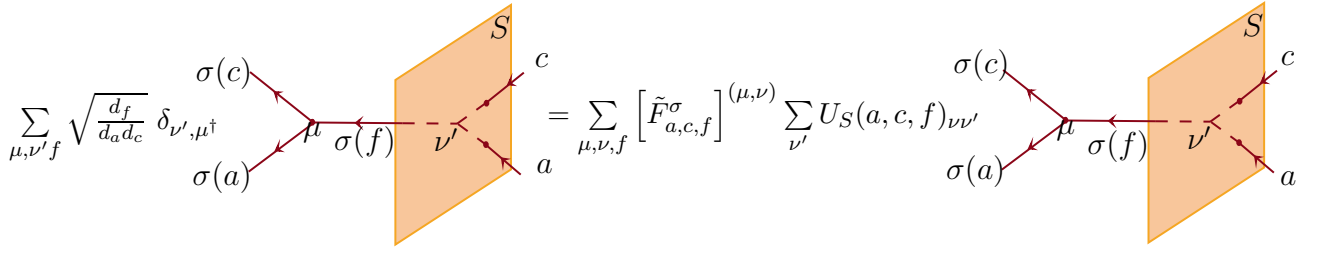
The L.H.S. of \eqref{invRel} is an invertible matrix and the R.H.S. is a product of two square matrices. Therefore, both $\tilde F$ and $U_S$ have to be invertible matrices. In particular, we see that:

\medskip
\noindent
{\bf Lemma 2:} {\it If a surface acts invertibly on lines, then it also acts invertibly on fusion spaces.}

\medskip
\noindent
As a result, the invertible ambiguity found by Davydov in \cite{Davydov_2014} does not generalize to a non-invertible ambiguity. Said differently, an invertible symmetry and a non-invertible symmetry cannot differ only by their action on fusion spaces.

Given these results, let us prove a partial generalization of theorem 1 to unitary non-Abelian TQFTs. To that end, suppose we have a TQFT without any bosonic line operators. Since the action of a surface operator preserves topological twists, it cannot map a non-trivial line operator in such a theory to the trivial line operator. By the same logic, a surface cannot map the trivial line operator to a non-trivial line. Therefore, we have
\be
\label{eq:nobosoncond}
n_{S \cdot1}^b = n_{S \cdot b}^{1} = \delta_{b,1}~.
\ee
Next, since $n_{S\cdot a}^{b}$ commutes with the modular $S$-matrix of the TQFT (it describes a modular invariant in the boundary CFT), we have
\be
\sum_{c} \mathcal{S}_{ac} n_{S\cdot c}^{b} = \sum_{c} n_{S\cdot a}^{c} \mathcal{S}_{cb}~,
\ee
where $\mathcal{S}$ is the modular S-matrix. Choosing $b=1$ and using \eqref{eq:nobosoncond}, we get
\be
\label{eq:qdimcond1}
d_{a}= \sum_c n_{S\cdot a}^c d_c~,
\ee
where $d_a$ are the quantum dimensions of the line operators. This logic shows that, when the TQFT does not have any bosons, the action of the surface operator preserves the quantum dimensions. On the other hand, choosing $a=1$ and using \eqref{eq:nobosoncond}, we get
\be
\label{eq:qdimcond2}
\sum_c n_{S\cdot c}^b d_c = d_{b}~.
\ee
In a unitary TQFT, the quantum dimensions satisfy $d_a \geq 1$ for all line operators $a$. Also, note that $n_{S\cdot a}^b$ are non-negative integers. Suppose $n_{S\cdot e}^f\neq 0$, then using \eqref{eq:qdimcond1} for $a=e$ we find that $d_{f}\leq d_e$. Also, using \eqref{eq:qdimcond2} for $b=f$, we have $d_f \geq d_e$. Therefore, $d_e=d_f$ and the only consistent solution is
\be
n_{S\cdot a}^b= \delta_{b,\sigma(a)}~,
\ee
for some permutation, $\sigma$, of the lines. That is, in a unitary TQFT without bosons, all symmetries act as permutations on the line operators. In the boundary CFT, the corresponding statement is that, after the maximal extension of the chiral algebra, the partition function is always given by a permutation modular invariant \cite{Moore:1988ss,Dijkgraaf:1988tf}. 

Therefore, using lemma 2, we arrive at:

\medskip
\noindent
{\bf Theorem 3:} {\it $2+1$d TQFTs without bosons only have invertible symmetries.} 

\medskip
In the Abelian case, we know from theorem 1 that the converse also holds. However, the converse does not hold more generally. Indeed, there are non-abelian TQFTs with bosonic line operators that do not have any non-invertible symmetries. To understand this statement, let us assume that a TQFT, $\CT$, has a non-invertible symmetry implemented by some surface operator, $S$. From the argument leading to theorem 3, we know that either $S$ should map a non-trivial line operator the trivial line operator, or $S$ should map the trivial line operator to a non-trivial line operator. That is,
\be
n_{S\cdot1}^{a}\neq 0 \text{ or } n_{S\cdot a}^{1}\neq 0~,
\ee
for some non-trivial line operator, $a$. Therefore, the TQFT, $\CT$, has a bosonic line operator, $a$. More precisely, let $\CA_L$ be the set of bosonic line operators such that $n_{S\cdot a}^1 \neq 0$, and similarly define $\CA_R$ to be the set of bosonic line operators such that $n_{S\cdot1}^a\ne0$. The authors of \cite{Fuchs:2002cm} showed that $\CA_L$ and $\CA_R$ admit the structure of a commutative Frobenius algebra. In other words, $\CA_L$ and $\CA_R$ are bosons that can be condensed.\footnote{As we will discuss below, the bosonic lines in $\CA_L/\CA_R$ can be condensed on a surface to define a surface operator. The action of the surface operator on the line operators correspond to a type 1 parent of the original partition function in the boundary CFT. Type 1 partition functions are sums of squares of linear combinations of characters \cite{ostrik2001module}.} Therefore, a necessary condition for a TQFT to have a non-invertible symmetry is that it must contain a bosonic line operator which can be condensed. 

For example, the Fibonacci TQFT does not contain any bosons. Therefore, it is clear that this theory does not have any non-invertible symmetries. If we stack five copies of Fibonacci TQFT, then this theory does contain a boson. However, the authors of \cite{davydov2011commutative,Neupert_2016} showed that this boson cannot be condensed. Therefore, this TQFT, despite having a boson, does not have any non-invertible symmetries. More generally, $SO(3)_k$ Chern-Simon theory at odd $k$, or any of its powers do not have non-invertible symmetries \cite{neupert2016no}.\footnote{A modular tensor category which does not contain condensable bosons is called completely anisotropic. This is a stronger condition than anisotropy, which only requires the TQFT does not contain a Rep$(G)$ subcategory for some non-trivial group $G$.}

To find a converse of theorem 3, we will need to introduce aspects of higher gauging in the next section. In other words, we will need to understand some details of how to construct topological surfaces from lines.

However, before moving on to the resolution of surfaces in terms of lines, let us consider an Abelian TQFT with pointed MTC, $\CC$, and show that our general non-Abelian symmetry definition reduces to our Abelian one. Then the conditions \ref{def:nonabnoninvsym1} and \ref{def:nonabnoninvsym2} reduce to corresponding conditions in the Abelian case. Moreover, condition \ref{def:nonabnoninvsym3} leads to \eqref{RFtRI} which, in turn, leads (for $b=d$ and $a=c$) to $\theta_a=\theta_b$, which reproduces the first condition in the Abelian case.

\section{Resolving surfaces into lines: higher gauging and applications}\label{resolution}
In this section we build on the general results of the previous two sections by specializing to the case of condensation surfaces. As alluded to above, these are defects that are constructed by summing over lines (higher gauging) on a two-dimensional submanifold of spacetime.

After reviewing basic properties of higher gauging in Abelian TQFTs, we generalize the \lq\lq universal" non-invertible surfaces we constructed in \eqref{eq:SBsym} and consider arbitrary condensation surfaces. We explicitly construct the action of these general surfaces on the lines of Abelian TQFTs and show how to reconstruct the higher-gauging data from the action on lines.

Then, we move on to condensation surfaces in non-Abelian TQFTs and give a construction of a non-Abelian analog of the condensation surface in \eqref{eq:SBsym}. This result leads to a generalization of theorem 1 to non-Abelian TQFTs.

Finally, we conclude this section by considering arbitrary condensation surfaces built out of Abelian lines in general (not necessarily Abelian) TQFTs. We characterize precisely when such surfaces are (non-)invertible.

\subsection{Abelian TQFTs}

\label{sec:noinv defects from condensation}

As in section \ref{sec: Non-inv sym abelian TQFTs}, we consider a non-spin $2+1$d TQFT with invertible topological line operators forming an abelian group, $A$, under fusion. The full data that describes the 1-form symmetry is the group, $A$, along with the fusion and braiding phases: $F(a,b,c)$ and $R(a,b)$ respectively. This data forms a pointed braided fusion category, $\CC$. 

Consider the higher gauging of a 1-form symmetry subgroup, $G \le A$, on a 2-dimensional submanifold, $\Sigma$ \cite{Roumpedakis:2022aik}. For this procedure to be consistent, the 1-anomaly of $G$ should vanish. In other words, $F(g,h,k)$ restricted to $G$ should be trivial in $H^3(G,U(1))$. Therefore, there exist phases, $\tau_{G}(a,b): G \times G \to \mathds{C}$, such that
\be
F(g,h,k)=\frac{\tau_{G}(h,k)\tau_{G}(g,h+k)}{\tau_{G}(g+h,k)\tau_{G}(g,h)}~,  ~ ~ g,h,k \in G~.
\ee
This condition ensures that the gauging of the $G$-lines on $\Sigma$ does not depend on a choice of a triangulation of $\Sigma$.
While higher gauging $G$ on $\Sigma$, we have the freedom to add an SPT phase (a.k.a. discrete torsion) labelled by a 2-cocycle, $\sigma\in H^2(G,U(1))$. Let us label the condensation surface operator obtained in this way as $S(\Sigma, G,\sigma)$. 

As a concrete example, consider the higher gauging of a subgroup, $\DZ_2 \le A$, on $\Sigma$ \cite{Roumpedakis:2022aik}. The line operator, $a$, generating this subgroup can be higher-gauged if and only if $\theta_a=\pm 1$. This condition is equivalent to the existence of a gauge in which the $\mathbb{Z}_2$ restriction of the $F$ matrices satisfy $F|_{\DZ_2}=1$. We can then define
\be
S(\Sigma,\DZ_2,1):=\frac{1}{\sqrt{|H_1(\Sigma,\DZ_2)}|} \sum_{\gamma \in H_1(\Sigma,\DZ_2)} a(\gamma)~.
\ee
This condensation surface has the following fusion rule depending on $\theta_a$:
\bea
\theta_a=1 &:& S(\Sigma, \DZ_2, 1) \times S(\Sigma,\DZ_2,1)=\sqrt{|H_1(\Sigma, \DZ_2)|} ~ S(\Sigma, \DZ_2, 1)~, \\
\theta_a=-1 &:& S(\Sigma, \DZ_2, 1) \times S(\Sigma, \DZ_2, 1)= \mathds{1} ~,
\eea
where $\mathds{1}$ is the trivial surface operator. Whether the condensation defect, $S(\Sigma,\DZ_2,1)$, acts non-trivially on $\CC$ or not depends crucially on how the line operator, $a$, is embedded in $\CC$. For example, if $\CC$ is $\DZ_2$ discrete gauge theory, $S(\Sigma,\DZ_2,1)$ implements electric-magnetic duality. On the other hand, if $\CC$ is the Ising category, then $S(\Sigma,\DZ_2,1)$ acts trivially. 

Note that the fusion rule for $S(\Sigma,G,\sigma)$ depends on the 2-manifold, $\Sigma$. For example, in the case of $S(\Sigma,\DZ_2,1)$, we get the fusion rule \cite{Roumpedakis:2022aik}
\be
S(\Sigma, \DZ_2, 1) \times S(\Sigma, \DZ_2, 1)= \sqrt{|H_1(\Sigma, \DZ_2)|} ~ S(\Sigma, \DZ_2, 1)~.
\ee
In particular, for $\Sigma=\mathds{T}^2$, we find
\be
S(\mathds{T}^2, \DZ_2, 1) \times S(\mathds{T}^2, \DZ_2, 1)= 2 ~ S(\mathds{T}^2, \DZ_2, 1)~.
\ee
From now on, we will fix $\Sigma=\mathds{T}^2$ and write the surface operator as $S(G,\sigma)$ without explicitly mentioning the dependence on the 2-manifold. We will be mainly interested in distinguishing invertible and non-invertible surface operators, and so there is no loss of generality in choosing $\Sigma=\mathds{T}^2$. Indeed, we have 
\be
H_1(\Sigma,M)\cong M^{2g}~,
\ee
where $\Sigma$ is a compact genus $g$ surface, and $M$ is an abelian group. Therefore, if the fusion rules are non-invertible over the torus, they are non-invertible on any higher genus surface.

\subsubsection{Action on line operators}
To describe the action of the condensation defect on line operators, we must consider the direction normal to the surface as well (since the surface acts on lines that pierce it) and therefore the braiding data of the lines naturally enters.\footnote{Since the surface itself is two dimensional, the obstruction to constructing it only depends on the fusion data (i.e., the $F$ matrices).} Moreover, even though the 2d SPT does not affect the 1-anomaly of the lines, it naturally enters the action of the surface on lines (and therefore also the surface fusion rules). In particular, it turns out to be useful to define the following phase
\be
\label{eq:defXi}
\Xi_{[G,\sigma]}(g,h):= R(g,h) \frac{\tau_G(g,h)}{\tau_G(h,g)}\frac{\sigma(g,h)}{\sigma(h,g)}~,
\ee
which explicitly contains braiding and SPT data. Indeed, we will show that the action of the condensation defect, $S(G,\sigma)$, on a line operator, $p$, is given by
\be
\label{eq:Saction}
S(G,\sigma) \cdot p = \sum_{g \in G} \Bigg [\frac{1}{|G|}  \sum_{h\in G}  \frac{\theta_{h \cdot p}}{\theta_h \theta_p} \Xi_{[G,\sigma]}(g,h) \Bigg ] g \cdot p~.
\ee
From this expression, note that $g \cdot p \in S(G,\sigma)\cdot p$ if 
\be
\frac{1}{|G|}  \sum_{h\in G}  \frac{\theta_{h \cdot p}}{\theta_h \theta_p} \Xi_{[G,\sigma]}(g,h)\neq 0~.
\ee
Since this sum is over $|G|$ phases, we have 
\be 
\frac{1}{|G|}  \sum_{h\in G}  \frac{\theta_{h \cdot p}}{\theta_h \theta_p} \Xi_{[G,\sigma]}(g,h) \in \{0,1\}~.
\ee
For this sum to be equal to $1$, $p$ should satisfy
\be
\label{eq:pconst}
 \frac{\theta_{h \cdot p}}{\theta_h \theta_p} \Xi_{[G,\sigma]}(g,h) =1~, ~ \forall ~ h \in G~.
\ee

Now, let us understand how the expression \eqref{eq:Saction} for the action of the surface operator on the line operators can be derived. Consider a CFT $\CT$ with charge conjugation partition function
\be
Z_{\CT}= \sum_p \chi_{p} \bar \chi_{\bar p}~,
\ee
where $p$ labels the representations of some chiral algebra, $V$. We will choose a chiral algebra such that Rep$(V)\simeq \CC$ as modular tensor categories. Then, this CFT has a 0-form symmetry implemented by Verlinde lines labelled by $a \in A$. Gauging a subgroup, $G \le A$, with discrete torsion, $\sigma \in H^2(G,U(1))$, results in a CFT, $\CT/(G,\sigma)$, with partition function  \cite{Fuchs_2004}
\be
\label{eq:partfunc}
Z_{\CT/(G,\sigma)}= \sum_{g \in G} \Bigg [\frac{1}{|G|}   \sum_{h\in G} \frac{\theta_{h \cdot p}}{\theta_h \theta_p} \Xi_{[G,\sigma]}(g,h) \Bigg ] \chi_{p} \bar \chi_{\overline{g \cdot p}}~.
\ee
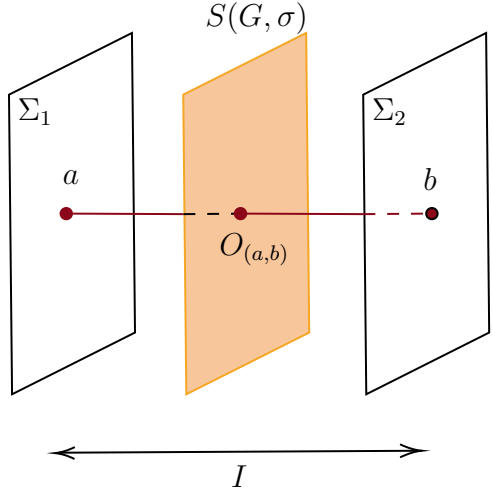
\begin{figure}
\centering

\tikzset{every picture/.style={line width=0.75pt}} %set default line width to 0.75pt        

\begin{tikzpicture}[x=0.75pt,y=0.75pt,yscale=-1,xscale=1]

%Shape: Parallelogram [id:dp10913265176241915] 
\draw   (444.58,203.85) -- (382.59,235) -- (381.04,83.79) -- (443.03,52.64) -- cycle ;
%Shape: Parallelogram [id:dp6742324505083032] 
\draw  [color={rgb, 255:red, 245; green, 166; blue, 35 }  ,draw opacity=1 ][fill={rgb, 255:red, 248; green, 198; blue, 155 }  ,fill opacity=1 ] (353.82,203.85) -- (291.83,235) -- (290.28,83.79) -- (352.27,52.64) -- cycle ;
%Shape: Parallelogram [id:dp4470417955949598] 
\draw   (265.95,203.85) -- (203.95,235) -- (202.4,83.79) -- (264.4,52.64) -- cycle ;
%Straight Lines [id:da057541279435106785] 
\draw [color={rgb, 255:red, 139; green, 6; blue, 24 }  ,draw opacity=1 ]   (234.17,143.82) -- (290.38,144.08) ;
%Straight Lines [id:da34835219849873] 
\draw  [dash pattern={on 4.5pt off 4.5pt}]  (290.38,144.08) -- (322.05,143.82) ;
%Straight Lines [id:da284360693972605] 
\draw [color={rgb, 255:red, 139; green, 6; blue, 24 }  ,draw opacity=1 ]   (322.05,143.82) -- (381.14,144.08) ;
%Straight Lines [id:da24766568848618786] 
\draw [color={rgb, 255:red, 139; green, 6; blue, 24 }  ,draw opacity=1 ] [dash pattern={on 4.5pt off 4.5pt}]  (381.14,144.08) -- (412.81,143.82) ;
%Shape: Ellipse [id:dp8557178762632961] 
\draw  [color={rgb, 255:red, 139; green, 6; blue, 24 }  ,draw opacity=1 ][fill={rgb, 255:red, 139; green, 6; blue, 24 }  ,fill opacity=1 ] (316.29,143.82) .. controls (316.29,142.25) and (317.58,140.98) .. (319.17,140.98) .. controls (320.76,140.98) and (322.05,142.25) .. (322.05,143.82) .. controls (322.05,145.39) and (320.76,146.66) .. (319.17,146.66) .. controls (317.58,146.66) and (316.29,145.39) .. (316.29,143.82) -- cycle ;
%Shape: Ellipse [id:dp23839756180614025] 
\draw  [color={rgb, 255:red, 139; green, 6; blue, 24 }  ,draw opacity=1 ][fill={rgb, 255:red, 139; green, 6; blue, 24 }  ,fill opacity=1 ] (228.41,143.82) .. controls (228.41,142.25) and (229.7,140.98) .. (231.29,140.98) .. controls (232.88,140.98) and (234.17,142.25) .. (234.17,143.82) .. controls (234.17,145.39) and (232.88,146.66) .. (231.29,146.66) .. controls (229.7,146.66) and (228.41,145.39) .. (228.41,143.82) -- cycle ;
%Shape: Ellipse [id:dp6809297828247807] 
\draw  [fill={rgb, 255:red, 139; green, 6; blue, 24 }  ,fill opacity=1 ] (412.81,143.82) .. controls (412.81,142.25) and (414.1,140.98) .. (415.69,140.98) .. controls (417.28,140.98) and (418.57,142.25) .. (418.57,143.82) .. controls (418.57,145.39) and (417.28,146.66) .. (415.69,146.66) .. controls (414.1,146.66) and (412.81,145.39) .. (412.81,143.82) -- cycle ;
%Straight Lines [id:da7500839289261838] 
\draw    (227.69,264.5) -- (408.58,263.51) ;
\draw [shift={(410.58,263.5)}, rotate = 179.69] [color={rgb, 255:red, 0; green, 0; blue, 0 }  ][line width=0.75]    (10.93,-3.29) .. controls (6.95,-1.4) and (3.31,-0.3) .. (0,0) .. controls (3.31,0.3) and (6.95,1.4) .. (10.93,3.29)   ;
\draw   (235.42,267.68) .. controls (232.16,266.01) and (228.88,265.02) .. (225.58,264.7) .. controls (228.9,264.36) and (232.22,263.33) .. (235.58,261.65) ;

% Text Node
\draw (228,120.4) node [anchor=north west][inner sep=0.75pt]    {$a$};
% Text Node
\draw (410,119.4) node [anchor=north west][inner sep=0.75pt]    {$b$};
% Text Node
\draw (307,152.4) node [anchor=north west][inner sep=0.75pt]    {$O_{( a,b)}$};
% Text Node
\draw (300,35.4) node [anchor=north west][inner sep=0.75pt]    {$S( G,\sigma )$};
% Text Node
\draw (205.4,84.19) node [anchor=north west][inner sep=0.75pt]    {$\Sigma _{1}$};
% Text Node
\draw (384.04,84.19) node [anchor=north west][inner sep=0.75pt]    {$\Sigma _{2}$};
% Text Node
\draw (313,269.4) node [anchor=north west][inner sep=0.75pt]    {$I$};

\end{tikzpicture}
\caption{The pairing of 2d CFT left and right movers on $\Sigma_1$ and $\Sigma_2$ respectively is specified by an abelian TQFT on $X\simeq\Sigma_1 \times I$ with a surface operator, $S(G,\sigma)$, inserted between \cite{Kapustin:2010if,Komargodski:2020mxz}.  Different $S(G,\sigma)$ lead to different modular invariants.}
\label{fig:3d2d}
\end{figure}
Consider the 3d-2d picture where the bulk theory is the TQFT described by the modular tensor category $\CC$, and the CFT left and right movers are on $\Sigma_1$ and $\Sigma_2$ respectively (see Fig. \ref{fig:3d2d}). The partition function of the CFT is described by how Wilson lines end on the surface operator $S(G,\sigma)$ of the bulk TQFT. More precisely, if the Wilson lines $p$ and $g \cdot p$ form a junction on the surface operator, then the CFT contains the primary operator $O_{(p,g \cdot p)}$. Conversely, if the CFT contains a primary operator $O_{(p,g \cdot p)}$, then the Wilson lines with labels $p$ and $g \cdot p$ can form a junction on the surface operator $S(G,\sigma)$. This logic shows that the action of $S(G,\sigma)$ on the Wilson line, $p$, contains the outcome $g \cdot p$ (possibly among others). Therefore, using the general expression in equation \eqref{eq:partfunc} for the partition function, we get the action of the surface operator on Wilson lines in equation \eqref{eq:Saction}.

Note that, in an Abelian TQFT, all non-trivial surface operators act non-trivially on the line operators. Indeed, suppose we have some condensation defect, $S(G,\sigma)$, which acts trivially on all line operators. Then, using \eqref{eq:pconst}, we get
\be
 \frac{\theta_{h \cdot p}}{\theta_h \theta_p} \Xi_{[G,\sigma]}(g,h) =1~,\ \forall\ h \in G~,\  \forall p \in \CC~,
\ee
if and only if $g$ is the trivial line operator. This logic implies
\be
\frac{\theta_{h \cdot p}}{\theta_h \theta_p}  = |A| \CS_{h,p}=1~,\  \forall h \in G~,\ \forall p \in \CC~.
\ee
To be consistent with modularity, $h$ must be the trivial line operator. Therefore, if $S(G,\sigma)$ acts trivially on all line operators, then $G$ must be the trivial subgroup. Note that $S(\DZ_1,1)$ is the trivial surface operator, and this is the only such surface operator which acts trivially on all line operators. 

Therefore, non-trivial surface operators of an abelian TQFT are completely characterised by their action on the line operators. As we discussed in section \ref{sec:noninv sym of non-abelian TQFTs}, this is in contrast with non-abelian TQFTs where surface operators can act non-trivially on fusion spaces but trivially on the lines \cite{Davydov_2014}.

\subsubsection{Reconstructing the higher-gauged symmetry and SPT}
\label{sec:Sactiontocond}
Given the construction of condensation defects, $S(G,\sigma)$, in the previous subsection, we would like to understand if we can reconstruct $G\le A$ and $\sigma\in H^2(G,U(1))$ from the symmetry action. Recall from \eqref{eq:Saction} that the action of this surface operator on the lines is given by 
\be
S(G,\sigma) \cdot p = \sum_{g \in G} \Bigg [\frac{1}{|G|}  \sum_{h\in G}  \frac{\theta_{h \cdot p}}{\theta_h \theta_p} \Xi_{[G,\sigma]}(g,h) \Bigg ] g \cdot p~.
\ee
Therefore, the elements of $G$ as a set are given by 
\be
\label{eq:GfromSact}
G=\{(S(G,\sigma) \cdot p) \cdot \bar p~|~ p \in \CC\}~.
\ee
Let us verify that fusions of the form $(S(G,\sigma) \cdot p) \cdot \bar p$ give us all line operators, $g$, that are condensed. In other words, we have to show that for every $g$ in the group being condensed, there is at least one line operator, say $p$, such that $S(G,\sigma) \cdot p=g\cdot p$. We know that the line operators $p$ and $g \cdot p$ are related by a symmetry action if and only if the condition \eqref{eq:pconst} is satisfied.  Let us show that, for a given $g$, there is at least one line operator, $p$, satisfying this condition. 

From the modularity of the S-matrix, we know that all 1-dimensional representations of the fusion rules of line operators are given by 
\be
\lambda_{x}(a)=\frac{\CS_{ax}}{\CS_{1x}}~.
\ee
where $x,a\in A$. For a given line operator $x$, the function $\lambda_{x}(a)$, when restricted to the abelian subgroup of line operators $G$, must be a character of $G$. Moreover,
\be
\lambda_x(a)=\lambda_{x'}(a)~,
\ee
when $x'=x \cdot e$, and $e \in Z_{A}(G)$. Here, $Z_{A}(G)$ is the set of line operators in $A$ which braid trivially with all line operators in $G$. Therefore, the number of distinct characters is
\be
\frac{|A|}{|Z_A(G)|}=|G|~,
\ee
where we have used the result that $|Z_A(G)|=\frac{|A|}{|G|}$ \cite{muger2003structure}. Therefore, restricting $\lambda_x(a)$ to the subgroup $G$ gives us all characters of the group $G$. Now, the constraint \eqref{eq:pconst} can be written in terms of the S-matrix as
\be
 \lambda_p(h)=\frac{\CS_{hp}}{\CS_{1p}} =(\Xi_{[G,\sigma]}(g,h))^{-1}, ~ \forall ~ h \in G~.
\ee
$\Xi_{[G,\sigma]}^{-1}$ is a bicharacter (and in particular a character) of the group $G$. Since the functions $\lambda_p(h)$ for $p \in A$ exhaust all characters of $G$, there must be some $p$ such that the above equation is satisfied. This finishes the proof that for a given $g$ there is always some $p$ such that $S(G,\sigma) \cdot p=g\cdot p$. 

Also, if $g\cdot p \in S\cdot p$, then, using \eqref{eq:pconst}, we get
\be
\label{eq:discrete tor from sym action}
\frac{\sigma(g,h)}{\sigma(h,g)}=\frac{\theta_h\theta_p}{\theta_{h\cdot p}} \frac{\tau_G(h,g)}{\tau_G(g,h)} \frac{1}{R(g,h)}~.
\ee
As a result, the choice of discrete torsion is also fixed by the symmetry action.\footnote{In a non-Abelian TQFT with a symmetry implemented by a surface operator, $S(G,\sigma)$, obtained from higher-gauging invertible line operators, $G$ can be fixed by choosing some invertible line operator $g \in \CC$ such that
\be
S(G,\sigma)\cdot p\ni g \cdot p~,
\ee
for each $p \in \CC$. The discrete torsion, $\sigma$, can then be fixed using \eqref{eq:discrete tor from sym action}. Note that there may not be a unique set of $G$ and $\sigma$ satisfying these constraints. Indeed, distinct algebra objects in non-abelian TQFTs can be Morita equivalent. For example, consider the surface operator, $S$, that acts trivially on the Ising model. We have
\be
S\cdot 1=1~, ~ S\cdot \psi=\psi~,~ S\cdot \sigma=\sigma~.
\ee
We can choose the group $G$ to be either the trivial group or a $\DZ_2$ group generated by $\psi$. Indeed the algebra objects
\be
1 \text{ and } 1+\psi~,
\ee
are Morita equivalent in the Ising category. This discussion illustrates that the embedding of the $G$-lines in the larger TQFT is important for understanding the action of the corresponding surface.}

Next, we consider a set of examples to illustrate the above points. The first example has non-trivial invertible and non-invertible surfaces, while the second and third examples only have non-trivial non-invertible and invertible surfaces respectively. Finally, in Appendix \ref{AppB}, we consider two infinite classes of theories and determine all their surface operators. In all the examples, we find that non-invertible surface operators are always of the form $S_{\rm inv}\times S_B$, where $S_B$ is a universal surface operator described in \eqref{eq:SBsym}, and $S_{\rm inv}$ is an invertible surface operator. It will be interesting to understand if this statement is true for non-invertible surface operators of an arbitrary Abelian TQFT. 

\subsubsection{Example 1: Toric code $\cong$ Spin$(16)_1$ Chern-Simons theory $\cong$ $\mathbb{Z}_2$ discrete gauge theory}
\label{subsec:toriccode}
Consider the Toric code (also known as Spin$(16)_1$ Chern-Simons Theory or $\mathbb{Z}_2$ discrete gauge theory) with lines operators $1$, $e$, $m$, and $\psi$ corresponding to topological spins
\be
\theta_1=1, ~ \theta_e=1,~ \theta_m=1,~ \theta_\psi=1~.
\ee
Let us use the characterisation of symmetries in terms of their action on line operators to determine the symmetries of this TQFT. Let us consider the action of the symmetry on the trivial line operator. $S \cdot 1$ cannot be a sum of more than two outcomes. This follows from applying \eqref{eq:noninvsymcond} 
\be
Nn=4~,
\ee
where $N$ is the number of outcomes for lines that are not annihilated by $S$, and $n$ is the number of lines that are not annihilated by $S$. Recall that $n\geq 2$ from modularity. Therefore, $N$ has to be $2$ or $1$. Let us consider these cases in turn:
\begin{outline}
\1 $N=1$. In this case, the trivial line operator is fixed by the symmetry action. Therefore, the symmetry is invertible. Invertible symmetries of the Toric code are the trivial symmetry and the electric-magnetic duality (which exchanges $e$ and $m$ while fixing $1$ and $\psi$).
\1 $N=2$. In this case, we have 
\be
S\cdot 1= 1+\ell~.
\ee
Preservation of topological spins implies that $\ell$ is $e$ or $m$. Also, if $S\cdot \psi$ is non-trivial, then it should be a sum of two fermionic lines, but $\psi$ is the only fermion. Therefore, $S\cdot \psi=0$. 

Now, let us consider the case with $\ell=e$ in more detail (the case of $\ell=m$ will be related to it by fusion with the electric-magnetic duality surface). If the action of $S$ on $e$ is non-trivial, it should be a sum of two line operators. Note that $m\not\in S\cdot e$ since then $\psi \in S\cdot e$, which is inconsistent with preserving spins. Therefore, we have two options. $S\cdot e=1+e$ or $S\cdot e=0$.
     \2[${\bf(a)}$] $S\cdot e=1+e$. To satisfy \eqref{eq:noninvsymcond} we must have $S \cdot m=0$. This is the universal $S_B$ symmetry for $B=\{1,e\}$ described in \eqref{eq:SBsym}.
     \2[${\bf(b)}$] $S\cdot e=0$. In this case, to satisfy \eqref{eq:noninvsymcond}, $S \cdot m$ must be non-trivial. The only consistent choice is $S\cdot m=1+e$.
\end{outline}
Including the surfaces related to the two non-invertible symmetries described above via electric-magnetic duality (i.e., via $e\leftrightarrow m$), we get $4$ non-invertible symmetries and $2$ invertible symmetries (including the trivial one). This result agrees with the known classification of symmetries in terms of condensation defects \cite{Roumpedakis:2022aik}.

Next, let us use the discussion in section \ref{sec:Sactiontocond} to find the lines that need to be condensed to get surface operators implementing the symmetries above. For the electric-magnetic duality symmetry discussed above, using \eqref{eq:GfromSact} we find that the lines to be condensed are
\be
1~,\ \psi~,
\ee
without any discrete torsion, as $H^2(\DZ_2,U(1))\cong \DZ_1$.

Now, let us consider the symmetry
\be
S \cdot 1= S\cdot e= 1 + e~, ~ S \cdot m = S \cdot \psi=0~.
\ee
Using \eqref{eq:GfromSact}, we find that the lines to be condensed are
\be
1~,\ e~.
\ee
Similarly, for the symmetry 
\be
S \cdot 1= S\cdot m= 1 + m~, ~ S \cdot e = S \cdot \psi=0~,
\ee
the lines to be condensed are
\be
1~,\ m~.
\ee

Now, let us consider the symmetry
\be
S \cdot 1= 1 + e~, ~ S \cdot m = 1+e~, ~ S \cdot e= S \cdot \psi=0~.
\ee
Using \eqref{eq:GfromSact}, we find that the lines to be condensed are
\be
1~,\ e~,\ m~,\ \psi~.
\ee
Since the 1-form symmetry to be condensed is $\DZ_2 \times \DZ_2$, we have a choice of discrete torsion: $H^2(\DZ_2\times \DZ_2,U(1))\cong \DZ_2$. To understand which discrete torsion corresponds to the above surface, we can choose a basis where the $F$ matrices are all trivial and the R-matrices given by \cite{rowell2009classification}
\bea
&& R_{e,e}^{1}=R_{m,m}^{1}=R_{\psi,m}^{e}=R_{e,m}^{\psi}=R_{e,\psi}^{m}=1~,\cr
&& R_{\psi,\psi}^{1}=R_{m,e}^{\psi}=R_{m,\psi}^{e}=R_{\psi,e}^{m}=-1~.
\eea
For the Toric code, \eqref{eq:discrete tor from sym action} simplifies to
\be
\frac{\sigma(g,h)}{\sigma(h,g)}= \frac{1}{R(h,p) R(p,h) R(h,g)}~,
\ee
and we get
\be
\frac{\sigma(e,\psi)}{\sigma(\psi,e)}=-1~, ~ \frac{\sigma(e,m)}{\sigma(m,e)}=-1~, ~ \frac{\sigma(m,\psi)}{\sigma(\psi,m)}=-1~.
\ee
Clearly, we have to choose the non-trivial discrete torsion in $H^2(\DZ_2\times \DZ_2,U(1))$. 

Finally, let us consider the symmetry
\be
S \cdot 1= 1 + m~, ~ S \cdot e = 1+m~, ~ S \cdot m= S \cdot \psi=0~.
\ee
Using \eqref{eq:GfromSact} we find that the lines to be condensed are
\be
1~,\ e~,\ m~,\ \psi~.
\ee
Using
\be
\frac{\sigma(g,h)}{\sigma(h,g)}= \frac{1}{R(h,p) R(p,h) R(h,g)}~,
\ee
we get
\be
\frac{\sigma(g,h)}{\sigma(h,g)}=1 ~ \forall g,h \in \{1,e,m,\psi\}~.
\ee
Therefore, this symmetry is obtained from higher-gauging the $\DZ_2 \times \DZ_2$ 1-form symmetry without discrete torsion. 

The set of condensable lines found from symmetry actions in this section agrees with the discussion in \cite[Section 6.3]{Roumpedakis:2022aik} up to a relabelling of the line operators, $e \leftrightarrow m$. 

\subsubsection{Example 2: Double semion Model $\cong$ twisted $\mathbb{Z}_2$ discrete gauge theory}

\label{subsec:doublesemionun}

Consider the double semion model (also known as the twisted $\mathbb{Z}_2$ discrete gauge theory) with line operators $1$, $s$, $\bar s$, and $d$ and topological spins
\be
\theta_1=1~, ~ \theta_s=i~,~ \theta_{\bar s}=-i~,~ \theta_{d}=1~.
\ee
To determine the symmetries, let us consider the action of the symmetry on the trivial line operator. As in the previous example, $S \cdot 1$ cannot be a sum of more than two outcomes, since \eqref{eq:noninvsymcond} implies
\be
Nn=4~.
\ee
We again know that $n\ge2$. Therefore, $N=1,2$. 
\begin{outline}
\1 $N=1$. In this case, the trivial line operator is fixed by the symmetry action. Therefore, the symmetry is invertible. The Double semion model does not have any non-trivial unitary invertible symmetries.
\1 $N=2$. In this case we have
\be
S\cdot 1=1+d~,~ S\cdot d=1+d~,~ S\cdot s=S \cdot \bar s=0~.
\ee
This is the universal $S_B$ symmetry, for $B=\{1,d\}$, described in \eqref{eq:SBsym}. From \eqref{eq:GfromSact}, $B$ is precisely the 1-form symmetry group that needs to be higher-gauged to get the surface, $S_B$. 
\end{outline}

\subsubsection{Example 3: 3-Fermion Model $\cong$ Spin$(8)_1$ Chern-Simons Theory}
As a final example, consider the 3-Fermion model with lines operators $1$, $\psi_1$, $\psi_2$, $\psi_3$ and topological twists
\be
\theta_1=1~, ~ \theta_{\psi_1}=\theta_{\psi_2}=\theta_{\psi_3}=-1~.
\ee
All line operators in this TQFT are higher-gaugeable. A priori one might imagine that there are non-invertible symmetries. However, from theorem 1, we know that this model cannot have any non-invertible symmetries since it lacks a non-anomalous 1-form symmetry. Therefore, the 3-fermion models only has invertible symmetries. Any permutation of the three fermionic lines is a symmetry of this theory. Therefore, the invertible symmetry group is $S_3$. 

Consider the order-two symmetry which acts as
\be
S\cdot \psi_1=\psi_2~,~ S\cdot \psi_2=\psi_1~,
\ee
and leaves the other lines invariant. Using \eqref{eq:GfromSact} we find that the surface operator which implements this symmetry is obtained from higher-gauging a $\DZ_2$ 1-form symmetry generated by the line operator $\psi_3$. In general, condensing the line operator $\psi_i$ gives the order two symmetry which fixes $\psi_i$ and permutes the other two fermionic lines. Finally, the surfaces implementing the two order three symmetries are obtained from higher-gauging the full $\DZ_2 \times \DZ_2$ 1-form symmetry, with and without discrete torsion. 

\subsection{Non-Abelian TQFTs}
\label{sec:TQFTs with bosonic lines}
In this section, we introduce basic aspects of higher gauging in the context of non-Abelian TQFTs. Our goal is to give a partial converse of theorem 3 and, in so doing, find a non-Abelian generalization of theorem 1.

To that end, recall from the discussion in section \ref{sec:noninv sym of non-abelian TQFTs} that a necessary condition to have a non-invertible symmetry is that the TQFT in question, $\CT$, must have condensible bosons living in commutative Frobenius algebras $\CA_{L}$ or $\CA_R$ \cite{Fuchs:2002cm}. These algebras consist of bosons satisfying $n_{S\cdot a}^1\ne0$ and $n_{S\cdot 1}^a\ne0$ respectively.

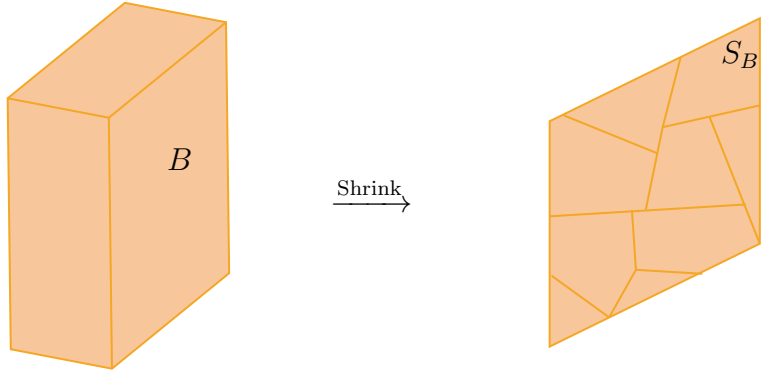
\begin{figure}[h!]
    \centering

\tikzset{every picture/.style={line width=0.75pt}} %set default line width to 0.75pt        

\begin{tikzpicture}[x=0.75pt,y=0.75pt,yscale=-1,xscale=1]

%Shape: Parallelogram [id:dp9547745150430316] 
\draw  [color={rgb, 255:red, 245; green, 166; blue, 35 }  ,draw opacity=1 ][fill={rgb, 255:red, 248; green, 198; blue, 155 }  ,fill opacity=1 ] (428.98,234.76) -- (429,121) -- (535,69.06) -- (534.98,182.82) -- cycle ;
%Straight Lines [id:da8607671975291249] 
\draw [color={rgb, 255:red, 245; green, 166; blue, 35 }  ,draw opacity=1 ]   (495,89) -- (486,124) ;
%Straight Lines [id:da38022251469841983] 
\draw [color={rgb, 255:red, 245; green, 166; blue, 35 }  ,draw opacity=1 ]   (535,113) -- (486,124) ;
%Straight Lines [id:da37118389736541924] 
\draw [color={rgb, 255:red, 245; green, 166; blue, 35 }  ,draw opacity=1 ]   (486,124) -- (477.5,165.5) ;
%Straight Lines [id:da8152343077303414] 
\draw [color={rgb, 255:red, 245; green, 166; blue, 35 }  ,draw opacity=1 ]   (436,118) -- (483.17,137.13) ;
%Straight Lines [id:da8230675583350993] 
\draw [color={rgb, 255:red, 245; green, 166; blue, 35 }  ,draw opacity=1 ]   (509.5,118.5) -- (534.98,182.82) ;
%Straight Lines [id:da6212696335117135] 
\draw [color={rgb, 255:red, 245; green, 166; blue, 35 }  ,draw opacity=1 ]   (472.5,196) -- (506,198) ;
%Straight Lines [id:da5128805953271293] 
\draw [color={rgb, 255:red, 245; green, 166; blue, 35 }  ,draw opacity=1 ]   (429,169) -- (528,163) ;
%Straight Lines [id:da793312586270952] 
\draw [color={rgb, 255:red, 245; green, 166; blue, 35 }  ,draw opacity=1 ]   (470.5,166) -- (472.5,196) ;
%Straight Lines [id:da9600969007505765] 
\draw [color={rgb, 255:red, 245; green, 166; blue, 35 }  ,draw opacity=1 ]   (472.5,196) -- (459,220) ;
%Straight Lines [id:da7430904857930009] 
\draw [color={rgb, 255:red, 245; green, 166; blue, 35 }  ,draw opacity=1 ]   (430,199) -- (459,220) ;
%Curve Lines [id:da12899901682248704] 
\draw [color={rgb, 255:red, 245; green, 166; blue, 35 }  ,draw opacity=1 ]   (165.95,142.99) .. controls (190.55,132.32) and (186.64,221.27) .. (204.52,232.84) ;
%Curve Lines [id:da5412711980867769] 
\draw [color={rgb, 255:red, 245; green, 166; blue, 35 }  ,draw opacity=1 ]   (160.6,192.81) .. controls (171.52,195.48) and (181.01,181.35) .. (184.41,176.02) ;
%Curve Lines [id:da5943711452742537] 
\draw [color={rgb, 255:red, 245; green, 166; blue, 35 }  ,draw opacity=1 ]   (188.53,170.79) .. controls (193.07,164.56) and (179.72,142.1) .. (184.18,125.2) ;
%Curve Lines [id:da9471882893526227] 
\draw [color={rgb, 255:red, 245; green, 166; blue, 35 }  ,draw opacity=1 ]   (184.18,125.2) .. controls (195.42,94.07) and (234.59,188.36) .. (257.34,161.67) ;
%Curve Lines [id:da49369154526007974] 
\draw [color={rgb, 255:red, 245; green, 166; blue, 35 }  ,draw opacity=1 ]   (226.91,159) .. controls (218.08,206.15) and (249.46,181.24) .. (248.45,199.03) ;
%Curve Lines [id:da8962385525270895] 
\draw [color={rgb, 255:red, 245; green, 166; blue, 35 }  ,draw opacity=1 ]   (190,190) .. controls (223.52,146.41) and (221.32,178.57) .. (227.92,216.82) ;
%Curve Lines [id:da5695376560266645] 
\draw [color={rgb, 255:red, 245; green, 166; blue, 35 }  ,draw opacity=1 ]   (201.62,79.83) .. controls (203.18,133.21) and (244.35,113.64) .. (229.7,149.22) ;
%Shape: Cube [id:dp8391455624003534] 
\draw  [color={rgb, 255:red, 245; green, 166; blue, 35 }  ,draw opacity=1 ][fill={rgb, 255:red, 248; green, 198; blue, 155 }  ,fill opacity=0.58 ] (208.24,245.9) -- (157.21,236.09) -- (155.64,109.42) -- (214.75,61.23) -- (265.78,71.04) -- (267.35,197.71) -- cycle ; \draw  [color={rgb, 255:red, 245; green, 166; blue, 35 }  ,draw opacity=1 ] (155.64,109.42) -- (206.67,119.23) -- (208.24,245.9) ; \draw  [color={rgb, 255:red, 245; green, 166; blue, 35 }  ,draw opacity=1 ] (206.67,119.23) -- (265.78,71.04) ;

% Text Node
\draw (514,79.4) node [anchor=north west][inner sep=0.75pt]    {$S_{B}{}$};
% Text Node
\draw (317,147.4) node [anchor=north west][inner sep=0.75pt]    {$\xrightarrow{\text{Shrink}}$};
% Text Node
\draw (234.6,133.21) node [anchor=north west][inner sep=0.75pt]    {$B$};

\end{tikzpicture}
    \caption{We can gauge the commutative algebra object, $B$, in a 2+1d slab and shrink the slab to get the surface operator, $S_B$.}
    \label{fig:surface from slab}
\end{figure}

Since the bosons are condensible, we can gauge them in a three-dimensional slab of spacetime. Then, as shown in Fig. \ref{fig:surface from slab}, we can compress this slab to form a non-Abelian generalization of the universal surfaces described in \eqref{eq:SBsym}. All that remains to show is that such surfaces are always non-invertible.

More explicitly, suppose a TQFT contains a condensable algebra, $B$, and that we compress a slab of $B$ condensate to form a surface operator, $S_B$. Let $\CC/B$ be the MTC obtained from condensing $B$. Under the condensation, a line operator, $a \in \CC$, becomes a superposition of line operators in $\CC/B$, and we have \cite{Neupert_2016}
\be
a \to \sum_{t \in \CC/B} n_{a}^t ~ t~,
\ee
where $n_{a}^t$ are non-negative integers. Using this condensation, we can define $S_B$, whose action on the line operators in $\CC$ is given by
\be
n_{S_B a}^{b}= \sum_{i\in \CC/B} n_{a}^i n_{b}^i~.
\ee
Note that, as promised, this is a generalization of the universal surface operator, $S_B$, defined in \eqref{eq:SBsym} for abelian TQFTs.

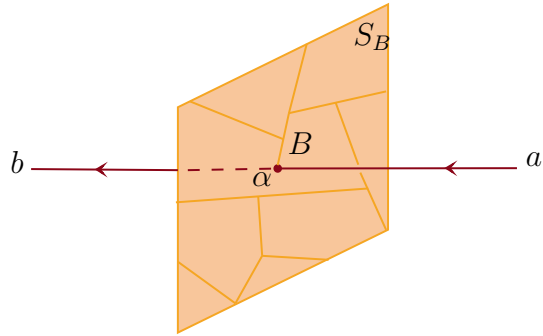
\begin{figure}[h!]
    \centering

\tikzset{every picture/.style={line width=0.75pt}} %set default line width to 0.75pt        

\begin{tikzpicture}[x=0.75pt,y=0.75pt,yscale=-1,xscale=1]

%Shape: Parallelogram [id:dp9547745150430316] 
\draw  [color={rgb, 255:red, 245; green, 166; blue, 35 }  ,draw opacity=1 ][fill={rgb, 255:red, 248; green, 198; blue, 155 }  ,fill opacity=1 ] (264.98,219.76) -- (265,106) -- (371,54.06) -- (370.98,167.82) -- cycle ;
%Straight Lines [id:da9423078429202264] 
\draw [color={rgb, 255:red, 139; green, 6; blue, 24 }  ,draw opacity=1 ]   (316.99,136.91) -- (436,136.76) ;
%Straight Lines [id:da8558727069000894] 
\draw [color={rgb, 255:red, 139; green, 6; blue, 24 }  ,draw opacity=1 ] [dash pattern={on 4.5pt off 4.5pt}]  (270,137.76) -- (316.99,136.91) ;
%Straight Lines [id:da8364306225426728] 
\draw [color={rgb, 255:red, 139; green, 6; blue, 24 }  ,draw opacity=1 ]   (190.97,137.21) -- (265,137.76) ;
\draw  [color={rgb, 255:red, 139; green, 6; blue, 24 }  ,draw opacity=1 ][fill={rgb, 255:red, 139; green, 6; blue, 24 }  ,fill opacity=1 ] (227.33,139.5) -- (223.98,137.39) -- (227.4,135.38) -- (225.67,137.42) -- cycle ;
%Shape: Circle [id:dp5656852943928684] 
\draw  [color={rgb, 255:red, 139; green, 6; blue, 24 }  ,draw opacity=1 ][fill={rgb, 255:red, 139; green, 6; blue, 24 }  ,fill opacity=1 ] (313.69,136.91) .. controls (313.69,136) and (314.43,135.26) .. (315.34,135.26) .. controls (316.25,135.26) and (316.99,136) .. (316.99,136.91) .. controls (316.99,137.82) and (316.25,138.56) .. (315.34,138.56) .. controls (314.43,138.56) and (313.69,137.82) .. (313.69,136.91) -- cycle ;
\draw  [color={rgb, 255:red, 139; green, 6; blue, 24 }  ,draw opacity=1 ][fill={rgb, 255:red, 139; green, 6; blue, 24 }  ,fill opacity=1 ] (403.67,138.83) -- (400.31,136.72) -- (403.73,134.72) -- (402,136.75) -- cycle ;
%Straight Lines [id:da8607671975291249] 
\draw [color={rgb, 255:red, 245; green, 166; blue, 35 }  ,draw opacity=1 ]   (330,74) -- (321,109) ;
%Straight Lines [id:da38022251469841983] 
\draw [color={rgb, 255:red, 245; green, 166; blue, 35 }  ,draw opacity=1 ]   (370,98) -- (321,109) ;
%Straight Lines [id:da37118389736541924] 
\draw [color={rgb, 255:red, 245; green, 166; blue, 35 }  ,draw opacity=1 ]   (321,109) -- (315.34,135.26) ;
%Straight Lines [id:da8152343077303414] 
\draw [color={rgb, 255:red, 245; green, 166; blue, 35 }  ,draw opacity=1 ]   (271,103) -- (318.17,122.13) ;
%Straight Lines [id:da8230675583350993] 
\draw [color={rgb, 255:red, 245; green, 166; blue, 35 }  ,draw opacity=1 ]   (344.5,103.5) -- (356,135) ;
%Straight Lines [id:da6932942582906912] 
\draw [color={rgb, 255:red, 245; green, 166; blue, 35 }  ,draw opacity=1 ]   (357,139) -- (369.98,167.82) ;
%Straight Lines [id:da6212696335117135] 
\draw [color={rgb, 255:red, 245; green, 166; blue, 35 }  ,draw opacity=1 ]   (307.5,181) -- (341,183) ;
%Straight Lines [id:da5128805953271293] 
\draw [color={rgb, 255:red, 245; green, 166; blue, 35 }  ,draw opacity=1 ]   (264,154) -- (361,147) ;
%Straight Lines [id:da793312586270952] 
\draw [color={rgb, 255:red, 245; green, 166; blue, 35 }  ,draw opacity=1 ]   (305.5,151) -- (307.5,181) ;
%Straight Lines [id:da9600969007505765] 
\draw [color={rgb, 255:red, 245; green, 166; blue, 35 }  ,draw opacity=1 ]   (307.5,181) -- (294,205) ;
%Straight Lines [id:da7430904857930009] 
\draw [color={rgb, 255:red, 245; green, 166; blue, 35 }  ,draw opacity=1 ]   (265,184) -- (294,205) ;

% Text Node
\draw (352,62.4) node [anchor=north west][inner sep=0.75pt]    {$S_{B}{}$};
% Text Node
\draw (439,127.4) node [anchor=north west][inner sep=0.75pt]    {$a$};
% Text Node
\draw (179,126.4) node [anchor=north west][inner sep=0.75pt]    {$b$};
% Text Node
\draw (319,117.4) node [anchor=north west][inner sep=0.75pt]    {$B$};
% Text Node
\draw (301,137.4) node [anchor=north west][inner sep=0.75pt]    {$\alpha $};

\end{tikzpicture}
    \caption{Action of $S_B$ on a line.}
    \label{fig:SBact}
\end{figure}

\begin{figure}[h!]
    \centering

\tikzset{every picture/.style={line width=0.75pt}} %set default line width to 0.75pt        

\begin{tikzpicture}[x=0.75pt,y=0.75pt,yscale=-1,xscale=1]

%Straight Lines [id:da003358239188612888] 
\draw [color={rgb, 255:red, 139; green, 6; blue, 24 }  ,draw opacity=1 ]   (257,172.79) -- (257,206.17) ;
%Straight Lines [id:da8525147337015229] 
\draw [color={rgb, 255:red, 139; green, 6; blue, 24 }  ,draw opacity=1 ]   (257,106.67) -- (257,168.3) ;
%Straight Lines [id:da6317826504360281] 
\draw [color={rgb, 255:red, 245; green, 166; blue, 35 }  ,draw opacity=1 ][fill={rgb, 255:red, 245; green, 166; blue, 35 }  ,fill opacity=1 ]   (206,107.31) -- (206,140.05) ;
%Curve Lines [id:da6954939048329682] 
\draw [color={rgb, 255:red, 139; green, 6; blue, 24 }  ,draw opacity=1 ]   (231,82.91) .. controls (240,87.41) and (256,91) .. (257,106.67) ;
%Curve Lines [id:da5795765318423646] 
\draw [color={rgb, 255:red, 245; green, 166; blue, 35 }  ,draw opacity=1 ]   (206,107.31) .. controls (207,95.11) and (216,90.62) .. (231,82.91) ;
%Straight Lines [id:da8997437953927581] 
\draw [color={rgb, 255:red, 139; green, 6; blue, 24 }  ,draw opacity=1 ]   (231,60.44) -- (231,82.91) ;
%Straight Lines [id:da737899802115924] 
\draw [color={rgb, 255:red, 245; green, 166; blue, 35 }  ,draw opacity=1 ]   (208,190.77) -- (208,206.17) ;
%Curve Lines [id:da4933363227730967] 
\draw [color={rgb, 255:red, 245; green, 166; blue, 35 }  ,draw opacity=1 ]   (261,123.36) .. controls (324,173.43) and (230,167.65) .. (208,190.77) ;
%Curve Lines [id:da5812044852352973] 
\draw [color={rgb, 255:red, 245; green, 166; blue, 35 }  ,draw opacity=1 ]   (211,109.88) .. controls (226,110.52) and (237,111.8) .. (253,118.86) ;
%Curve Lines [id:da892290891716338] 
\draw [color={rgb, 255:red, 245; green, 166; blue, 35 }  ,draw opacity=1 ]   (206,140.05) .. controls (167,140.69) and (177,107.31) .. (202,109.88) ;
%Curve Lines [id:da09451927514943292] 
\draw [color={rgb, 255:red, 245; green, 166; blue, 35 }  ,draw opacity=1 ]   (208,190.77) .. controls (155,176) and (256,153.53) .. (206,140.05) ;
%Curve Lines [id:da6869240489720657] 
\draw [color={rgb, 255:red, 139; green, 6; blue, 24 }  ,draw opacity=1 ]   (439,82.27) .. controls (448,86.76) and (466,94.47) .. (465,106.02) ;
%Curve Lines [id:da8815422307894785] 
\draw [color={rgb, 255:red, 245; green, 166; blue, 35 }  ,draw opacity=1 ]   (414,106.67) .. controls (415,94.47) and (424,89.97) .. (439,82.27) ;
%Straight Lines [id:da9425219583796125] 
\draw [color={rgb, 255:red, 139; green, 6; blue, 24 }  ,draw opacity=1 ]   (439,59.8) -- (439,82.27) ;
%Straight Lines [id:da4289194783383793] 
\draw [color={rgb, 255:red, 139; green, 6; blue, 24 }  ,draw opacity=1 ]   (465,106.02) -- (465,206.82) ;
%Straight Lines [id:da5778111962386638] 
\draw [color={rgb, 255:red, 245; green, 166; blue, 35 }  ,draw opacity=1 ][fill={rgb, 255:red, 245; green, 166; blue, 35 }  ,fill opacity=1 ]   (414,106.67) -- (414,206.82) ;

% Text Node
\draw (252,205.71) node [anchor=north west][inner sep=0.75pt]    {$a$};
% Text Node
\draw (227,43.5) node [anchor=north west][inner sep=0.75pt]    {$b$};
% Text Node
\draw (201,207.35) node [anchor=north west][inner sep=0.75pt]    {$B$};
% Text Node
\draw (435,42.86) node [anchor=north west][inner sep=0.75pt]    {$b$};
% Text Node
\draw (408,207.35) node [anchor=north west][inner sep=0.75pt]    {$B$};
% Text Node
\draw (460,205.71) node [anchor=north west][inner sep=0.75pt]    {$a$};
% Text Node
\draw (330,138.87) node [anchor=north west][inner sep=0.75pt]    {$=$};
% Text Node
\draw (215,67.4) node [anchor=north west][inner sep=0.75pt]    {$\alpha $};
% Text Node
\draw (425,66.4) node [anchor=north west][inner sep=0.75pt]    {$\alpha $};

\end{tikzpicture}
    \caption{A consistency condition on $\alpha \in \text{Hom}(B\times a,b)$ from a change of triangulation.}
    \label{fig:localmorphismdef}
\end{figure}
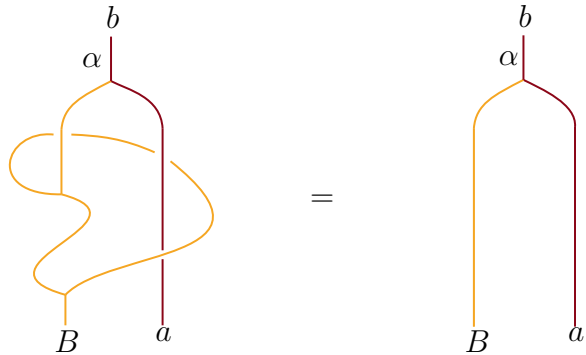

Let us show that the action of the surface $S_B$ satisfies
\be
S\cdot 1= \sum_{m\in B} m~,
\ee
where $1$ is the trivial line operator. To understand this claim, note that $S_B$ is made up of a network of the algebra object, $B$. A line operator, $b$, is in the output of the $S_B$-action on a line $a$ if $a$, $B$, and $b$ form a junction as in Fig. \ref{fig:SBact}. 
However, not all junctions, $\alpha \in$Hom$(B \times a,b)$, are consistent with changing the triangulation of $B$ lines on the surface. Indeed, only those junctions that satisfy the condition in Fig. \ref{fig:localmorphismdef} are allowed \cite{Fuchs:2002cm}.\footnote{We thank I.~Runkel for a discussion of this point.}

\begin{figure}
    \centering

\tikzset{every picture/.style={line width=0.75pt}} %set default line width to 0.75pt        

\begin{tikzpicture}[x=0.75pt,y=0.75pt,yscale=-1,xscale=1]

%Straight Lines [id:da003358239188612888] 
\draw [color={rgb, 255:red, 139; green, 6; blue, 24 }  ,draw opacity=1 ] [dash pattern={on 4.5pt off 4.5pt}]  (190,168.79) -- (190,202.17) ;
%Straight Lines [id:da8525147337015229] 
\draw [color={rgb, 255:red, 139; green, 6; blue, 24 }  ,draw opacity=1 ] [dash pattern={on 4.5pt off 4.5pt}]  (190,102.67) -- (190,164.3) ;
%Straight Lines [id:da6317826504360281] 
\draw [color={rgb, 255:red, 245; green, 166; blue, 35 }  ,draw opacity=1 ][fill={rgb, 255:red, 245; green, 166; blue, 35 }  ,fill opacity=1 ]   (139,103.31) -- (139,136.05) ;
%Curve Lines [id:da6954939048329682] 
\draw [color={rgb, 255:red, 139; green, 6; blue, 24 }  ,draw opacity=1 ] [dash pattern={on 4.5pt off 4.5pt}]  (164,78.91) .. controls (173,83.41) and (189,87) .. (190,102.67) ;
%Curve Lines [id:da5795765318423646] 
\draw [color={rgb, 255:red, 245; green, 166; blue, 35 }  ,draw opacity=1 ]   (139,103.31) .. controls (140,91.11) and (149,86.62) .. (164,78.91) ;
%Straight Lines [id:da8997437953927581] 
\draw [color={rgb, 255:red, 139; green, 6; blue, 24 }  ,draw opacity=1 ]   (164,56.44) -- (164,78.91) ;
%Straight Lines [id:da737899802115924] 
\draw [color={rgb, 255:red, 245; green, 166; blue, 35 }  ,draw opacity=1 ]   (141,186.77) -- (141,202.17) ;
%Curve Lines [id:da4933363227730967] 
\draw [color={rgb, 255:red, 245; green, 166; blue, 35 }  ,draw opacity=1 ]   (194,119.36) .. controls (257,169.43) and (163,163.65) .. (141,186.77) ;
%Curve Lines [id:da5812044852352973] 
\draw [color={rgb, 255:red, 245; green, 166; blue, 35 }  ,draw opacity=1 ]   (144,105.88) .. controls (159,106.52) and (170,107.8) .. (186,114.86) ;
%Curve Lines [id:da892290891716338] 
\draw [color={rgb, 255:red, 245; green, 166; blue, 35 }  ,draw opacity=1 ]   (139,136.05) .. controls (100,136.69) and (110,103.31) .. (135,105.88) ;
%Curve Lines [id:da09451927514943292] 
\draw [color={rgb, 255:red, 245; green, 166; blue, 35 }  ,draw opacity=1 ]   (141,186.77) .. controls (88,172) and (189,149.53) .. (139,136.05) ;
%Curve Lines [id:da5098896784171582] 
\draw [color={rgb, 255:red, 139; green, 6; blue, 24 }  ,draw opacity=1 ] [dash pattern={on 4.5pt off 4.5pt}]  (346,80.27) .. controls (355,84.76) and (373,92.47) .. (372,104.02) ;
%Curve Lines [id:da5898222887861254] 
\draw [color={rgb, 255:red, 245; green, 166; blue, 35 }  ,draw opacity=1 ]   (321,104.67) .. controls (322,92.47) and (331,87.97) .. (346,80.27) ;
%Straight Lines [id:da40600808778324515] 
\draw [color={rgb, 255:red, 139; green, 6; blue, 24 }  ,draw opacity=1 ]   (346,57.8) -- (346,80.27) ;
%Straight Lines [id:da28862116975372953] 
\draw [color={rgb, 255:red, 139; green, 6; blue, 24 }  ,draw opacity=1 ] [dash pattern={on 4.5pt off 4.5pt}]  (372,104.02) -- (372,204.82) ;
%Straight Lines [id:da6141343443679683] 
\draw [color={rgb, 255:red, 245; green, 166; blue, 35 }  ,draw opacity=1 ][fill={rgb, 255:red, 245; green, 166; blue, 35 }  ,fill opacity=1 ]   (321,104.67) -- (321,204.82) ;
%Curve Lines [id:da2953471132739163] 
\draw [color={rgb, 255:red, 245; green, 166; blue, 35 }  ,draw opacity=1 ]   (321,165) .. controls (287,166) and (289,121) .. (321,122) ;
%Curve Lines [id:da6476885394049092] 
\draw [color={rgb, 255:red, 139; green, 6; blue, 24 }  ,draw opacity=1 ] [dash pattern={on 4.5pt off 4.5pt}]  (476,81.27) .. controls (485,85.76) and (503,93.47) .. (502,105.02) ;
%Curve Lines [id:da6885994925145102] 
\draw [color={rgb, 255:red, 245; green, 166; blue, 35 }  ,draw opacity=1 ]   (451,105.67) .. controls (452,93.47) and (461,88.97) .. (476,81.27) ;
%Straight Lines [id:da11069477043164688] 
\draw [color={rgb, 255:red, 139; green, 6; blue, 24 }  ,draw opacity=1 ]   (476,58.8) -- (476,81.27) ;
%Straight Lines [id:da26455291610531007] 
\draw [color={rgb, 255:red, 139; green, 6; blue, 24 }  ,draw opacity=1 ] [dash pattern={on 4.5pt off 4.5pt}]  (502,105.02) -- (502,205.82) ;
%Straight Lines [id:da11207968005142832] 
\draw [color={rgb, 255:red, 245; green, 166; blue, 35 }  ,draw opacity=1 ][fill={rgb, 255:red, 245; green, 166; blue, 35 }  ,fill opacity=1 ]   (451,105.67) -- (451,205.82) ;

% Text Node
\draw (185,203.71) node [anchor=north west][inner sep=0.75pt]    {$1$};
% Text Node
\draw (160,39.5) node [anchor=north west][inner sep=0.75pt]    {$b$};
% Text Node
\draw (134,203.35) node [anchor=north west][inner sep=0.75pt]    {$B$};
% Text Node
\draw (243,134.87) node [anchor=north west][inner sep=0.75pt]    {$=$};
% Text Node
\draw (148,63.4) node [anchor=north west][inner sep=0.75pt]    {$\alpha $};
% Text Node
\draw (342,40.86) node [anchor=north west][inner sep=0.75pt]    {$b$};
% Text Node
\draw (315,205.35) node [anchor=north west][inner sep=0.75pt]    {$B$};
% Text Node
\draw (367,205.71) node [anchor=north west][inner sep=0.75pt]    {$1$};
% Text Node
\draw (332,64.4) node [anchor=north west][inner sep=0.75pt]    {$\alpha $};
% Text Node
\draw (402,134.87) node [anchor=north west][inner sep=0.75pt]    {$=$};
% Text Node
\draw (472,41.86) node [anchor=north west][inner sep=0.75pt]    {$b$};
% Text Node
\draw (445,206.35) node [anchor=north west][inner sep=0.75pt]    {$B$};
% Text Node
\draw (497,206.71) node [anchor=north west][inner sep=0.75pt]    {$1$};
% Text Node
\draw (462,65.4) node [anchor=north west][inner sep=0.75pt]    {$\alpha $};

\end{tikzpicture}
    \caption{Consistency of the $S_B$ action on the trivial line. In the first equality, we have used the fact that $B$ is a commutative algebra. Therefore, $B$ lines can be braided freely. In the second equality, we use the fact that bubbles in a network of $B$ lines can be removed, since $B$ is a Frobenius algebra.}
    \label{fig:localmorphismtrivial}
\end{figure}
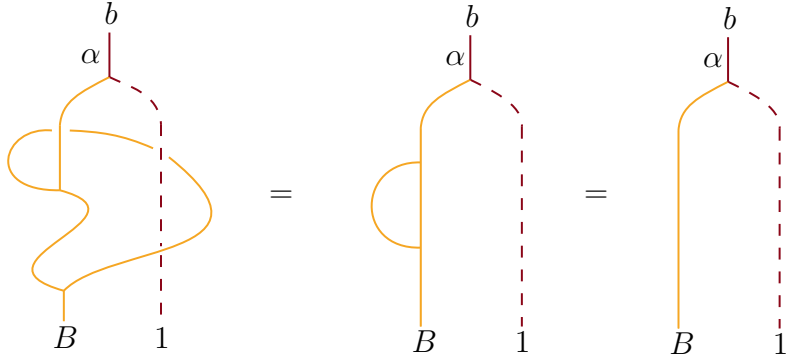

Consider the case when $a$ is the trivial line operator. Then, the L.H.S. of the equality in Fig. \ref{fig:localmorphismdef} can be simplified as in Fig. \ref{fig:localmorphismtrivial}.
Therefore, if $B$ is a commutative Frobenius algebra, any morphism, $\alpha \in \text{Hom}(B\times a,b)$, satisfies the constraint in Fig. \ref{fig:localmorphismdef}. As a result, any element $m \in B$ is a consistent outcome for the action of $S_B$ on the trivial line operator. Hence, for any condensable algebra, $B$, the surface, $S_B$, obtained from higher-gauging $B$ implements a non-invertible symmetry. In summary, we have the following result:

\bigskip
\noindent
{\bf Theorem 4:} {\it A $2+1$d TQFT has a non-invertible symmetry if and only if it contains condensable bosonic line operators.}

\bigskip
\noindent
This is our promised generalization of the Abelian theorem 1 to general (non-spin) non-Abelian TQFTs. In particular, we have learned that:

\medskip
\noindent
{\it All non-invertible symmetries in non-spin TQFTs are emergent: they come from sequentially gauging symmetries starting from a theory with no condensable bosons. In modern language, non-invertible symmetries in these theories are non-intrinsic.}

\medskip
As in the Abelian case, non-Abelian TQFTs typically have a plethora of non-universal surface operators. Indeed, given a modular tensor category, $\CC$, let $\CA$ be a special Frobenius algebra in $\CC$. The lines forming $\CA$ can be higher-gauged on a surface to get an operator, $S(\CA)$ \cite{Roumpedakis:2022aik}. The action of this surface operator can then be deduced from the partition function of the boundary RCFT defined by $\CA$. Note that, unlike the universal case described above, unless $\CA$ is commutative, the surface cannot be \lq\lq expanded" into a slab.\footnote{In general, we can imagine trying to extend the surface into a slab by summing over the condensable algebras $\CA_L$ and $\CA_R$ in a three-dimensional slab ending on $S$. Consider performing this operation with $\CA_L$ \lq\lq to the right of $S$" (where \lq\lq right" can be defined from the configuration in Fig. \ref{fig:3d2d}). This extension  involves summing over insertions of the identity line ending on the defect and producing a three-dimensional network of the object $\CA_L$ upon the action of $S$. When $\CA_L=\CA$, this maneuver amounts to expanding a universal surface into a slab (i.e., the inverse procedure of Fig. \ref{fig:surface from slab}). On the other hand, when $\CA\ne\CA_L$, this process instead amounts to defining a transition of the non-invertible surface to a new non-trivial surface.}

Assuming that the fusion rules of $\CC$ are multiplicity-free, and that the algebra object $\CA$ contains at most one copy of any given simple line operator, the action of the surface operator is given by \cite{Fuchs:2002cm}
\be
S(\CA) \cdot a = \sum_{b\in \CC} n_{S \cdot a}^b b~,
\ee
where 
\be
n_{S(\CA)\cdot a}^b= \sum_{i,j,k \in \CA} m_{jk}^i \Delta_{i}^{kj} \sum_{ c \in \CC} (F_{kjb}^{\bar a})^{-1}_{ic} R_{kj}^i \frac{\theta_{c}}{\theta_{b}} (F_{kjb}^{\bar a})_{ci}~.
\ee
In this latter expression, $m_{jk}^i$ and $\Delta_{i}^{kj}$ specify the multiplication and co-multiplication of $\CA$ respectively. For example, all special Frobenius algebras in an MTC describing a discrete gauge theory have been classified in \cite{ostrik2006module}. Plugging in the data of these algebras into the formula above (or its generalization for the case of theories with multiplicity in the fusion rule), we can compute the action of all surface operators on the lines. 

Our definition of non-invertible symmetries was directly in terms of the consistent ways in which a surface operator can act on the line operators and fusion spaces. Given a set of integers, $n_{Sa}^b$, satisfying this definition and an action of $S$ on the fusion spaces, it would be interesting to derive the full data of the algebra $\CA$ (the mathematical machinery in \cite{Kong_2008,Kong_2009} will likely be of use). We leave this generalization of section \ref{sec:Sactiontocond} for future work.

\subsubsection{Faithfullness of condensation defects}
\label{sec:faithfullness}

Condensation defects obtained from higher-gauging a non-Morita trivial algebra object act faithfully on the untwisted sector of the TQFT. To understand this statement, note that condensation defects of a TQFT with MTC, $\CC$, are in one-to-one correspondence with Lagrangian algebras of $\CC \boxtimes \bar \CC$ \cite{davydov2011witt, Roumpedakis:2022aik}. The condensation defects which act trivially on all line operators correspond to Lagrangian algebras obtained from soft braided tensor auto-equivalences of $\CC$ \cite{Davydov_2014}. In particular, the only soft braided tensor auto-equivalence which acts trivially on the fusion spaces is the trivial one. This symmetry corresponds to the canonical Lagrangian algebra in $\CC \boxtimes \bar \CC$. Therefore, the only condensation defect that acts trivially on both the line operators and fusion spaces is the trivial surface operator. 

\subsubsection{Non-invertibility from $\CA$}
From lemma 2 in section \ref{sec:noninv sym of non-abelian TQFTs}, we know that the invertibility of a surface operator can be determined by studying its action on line operators. Let $\CA$ be a special Frobenius algebra. When is the surface operator $S$ obtained from higher-gauging $\CA$ on a surface non-invertible?

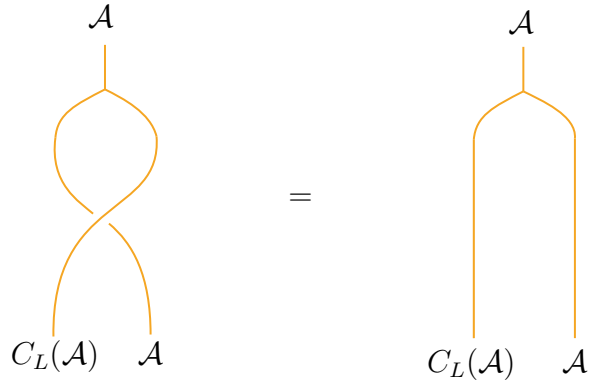
\begin{figure}[h!]
    \centering

\tikzset{every picture/.style={line width=0.75pt}} %set default line width to 0.75pt        

\begin{tikzpicture}[x=0.75pt,y=0.75pt,yscale=-1,xscale=1]

%Curve Lines [id:da032744426288151884] 
\draw [color={rgb, 255:red, 245; green, 166; blue, 35 }  ,draw opacity=1 ]   (453,96.27) .. controls (462,100.76) and (480,108.47) .. (479,120.02) ;
%Curve Lines [id:da05472227269127916] 
\draw [color={rgb, 255:red, 245; green, 166; blue, 35 }  ,draw opacity=1 ]   (428,120.67) .. controls (429,108.47) and (438,103.97) .. (453,96.27) ;
%Straight Lines [id:da17351413254591552] 
\draw [color={rgb, 255:red, 245; green, 166; blue, 35 }  ,draw opacity=1 ]   (453,73.8) -- (453,96.27) ;
%Straight Lines [id:da4257074023439641] 
\draw [color={rgb, 255:red, 245; green, 166; blue, 35 }  ,draw opacity=1 ]   (479,120.02) -- (479,220.82) ;
%Straight Lines [id:da9973120059685249] 
\draw [color={rgb, 255:red, 245; green, 166; blue, 35 }  ,draw opacity=1 ][fill={rgb, 255:red, 245; green, 166; blue, 35 }  ,fill opacity=1 ]   (428,120.67) -- (428,220.82) ;
%Curve Lines [id:da2538700245753329] 
\draw [color={rgb, 255:red, 245; green, 166; blue, 35 }  ,draw opacity=1 ]   (242,95.27) .. controls (251,99.76) and (265,107) .. (268,119.02) ;
%Curve Lines [id:da9393892688570242] 
\draw [color={rgb, 255:red, 245; green, 166; blue, 35 }  ,draw opacity=1 ]   (217,119.67) .. controls (218,107.47) and (227,102.97) .. (242,95.27) ;
%Straight Lines [id:da5810838496548381] 
\draw [color={rgb, 255:red, 245; green, 166; blue, 35 }  ,draw opacity=1 ]   (242,72.8) -- (242,95.27) ;
%Curve Lines [id:da009248254396189814] 
\draw [color={rgb, 255:red, 245; green, 166; blue, 35 }  ,draw opacity=1 ]   (216,220) .. controls (215,148) and (271,160) .. (268,119.02) ;
%Curve Lines [id:da012457160800589229] 
\draw [color={rgb, 255:red, 245; green, 166; blue, 35 }  ,draw opacity=1 ]   (217,119.67) .. controls (214,143) and (229,152) .. (236,158) ;
%Curve Lines [id:da03972340604752145] 
\draw [color={rgb, 255:red, 245; green, 166; blue, 35 }  ,draw opacity=1 ]   (244,163) .. controls (256,173) and (265,189) .. (265,219) ;

% Text Node
\draw (471,225.35) node [anchor=north west][inner sep=0.75pt]    {$\mathcal{A}$};
% Text Node
\draw (403,224.4) node [anchor=north west][inner sep=0.75pt]    {$C_{L}(\mathcal{A})$};
% Text Node
\draw (444,54.35) node [anchor=north west][inner sep=0.75pt]    {$\mathcal{A}$};
% Text Node
\draw (193,220.4) node [anchor=north west][inner sep=0.75pt]    {$C_{L}(\mathcal{A})$};
% Text Node
\draw (257,221.35) node [anchor=north west][inner sep=0.75pt]    {$\mathcal{A}$};
% Text Node
\draw (232,51.35) node [anchor=north west][inner sep=0.75pt]    {$\mathcal{A}$};
% Text Node
\draw (333,146.4) node [anchor=north west][inner sep=0.75pt]    {$=$};

\end{tikzpicture}
    \caption{Left center of $\CA$.}
    \label{fig:leftcenter}
\end{figure}

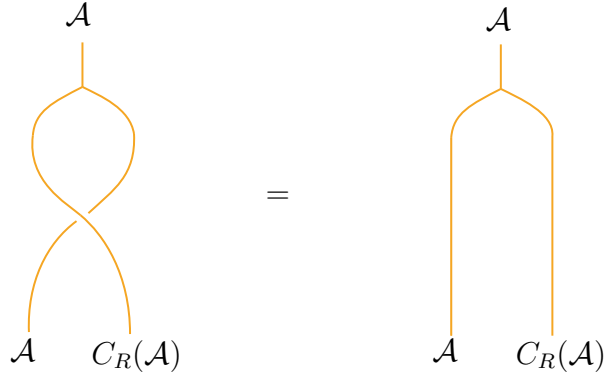
\begin{figure}[h!]
    \centering

\tikzset{every picture/.style={line width=0.75pt}} %set default line width to 0.75pt        

\begin{tikzpicture}[x=0.75pt,y=0.75pt,yscale=-1,xscale=1]

%Curve Lines [id:da032744426288151884] 
\draw [color={rgb, 255:red, 245; green, 166; blue, 35 }  ,draw opacity=1 ]   (453,96.27) .. controls (462,100.76) and (480,108.47) .. (479,120.02) ;
%Curve Lines [id:da05472227269127916] 
\draw [color={rgb, 255:red, 245; green, 166; blue, 35 }  ,draw opacity=1 ]   (428,120.67) .. controls (429,108.47) and (438,103.97) .. (453,96.27) ;
%Straight Lines [id:da17351413254591552] 
\draw [color={rgb, 255:red, 245; green, 166; blue, 35 }  ,draw opacity=1 ]   (453,73.8) -- (453,96.27) ;
%Straight Lines [id:da4257074023439641] 
\draw [color={rgb, 255:red, 245; green, 166; blue, 35 }  ,draw opacity=1 ]   (479,120.02) -- (479,220.82) ;
%Straight Lines [id:da9973120059685249] 
\draw [color={rgb, 255:red, 245; green, 166; blue, 35 }  ,draw opacity=1 ][fill={rgb, 255:red, 245; green, 166; blue, 35 }  ,fill opacity=1 ]   (428,120.67) -- (428,220.82) ;
%Curve Lines [id:da2538700245753329] 
\draw [color={rgb, 255:red, 245; green, 166; blue, 35 }  ,draw opacity=1 ]   (242,95.27) .. controls (251,99.76) and (266,106) .. (268,119.02) ;
%Curve Lines [id:da9393892688570242] 
\draw [color={rgb, 255:red, 245; green, 166; blue, 35 }  ,draw opacity=1 ]   (217,119.67) .. controls (218,107.47) and (227,102.97) .. (242,95.27) ;
%Straight Lines [id:da5810838496548381] 
\draw [color={rgb, 255:red, 245; green, 166; blue, 35 }  ,draw opacity=1 ]   (242,72.8) -- (242,95.27) ;
%Curve Lines [id:da009248254396189814] 
\draw [color={rgb, 255:red, 245; green, 166; blue, 35 }  ,draw opacity=1 ]   (245,159) .. controls (258,147) and (269,140) .. (268,119.02) ;
%Curve Lines [id:da012457160800589229] 
\draw [color={rgb, 255:red, 245; green, 166; blue, 35 }  ,draw opacity=1 ]   (217,119.67) .. controls (214,143) and (234,154) .. (241,160) ;
%Curve Lines [id:da03972340604752145] 
\draw [color={rgb, 255:red, 245; green, 166; blue, 35 }  ,draw opacity=1 ]   (241,160) .. controls (253,170) and (266,190) .. (266,220) ;
%Curve Lines [id:da9066201079033632] 
\draw [color={rgb, 255:red, 245; green, 166; blue, 35 }  ,draw opacity=1 ]   (215,219) .. controls (214,194) and (224,173.98) .. (239,163) ;

% Text Node
\draw (417,221.35) node [anchor=north west][inner sep=0.75pt]    {$\mathcal{A}$};
% Text Node
\draw (459,221.4) node [anchor=north west][inner sep=0.75pt]    {$C_{R}(\mathcal{A})$};
% Text Node
\draw (444,54.35) node [anchor=north west][inner sep=0.75pt]    {$\mathcal{A}$};
% Text Node
\draw (245,221.4) node [anchor=north west][inner sep=0.75pt]    {$C_{R}(\mathcal{A})$};
% Text Node
\draw (204,221.35) node [anchor=north west][inner sep=0.75pt]    {$\mathcal{A}$};
% Text Node
\draw (232,51.35) node [anchor=north west][inner sep=0.75pt]    {$\mathcal{A}$};
% Text Node
\draw (333,146.4) node [anchor=north west][inner sep=0.75pt]    {$=$};

\end{tikzpicture}
    \caption{Right center of $\CA$.}
    \label{fig:rightcenter}
\end{figure}

To answer this question, let us review the definition of left and right center of the algebra $\CA$ \cite{Fuchs:2002cm}. The left center, $C_L(\CA)$ of $A$, is the maximal sub-object of $\CA$ such that 
\be
m_{a,i}^{b} ~ R_{i,a}^{b}= m_{i,a}^b~,
\ee
where $m$ is the multiplication in $\CA$, and $R_{i,a}^{b}$ is the braiding of the lines $a \in \CA$ and $i\in C_{L}(\CA)$. Graphically, we have Fig. \ref{fig:leftcenter}.
Similarly, the right center, $C_R(\CA)$, is defined as the maximal sub-object of $\CA$ such that
\be
m_{j,a}^{b} ~ R_{a,j}^{b}= m_{a,j}^{b}~,
\ee
where $a \in \CA$ and $j\in C_{R}(\CA)$. 
Graphically, we have Fig. \ref{fig:rightcenter}.

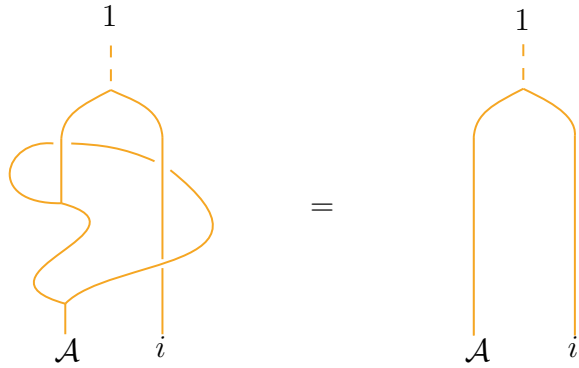
\begin{figure}[h!]
    \centering

\tikzset{every picture/.style={line width=0.75pt}} %set default line width to 0.75pt        

\begin{tikzpicture}[x=0.75pt,y=0.75pt,yscale=-1,xscale=1]

%Straight Lines [id:da003358239188612888] 
\draw [color={rgb, 255:red, 245; green, 166; blue, 35 }  ,draw opacity=1 ]   (257,172.79) -- (257,206.17) ;
%Straight Lines [id:da8525147337015229] 
\draw [color={rgb, 255:red, 245; green, 166; blue, 35 }  ,draw opacity=1 ]   (257,106.67) -- (257,168.3) ;
%Straight Lines [id:da6317826504360281] 
\draw [color={rgb, 255:red, 245; green, 166; blue, 35 }  ,draw opacity=1 ][fill={rgb, 255:red, 245; green, 166; blue, 35 }  ,fill opacity=1 ]   (206,107.31) -- (206,140.05) ;
%Curve Lines [id:da6954939048329682] 
\draw [color={rgb, 255:red, 245; green, 166; blue, 35 }  ,draw opacity=1 ]   (231,82.91) .. controls (240,87.41) and (256,91) .. (257,106.67) ;
%Curve Lines [id:da5795765318423646] 
\draw [color={rgb, 255:red, 245; green, 166; blue, 35 }  ,draw opacity=1 ]   (206,107.31) .. controls (207,95.11) and (216,90.62) .. (231,82.91) ;
%Straight Lines [id:da8997437953927581] 
\draw [color={rgb, 255:red, 245; green, 166; blue, 35 }  ,draw opacity=1 ] [dash pattern={on 4.5pt off 4.5pt}]  (231,60.44) -- (231,82.91) ;
%Straight Lines [id:da737899802115924] 
\draw [color={rgb, 255:red, 245; green, 166; blue, 35 }  ,draw opacity=1 ]   (208,190.77) -- (208,206.17) ;
%Curve Lines [id:da4933363227730967] 
\draw [color={rgb, 255:red, 245; green, 166; blue, 35 }  ,draw opacity=1 ]   (261,123.36) .. controls (324,173.43) and (230,167.65) .. (208,190.77) ;
%Curve Lines [id:da5812044852352973] 
\draw [color={rgb, 255:red, 245; green, 166; blue, 35 }  ,draw opacity=1 ]   (211,109.88) .. controls (226,110.52) and (237,111.8) .. (253,118.86) ;
%Curve Lines [id:da892290891716338] 
\draw [color={rgb, 255:red, 245; green, 166; blue, 35 }  ,draw opacity=1 ]   (206,140.05) .. controls (167,140.69) and (177,107.31) .. (202,109.88) ;
%Curve Lines [id:da09451927514943292] 
\draw [color={rgb, 255:red, 245; green, 166; blue, 35 }  ,draw opacity=1 ]   (208,190.77) .. controls (155,176) and (256,153.53) .. (206,140.05) ;
%Curve Lines [id:da6869240489720657] 
\draw [color={rgb, 255:red, 245; green, 166; blue, 35 }  ,draw opacity=1 ]   (439,82.27) .. controls (448,86.76) and (466,94.47) .. (465,106.02) ;
%Curve Lines [id:da8815422307894785] 
\draw [color={rgb, 255:red, 245; green, 166; blue, 35 }  ,draw opacity=1 ]   (414,106.67) .. controls (415,94.47) and (424,89.97) .. (439,82.27) ;
%Straight Lines [id:da9425219583796125] 
\draw [color={rgb, 255:red, 245; green, 166; blue, 35 }  ,draw opacity=1 ] [dash pattern={on 4.5pt off 4.5pt}]  (439,59.8) -- (439,82.27) ;
%Straight Lines [id:da4289194783383793] 
\draw [color={rgb, 255:red, 245; green, 166; blue, 35 }  ,draw opacity=1 ]   (465,106.02) -- (465,206.82) ;
%Straight Lines [id:da5778111962386638] 
\draw [color={rgb, 255:red, 245; green, 166; blue, 35 }  ,draw opacity=1 ][fill={rgb, 255:red, 245; green, 166; blue, 35 }  ,fill opacity=1 ]   (414,106.67) -- (414,206.82) ;

% Text Node
\draw (252,205.71) node [anchor=north west][inner sep=0.75pt]    {$i$};
% Text Node
\draw (201,207.35) node [anchor=north west][inner sep=0.75pt]    {$\mathcal{A}$};
% Text Node
\draw (408,207.35) node [anchor=north west][inner sep=0.75pt]    {$\mathcal{A}$};
% Text Node
\draw (460,205.71) node [anchor=north west][inner sep=0.75pt]    {$i$};
% Text Node
\draw (330,138.87) node [anchor=north west][inner sep=0.75pt]    {$=$};
% Text Node
\draw (225,38.71) node [anchor=north west][inner sep=0.75pt]    {$1$};
% Text Node
\draw (433,40.71) node [anchor=north west][inner sep=0.75pt]    {$1$};

\end{tikzpicture}
    \caption{Condition for $n_{Si}^{1}\neq 0$.}
    \label{fig:leftcenterlocalmorphism}
\end{figure}

Now, we will show that if $\CA$ has a non-trivial left center, then 
\be
n_{S\cdot i}^{1}=1 ~,\ \forall i \in C_L(\CA)~.
\ee
In other words, $C_L(\CA)$ is precisely the $\CA_L$ algebra we discussed in the previous part of section \ref{sec:TQFTs with bosonic lines}. To see this, note that $n_{S\cdot a}^{1}\neq 0$, if and only if the line $a$ can end on the surface operator $S$. For the end-point to be consistent with a change of triangulation of the network of $\CA$ defects on the surface, it should satisfy Fig. \ref{fig:leftcenterlocalmorphism}. This figure does indeed hold, since $i$ belongs to the left center of $\CA$, and therefore the $i$ and $\CA$ lines can be unbraided freely.
Similarly, we can show that if the algebra $\CA$ has a non-trivial right center, then 
\be
n_{S\cdot 1}^{a}=1 ~,\ \forall a \in C_R(\CA)~.
\ee
Clearly, $C_R(\CA)$ is just the $\CA_R$ algebra we discussed before.

In conclusion, if the algebra $\CA$ has a non-trivial left or right center, then the corresponding surface operator acts non-invertibly on the line operators. See \cite[Section 5.5]{Fuchs:2002cm} for a detailed proof of this in the context of extending chiral algebras of the boundary CFT. 

In the following section, we will derive an explicit necessary and sufficient criterion to determine invertibility of surface operators obtained from higher-gauging an invertible 1-form symmetry in both Abelian and non-Abelian TQFTs. 

\subsubsection{A precise condition for invertibility of condensation defects from Abelian lines}\label{preciseCond}
\begin{table}[h!]
\centering
 \begin{tabular}{||c | c||} 
 \hline
 Property & Determined by \\ [0.5ex] 
 \hline\hline
 Condensability  & $F$ matrices of $\CA$.\\
Fusion Rules & $R$ matrices of $\CC$ and $\sigma$\\
Action on lines & $\CA \hookrightarrow \CC$ and $\sigma$ 
\\ [1ex] 
 \hline
 \end{tabular}
 \caption{This table summarizes the specific data in $\CC$ that  determines various properties of condensation defects. Here, $\sigma \in H^2(\CA,U(1))$ is the discrete torsion.}
 \label{tb:propofcond}
\end{table}

Recall the notation for a condensation defect, $S(G,\sigma)$, obtained from higher-gauging a 1-form symmetry, $G$, with discrete torsion, $\sigma$ introduced in Sec. \ref{sec:noinv defects from condensation}. The important data is summarized in table \ref{tb:propofcond}. From Sec. \ref{sec:invertibility_basic}, we know that $S(G,\sigma)$ implements an invertible symmetry if and only if its action on the lines is a permutation. Therefore, the action of the surface operator on a line, $p$, should result in a unique outcome. In the boundary CFT, this means that the CFT partition function should be a permutation modular invariant.

For an invertible action, using \eqref{eq:Saction} and \eqref{eq:pconst}, for a given $p$ the condition
\be
\label{eq:invconst}
 \frac{\theta_{h \cdot p}}{\theta_h \theta_p} \Xi_{[G,\sigma]}(g,h) =1 ~,\ \forall ~ h \in G~,
\ee
should not be satisfied for all but one value of $g \in G$. Moreover, we should not have 
\be
S(G,\sigma) \cdot p= S(G,\sigma) \cdot q~,
\ee
for $p\ne q$.

Note that the last two equations equations along with \eqref{eq:Saction} and the results in \cite{Fuchs_2004} upon which this discussion is based apply to any non-spin TQFT (including non-abelian ones) with a $G$ invertible 1-form symmetry subgroup. We are therefore ready to prove the following lemma:

\vspace{0.2cm}

\noindent {\bf Lemma 5:} {\it In a $2+1$d TQFT with invertible 1-form symmetry subgroup $G$, the condensation defect, $S(G,\sigma)$, is invertible if and only if $\Xi_{[G,\sigma]}(g,h)$ is a non-degenerate bicharacter.}

\vspace{0.2cm}

\noindent {\bf Proof:} {\bf If:} Assume that the condensation defect, $S(G,\sigma)$, is invertible, and suppose the bicharacter, $\Xi_{(G,\sigma)}(g,h)$, is degenerate. That is, there exists $g$ and $\ell$ such that\footnote{Since $\Xi_{[G,\sigma]}(g,h)$ is a bicharacter, for a fixed $g$ (or fixed $h$) it must be a linear character of the group $G$. As a matrix, a non-degenerate $\Xi_{[G,\sigma]}$ has distinct characters of the group $G$ as its rows and columns. A degenerate $\Xi_{[G,\sigma]}$ has some repeated rows or columns. Since row rank and column rank agree, we will only consider the case of repeated columns.} 
\be
\label{eq:lemma11}
\Xi_{(G,\sigma)}(g,k)=\Xi_{(G,\sigma)}(\ell,k) ~,\  \forall ~ k \in G~.
\ee
This expression can be rearranged to get
\bea
\Xi_{(G,\sigma)}(g,k)\Xi^{-1}_{(G,\sigma)}(\ell,k) &=&\Xi_{(G,\sigma)}(g,k)\Xi_{(G,\sigma)}(-\ell,k)\cr
 &=&\Xi_{(G,\sigma)}(g-\ell,k)=\Xi_{(G,\sigma)}(u,k)=1 ~,\ \forall ~ k \in G~.
\eea
We have used the bicharacter property of $\Xi_{(G,\sigma)}$ and defined $u:=g-\ell$ to be some non-identity element of $G$. This argument implies that the action of $S(G,\sigma)$ on the trivial line, $1$, of the MTC, $\CC$, gives the outcome
\be
1 + u\in S(G,\sigma)\cdot1~.
\ee
This result contradicts the assumption that $S(G,\sigma)$ is invertible. Hence, for an invertible action, the bicharacter, $\Xi_{(G,\sigma)}(g,h)$, is non-degenerate and satisfies 
\be
\Xi_{[G,\sigma]}(g,k)=\Xi_{[G,\sigma]}(l,k) ~,\ \forall ~ k \in G \implies g=l~.
\ee

\medskip
\noindent
{\bf Only if:} Assume that $\Xi_{[G,\sigma]}(g,h)$ is non-degenerate. Then, for a given $p \in \CC$, there is always some $g \in G$ which satisfies 
\be
\label{invconst2}
\frac{\theta_{h \cdot p}}{\theta_h \theta_p} \Xi_{[G,\sigma]}(g,h) =1 ~,\ \forall ~ h \in G~.
\ee
To see this, let us rewrite this equation in terms of the S-matrix
\be
\frac{\CS_{hp}}{\CS_{1p}} = (\Xi_{[G,\sigma]}(g,h))^{-1} ~,\ \forall ~ h \in G~.
\ee
Recall that $\lambda_p(h):=\frac{\CS_{hp}}{\CS_{1p}}$ is a character of the group $G$. By assumption, $\Xi_{[G,\sigma]}(g,h)$ is non-degenerate which implies that the rows of $\Xi_{[G,\sigma]}$ exhaust all characters of $G$. Therefore, there exists some $g$ such that the above equation is satisfied. Moreover, this $g$ must be unique. If this were not true, then the bicharacter, $\Xi_{[G,\sigma]}(g,k)$, would not be non-degenerate. Indeed, suppose \eqref{invconst2} held for distinct $g,\ell\in G$, then
\be
 \Xi_{[G,\sigma]}(g,k)=\Xi_{[G,\sigma]}(\ell,k) ~,\ \forall ~ k \in G~,
\ee
contradicting non-degeneracy of $ \Xi_{[G,\sigma]}$. Moreover, the non-degeneracy of $\Xi_{[G,\sigma]}(g,h)$ implies that 
\be
S(G,\sigma) \cdot p \neq S(G,\sigma) \cdot q~,
\ee
for any $p,q\in \CC$ with $p\ne q$. To see this statement holds, suppose we have
\be
S(G,\sigma) \cdot p = S(G,\sigma) \cdot q~.
\ee 
Let $S(G,\sigma) \cdot p= g \cdot p$ and $S(G,\sigma)\cdot q =\ell\cdot q$ for some $g ,\ell \in G$. We will show that the non-degeneracy of $\Xi_{[G,\sigma]}$  implies that $p=q$. Note that $g,\ell$ are invertible line operators while $p,q$ are non-invertible in general. From the assumptions, we have 
\be 
g \cdot p= \ell \cdot q \implies p= \bar g \cdot \ell \cdot q~,
\ee
and 
\be
\label{eq:onlyifsimp}
\frac{\theta_{h \cdot p}}{\theta_h \theta_p} \Xi_{[G,\sigma]}(g,h) =1 ~,\ \forall ~ h \in G~, ~~~ \frac{\theta_{h \cdot q}}{\theta_h \theta_q} \Xi_{[G,\sigma]}(\ell,h) =1 ~,\ \forall ~ h \in G~.
\ee

Now, let us write the topological twists in terms of $\CS$-matrix elements. We have
\be
\frac{\theta_{h \cdot p}}{\theta_h \theta_p} = \frac{D}{d_{h\cdot p} }\CS_{h,p} = \frac{D}{d_{h\cdot \bar g \cdot \ell \cdot q} }\CS_{h,\bar g \cdot \ell \cdot q}= \frac{D}{d_{h\cdot \bar g \cdot \ell\cdot q}} \frac{\CS_{h,\bar g \cdot \ell}\CS_{h,q}}{\CS_{h,1}}= \frac{1}{d_{\bar g \cdot \ell}}\frac{\CS_{h,\bar g \cdot \ell}}{\CS_{h,1}}  \frac{D}{d_{h\cdot q}} \CS_{h,q} = \frac{1}{d_{\bar g \cdot \ell}}\frac{\CS_{h,\bar g \cdot \ell}}{\CS_{h,1}}  \frac{\theta_{h \cdot q}}{\theta_h \theta_q}~,
\ee
where $D$ is the total quantum dimension, $1$ is the trivial line operator, and, in writing the third equality, we have used
\be
\frac{\CS_{ax}}{\CS_{1x}}\frac{\CS_{bx}}{\CS_{1x}}=\sum_c N_{ab}^c \frac{\CS_{cx}}{\CS_{1x}}~,
\ee
for $a=\bar g\cdot \ell$, $b=q$, and $x=h$. We can substitute the expression obtained above into the first equation in \eqref{eq:onlyifsimp} to get
\be
\frac{1}{d_{\bar g \cdot \ell}}\frac{S_{h,\bar g \cdot \ell}}{\CS_{h,1}}  \frac{\theta_{h \cdot q}}{\theta_h \theta_q}  \Xi_{[G,\sigma]}(g,h) =1 ~,\ \forall ~ h \in G~.
\ee
Using the second equation in \eqref{eq:onlyifsimp} and rearranging the terms, we get
\be
\Xi_{[G,\sigma]}(l,h)=\frac{1}{d_{\bar g \cdot l}}\frac{\CS_{h,\bar g \cdot l}}{\CS_{h,1}}   \Xi_{[G,\sigma]}(g,h)  ~,\ \forall ~ h \in G~.
\ee
Note that invertibility of $\bar g\cdot \ell$ implies $d_{\bar g\cdot \ell}=1$ and that therefore $\CS$ in the expression above can be written as
\be
\frac{\CS_{h,\bar g \cdot \ell}}{\CS_{h,1}}  = \Xi_{[G,\sigma]}(h,\bar g \cdot \ell)\Xi_{[G,\sigma]}(\bar g \cdot \ell, h)~.
\ee
This statement follows from the definition of $\Xi$ given in \eqref{eq:defXi}. Therefore, we get
\be
\Xi_{[G,\sigma]}(\ell,h)=\Xi_{[G,\sigma]}(h,\bar g \cdot \ell)\Xi_{[G,\sigma]}(\bar g \cdot \ell, h) \Xi_{[G,\sigma]}(g,h)=  \Xi_{[G,\sigma]}(h,\bar g \cdot \ell)\Xi_{[G,\sigma]}(\ell, h) ~,\ \forall ~ h \in G~,
\ee
where, in the last equality, we used the bicharacter property of $\Xi_{[G,\sigma]}$. This argument implies
\be
 \Xi_{[G,\sigma]}(h,\bar g \cdot \ell)=1 ~, ~~~ \forall ~ h \in G~.
\ee
Since $ \Xi_{[G,\sigma]}$ is non-degenerate, we get $\bar g \cdot \ell=1 \implies \ell=g$. Then, from $p= \bar g \cdot \ell \cdot q$ we have $p =q$. $\square$

\medskip
For a non-invertible surface operator, $S(G,\sigma)$, let $U$ be the set of elements in $G$ such that any $u \in U$ satisfies
\be
\Xi_{[G,\sigma]}(u,k)=1 ~ \forall ~ k \in G~.
\ee
From the bicharacter property of $\Xi_{[G,\sigma]}$, $U$ is in fact a subgroup of $G$. The order of $U$ can be seen as the ``amount of non-invertibility" of the surface operator. In particular, the action of $S(G,\sigma)$ on the trivial line operator is given by 
\be
S(G,\sigma) \cdot 1= \sum_{u\in U} u~.
\ee
Note that, in general, the line operators being condensed cannot all end on the surface operator. For example, if the condensation of a non-trivial group $G$ of lines results in an invertible surface operator, then only the trivial line operator can end on this surface.  

For example, let us consider the case of higher-gauging a $\DZ_2$ 1-form symmetry generated by a line $a$. The expression \eqref{eq:defXi} for $\Xi$ reduces to
\be
\Xi_{[\DZ_2,1]}(g,h)=R(g,h)~,
\ee
where we have used $\tau_G(g,h)=1$ since there is a gauge in which the $F$ matrix is trivial for the $\DZ_2$ to be higher-gaugeable. Moreover, there is no non-trivial choice of discrete torsion in this case. For the invertibility of the condensation defect, we want $\Xi_{[\DZ_2,1]}$ to be non-degenerate. 
\be
\Xi_{[\DZ_2,1]}(a,a) \neq 1 \implies R(a,a)\neq 1~.
\ee
For an order two line operator $a$, $R(a,a)=\pm 1$. Therefore, for the invertibility of the condensation defect we require the line operator being condensed to be fermionic. This agrees with the discussion in \cite{Roumpedakis:2022aik}.

\section{From TQFTs to more general QFTs}\label{sec:QFTembedd}
In this section, our goal is to first give a simple pictorial \lq\lq derivation" of the result from \cite{thibault2022drinfeld} discussed around \eqref{introMorita}. Our reason for doing this is to more precisely show how our above results (in particular theorem 4; in the final part of this section we will also discuss lemma 5) generalize to non-topological $2+1$-dimensional QFTs.

As discussed in the introduction, it is generally interesting to understand relations between different QFTs and to encode these relations in domain walls between the theories being compared. Here we will consider gapped interfaces between QFTs related by \lq\lq topological manipulations." More specifically, suppose we have a QFT, $\CQ$, with a non-invertible symmetry, $\CC_{\CQ}$. If we perform a topological manipulation, $\phi$, on $\CQ$ to obtain another QFT, $\CQ'$, such that $\phi(\CC_{\CQ})$ is an invertible symmetry of $\CQ'$, then $\CC_{\CQ}$ is called a \lq\lq non-intrinsically non-invertible symmetry" \cite{Kaidi:2022cpf,Kaidi:2023maf}. This statement amounts to being able to construct gapped domain walls between the two theories hosting $\CC_{\CQ}$ and $\CC_{\CQ'}$.

We will look at the structure of non-invertible symmetries in a 2+1D QFT and show that non-invertible symmetries implemented by surface operators are non-intrinsic in 2+1D QFTs without transparent fermionic line operators.

Before explaining this statement further, let us briefly recapitulate our results leading up to theorem 4. First, recall that in a TQFT, $\CT_1$, with a set of condensable bosons, $B_1$, we can condense $B_1$ to get a TQFT $\CT_2$. If $\CT_2$ has condensable bosons, $B_2$, we can further condense these bosons to get a new TQFT, $\CT_3$. Continuing this way, we end up with a TQFT, $\CT_n$, which does not contain any condensable bosons. In other words, $\CT_n$ is described by a strongly anisotropic modular tensor category
\be
\label{eq:TQFTseqG}
\CT_1 \xrightarrow{\text{condense } B_1} \CT_2 \xrightarrow{\text{condense } B_2} \CT_3 \cdots \xrightarrow{\text{condense }B_{n-1}} \CT_n~.
\ee
From our results, we know that $\CT_n$ only has invertible surface operators. Therefore, all non-invertible surface operators in $\CT_1$ arise from starting from $\CT$, and sequentially gauging symmetries (implemented by surface operators) to move in the reverse direction in the sequence of TQFTs \eqref{eq:TQFTseqG}. This logic shows that there is a topological manipulation on the TQFT $\CT_1$ such that all non-invertible surface operators become invertible. Therefore, non-invertible symmetries implemented by surface operaors in a 2+1D TQFT are non-intrinsic.  

\begin{figure}
    \centering

\tikzset{every picture/.style={line width=0.75pt}} %set default line width to 0.75pt        

\begin{tikzpicture}[x=0.75pt,y=0.75pt,yscale=-1,xscale=1]

%Straight Lines [id:da1516567082469471] 
\draw [color={rgb, 255:red, 139; green, 6; blue, 24 }  ,draw opacity=1 ]   (427.27,117.62) -- (426.82,203) ;
%Straight Lines [id:da0929646539369614] 
\draw [color={rgb, 255:red, 139; green, 6; blue, 24 }  ,draw opacity=1 ]   (388.48,58.32) -- (415.46,100) ;
%Straight Lines [id:da5529385853542841] 
\draw [color={rgb, 255:red, 139; green, 6; blue, 24 }  ,draw opacity=1 ]   (468.93,60.93) -- (442.54,98) ;
\draw  [color={rgb, 255:red, 139; green, 6; blue, 24 }  ,draw opacity=1 ][fill={rgb, 255:red, 139; green, 6; blue, 24 }  ,fill opacity=1 ] (406.49,82.19) -- (406.9,86.21) -- (402.67,85.18) -- (405.74,84.95) -- cycle ;
\draw  [color={rgb, 255:red, 139; green, 6; blue, 24 }  ,draw opacity=1 ][fill={rgb, 255:red, 139; green, 6; blue, 24 }  ,fill opacity=1 ] (454.18,84.69) -- (449.95,86.14) -- (450.25,82.12) -- (451.08,84.77) -- cycle ;
%Shape: Ellipse [id:dp2513994905128194] 
\draw  [color={rgb, 255:red, 139; green, 6; blue, 24 }  ,draw opacity=1 ][fill={rgb, 255:red, 139; green, 6; blue, 24 }  ,fill opacity=1 ] (427.29,115.96) .. controls (428.36,115.99) and (429.22,116.75) .. (429.21,117.67) .. controls (429.19,118.59) and (428.31,119.31) .. (427.24,119.29) .. controls (426.17,119.26) and (425.31,118.5) .. (425.32,117.58) .. controls (425.34,116.66) and (426.22,115.94) .. (427.29,115.96) -- cycle ;
\draw  [color={rgb, 255:red, 139; green, 6; blue, 24 }  ,draw opacity=1 ][fill={rgb, 255:red, 139; green, 6; blue, 24 }  ,fill opacity=1 ] (429.18,157.65) -- (426.86,161.32) -- (424.79,157.46) -- (426.92,159.44) -- cycle ;
%Curve Lines [id:da31446717524243606] 
\draw    (460.87,86) .. controls (482.71,112) and (365.69,113) .. (394.51,83) ;
%Straight Lines [id:da12031923347450457] 
\draw [color={rgb, 255:red, 139; green, 6; blue, 24 }  ,draw opacity=1 ]   (437.86,106.46) -- (427.24,119.29) ;
%Straight Lines [id:da39701863580509844] 
\draw [color={rgb, 255:red, 139; green, 6; blue, 24 }  ,draw opacity=1 ]   (418.96,107) -- (427.24,119.29) ;
%Curve Lines [id:da960560674073757] 
\draw    (406.73,74) .. controls (421.58,66) and (438.17,69) .. (450.39,75) ;
%Straight Lines [id:da5164136273130279] 
\draw [color={rgb, 255:red, 139; green, 6; blue, 24 }  ,draw opacity=1 ]   (602.75,120.62) -- (602.3,206) ;
%Straight Lines [id:da5314122540947187] 
\draw [color={rgb, 255:red, 139; green, 6; blue, 24 }  ,draw opacity=1 ]   (563.96,61.32) -- (594.44,110) ;
%Straight Lines [id:da9745291298652214] 
\draw [color={rgb, 255:red, 139; green, 6; blue, 24 }  ,draw opacity=1 ]   (644.42,63.93) -- (603.65,119.96) ;
\draw  [color={rgb, 255:red, 139; green, 6; blue, 24 }  ,draw opacity=1 ][fill={rgb, 255:red, 139; green, 6; blue, 24 }  ,fill opacity=1 ] (581.98,85.19) -- (582.38,89.21) -- (578.16,88.18) -- (581.22,87.95) -- cycle ;
\draw  [color={rgb, 255:red, 139; green, 6; blue, 24 }  ,draw opacity=1 ][fill={rgb, 255:red, 139; green, 6; blue, 24 }  ,fill opacity=1 ] (629.66,87.69) -- (625.43,89.14) -- (625.73,85.12) -- (626.57,87.77) -- cycle ;
%Shape: Ellipse [id:dp22877281808766048] 
\draw  [color={rgb, 255:red, 139; green, 6; blue, 24 }  ,draw opacity=1 ][fill={rgb, 255:red, 139; green, 6; blue, 24 }  ,fill opacity=1 ] (602.77,118.96) .. controls (603.84,118.99) and (604.7,119.75) .. (604.69,120.67) .. controls (604.68,121.59) and (603.8,122.31) .. (602.72,122.29) .. controls (601.65,122.26) and (600.79,121.5) .. (600.81,120.58) .. controls (600.82,119.66) and (601.7,118.94) .. (602.77,118.96) -- cycle ;
\draw  [color={rgb, 255:red, 139; green, 6; blue, 24 }  ,draw opacity=1 ][fill={rgb, 255:red, 139; green, 6; blue, 24 }  ,fill opacity=1 ] (604.67,160.65) -- (602.34,164.32) -- (600.27,160.46) -- (602.41,162.44) -- cycle ;
%Straight Lines [id:da9763225068375294] 
\draw [color={rgb, 255:red, 139; green, 6; blue, 24 }  ,draw opacity=1 ]   (594.44,110) -- (602.72,122.29) ;
%Straight Lines [id:da45471725608440994] 
\draw [color={rgb, 255:red, 139; green, 6; blue, 24 }  ,draw opacity=1 ]   (256.13,114.96) -- (255.83,172) ;
%Straight Lines [id:da297164798937534] 
\draw [color={rgb, 255:red, 139; green, 6; blue, 24 }  ,draw opacity=1 ]   (217.32,57.32) -- (247.79,106) ;
%Straight Lines [id:da5249292877671983] 
\draw [color={rgb, 255:red, 139; green, 6; blue, 24 }  ,draw opacity=1 ]   (297.77,59.93) -- (257,115.96) ;
\draw  [color={rgb, 255:red, 139; green, 6; blue, 24 }  ,draw opacity=1 ][fill={rgb, 255:red, 139; green, 6; blue, 24 }  ,fill opacity=1 ] (235.33,81.19) -- (235.74,85.21) -- (231.51,84.18) -- (234.58,83.95) -- cycle ;
\draw  [color={rgb, 255:red, 139; green, 6; blue, 24 }  ,draw opacity=1 ][fill={rgb, 255:red, 139; green, 6; blue, 24 }  ,fill opacity=1 ] (283.01,83.69) -- (278.79,85.14) -- (279.08,81.12) -- (279.92,83.77) -- cycle ;
%Shape: Ellipse [id:dp39247114709868036] 
\draw  [color={rgb, 255:red, 139; green, 6; blue, 24 }  ,draw opacity=1 ][fill={rgb, 255:red, 139; green, 6; blue, 24 }  ,fill opacity=1 ] (256.13,114.96) .. controls (257.2,114.99) and (258.06,115.75) .. (258.04,116.67) .. controls (258.03,117.59) and (257.15,118.31) .. (256.08,118.29) .. controls (255.01,118.26) and (254.15,117.5) .. (254.16,116.58) .. controls (254.17,115.66) and (255.05,114.94) .. (256.13,114.96) -- cycle ;
\draw  [color={rgb, 255:red, 139; green, 6; blue, 24 }  ,draw opacity=1 ][fill={rgb, 255:red, 139; green, 6; blue, 24 }  ,fill opacity=1 ] (258.02,156.65) -- (255.7,160.32) -- (253.62,156.46) -- (255.76,158.44) -- cycle ;
%Curve Lines [id:da11704214656765943] 
\draw    (288.84,156) .. controls (300.36,179) and (212.17,186) .. (222.47,153) ;
%Straight Lines [id:da5673230498414582] 
\draw [color={rgb, 255:red, 139; green, 6; blue, 24 }  ,draw opacity=1 ]   (247.79,106) -- (256.08,118.29) ;
%Straight Lines [id:da6721166373594945] 
\draw [color={rgb, 255:red, 139; green, 6; blue, 24 }  ,draw opacity=1 ]   (255.25,181.96) -- (254.95,203) ;
%Curve Lines [id:da2682379626094322] 
\draw    (222.47,153) .. controls (227.88,144) and (241.86,143) .. (251.46,144) ;
%Curve Lines [id:da6690884335445441] 
\draw    (261.94,144) .. controls (285.52,145) and (287.27,152) .. (288.84,156) ;
%Straight Lines [id:da47066590904285166] 
\draw [color={rgb, 255:red, 139; green, 6; blue, 24 }  ,draw opacity=1 ]   (90.74,117.62) -- (90.29,203) ;
%Straight Lines [id:da269275246620407] 
\draw [color={rgb, 255:red, 139; green, 6; blue, 24 }  ,draw opacity=1 ]   (51.95,58.32) -- (82.43,107) ;
%Straight Lines [id:da640460708793363] 
\draw [color={rgb, 255:red, 139; green, 6; blue, 24 }  ,draw opacity=1 ]   (132.41,60.93) -- (91.63,116.96) ;
\draw  [color={rgb, 255:red, 139; green, 6; blue, 24 }  ,draw opacity=1 ][fill={rgb, 255:red, 139; green, 6; blue, 24 }  ,fill opacity=1 ] (69.96,82.19) -- (70.37,86.21) -- (66.14,85.18) -- (69.21,84.95) -- cycle ;
\draw  [color={rgb, 255:red, 139; green, 6; blue, 24 }  ,draw opacity=1 ][fill={rgb, 255:red, 139; green, 6; blue, 24 }  ,fill opacity=1 ] (117.65,84.69) -- (113.42,86.14) -- (113.72,82.12) -- (114.55,84.77) -- cycle ;
%Shape: Ellipse [id:dp050904081742344065] 
\draw  [color={rgb, 255:red, 139; green, 6; blue, 24 }  ,draw opacity=1 ][fill={rgb, 255:red, 139; green, 6; blue, 24 }  ,fill opacity=1 ] (90.76,115.96) .. controls (91.83,115.99) and (92.69,116.75) .. (92.68,117.67) .. controls (92.66,118.59) and (91.78,119.31) .. (90.71,119.29) .. controls (89.64,119.26) and (88.78,118.5) .. (88.8,117.58) .. controls (88.81,116.66) and (89.69,115.94) .. (90.76,115.96) -- cycle ;
\draw  [color={rgb, 255:red, 139; green, 6; blue, 24 }  ,draw opacity=1 ][fill={rgb, 255:red, 139; green, 6; blue, 24 }  ,fill opacity=1 ] (92.65,157.65) -- (90.33,161.32) -- (88.26,157.46) -- (90.39,159.44) -- cycle ;
%Straight Lines [id:da1352664080186603] 
\draw [color={rgb, 255:red, 139; green, 6; blue, 24 }  ,draw opacity=1 ]   (82.43,107) -- (90.71,119.29) ;

% Text Node
\draw (465.53,37.93) node [anchor=north west][inner sep=0.75pt]    {$b$};
% Text Node
\draw (380.59,37.83) node [anchor=north west][inner sep=0.75pt]    {$a$};
% Text Node
\draw (420.3,205.07) node [anchor=north west][inner sep=0.75pt]    {$c$};
% Text Node
\draw (641.01,40.93) node [anchor=north west][inner sep=0.75pt]    {$b$};
% Text Node
\draw (556.08,40.83) node [anchor=north west][inner sep=0.75pt]    {$a$};
% Text Node
\draw (595.78,208.07) node [anchor=north west][inner sep=0.75pt]    {$c$};
% Text Node
\draw (294.36,36.93) node [anchor=north west][inner sep=0.75pt]    {$b$};
% Text Node
\draw (209.43,36.83) node [anchor=north west][inner sep=0.75pt]    {$a$};
% Text Node
\draw (249.13,204.07) node [anchor=north west][inner sep=0.75pt]    {$c$};
% Text Node
\draw (129,37.93) node [anchor=north west][inner sep=0.75pt]    {$b$};
% Text Node
\draw (44.07,37.83) node [anchor=north west][inner sep=0.75pt]    {$a$};
% Text Node
\draw (84.77,203.07) node [anchor=north west][inner sep=0.75pt]    {$c$};
% Text Node
\draw (17,106.4) node [scale=1.5,anchor=north west][inner sep=0.75pt]    {$\frac{S_{xc}}{S_{1c}}$};
% Text Node
\draw (158,119.4) node [anchor=north west][inner sep=0.75pt]    {$=$};
% Text Node
\draw (329,120.4) node [anchor=north west][inner sep=0.75pt]    {$=$};
% Text Node
\draw (502,121.4) node [anchor=north west][inner sep=0.75pt]    {$=$};
% Text Node
\draw (207,154.4) node [anchor=north west][inner sep=0.75pt]    {$x$};
% Text Node
\draw (375,81.4) node [anchor=north west][inner sep=0.75pt]    {$x$};

\end{tikzpicture}
    \caption{Suppose the line operator $a$ and $b$ are transparent. Then, $c$ is also transparent.}
    \label{fig:transparent lines}
\end{figure}
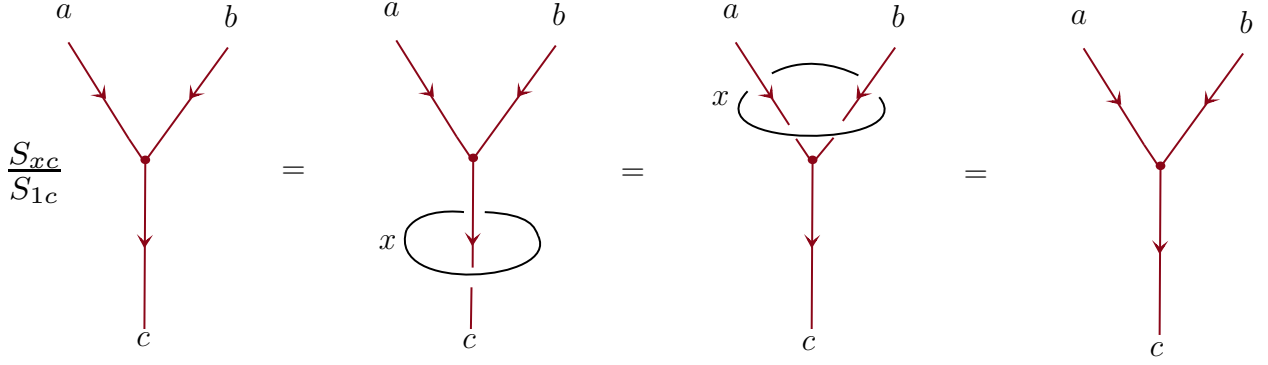

Now, consider the set of topological surface and line operators in a 2+1D QFT, $\CQ$. In general, these operators may form an infinite family. We will focus on the subset of topological operators which form a fusion 2-category. The set of topological line operators form a braided fusion category, $\CC_{\CQ}$. We will assume that there are no fermionic line operators which braided trivially with all the other line operators. Therefore, the set of line operators which braid trivially with all lines in $\CC_{\CQ}$ are bosonic. Moreover, the pictorial equalities in Fig. \ref{fig:transparent lines} show that transparent lines are closed under fusion. 
As a result, they form a symmetric fusion category, Rep$(K)$, for some finite group, $K$ \cite{deligne2002categories}. Condensing Rep$(K)$ gives a QFT, $\CQ'$, with no transparent line operators. \cite[Theorem 2]{Sawin:1999xt}. In other words, $\CC_{\CQ'}$ is an MTC. Let us denote the full fusion 2-category of surfaces and lines in $\CQ'$ as $\CC_{\CQ'}^{(2)}$.

One can show that it is possible to perform further topological manipulations on $\CQ'$ to get a QFT, $\CQ''$, such that, as alluded to in the introduction, the fusion 2-category of symmetries of $\CQ''$ is of the form \cite{thibault2022drinfeld}
\be
\label{eq:every 2fusion cat}
\text{2Vec}_{G}^{\omega} \boxtimes \text{ Mod}(\CC_{\CQ'})~,
\ee
for some finite group, $G$, and 4-cocycle $\omega \in Z^4(G,U(1))$. Here, $\text{2Vec}_{G}^{\omega}$ denotes a set of invertible surface operators which form the group $G$ under fusion with anomaly $\omega$. The second factor, $\text{ Mod}(\CC_{\CQ'})$, denotes the line operators which form the MTC, $\CC_{\CQ'}$, and their condensation defects.

\begin{figure}[h!]
    \centering

\tikzset{every picture/.style={line width=0.75pt}} %set default line width to 0.75pt        

\begin{tikzpicture}[x=0.75pt,y=0.75pt,yscale=-0.8,xscale=0.8]

%Shape: Cube [id:dp5414738226532101] 
\draw  [color={rgb, 255:red, 245; green, 166; blue, 35 }  ,draw opacity=1 ][fill={rgb, 255:red, 248; green, 198; blue, 155 }  ,fill opacity=0.14 ] (144.52,112.16) -- (206.42,49.92) -- (364.42,51.93) -- (366.78,185.84) -- (304.87,248.08) -- (146.88,246.07) -- cycle ; \draw  [color={rgb, 255:red, 245; green, 166; blue, 35 }  ,draw opacity=1 ] (364.42,51.93) -- (302.51,114.17) -- (144.52,112.16) ; \draw  [color={rgb, 255:red, 245; green, 166; blue, 35 }  ,draw opacity=1 ] (302.51,114.17) -- (304.87,248.08) ;
%Shape: Cube [id:dp9435223638820807] 
\draw  [color={rgb, 255:red, 245; green, 166; blue, 35 }  ,draw opacity=1 ][fill={rgb, 255:red, 248; green, 198; blue, 155 }  ,fill opacity=0.58 ] (300.62,114.28) -- (362.53,52.04) -- (520.53,54.05) -- (522.89,187.96) -- (460.98,250.2) -- (302.98,248.18) -- cycle ; \draw  [color={rgb, 255:red, 245; green, 166; blue, 35 }  ,draw opacity=1 ] (520.53,54.05) -- (458.62,116.29) -- (300.62,114.28) ; \draw  [color={rgb, 255:red, 245; green, 166; blue, 35 }  ,draw opacity=1 ] (458.62,116.29) -- (460.98,250.2) ;
%Shape: Polygon [id:ds7901248975533798] 
\draw  [color={rgb, 255:red, 245; green, 166; blue, 35 }  ,draw opacity=1 ][fill={rgb, 255:red, 248; green, 198; blue, 155 }  ,fill opacity=1 ] (321.55,92.21) -- (362.91,51.97) -- (365.27,185.85) -- (303.13,248.05) -- (300.64,114.26) -- cycle ;
%Straight Lines [id:da5986605726902466] 
\draw [color={rgb, 255:red, 245; green, 166; blue, 35 }  ,draw opacity=1 ][fill={rgb, 255:red, 248; green, 198; blue, 155 }  ,fill opacity=1 ]   (300.64,114.26) -- (364.2,115.12) ;

% Text Node
\draw (203.47,169.83) node [anchor=north west][inner sep=0.75pt]    {2Vec};
% Text Node
\draw (312.2,169.35) node [anchor=north west][inner sep=0.75pt]  [font=\small]  {$\text{Mod}(\mathcal{C}_{\CQ'}) \boxtimes \text{Mod}(\overline{\mathcal{C}}_{\CQ'})$};
\end{tikzpicture}
    \caption{Gapped interface between 2Vec and $\text{Mod}(\CC_{\CQ'}) \boxtimes \text{Mod}(\bar \CC_{\CQ'})$.}
    \label{fig:gapped interface 1}
\end{figure}
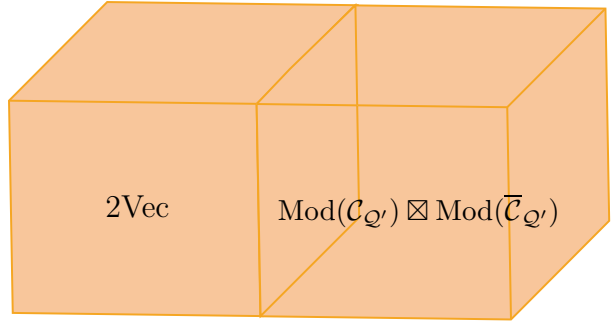

\begin{figure}[h!]
    \centering

\tikzset{every picture/.style={line width=0.75pt}} %set default line width to 0.75pt        

\begin{tikzpicture}[x=0.75pt,y=0.75pt,yscale=-0.8,xscale=0.8]

%Shape: Cube [id:dp5414738226532101] 
\draw  [color={rgb, 255:red, 245; green, 166; blue, 35 }  ,draw opacity=1 ][fill={rgb, 255:red, 248; green, 198; blue, 155 }  ,fill opacity=0.55 ] (144.52,112.16) -- (206.42,49.92) -- (364.42,51.93) -- (366.78,185.84) -- (304.87,248.08) -- (146.88,246.07) -- cycle ; \draw  [color={rgb, 255:red, 245; green, 166; blue, 35 }  ,draw opacity=1 ] (364.42,51.93) -- (302.51,114.17) -- (144.52,112.16) ; \draw  [color={rgb, 255:red, 245; green, 166; blue, 35 }  ,draw opacity=1 ] (302.51,114.17) -- (304.87,248.08) ;
%Shape: Cube [id:dp9435223638820807] 
\draw  [color={rgb, 255:red, 245; green, 166; blue, 35 }  ,draw opacity=1 ][fill={rgb, 255:red, 248; green, 198; blue, 155 }  ,fill opacity=0.58 ] (300.62,114.28) -- (362.53,52.04) -- (520.53,54.05) -- (522.89,187.96) -- (460.98,250.2) -- (302.98,248.18) -- cycle ; \draw  [color={rgb, 255:red, 245; green, 166; blue, 35 }  ,draw opacity=1 ] (520.53,54.05) -- (458.62,116.29) -- (300.62,114.28) ; \draw  [color={rgb, 255:red, 245; green, 166; blue, 35 }  ,draw opacity=1 ] (458.62,116.29) -- (460.98,250.2) ;
%Shape: Polygon [id:ds7901248975533798] 
\draw  [color={rgb, 255:red, 245; green, 166; blue, 35 }  ,draw opacity=1 ][fill={rgb, 255:red, 248; green, 198; blue, 155 }  ,fill opacity=1 ] (321.55,92.21) -- (362.91,51.97) -- (365.27,185.85) -- (303.13,248.05) -- (300.64,114.26) -- cycle ;
%Straight Lines [id:da5986605726902466] 
\draw [color={rgb, 255:red, 245; green, 166; blue, 35 }  ,draw opacity=1 ][fill={rgb, 255:red, 248; green, 198; blue, 155 }  ,fill opacity=1 ]   (300.64,114.26) -- (364.2,115.12) ;

% Text Node
\draw (204.47,167.83) node [anchor=north west][inner sep=0.75pt]    {$\mathcal{C}_{\CQ'}^{(2)}$};
% Text Node
\draw (399.2,171.35) node [anchor=north west][inner sep=0.75pt]  [font=\normalsize]  {$\mathcal{B}$};

\end{tikzpicture}
    \caption{Gapped interface between $\CC_{\CQ'}^{(2)}$ and $\CB:=\text{Mod}(\CC_{\CQ'}) \boxtimes \text{Mod}(\bar \CC_{\CQ'}) \boxtimes \CC_{\CQ'}^{(2)}$}
    \label{fig:gapped interface 2}
\end{figure}
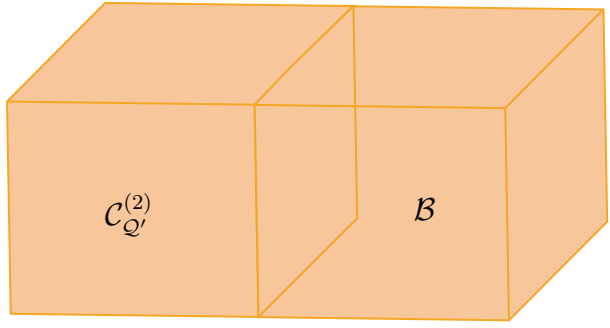

\begin{figure}[h!]
    \centering

\tikzset{every picture/.style={line width=0.75pt}} %set default line width to 0.75pt        

\begin{tikzpicture}[x=0.75pt,y=0.75pt,yscale=-1,xscale=1]

%Shape: Cube [id:dp5414738226532101] 
\draw  [color={rgb, 255:red, 245; green, 166; blue, 35 }  ,draw opacity=1 ][fill={rgb, 255:red, 248; green, 198; blue, 155 }  ,fill opacity=0.55 ] (114.31,126.23) -- (165.5,74.79) -- (293.76,76.45) -- (295.69,187.14) -- (244.51,238.58) -- (116.25,236.92) -- cycle ; \draw  [color={rgb, 255:red, 245; green, 166; blue, 35 }  ,draw opacity=1 ] (293.76,76.45) -- (242.58,127.89) -- (114.31,126.23) ; \draw  [color={rgb, 255:red, 245; green, 166; blue, 35 }  ,draw opacity=1 ] (242.58,127.89) -- (244.51,238.58) ;
%Shape: Cube [id:dp9435223638820807] 
\draw  [color={rgb, 255:red, 245; green, 166; blue, 35 }  ,draw opacity=1 ][fill={rgb, 255:red, 248; green, 198; blue, 155 }  ,fill opacity=0.58 ] (241.71,127.98) -- (292.89,76.54) -- (421.16,78.2) -- (423.09,188.89) -- (371.91,240.33) -- (243.64,238.67) -- cycle ; \draw  [color={rgb, 255:red, 245; green, 166; blue, 35 }  ,draw opacity=1 ] (421.16,78.2) -- (369.97,129.64) -- (241.71,127.98) ; \draw  [color={rgb, 255:red, 245; green, 166; blue, 35 }  ,draw opacity=1 ] (369.97,129.64) -- (371.91,240.33) ;
%Shape: Polygon [id:ds7901248975533798] 
\draw  [color={rgb, 255:red, 245; green, 166; blue, 35 }  ,draw opacity=1 ][fill={rgb, 255:red, 248; green, 198; blue, 155 }  ,fill opacity=1 ] (258.78,109.74) -- (292.54,76.47) -- (294.46,187.15) -- (243.75,238.56) -- (241.72,127.96) -- cycle ;
%Straight Lines [id:da5986605726902466] 
\draw [color={rgb, 255:red, 245; green, 166; blue, 35 }  ,draw opacity=1 ][fill={rgb, 255:red, 248; green, 198; blue, 155 }  ,fill opacity=1 ]   (241.72,127.96) -- (293.59,128.68) ;
%Shape: Cube [id:dp8695464619573078] 
\draw  [color={rgb, 255:red, 245; green, 166; blue, 35 }  ,draw opacity=1 ][fill={rgb, 255:red, 248; green, 198; blue, 155 }  ,fill opacity=0.55 ] (371.31,129.23) -- (422.5,77.79) -- (550.76,79.45) -- (552.69,190.14) -- (501.51,241.58) -- (373.25,239.92) -- cycle ; \draw  [color={rgb, 255:red, 245; green, 166; blue, 35 }  ,draw opacity=1 ] (550.76,79.45) -- (499.58,130.89) -- (371.31,129.23) ; \draw  [color={rgb, 255:red, 245; green, 166; blue, 35 }  ,draw opacity=1 ] (499.58,130.89) -- (501.51,241.58) ;
%Straight Lines [id:da8045277259185671] 
\draw [color={rgb, 255:red, 245; green, 166; blue, 35 }  ,draw opacity=1 ][fill={rgb, 255:red, 248; green, 198; blue, 155 }  ,fill opacity=1 ]   (498.72,130.96) -- (550.59,131.68) ;
%Shape: Polygon [id:ds027960716320158174] 
\draw  [color={rgb, 255:red, 245; green, 166; blue, 35 }  ,draw opacity=1 ][fill={rgb, 255:red, 248; green, 198; blue, 155 }  ,fill opacity=1 ] (387.41,111.48) -- (421.17,78.21) -- (423.09,188.89) -- (372.38,240.3) -- (370.35,129.7) -- cycle ;
%Shape: Polygon [id:ds9525817520218359] 
\draw  [color={rgb, 255:red, 245; green, 166; blue, 35 }  ,draw opacity=1 ][fill={rgb, 255:red, 248; green, 198; blue, 155 }  ,fill opacity=1 ] (515.78,112.74) -- (549.54,79.47) -- (551.46,190.15) -- (500.75,241.56) -- (498.72,130.96) -- cycle ;
%Straight Lines [id:da3795818663699583] 
\draw [color={rgb, 255:red, 245; green, 166; blue, 35 }  ,draw opacity=1 ][fill={rgb, 255:red, 248; green, 198; blue, 155 }  ,fill opacity=1 ]   (371.31,129.23) -- (423.18,129.95) ;

% Text Node
\draw (160.66,166.76) node [anchor=north west][inner sep=0.75pt]    {$\mathcal{\CC}_{\CQ'}^{(2)}$};
% Text Node
\draw (308.2,170.35) node [anchor=north west][inner sep=0.75pt]  [font=\normalsize]  {$\mathcal{B}$};
% Text Node
\draw (378,169.4) node [anchor=north west][inner sep=0.75pt]  [font=\small]  {$\text{2Vec}_{G}^{\alpha } \boxtimes \text{Mod}(\mathcal{\CC}_{\CQ'})$};

\end{tikzpicture}
    \caption{A set of line operators in $\CB$ can be gauged to get a gapped interface between it and $\text{2Vec}_{G}^{\alpha } \boxtimes \text{Mod}(\mathcal{\CC}_{\CQ'})$}
    \label{fig:gapped interface 3}
\end{figure}
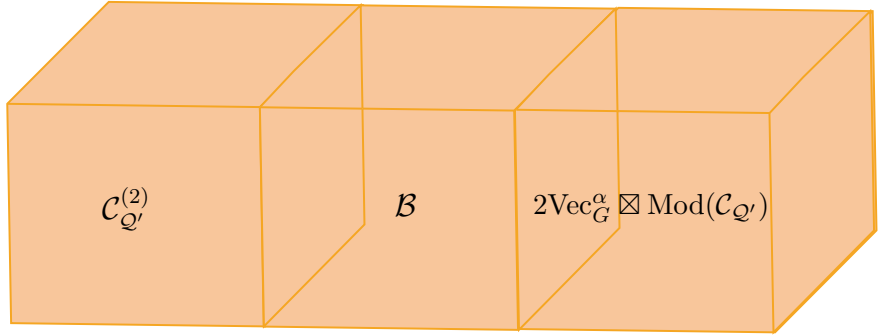

Mathematically, the above result is the statement that any fusion 2-category is Morita equivalent to a fusion 2-category of the form \eqref{eq:every 2fusion cat}. The argument in \cite{thibault2022drinfeld} is as follows. We start by recalling the fact that the MTC, $\CC_{\CQ'} \boxtimes \bar \CC_{\CQ'}$, admits a canonical gapped boundary. Including surface operators, the resulting structure can be written in terms of fusion 2-categories as the statement that $\text{Mod}(\CC_{\CQ'}) \boxtimes \text{Mod}(\bar \CC_{\CQ'})$ admits a gapped boundary. In other words, there is a gapped interface between $\text{Mod}(\CC_{\CQ'}) \boxtimes \text{Mod}(\bar \CC_{\CQ'})$  and 2Vec (see Fig. \ref{fig:gapped interface 1}). 
Stacking the 2-category, $\CC_{\CQ'}^{(2)}$, on both sides of the interface, we find that there is a gapped interface between $\CB:=\text{Mod}(\CC_{\CQ'}) \boxtimes \text{Mod}(\bar \CC_{\CQ'}) \boxtimes \CC_{\CQ'}^{(2)}$ and $\CC_{\CQ'}^{(2)}$ (see Fig. \ref{fig:gapped interface 2}).   
The line operators in $\CB:=\text{Mod}(\CC_{\CQ'}) \boxtimes \text{Mod}(\bar \CC_{\CQ'}) \boxtimes \CC_{\CQ'}^{(2)}$ form the MTC $\CC_{\CQ'}\boxtimes \bar \CC_{ \CQ'}\boxtimes  \CC_{\CQ'}$.  Condensing the canonical Lagrangian algebra in $\bar \CC_{ \CQ'}\boxtimes  \CC_{\CQ'}$ we reduce the fusion 2-category $\text{Mod}(\bar \CC_{\CQ'}) \boxtimes \CC_{\CQ'}^{(2)}$ to a fusion 2-category with only surface operators. As explained pictorially in the introduction, such a fusion 2-category must necessarily have only invertible surface operators and is of the form 2Vec$_{G}^{\omega}$ for some finite group $G$ and 4-cocycle $\omega$ \cite{JOHNSON_FREYD_2021}. Therefore, under this condensation, the fusion 2-category $\CB$ becomes $\text{Mod}(\CC_{{\CQ'}}) \boxtimes \text{2Vec}_G^{\omega}$ (see Fig. \ref{fig:gapped interface 3}). 
This completes the proof in \cite{thibault2022drinfeld}. 

\begin{figure}

\tikzset{every picture/.style={line width=0.75pt}} %set default line width to 0.75pt        

\begin{tikzpicture}[x=0.75pt,y=0.75pt,yscale=-1,xscale=1]

%Shape: Cube [id:dp5414738226532101] 
\draw  [color={rgb, 255:red, 245; green, 166; blue, 35 }  ,draw opacity=1 ][fill={rgb, 255:red, 248; green, 198; blue, 155 }  ,fill opacity=0.55 ] (45.31,121.23) -- (96.5,69.79) -- (224.76,71.45) -- (226.69,182.14) -- (175.51,233.58) -- (47.25,231.92) -- cycle ; \draw  [color={rgb, 255:red, 245; green, 166; blue, 35 }  ,draw opacity=1 ] (224.76,71.45) -- (173.58,122.89) -- (45.31,121.23) ; \draw  [color={rgb, 255:red, 245; green, 166; blue, 35 }  ,draw opacity=1 ] (173.58,122.89) -- (175.51,233.58) ;
%Shape: Cube [id:dp9435223638820807] 
\draw  [color={rgb, 255:red, 245; green, 166; blue, 35 }  ,draw opacity=1 ][fill={rgb, 255:red, 248; green, 198; blue, 155 }  ,fill opacity=0.58 ] (172.71,122.98) -- (223.89,71.54) -- (352.16,73.2) -- (354.09,183.89) -- (302.91,235.33) -- (174.64,233.67) -- cycle ; \draw  [color={rgb, 255:red, 245; green, 166; blue, 35 }  ,draw opacity=1 ] (352.16,73.2) -- (300.97,124.64) -- (172.71,122.98) ; \draw  [color={rgb, 255:red, 245; green, 166; blue, 35 }  ,draw opacity=1 ] (300.97,124.64) -- (302.91,235.33) ;
%Shape: Polygon [id:ds7901248975533798] 
\draw  [color={rgb, 255:red, 245; green, 166; blue, 35 }  ,draw opacity=1 ][fill={rgb, 255:red, 248; green, 198; blue, 155 }  ,fill opacity=1 ] (189.78,104.74) -- (223.54,71.47) -- (225.46,182.15) -- (174.75,233.56) -- (172.72,122.96) -- cycle ;
%Straight Lines [id:da5986605726902466] 
\draw [color={rgb, 255:red, 245; green, 166; blue, 35 }  ,draw opacity=1 ][fill={rgb, 255:red, 248; green, 198; blue, 155 }  ,fill opacity=1 ]   (172.72,122.96) -- (224.59,123.68) ;
%Shape: Cube [id:dp8695464619573078] 
\draw  [color={rgb, 255:red, 245; green, 166; blue, 35 }  ,draw opacity=1 ][fill={rgb, 255:red, 248; green, 198; blue, 155 }  ,fill opacity=0.55 ] (302.31,124.23) -- (353.5,72.79) -- (481.76,74.45) -- (483.69,185.14) -- (432.51,236.58) -- (304.25,234.92) -- cycle ; \draw  [color={rgb, 255:red, 245; green, 166; blue, 35 }  ,draw opacity=1 ] (481.76,74.45) -- (430.58,125.89) -- (302.31,124.23) ; \draw  [color={rgb, 255:red, 245; green, 166; blue, 35 }  ,draw opacity=1 ] (430.58,125.89) -- (432.51,236.58) ;
%Straight Lines [id:da8045277259185671] 
\draw [color={rgb, 255:red, 245; green, 166; blue, 35 }  ,draw opacity=1 ][fill={rgb, 255:red, 248; green, 198; blue, 155 }  ,fill opacity=1 ]   (429.72,125.96) -- (481.59,126.68) ;
%Shape: Polygon [id:ds027960716320158174] 
\draw  [color={rgb, 255:red, 245; green, 166; blue, 35 }  ,draw opacity=1 ][fill={rgb, 255:red, 248; green, 198; blue, 155 }  ,fill opacity=1 ] (318.41,106.48) -- (352.17,73.21) -- (354.09,183.89) -- (303.38,235.3) -- (301.35,124.7) -- cycle ;
%Shape: Polygon [id:ds9525817520218359] 
\draw  [color={rgb, 255:red, 245; green, 166; blue, 35 }  ,draw opacity=1 ][fill={rgb, 255:red, 248; green, 198; blue, 155 }  ,fill opacity=1 ] (446.78,107.74) -- (480.54,74.47) -- (482.46,185.15) -- (431.75,236.56) -- (429.72,125.96) -- cycle ;
%Straight Lines [id:da3795818663699583] 
\draw [color={rgb, 255:red, 245; green, 166; blue, 35 }  ,draw opacity=1 ][fill={rgb, 255:red, 248; green, 198; blue, 155 }  ,fill opacity=1 ]   (302.31,124.23) -- (354.18,124.95) ;
%Shape: Cube [id:dp7523811897052896] 
\draw  [color={rgb, 255:red, 245; green, 166; blue, 35 }  ,draw opacity=1 ][fill={rgb, 255:red, 248; green, 198; blue, 155 }  ,fill opacity=0.55 ] (430.31,126.23) -- (481.5,74.79) -- (609.76,76.45) -- (611.69,187.14) -- (560.51,238.58) -- (432.25,236.92) -- cycle ; \draw  [color={rgb, 255:red, 245; green, 166; blue, 35 }  ,draw opacity=1 ] (609.76,76.45) -- (558.58,127.89) -- (430.31,126.23) ; \draw  [color={rgb, 255:red, 245; green, 166; blue, 35 }  ,draw opacity=1 ] (558.58,127.89) -- (560.51,238.58) ;
%Straight Lines [id:da10630621541718355] 
\draw [color={rgb, 255:red, 245; green, 166; blue, 35 }  ,draw opacity=1 ][fill={rgb, 255:red, 248; green, 198; blue, 155 }  ,fill opacity=1 ]   (557.72,127.96) -- (609.59,128.68) ;
%Shape: Polygon [id:ds22098261691007715] 
\draw  [color={rgb, 255:red, 245; green, 166; blue, 35 }  ,draw opacity=1 ][fill={rgb, 255:red, 248; green, 198; blue, 155 }  ,fill opacity=1 ] (574.78,109.74) -- (608.54,76.47) -- (610.46,187.15) -- (559.75,238.56) -- (557.72,127.96) -- cycle ;

% Text Node
\draw (95.66,162.76) node [anchor=north west][inner sep=0.75pt]    {$\mathcal{C}_{\CQ'}$};
% Text Node
\draw (239.2,165.35) node [anchor=north west][inner sep=0.75pt]  [font=\normalsize]  {$\mathcal{B}$};
% Text Node
\draw (309,164.4) node [anchor=north west][inner sep=0.75pt]  [font=\small]  {$\text{2Vec}_{G}^{\alpha } \boxtimes \text{Mod}(\mathcal{C}_{\CQ'})$};
% Text Node
\draw (436,166.4) node [anchor=north west][inner sep=0.75pt]  [font=\small]  {$\text{2Vec}_{G}^{\alpha } \boxtimes \text{Mod}(\mathcal{C})$};
\end{tikzpicture}
    \caption{Condensing all condensable line operators in the fusion 2-category $\text{2Vec}_{G}^{\alpha } \boxtimes \text{Mod}(\mathcal{C}_{\CQ'})$ leads to a fusion 2-category $\text{2Vec}_{G}^{\alpha } \boxtimes Mod(\mathcal{C})$ which does not contain any non-invertible surface operators where $\CC$ is some strongly-anisotropic modular tensor category.}
    \label{fig:gapped interface 4}
\end{figure}
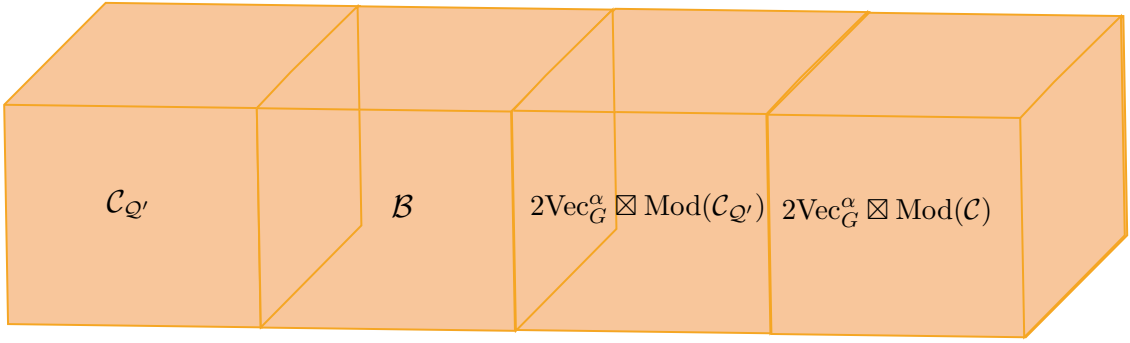

The only source of non-invertible surface operators in $\text{2Vec}_{G}^{\alpha } \boxtimes \text{Mod}(\mathcal{\CC}_{\CQ'})$ are the condensation defects denoted by objects in the fusion 2-category $\text{Mod}(\mathcal{\CC}_{\CQ'})$. Note that $\text{Mod}(\mathcal{\CC}_{\CQ'})$ describes the surface and line operators of a 2+1D TQFT. The fusion 2-category is completely determined by the MTC $\CC_{\CQ'}$. From our theorem 4, we know that there are non-invertible condensation defects if and only if there are condensable bosonic line operators in $\CC_{\CQ'}$. Therefore, we know that sequentially condensing all the condensable line operators in $\CC_{\CQ'}$ gives a strongly anisotropic MTC, say $\CC$, which does not contain any condensable bosons. Using this discussion, we can extend Fig. \ref{fig:gapped interface 3} to get Fig. \ref{fig:gapped interface 4}. 

In summary, we have shown that, given the symmetries of a 2+1D QFT described by a fusion 2-category, there is a sequence of topological manipulations leading to a QFT which has only invertible topological surface operators.\footnote{Note that to invert the sequence of topological manipulations (gauging line operators) in Fig. \ref{fig:gapped interface 4} we may have to keep track of dual symmetries implemented by surfaces which act trivially on the topological lines and their fusion spaces. For example, gauging the boson in the Toric code results in the trivial TQFT. To invert this gauging procedure we have to note that the resulting trivial TQFT is in fact a $\DZ_2$-SPT.} Therefore, we have arrived at the following result:

\medskip
\noindent
{\bf Theorem 6:} {\it In a 2+1d QFT without transparent fermionic lines, non-invertible symmetries implemented by non-invertible surface operators are non-intrinsic.} 

\subsec{Generalizing lemma 5 from TQFT to QFT}
In a similar spirit to our generalization of theorem 4 from TQFT to more general QFTs and theorem 5 in the previous subsection, let us discuss how to extend lemma 5 in Sec. \ref{preciseCond} to any $2+1$d QFT that has an invertible 1-form symmetry.

To that end, let $\CC$ be the braided fusion category of line operators in a $2+1$D TQFT. The condensation defects that can be constructed by higher-gauging these line operators are described by module categories of $\CC$. Consider a set of invertible line operators which form a group, $A$, under fusion. In this case, $\CC \cong \text{Vec}_A^{\omega}$ for some $\omega \in H^3(A,U(1))$ (we have suppressed the braiding). A module category, $\CM(G,\sigma)$ over $\CC$, is labelled by a subgroup $G\le A$ such that $\omega|_{G}$ is trivial in cohomology, and $\sigma \in H^2(G,U(1))$. In the notation used in other sections of this paper, we have $\CM(G,\sigma) =S(G,\sigma)$. In \cite{davydov2012picard}, the authors study fusion rules of module categories of $\text{Vec}_A^{\omega}$ and their invertibility under fusion. Rephrasing Proposition 5.5 in \cite{davydov2012picard} in our language, we get the following statement noting that this proposition holds even with a degenerate $\CS$ matrix

\bigskip
\noindent
{\bf Theorem 7:} \textit{In any $2+1$d QFT with abelian 1-form symmetry subgroup, $G$, the condensation defect $S(G,\sigma)$ is invertible if and only if the phase $\Xi_{[G,\sigma]}(g,h)$ is non-degenerate. }

\section{Non-invertible time-reversal symmetries of TQFTs}\label{non-inv-TR}
In previous sections, we considered non-invertible generalizations of unitary symmetries (which we will continue to refer to as \lq\lq non-invertible symmetries"). In this section, we consider non-invertible generalizations of anti-unitary / time-reversal symmetries, which we will colloquially call \lq\lq non-invertible time-reversal symmetries."

As we have seen, unitary symmetries give rise to automorphisms of the fusion group that preserve the topological spins. On the other hand, anti-unitary / time-reversal symmetries give rise to automorphisms of the fusion group that conjugate the topological spins
\begin{equation}\label{UTprod}
\theta_{S_U\cdot a}=\theta_a~,\ \ \ \theta_{S_T\cdot a}=\theta^*_a~,
\end{equation}
where the \lq\lq $U$" subscript stands for unitary, and the \lq\lq $T$" subscript stands for \lq\lq time reversal."

At the level of an Abelian MTC, we require at least one line with complex topological spin, $\theta_a\ne\theta_a^*$, in order to have an anti-unitary symmetry (otherwise all the data of the MTC can be gauge-transformed to be real, and the symmetry is unitary). Toric code / $\mathbb{Z}_2$ Dijkgraaf-Witten theory (see section \ref{subsec:toriccode}) is an example of a theory with a real MTC that lacks a time-reversal symmetry, while double semion / twisted $\mathbb{Z}_2$ Dijkgraaf-Witten theory (see section \ref{subsec:doublesemionun}) has an MTC with complex topological spins and a time-reversal symmetry. In the non-Abelian case, we can a priori imagine that there are non-trivial time reversal symmetries even if all spins are real (since the spins do not determine the MTC).\footnote{On the other hand, up to rank eleven, there are no non-Abelian unitary MTCs consisting solely of bosons and fermions (i.e., all examples with real spins up to this rank are Abelian MTCs) \cite{ng2023classification}. It would be interesting to understand if this statement is true more generally.}

When we have a complex topological spin for some line, $a$, then \eqref{UTprod} implies that unitary and time-reversal symmetries have different actions on $a$. In particular
\begin{equation}\label{distinguish}
\exists\ b\ {\rm s.t.}\ n_{S_U\cdot a}^b\ne n_{S_T\cdot a}^b~.
\end{equation}
In other words, it is always possible to distinguish a unitary and time-reversal symmetry of an MTC via symmetry actions on the complex-spin topological lines. On the other hand, we claim that:

\medskip
\noindent
{\it Non-invertible time-reversal symmetries of $2+1$d TQFTs and non-invertible symmetries (i.e., non-invertible generalizations of unitary symmetries) can have identical actions on lines, even when the topological spins of those lines are complex.}

\medskip
One way to understand this claim is to recall from our discussion in section \ref{sec:char sym of abelian TQFTs} that, in an Abelian TQFT, a non-invertible symmetry, $S$, has a non-trivial space of lines it annihilates (see \eqref{nullS}): ${\rm Null}(S)\ne\emptyset$. Now, let us suppose that $S$ annihilates all lines, $a_i$, with complex spin, $\theta_{a_i}\ne\theta_{a_i}^*$. If the theory has an invertible time-reversal symmetry, $S_T$, then the product
\begin{equation}\label{noninvT}
S\times S_T=S_{T'}~,
\end{equation}
is a non-invertible time-reversal symmetry.\footnote{Consider a configuration in which the $S$ and $S_T$ defects are close together but have not fused. This configuration gives rise to a time-reversal action on the theory and therefore so to does the fusion product.} However, it is clear that ${\rm Null}(S_{T'})={\rm Null}(S)$ and that, unlike the invertible case in \eqref{distinguish}, here
\begin{equation}\label{equivUT}
n_{S_{T'}\cdot a}^b=n_{S\cdot a}^b~,\ \forall\ a,b\in\CC~,
\end{equation}
even though $S_{T'}$ is a time-reversal symmetry, and $S$ is not.

As we will explain in our example subsection below, the phenomenon discussed in \eqref{equivUT} manifests itself in the simplest theory with time reversal: the double semion. In this theory there is a universal non-invertible surface, $S_B$, playing the role of $S$ and an invertible time reversal symmetry, $S_T$, exchanging the semion and anti-semion. Moreover, $S_B$ annihilates the complex spin lines (i.e., the semion and anti-semion lines).

We can explicitly describe this scenario more generally in Abelian TQFT. The conditions for the existence of an $S_B$ surface and an $S_T$ surface are:
\begin{enumerate}
    \item $\theta_b=1$ for some non-trivial $b\in \CC$.\label{cond1au}
    \item $\sum_{a\in \CC} \theta_a^n \in \mathds{R}$ for all positive integer $n$ \cite{Geiko:2022qjy}.\label{cond2au}
\end{enumerate}
We then have a non-invertible time-reversal symmetry, $S_{T'}$, via \eqref{noninvT}. If all lines, $a$, with $\theta_a\ne\theta_a^*$ braid non-trivially with some of the bosons, $B$, we are condensing, then we are precisely in the situation of \eqref{equivUT}, where the non-invertible time reversal symmetry is indistinguishable from the non-invertible symmetry. On the other hand, if a line, $a$, with complex spin braids trivially with $B$, then we can distinguish the non-invertible time reversal symmetry from the non-invertible symmetry via the action on lines.

Our goal in the next subsection will be to give a generalization of the definition of (non-invertible) symmetries of 2+1D TQFTs to the time reversal case. After this general discussion, we give a set of explicit examples. All the non-invertible time reversal symmetries in these theories are of the form $S_T\times S_B$.\footnote{It will be interesting to understand if all non-invertible time-reversal symmetries are of this form.} 

\subsection{A general definition of a (non-invertible) time reversal symmetry}
To motivate our general definition of (non-invertible) time-reversal symmetries, we first return to condition \ref{cond2au} above for the existence of an invertible time reversal symmetry. This condition implies that
\be
\frac{\sum_{(a,b)\in \CC \times \CC} \theta_{(a,b)}^n}{|\sum_{(a,b)\in \CC \times \CC} \theta_{(a,b)}^n|}=\frac{\sum_{a,b\in \CC} \theta_a^n \theta_b^n}{|\sum_{a,b\in \CC} \theta_a^n \theta_b^n|}=1~,
\ee
for all $n$, which, in turn, implies that $\CC \boxtimes \CC$ admits a gapped boundary \cite{Kaidi:2021gbs}. Therefore, the TQFT described by the category $\CC \boxtimes \CC$ admits a gapped boundary if  $\CC$ has a time-reversal symmetry. 

Generalizing the discussion in the case of Abelian TQFTs, we find that a general TQFT admits a time-reversal symmetry only if there is a gapped domain wall between itself and the complex-conjugated TQFT that arises from unfolding the $\CC\boxtimes\CC$ theory with gapped boundary. In terms of MTCs, this is a domain wall between $\CC$ and $\bar \CC$. 

\begin{figure}[h!]
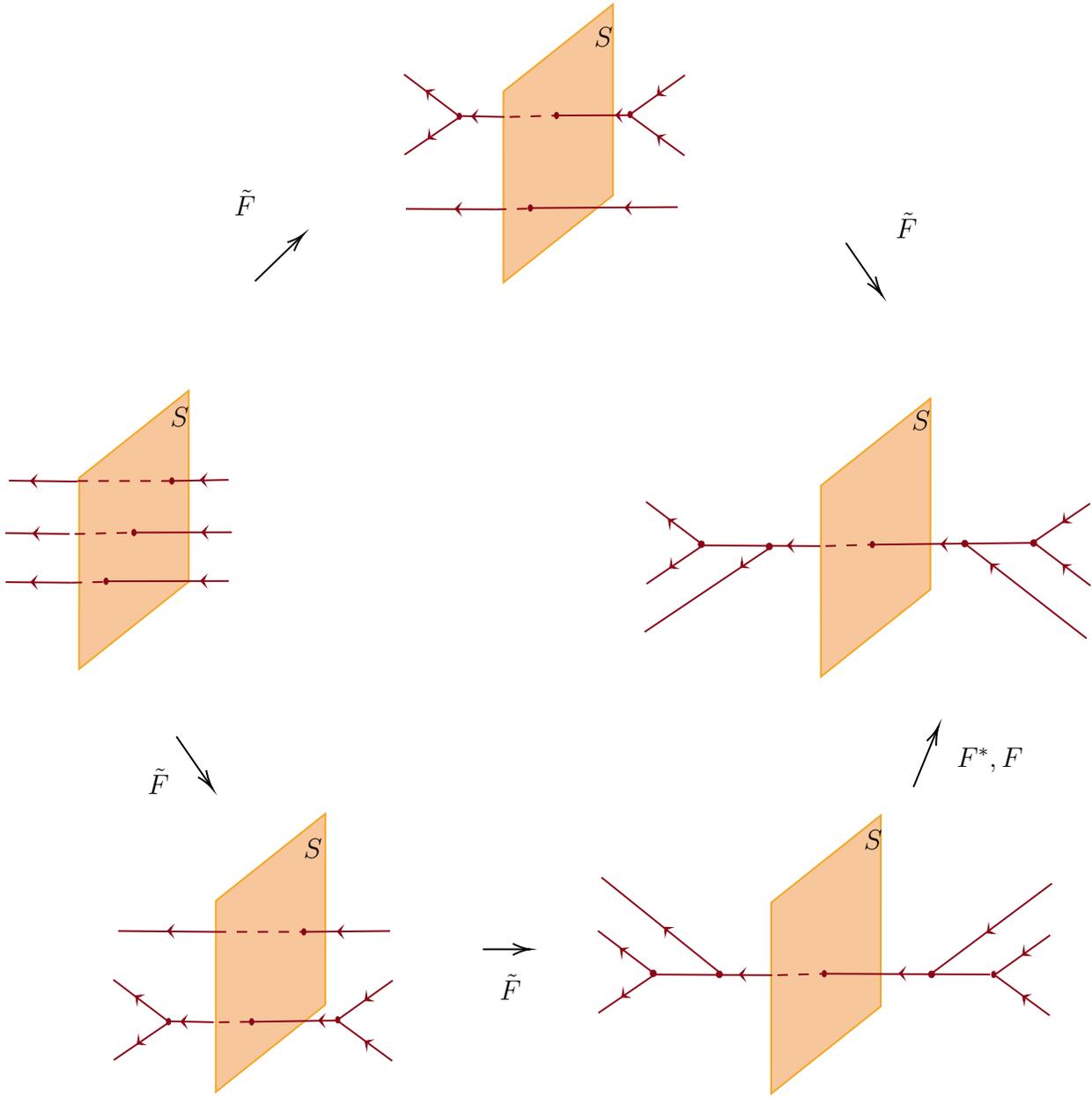

    \centering
\tikzset{every picture/.style={line width=0.75pt}} %set default line width to 0.75pt        


    \caption{The constraints on $\tilde F$ matrices from associativity of fusion. On the right bottom of the diagram, the $F^{*},F$ action indicates the $F$-move on either side of the surface operator. Note that, unlike in the case of unitary symmetry and its non-invertible generalization, in this case we use the complex conjugate $F$-move on the L.H.S. of the surface operator since the TQFT on the L.H.S. is the time-reversal of the TQFT on the R.H.S. of the surface.}
    \label{fig:tildeFconsttimerev}
\end{figure}

We can generalize the definition of non-invertible symmetries to accommodate time-reversal as follows. We have a surface operator implementing the time-reversal symmetry with $\tilde F$ as defined in Fig. \ref{fig:SactFdef}. However, now the $\tilde F$ are anti-linear maps. The constraints on $\tilde F$ from associativity of fusion is as in Fig. \ref{fig:tildeFconsttimerev}.
Moreover, the constraint on $\tilde F$ from braiding is
\be
\label{eq:tildeFRconsttimerev}
(R_{bd}^{e})^{*} \tilde F_{(c,d),(a,b),(f,e)} (R_{ac}^f)^{-1}= \tilde F_{(a,b),(c,d),(f,e)}~,
\ee
where $S\cdot a\ni b$ and $S\cdot c\ni d$. With these conditions in place, we can define a time-reversal symmetry as follows:

\medskip
\noindent
{\bf Definition:} A (non-invertible) time-reversal symmetry of a non-Abelian TQFT, $\CT$, with MTC, $\CC$, is a map, $S$, on the ring of line operators such that 
\begin{enumerate}
    \item \label{def:nonabnoninvsymt3} The $\tilde F$ matrices defined in Fig. \ref{fig:SactFdef} satisfying the conditions in Fig. \ref{fig:tildeFconsttimerev} and \eqref{eq:tildeFRconsttimerev} exist.
    \item \label{def:nonabnoninvsymt1} $n_{Sa}^bn_{Sc}^d \leq  \sum_{e,f} N_{ac}^fN_{bd}^e n_{Sf}^e$ and $n_{S1}^1=1$. 
    \item \label{def:nonabnoninvsymt2} $\sum_{a,b\in \CC} n_{Sa}^b d_a d_b =\text{dim}(\CC)$.
\end{enumerate}

The action of the non-invertible time-reversal symmetry on the fusion spaces is given by an anti-linear map $U_{T}( a,b,c,f)_{e,d}$, acting on the fusion space, $V_{ed}^f$. At the level of fusion spaces, this action can be written as
\be
\label{eq:Sact fusion space time rev}
U_T(a,b,c): \bigoplus_{d,e,f} N_{de}^f ~ n_{Sa}^{d} ~ n_{Sb}^{e} ~ n_{Sc}^{f} ~ V_{de}^{f} \rightarrow V_{ab}^c~, \ee
where $U_T(a,b,c)$ is a block-diagonal matrix, with the block $U_T(a,b,c,f)_{d,e}$ acting on $V_{de}^f$. Note that in general $U_T$ is a rectangular matrix, and is therefore non-invertible.

When the time-reversal symmetry is invertible, then the constraints on $\tilde F$ discussed above imply that $F_{abc}^d$ and $R_{ab}^c$ are related by symmetry action on the fusion spaces to $(F_{S\cdot a,S\cdot b,S\cdot c}^{S\cdot d})^{*}$ and $(R_{S\cdot a,S\cdot b}^{S\cdot c})^{*}$ respectively. This statement agrees with the discussion of invertible time-reversal symmetries in \cite{barkeshli2019symmetry}.\footnote{Non-invertible time-reversal symmetries in QFTs including Maxwell theory and massive QCD were first discussed in \cite{Choi:2022rfe}.}

\label{sec:abelianCStime-reversal}

Many results that we proved for non-invertible generalizations of unitary symmetries also hold for non-invertible time-reversal symemtries. For example, 
if a TQFT admits a non-invertible time-reversal symmetry, then it necessarily contains bosonic line operators. For abelian TQFTs, this can be seen by noting that the argument we used to prove the corresponding statement for abelian TQFTs in section \ref{sec: Non-inv sym abelian TQFTs} (specifically around \eqref{eq:propnoninv}) also applies to non-invertible time-reversal symmetries.

Consider a (non-abelian) TQFT without bosonic line operators. Let $S_T$ be a non-invertible time-reversal symmetry with the action on the line operators specified by the non-negative integers $n_{S_T \cdot a}^b$. Since there are no bosonic line operators, we have
\be
\label{eq:nobosoncond time-rev}
n_{S_T \cdot1}^b = n_{S_T \cdot b}^{1} = \delta_{b,1}~.
\ee
Also, note that the matrix of integers $n_{S_T\cdot a}^b$ satisfies \cite{Lan:2014uaa}\cite[Theorem 3.3]{Kawahigashi_2015}
\be
\sum_{c} \CS_{ac} n_{S_T \cdot c}^{b}= \sum_{c} n_{S_T \cdot a}^c \bar\CS_{cb}~,
\ee
where $ \bar\CS_{cb}$ is the complex-conjugate of the S-matrix. Choosing $b=1$, using \eqref{eq:nobosoncond time-rev} and noting that quantum dimensions of line operators are real numbers, we get
\be
\label{eq:qdimcond1 time-rev}
d_{a}= \sum_c n_{S\cdot a}^c d_c~,
\ee
where $d_a$ are the quantum dimensions of the line operators. On the other hand, choosing $a=1$ and using \eqref{eq:nobosoncond time-rev}, we get
\be
\label{eq:qdimcond2 time-rev}
\sum_c n_{S\cdot c}^b d_c = d_{b}~.
\ee
In a unitary TQFT, the quantum dimensions satisfy $d_a \geq 1$ for all line operators $a$. Therefore, the constraints \eqref{eq:qdimcond1 time-rev} and \eqref{eq:qdimcond2 time-rev} imply that 
\be
n_{S\cdot a}^b= \delta_{b,\sigma(a)}~,
\ee
for some permutation, $\sigma$, of the lines. That is, in a unitary TQFT without bosons, (non-invertible) time-reversal symmetries act as permutations on the line operators. In fact, if a time-reversal symmetry acts as permutations on the line operators, then its action on fusion spaces must be invertible. This statement follows from noting that the corresponding argument for non-invertible symmetries around \eqref{invRel} can be readily generalized for non-invertile time-reversal symmetries using the anti-linear maps $\tilde F$ and $U_T$. 

In summary, if the TQFT does not contain any bosonic line operators, it cannot have any non-invertible time-reversal symmetries. In the Abelian case, the existence of bosonic line operators is enough to guarantee the existence of a non-invertible generalization of a unitary symmetry and so, if an Abelian TQFT has a non-invertible time-reversal symmetry, then that TQFT also has a non-invertible symmetry.\footnote{This is a necessary but not sufficient condition to prove that all time reversal symmetries in Abelian TQFTs are of the form $S_B\times S_T$.} 

Non-invertible symmetries have a description in terms of Lagrangian algebras in the $\CC \boxtimes \bar \CC$ MTC (see Appendix \ref{AppA}). For time-reversal symmetries, the analogous structure is a Lagrangian subalgebra in $\CC \boxtimes \CC$. In other words, if $\CC$ has a time-reversal symmetry, then $\CC \boxtimes \CC$ has a non-anomalous symmetry implemented by topological line operators, which when gauged gives the trivial TQFT. That is, $\CC \boxtimes \CC$ must admit a Lagrangian algebra. Note that while $\CC \boxtimes \bar \CC$ always has at least one Lagrangian algebra, $\CC \boxtimes \CC$ may not have any. This statement corresponds to the fact that the any TQFT has the trivial (unitary) symmetry while it may not have any time-reversal symmetries. 

Note that even if $\CC \boxtimes \CC$ has a Lagrangian algebra object, it does not imply that the $\CC$ is time-reversal invariant. For example, $SU(9)_1$  MTC does not have a non-trivial time-reversal symmetry even though $SU(9)_{1} \boxtimes SU(9)_{1}$ has a Lagrangian subgroup.

\subsection{Example 1: Double semion model $\cong$ twisted $\mathbb{Z}_2$ discrete gauge theory}

Consider the double semion model with line operators $1$, $s$, $\bar s$, $d$ and topological spins
\be
\theta_1=1, ~ \theta_s=i,~ \theta_{\bar s}=-i,~ \theta_{d}=1~.
\ee
This theory has an invertible time-reversal symmetry implemented by a surface, $S_T$, satisfying
\be
S_T\cdot 1 =1~,\  S_T \cdot d=d,~ S_T \cdot s =\bar s~.
\ee
$S_T$ corresponds to the Lagrangian subgroup
\be
\left\{(1,1)~,\ (d,d)~,\ (s,\bar s)~,\ (\bar s, s)\right\}\in\CC\boxtimes\CC~,
\ee
where $\CC$ is the MTC of the double semion model. 

Next, recall that, in section \ref{subsec:doublesemionun}, we found a non-invertible symmetry in this TQFT implemented by the surface operator, $S_B$. We showed that that this surface operator acts as
\be
S_B\cdot 1=S_B \cdot d=1+d~, ~ S_B \cdot s=S_B \cdot \bar s=0~.
\ee
From this action, and the action of $S_T$, we can construct the symmetry
\be
S_{T'}:=S_B \times S_T~.
 \ee
Note that $S_{T'}$ acts on lines in the same way as $S_B$
\begin{equation}
S_{T'}\cdot1=S_{T'}\cdot d=1+d~,\ S_{T'}\cdot s=S_{T'}\cdot \bar s=0~.
\end{equation}
This is the simplest example of the general phenomenon discussed in the introduction to this section: at the level of the action on line operators, non-invertible symmetries and their time-reversal counterparts need not be distinguishable. 

Let us look at the action of $S_{T'}$ on the fusion spaces. This is determined by the action of $S_T$ and $S_B$ on the fusion spaces. $S_T$ acts on the fusion spaces as
\be
U_T(a,b): V_{ab}^c \to V_{S_T(a),S_T(b)}^{S_T(c)}~.
\ee
We can choose a gauge in which $F(a,b,c)=1$ $,~ a,b,c \in \CC$ \cite{rowell2009classification}. Also, $1,d$ are bosons. Therefore, it is consistent to choose 
\be
U_T(a,b,c)=1 ~ \forall ~ a,b,c ~ \in \{1,d\}~.
\ee

Now, let us consider the $S_B$ symmetry. Since both $F$ and $R$ matrices of the line operators $1,d$ are trivial, we can choose
\be
\tilde F^{S_B}_{a,b,c} =1~,
\ee
where $\tilde F^{S_B}$ is only defined when $a,b,c \in \{1,d\}$. Since the action of $S_B$ on $s$ and $\bar s$ is zero, the action of $S_B$ on fusion spaces involving these line operators is ill-defined. Using \eqref{invRel} and the choice of $\tilde F^{S_B}$ above, the action of $S_B$ on fusion spaces $V_{ab}^c, ~ a,b,c \in \{1.d\}$ is given by
\be
U_{S_B}(a,b,c)=1~.
\ee
The action of the symmetry $S_{T'}$ on the fusion spaces is given by 
\be
U_{T'}(a,b,c)=U_T(a,b,c)U_{S_B}(a,b,c)=1 ~ \forall ~ a,b,c \in \{1,d\}~,
\ee
and the action on other fusion spaces is ill-defined. Therefore, we conclude that the data $\{n_{S_B \cdot a}^b,U_{S_B}(a,b,c)\}$ and $\{n_{S_{T'} \cdot a}^b,U_{T'}(a,b,c)\}$ is identical. Therefore, this data cannot be used to distinguish between the symmetries $S_{T'}$ and $S_B$. 

We expect that the above ambiguity can be resolved by studying the action of $S_{T'}$ on non-genuine line operators. In particular, we expect that once we include twisted sectors, the action on $S_{T'}$ on the lines $s,\bar s$ will map them to non-genuine lines. Therefore, in this description the action of $S_{T'}$ on, say $V_{ss}^{1}$, will map it to a fusion space for non-genuine lines. This extra data could then be used to distinguish between $S_{T'}$ and $S_B$ symmetries. We leave this study for future work.

\subsection{Example 2: $B_5$ TQFT (Galois conjugate of $SU(5)_1$ Chern-Simons Theory)}

$U(1)_n$ Chern-Simons theories for even $n$ do not have any invertible time-reversal symmetries \cite{Delmastro:2019vnj}. On the other hand, abelian TQFTs with an odd number of line operators often have time-reversal symmtries. The simplest example is the $B_5$ prime abelian TQFT (which is a Galois conjugate of $SU(5)_1$ Chern-Simons theory). It has anyons that we label as $\ell_i$ for $i=0,\cdots,4$, with topological spins
\be
\theta_{\ell_0}=1~,~ \theta_{\ell_1}=\theta_{\ell_4}=e^{\frac{2\pi i}{5}}~, ~ \theta_{\ell_2}=\theta_{\ell_3}=e^{-\frac{2\pi i}{5}}~.
\ee
Since there are no non-trivial bosons, we know from our general discussion that this theory only has invertible time reversal symmetries. 

We clearly have the invertible time-reversal symmetry that acts as
\be
S_T \cdot \ell_0=\ell_0~,~ S_T \cdot \ell_1=\ell_2~,~ S_T \cdot \ell_2=\ell_4~,~ S_T \cdot \ell_4=\ell_3~,~ S_T \cdot \ell_3=\ell_1~,
\ee
but there are other time reversal symmetries as well since we have the fusion
\be
S_T \times S_T= S_C~,
\ee
where $S_C$ is the surface implementing the charge conjugation symmetry. Therefore, we also have
\begin{equation}
S_T^3\ne\mathds{1}~.
\end{equation}

The non-anomalous 1-form symmetry in $B_5 \boxtimes B_5$ corresponding to $S_T$ is a non-anomalous $\mathbb{Z}_5$ 1-form symmetry
\be
\left\{(\ell_0,\ell_0)~, (\ell_1,\ell_2)~,(\ell_2,\ell_4)~, (\ell_4,\ell_3)~, (\ell_3,\ell_1)\right\}\in B_5\boxtimes B_5~.
\ee
On the other hand, the $S_T^3$ time reversal symmetry corresponds to the distinct non-anomalous $\DZ_5$ 1-form symmetry
\be
\left\{(\ell_0,\ell_0)~,~ (\ell_2,\ell_1)~,~ (\ell_4,\ell_2)~,~ (\ell_3,\ell_4)~,~ (\ell_1,\ell_3)\right\}\in B_5\boxtimes B_5~.
\ee
Note that $B_5 \boxtimes B_5$ is dual to the $\DZ_5$ Dijkgraaf-Witten theory. The non-anomalous $\DZ_5$ subgroups found above correspond to the two Rep$(\DZ_5)$ subcategories in the $\DZ_5$ gauge theory.

\subsection{Example 3: $B_{25}$ TQFT (Galois conjugate of $SU(25)_1$ Chern-Simons theory)}

The $B_{25}$ TQFT has line operators, $\ell_a$, labelled by integers $0\leq a\leq 24$ with topological spins $e^{\frac{2\pi i a^2}{25}}$. This theory has all the cast of characters we have been introduced to above: a unitary symmetry, anti-unitary symmetries, and non-invertible generalizations of both.

To understand this statement, note that we have the order four time-reversal symmetry given by 
\be
S_T \cdot \ell_a = \ell_{7a \text{ mod } 25}~.
\ee
Moreover, it is easy to see that
\be
S_T \times S_T = S_C
\ee
where $S_C$ is the charge conjugation surface operator. Therefore, there is also an $S_T^3$ invertible time-reversal symmetry.

Additionally, since we have non-trivial bosons, we have a non-invertible symmetry implemented by the surface operator, $S_B$, where $B=\{\ell_0,~\ell_5,~\ell_{10},~\ell_{15},~\ell_{20}\}$. We have
\be
S_B \cdot \ell_a = \ell_0 + \ell_5 + \ell_{10} + \ell_{15} + \ell_{20}~,
\ee
when $a\in \{0,~5,~10,~15,~20\}$, and zero otherwise. It is clear that the surface operator
\be
S_B \times S_T=S_{T'}
\ee
has the same action on the lines as $S_B$. Note that the $B_{25} \boxtimes B_{25}$ TQFT is dual to the $\DZ_{25}$ Dijkgraaf-Witten theory. The $\DZ_{25}$ gauge theory has three Rep$(\DZ_{25})$ subgroups coming from the three normal subgroups of $\DZ_{25}$ \cite{naidu2006categorical}. These non-anomalous 1-form symmetries correspond to the three surface operators, $S_T$, $S_T^3$, and $S_B$. This example shows that for a TQFT described by an MTC $\CC$, some of the symmetries specified by Lagrangian subgroups of $\CC \boxtimes \CC$ correspond to symmetries that are not of time-reversal type. 

\subsection{Example 4: $B_{125}$ TQFT (Galois conjugate of $SU(125)_1$ Chern-Simons Theory)}

Finally, let us look at an example where the non-invertible time-reversal symmetry action can be observed at the level of the action on the line operators. The $B_{125}$ TQFT has a time-reversal symmetry given by
\be
S_T \cdot \ell_a = \ell_{57a \text{ mod } 125}~,
\ee
where $0\leq a \leq 124$. We have the fusion rules $S_T \times S_T=S_C$ where $S_C$ is the charge conjugation surface operator. We also have the non-invertible symmetry $S_B$ where $B=\{\ell_0,\ell_{25},\ell_{50},\ell_{75},\ell_{100}\}$. The surface operator, $S_B \times S_T$, is a non-invertible time-reversal symmetry. Its action on the line operators is given by
\be
(S_B \times S_T) \cdot \ell_a= B\text{-orbit of } \ell_{57 a \text{ mod }125}~,
\ee
if $\ell_{57 a \text{ mod }125}$ braids trivially with $B$ and zero otherwise. For example, we have
\be
(S_B \times S_T)\cdot \ell_5= \ell_{35} + \ell_{60} + \ell_{85}+ \ell_{110} + \ell_{15}~.
\ee
Note that $\theta_{\ell_5}=e^{\frac{2 \pi i}{5}}$ and $\theta_{\ell_{35}}=\theta_{\ell_{60}}=\theta_{\ell_{85}}=\theta_{\ell_{110}}=\theta_{\ell_{15}}=e^{\frac{2 \pi i 4}{5}}$. 

\section{Conclusion}

\label{sec:conclusion}
In this paper, we characterised non-invertible symmetries of $2+1$d TQFTs in terms of their action on line operators and fusion spaces. We learned that in these theories, non-invertible symmetries appear if and only if there are condensable bosons (in the case of Abelian TQFTs, this condition simplifies to the mere existence of bosons since all bosons are condensable in this case). As a result, all non-invertible symmetries are emergent or non-intrinsic in the sense that they appear via gauging 0-form symmetries in theories that have only invertible symmetries.

Indeed, given a TQFT, $\CT_1$, with a set of condensable bosons, $B_1$, we can condense $B_1$ to get a TQFT, $\CT_2$. If $\CT_2$ has condensable bosons, $B_2$, we can further condense these bosons to get a new TQFT, $\CT_3$. Continuing this way, we end up with a TQFT, $\CT_n$, which does not contain any condensable bosons. In other words, $\CT_n$ is discribed by a strongly-anisotropic modular tensor category.
\be
\label{eq:TQFTseq}
\CT_1 \xrightarrow{\text{condense } B_1} \CT_2 \xrightarrow{\text{condense } B_2} \CT_3 \cdots \xrightarrow{\text{condense }B_{n-1}} \CT_n~.
\ee
Our results show that $\CT_n$ only has invertible surface operators. Therefore, all non-invertible surface operators in the TQFT $\CT_1$ arise starting from $\CT_n$, and then sequentially gauging symmetries (implemented by surface operators) to move in the reverse direction in the sequence of TQFTs \eqref{eq:TQFTseq}.

Using \cite{thibault2022drinfeld}, we argued that our arguments above can be used to show that non-invertible symmetries in 2+1d QFTs are non-intrinsic as well. 

We reviewed the construction of surface operators by higher-gauging line operators and related it to the classification of surface operators in terms of surface action on the TQFT. In particular, we found an explicit criterion to determine the 1-form symmetry group and the discrete torsion data of the condensation defect from the action of the surface operator on the lines. Moreover, we derived an explicit criterion to determine the invertibility of a surface operator obtained from higher-gauging an invertible 1-form symmetry.

We reviewed the definition of time-reversal symmetry of a $2+1$d TQFT, and generalized it to include non-invertible time-reversal symmetries. We argued that non-invertible time-reversal symmetries of a TQFT described by an MTC, $\CC$, can be classified by studying the Lagrangian algebras in the category $\CC \boxtimes \CC$. We analyzed various explicit examples of Abelian TQFTs admitting invertible and non-invertible time-reversal symmetries and used our definition to understand the curious fact that non-invertible time reversal symmetries can sometimes act on lines in the same way as standard non-invertible symmetries, even when there are lines with complex topological spin. 

The results in this paper point to several questions to further understand non-invertible symmetries in 2+1d:
\begin{itemize}
\item SymTFTs are powerful tools to study possible topological manipulations of a QFT \cite{Freed:2012bs,Freed:2022qnc,Kaidi:2022cpf,Kaidi:2023maf}. As discussed above, any $2+1$d QFT with a non-invertible symmetry can be related to a QFT with only invertible surface operators through sequential gaugings. Therefore, the SymTFT of any $2+1$d QFT can be related to that of a QFT with only invertible surface operators. It will be very interesting to explore this direction further, and, in particular, to understand how taking the Drinfeld center of \eqref{introMorita} relates to constructing the SymTFT.
\item It will be interesting to understand gauging 0-form symmetries at the level of surface operators of a TQFT. This will make explicit how gauging 0-form symmetries of strongly anisotropic TQFTs lead to TQFTs with non-invertible symmetries. Better understanding this gauging may yield interesting insights into the classification of topological phases in $2+1$d.
\item  Motivated by the action on non-invertible symmetries in 1+1D, for example in the Ising model, one would expect that the genuine line operators that are set to zero by the non-invertible symmetry action are actually mapped to a twisted-sector line. It will be very interesting to show this explicitly, and more generally, to work out the action of the non-invertible symmetries discussed in this paper on the non-genuine lines bounding surface operators.
\item Many results discussed in this paper can be applied to spin TQFTs. It will be useful to systematically study possible condensations in this case.
\item We have seen glimmers that Galois theory has a role to play in our discussion. For example, some of the anisotropic TQFTs that play a role in the questions above also play a role in constructing Galois invariant theories. Moreover, time reversal symmetries are clearly related to Galois actions. It is tempting to try to understand this interplay even more generally.
\end{itemize}

\ack{We would like to thank Anuj Apte, Mahesh Balasubramanian, Francesco Benini, Lakshya Bhardwaj, Mathew Bullimore, Yichul Choi, Christian Copetti, Iñaki García Etxebarria, David Jordan, Zohar Komargodski, Kantaro Ohmori, Adrian Padellaro, Pavel Putrov, Shu-Heng Shao, Ingo Runkel and Yunqin Zheng for discussions and comments. M.~B. would like to thank SISSA, Nordita, and Durham University where part of this work was completed. R.~R. would like to thank Durham University, International Centre for Mathematical Sciences, International Centre for Theoretical Sciences, Nordita, Queen Mary University of London, Simons Centre for Geometry and Physics and SwissMap Research Station for hospitality during various stages of this project. Some of the results in this paper were first announced at Durham University. M.~B. is supported by the Royal Society under the grant, \lq\lq Relations, Transformations, and Emergence in Quantum Field Theory" and by STFC under the grant, \lq\lq Amplitudes, Strings and Duality." No new data were generated or analyzed in this study.}

\newpage
\begin{appendices}
\section{Relating our definitions to Lagrangian subalgebras in folded theories}
\label{AppA}
In this appendix, we explain how to relate our \lq\lq active" definition of symmetries on lines and fusion spaces to a more \lq\lq passive" definition in terms of boundaries and Lagrangian algebras in folded theories. We start with the case of Abelian TQFTs (see Sec. \ref{sec: Non-inv sym abelian TQFTs}) and then move on to non-Abelian TQFTs (see Sec. \ref{sec:noninv sym of non-abelian TQFTs}).

\subsec{The Abelian case}
Let us show that our definition of (non-invertible) symmetries of Abelian TQFTs is equivalent to the classification of (non-invertible) symmetries in terms of Lagrangian subgroups of the fusion group of the folded TQFT, $\CT \times \bar \CT$. If the original theory, $\CT$, has fusion group, $A$, the folded theory has fusion group, $A\times A$. Using $n_{S\cdot a}^b, n_{S\cdot c}^d\in \{0,1\}$, for non-zero $n_{S\cdot a}^{b}$ and $n_{S\cdot c}^d$ we can write condition \ref{def:noninvsym2} as 
\be
\label{eq:nnoninvsimpApp}
n_{S\cdot a}^bn_{S\cdot c}^d = n_{S\cdot(a\times c)}^{b\times d}~.
\ee
Given a set of non-zero integers, $n_{S\cdot a}^{b}$, satisfying \eqref{eq:nnoninvsimpApp} along with conditions \ref{def:noninvsym1} and \ref{def:noninvsym3}, the map
\be
n_{S\cdot a}^{b} \to (a,\bar b) \in \CT \times \bar \CT~,
\ee
defines a Lagrangian subgroup of the fusion group $A \times A$ of $\CT \times \bar \CT$. Conversely, given a Lagrangian subgroup, $L<A \times A$, we can define a non-invertible surface, $S_{L}$, whose action on the line operators are determined by the integers
\be
n_{S\cdot a}^{b}=\begin{cases} 
      1  & \text{ if } (a,\bar b) \in L~, \\
      0 & \text{ otherwise }~.
   \end{cases}
\ee
Therefore, our definition of non-invertible symmetries of $\CT$ is equivalent to the classification of (non-invertible) symmetries in terms of Lagrangian subgroups of the fusion group, $A \times A$, of the folded TQFT, $\CT \times \bar \CT$. 

\subsec{The non-Abelian case}
Let us show that our definition of (non-invertible) symmetries in non-Abelian TQFTs is equivalent to the description of surface operators in terms of Lagrangian algebras of $\CC \times \bar \CC$. To that end, suppose we have a map, $S$, satisfying the conditions in our non-Abelian definition. If $n_{S\cdot a}^b$ is non-zero, consider the algebra object
\be
\label{eq:defL}
L:=\sum_{a,b \in \CC} n_{Sa}^b (a,\bar b)\in\CC\times\bar\CC~.
\ee
Conditions \ref{def:nonabnoninvsym3}, \ref{def:nonabnoninvsym1}, and \ref{def:nonabnoninvsym2} imply that we can define a Lagrangian algebra structure on $L$. Indeed, using the ``folding trick" on Fig. \ref{fig:SactFdef}, we can convert the surface operator, $S$, into a boundary condition. $\tilde F$ is then related to the multiplication in the Lagrangian algebra, $L$,  describing this boundary condition \cite{Cong_2017}\cite[Appendix D]{Cheng_2020}.\footnote{We refer the reader to \cite[Appendix A]{choi2023comments} for a review of these constraints and the relation of $\tilde F$ to multiplication in the algebra in the context of 2d QFTs and their boundary conditions.} The constraint in Fig. \ref{fig:Sactioncond} ensures that $\tilde F$ defines an associative algebra structure on $L$. Moreover, the constraint in Fig. \ref{fig:constfrombraiding} ensures that this algebra is commutative. Using condition \ref{def:nonabnoninvsym2}, we have
\be
\text{dim}(L)=\sum_{a,b\in \CC} n_{Sa}^b d^2_{(a,\bar b)}=   \sum_{a,b\in \CC} n_{Sa}^b d_a^2d_b^2=\text{dim}(\CC)=\sqrt{\text{dim}(\CC\times \bar \CC)}~.
\ee
Also, since $n_{S1}^1=1$, the trivial line, $(1,1)\in \CC \times \bar \CC$, is an element of $L$ with multiplicity 1. Therefore, $L$ is a connected commutative algebra. Corollary 3.8 in \cite{Cong_2017} states that a connected commutative algebra, say $V=\sum_v n_v v$, in a unitary MTC,  $\CD$, with $\text{dim}(V)^2=\text{dim}(\CD)$ is a Lagrangian algebra if and only if the following inequality holds for all $v,w \in \CD$
\be
n_v n_w \leq \sum_u N_{vw}^u n_u~.
\ee
Using this theorem and condition \ref{def:nonabnoninvsym1}, we see that the algebra, $L$, defined above using the symmetry action, $S$, is indeed a Lagrangian algebra. 

Conversely, given a Lagrangian algebra 
\be
L=\sum_{(a,\bar b)\in \CC \times \bar \CC}n_{(a,\bar b)} (a,\bar b)~,
\ee
in $\CC \times \bar \CC$, we can define a non-invertible symmetry of the non-abelian TQFT with MTC, $\CC$, as
\be
n_{S\cdot a}^b:=n_{(a,\bar b)}~.
\ee
From the Lagrangian property of $L$, the integers $n_{Sa}^b$ satisfy the definition of a non-invertible symmetry. 

\section{Determining all symmetries in further Abelian examples}
\label{AppB}
In this section we determine all symmetries of several infinite classes of Abelian TQFTs. All non-invertible symmetries in these theories are of the form $S_B\times S_U$, where $S_B$ is a universal symmetry described in \eqref{eq:SBsym} and $S_U$ is an (invertible) unitary symmetry. 

\subsection{$B_{p^r}$ prime abelian TQFTs for odd prime $p$}
Consider the $B_{p^r}$ family of prime abelian TQFTs \cite{Wang_2020} with fusion rules given by the group, $\DZ_{p^r}$, and topological twists, $\theta_{\ell_a}=e^{\frac{2 \pi i a^2}{p^r}}$, where we have labelled the line operators as $\ell_a, a \in \{0,1,\dots,p^r\}$. Before discussing the general case, let us explicitly consider some theories with low rank: 
\begin{itemize}
\item Let us first consider the case $r=1$. We get a TQFT with $\DZ_p$ fusion rules  which does not have any non-trivial bosons. Therefore, this TQFT does not have any non-invertible symmetries. There is an invertible surface operator, $S_C$, which implements the charge conjugation symmetry. 
\item  Now, let us consider $r=2$. We get a TQFT with lines $\ell_0,\ell_1,\cdots, \ell_{p^2-1}$. We have the non-anomalous 1-form symmetry $B\cong \DZ_p=\{\ell_0,\ell_p\}$. We get a universal non-invertible surface operator, $S_B$, of the type \eqref{eq:SBsym}. We also have an invertible charge conjugation surface operator, $S_C$. We have the fusion rule $S_C \times S_B=S_B$. 
\item Now, let us consider $r=3$. We get a TQFT with lines $\ell_0,\ell_1,\cdots, \ell_{p^3-1}$. We have the non-anomalous 1-form symmetry $B\cong \DZ_p=\langle \ell_{p^2} \rangle$. We get a non-invertible surface operator, $S_B$, of the type \eqref{eq:SBsym}. We have an invertible charge conjugation surface operator, $S_C$. We also have the non-invertible surface operator $S_C \times S_B$.
\item Now, let us consider $r=4$. We get a TQFT with lines $\ell_0,\ell_1,\dots, \ell_{p^4-1}$. We have the non-anomalous 1-form symmetry $B\cong \DZ_{p^2}=\langle \ell_{p^2} \rangle$. We get a non-invertible surface operator, $S_B$, of the type \eqref{eq:SBsym}. We also get a non-invertible surface operator, $S_K$, for $K=\langle \ell_{p^{3}} \rangle \cong \DZ_{p} < B$. We also have an invertible charge conjugation surface operator, $S_C$. We have $S_C \times S_B =S_B$, while $S_C \times S_K$ is a new surface operator.   
\end{itemize}

For general $r$, from the expression for the topological spin of the line operators
\be
\theta_{\ell_a}=e^{\frac{2 \pi i a^2}{p^r}}~,
\ee
it is clear that the line operator, $\ell_{p^{\frac{r+1}{2}}}$, for odd $r$ and the line operator, $\ell_{p^{\frac{r}{2}}}$, for even $r$ are bosons. These line operators generate a $B\cong \DZ_{p^{\frac{r+1}{2}}}$ for odd $r$ and  $B\cong \DZ_{p^{\frac{r}{2}}}$ non-anomalous 1-form symmetry for even $r$. We get the non-invertible surface operator $S_B$ of type \eqref{eq:SBsym}. In fact, we get non-invertible surface operators for every subgroup $K < B$. We also have the invertible charge conjugation surface operator, $S_C$. Fusing the two types of surface operators we get new non-invertible surfaces $S_C \times S_K$. Suppose
\be
S_C \times S_K=S_K~.
\ee
For this to be true, the $K$-orbit of $\ell_a$ must be equal to the $K$-orbit of $\bar \ell_a$ for all $\ell_a$ that braids trivially with $K$. In other words, 
\be
\label{eq:fixedpointfusioncondp}
\ell_a + \ell_a \in K~,
\ee
for all $\ell_a$ that braid trivially with $K$. 

Suppose $r$ is even. Consider the group $K=\langle \ell_{p^{\frac{r}{2}+q}} \rangle$ for some $0\leq q \leq \frac{r}{2}$. For different values of $q$, we get all the subgroups of $B$. The line operators which braid trivially with $K$ satisfy
\be
e^{2 \pi i \frac{a p^{\frac{r}{2}+q}}{p^r}}=1 \implies a=0 \text{ mod }p^{\frac{r}{2}-q}~.
\ee
Imposing the condition \eqref{eq:fixedpointfusioncondp} we get
\be
l_{2 p^{\frac{r}{2}-q}} \in K~.
\ee
Since $l_{p^{\frac{r}{2}+q}} $ is a generator of $K$, there should be an integer $k$ such that
\be
kp^{\frac{r}{2}+q} =2 p^{\frac{r}{2}-q} \implies k= 2 p^{-2q}~.
\ee
Here $k$ is an integer only for $q=0$. Therefore, we get the fusion rule
\be
S_C \times S_B =S_B~.
\ee
For all proper subgroups $K < B$, we have 
\be
S_C \times S_K \neq S_K~.
\ee
Therefore, in total, we get $r+1$ surfaces. This counting agrees with the classification of surface operators of the $B_{p^r}$ TQFT in terms of condensation defects. Indeed, any subgroup of the $\DZ_{p^r}$ fusion group can be higher gauged. There are $r+1$ distinct subgroups leading to $r+1$ surface operators. The exact map between the surface operators discussed above and condensation defects can be explicitly worked out using the discussion in section \ref{sec:Sactiontocond}.

Suppose $r$ is odd. Consider the group $K=\langle \ell_{p^{\frac{r+1}{2}+q}} \rangle$ for some $0\leq q \leq \frac{r+1}{2}$. For different values of $q$, we get all the subgroups of $B$. The line operators which braid trivially with $K$ satisfy
\be
e^{2 \pi i \frac{p^{\frac{r+1}{2}+q}  a}{p^r}}=1 \implies a=0 \text{ mod }p^{\frac{r-1}{2}-q}~.
\ee
Imposing the condition \eqref{eq:fixedpointfusioncondp} we get
\be
\ell_{2 p^{\frac{r-1}{2}-q}} \in K~.
\ee
Since $\ell_{p^{\frac{r+1}{2}+q}} $ is a generator of $K$, there should be an integer $k$ such that
\be
kp^{\frac{r+1}{2}+q} =2 p^{\frac{r-1}{2}-q} \implies k= 2 p^{-2q-1}~.
\ee
Therefore, there is no integer solution for $k$ for any $q$. This argument shows that when $r$ is odd, we have $S_C \times S_K \neq S_K$ for all $K$. Again, we find that there are $r+1$ surface operators in total, which agrees with the classification of surface operators in terms of condensation defects. 

$A_{p^r}$ TQFTs is another class of prime abelian TQFTs which have the same fusion group as $B_{2^r}$ TQFTs but different topological spins. The classification of surface operators of $A_{p^r}$ is identical to that of $B_{p^r}$ since these two classes of TQFTs are related by a Galois conjugation \cite{Buican:2021axn}. 

\subsection{$A_{2^r}$ prime abelian TQFTs $\cong U(1)_{2^r}$ Chern-Simons Theory}

Next, consider the $A_{2^r}$ family of prime abelian TQFTs with fusion rules given by the group, $\DZ_{2^r}$, and topological spins, $\theta_{\ell_a}=e^{\frac{2 \pi i a^2}{2^{r+1}}}$. Before discussing the general case, let us explicitly consider some theories with low rank: 
\begin{itemize}
\item Let us first consider the case $r=1$. We get the Semion model, which does not have any non-trivial bosons. Therefore, this TQFT does not have any non-invertible symmetries. 
\item  Now, let us consider $r=2$. We get a TQFT with lines $\ell_0,\ell_1,\ell_2,\ell_3$ and topological twists $\theta_{\ell_0}=1,\theta_{\ell_1}=e^{\frac{2 \pi i}{8}},\theta_{\ell_2}=-1,\theta_{\ell_3}=e^{\frac{2 \pi i}{8}}$. Once again, there are no bosons, so this theory does not have any non-invertible symmetries. However, there is clearly an invertible symmetries which permutes $\ell_1\leftrightarrow \ell_3$. This invertible surface operator can be obtained from higher-gauging the fermionic line operator, $\ell_2$. 
\item Now, let us consider $r=3$. We get the line operators $\ell_0,\ell_1,\ell_2,\ell_3,\ell_4,\ell_5,\ell_6,\ell_7$ and topological spins $\theta_{\ell_0}=1,\theta_{\ell_1}=e^{\frac{2 \pi i}{16}},\theta_{\ell_2}=e^{\frac{2 \pi i}{4}},\theta_{\ell_3}=e^{\frac{2 \pi i 9}{16}},\theta_{\ell_4}=1,\theta_{\ell_5}=e^{\frac{2 \pi i 9}{16}},\theta_{\ell_6}=e^{\frac{2 \pi i}{4}},\theta_{\ell_7}=e^{\frac{2 \pi i 1}{16}}$. We have a boson which forms the non-anomalous 1-form symmetry group $B=\{\ell_0,\ell_4\}$. We have a universal non-invertible symmetry of the type \eqref{eq:SBsym} implemented by the surface $S_{B}$. The action of this surface on the line operators is
\be
S_B \cdot \ell_0 = \ell_0+\ell_4, ~ S_B \cdot \ell_2 = \ell_2+\ell_6, ~ S_B \cdot \ell_4 = \ell_0+\ell_4,~ S_B \cdot \ell_6 = \ell_2+\ell_6~,
\ee
and the action on $\ell_1,\ell_3,\ell_5,\ell_7$ are all zero. We also have the charge conjugation symmetry implemented by an invertible surface, say $S_C$. Moreover, $S_B \times S_C=S_B$. This exhausts all non-invertible symmetries of this TQFT. 
\item Let $r=4$. In this case, we have the bosonic line operators $B=\{\ell_0,\ell_8,\ell_{16},\ell_{24}\}\cong \DZ_4$. We get the non-invertible surface operators, $S_B$, as well as $S_{K}$, where $K=\{\ell_0,\ell_{16}\}\subset B$. Moreover, we have the charge conjugation surface operator $S_C$. By fusing these surface operators, we get the non-invertible surface $S_B \times S_C=S_C \times S_B=S_B$ and  $S_K \times S_C=S_C \times S_K$. In total, we get four non-trivial surface operators. 
\end{itemize}

Now let us consider the case of general $r$. To identity the non-invertible symmetries, we need to find the structure of bosons in the theory. Suppose $n$ is an integer such that $r=2n+1$ or $r=2n+2$. From the expression for the topological spin
\be
\theta_{\ell_a}=e^{\frac{2\pi i a^2}{2^{r+1}}}~,
\ee
it is clear that the line operator $\ell_{2^{r-n}}$ is a boson that generates a non-anomalous $B\cong \DZ_{2^n}$ subgroup of the full 1-form symmetry group. Let $S_B$ be the non-invertible surface operator defined by $B$. There are also non-invertible surface operators $S_K$ defined by a subgroup $K<B$. In total, we get $n$ such non-invertible surface operators. Also, we have the charge conjugation invertible surface operator, $S_C$. All other non-invertible surface operators of this TQFT are obtained from fusing $S_K$ and $S_C$. We have,
\be
(S_C \times S_K) \cdot \ell_a= S_C \cdot (\text{$K$-orbit of } \ell_a) =\text{ $K$-orbit of } \bar \ell_a= (S_K \times S_C) \cdot \ell_a~,
\ee
if $\ell_a$ braids trivially with $K$ and zero otherwise. Suppose we have
\be
S_C \times S_K= S_C \times S_{K'}~.
\ee
Then the $K$-orbit of $\ell_a$ must be equal to the $K'$-orbit of $\ell_a$, which is possible only if $K=K'$. Now, suppose
\be
S_C \times S_K=S_K~.
\ee
For this equality to hold, the $K$-orbit of $\ell_a$ must be equal to the $K$-orbit of $\bar \ell_a$ for all $\ell_a$ that braids trivially with $K$. In other words, 
\be
\label{eq:fixedpointfusioncond}
\ell_a + \ell_a \in K~,
\ee
for all $\ell_a$ that braids trivially with $K$. We have $B=\langle \ell_{2^{r-n}}\rangle$. Then the line operators which braid with $B$ satisfy
\be 
e^{2 \pi i \frac{2^{r-n} a}{2^r}}=1 \implies a=0 \text{ mod }2^n~.
\ee
Imposing the condition \eqref{eq:fixedpointfusioncond} we get
\be
\ell_{2^{n+1}} \in B~.
\ee
Since $\ell_{2^{r-n}}$ is a generator of $K$, there should be an integer, $k$, such that
\be
k2^{(r-n)}=2^{n+1} \implies k= 2^{2n+1-r}~.
\ee
Here, $r$ is either $2n+1$ or $2n+2$. In the former case $k=1$, and in the later case we don't have an integer solution for $k$. Therefore, if $r=2n+1$ we have the fusion rules 
\be
S_C \times S_B = S_B~,
\ee
and, if $r=2n+2$, then $S_C \times S_B$ is a new non-invertible surface operator. 

Now let us consider the group $K=\langle \ell_{2^{r-n+q}} \rangle$ for some $0\leq q \leq n$. For different values of $q$, we get all the subgroups of $B$. The line operators which braid trivially with $K$ satisfy
\be
e^{2 \pi i \frac{2^{r-n+q} a}{2^r}}=1 \implies a=0 \text{ mod }2^{n-q}~.
\ee
Imposing the condition \eqref{eq:fixedpointfusioncond} we get
\be
\ell_{2^{n-q+1}} \in K~.
\ee
Since $\ell_{2^{r-n-q}}$ is a generator of $K$, there should be an integer $k$ such that
\be
k2^{(r-n+q)}=2^{n-q+1} \implies k= 2^{2n+1-r-2q}~.
\ee
For $r=2n+1$ we get $k={-2q}$, which is an integer only for $q=0$. For $r=2n+2$, we get $k=2^{-1-2q}$ which is not an integer for any $q$. Therefore, the fusions
\be
S_C \times S_K \neq S_K~,
\ee
for $q>0$ and any value of $r$. 

$A_{2^r}$ TQFTs can be realized as $U(1)_{2^r}$ Chern-Simons theory. The surface operators in this Chern-Simons theory were studied in \cite{Kapustin_2011, Roumpedakis:2022aik}. Our classification of surface operators above agrees with the classification in these papers. There are other prime abelian TQFTs labelled as $B_{2^r}, C_{2^r}$ and $D_{2^r}$ TQFTs which have the same fusion group as $A_{2^r}$ but different topological spins \cite{Wang_2020}. The classification of surface operators in these latter theories are identical to that in $A_{2^r}$ since these theories are related by Galois conjugation \cite{Buican:2021axn}.  

Note that in the examples studied so far, the full set of surface operators can be constructed by first finding the invertible symmetries, then finding the $S_B$ type surface operators defined in \eqref{eq:SBsym} for every non-anomalous 1-form symmetry group $B$ of the TQFT and then studying their fusions. It would be interesting to understand if this is a general property of symmetries in Abelian TQFT.

\section{Action on twisted sectors}

\label{sec:twisted}

Throughout this paper we have been concerned with the action of symmetries on genuine lines. That is to say, we only considered the symmetry action of surfaces on lines that are not attached to surfaces. In this appendix, we begin to generalize this study (leaving a more complete picture for future work).

Indeed, the action of a surface operator on the genuine line operators of a QFT can be extended to an action on the non-genuine line operators that live at the end of surface operators (a.k.a. \lq\lq twisted sector" lines). Similar to the various conditions the action of a surface operator on genuine lines satisfy, we get various constraints on the surface action on twisted sectors. 

\begin{figure}[h!]
    \centering

\tikzset{every picture/.style={line width=0.75pt}} %set default line width to 0.75pt        

\begin{tikzpicture}[x=0.75pt,y=0.75pt,yscale=-1,xscale=1]

%Shape: Rectangle [id:dp6960540634213839] 
\draw  [color={rgb, 255:red, 96; green, 116; blue, 133 }  ,draw opacity=1 ][fill={rgb, 255:red, 152; green, 189; blue, 214 }  ,fill opacity=1 ] (202.45,85.7) -- (354.1,85.7) -- (354.1,177.62) -- (202.45,177.62) -- cycle ;
%Shape: Parallelogram [id:dp6562840942953989] 
\draw  [color={rgb, 255:red, 245; green, 166; blue, 35 }  ,draw opacity=1 ][fill={rgb, 255:red, 248; green, 198; blue, 155 }  ,fill opacity=1 ] (297.19,250) -- (297.22,115.92) -- (416.3,54.7) -- (416.28,188.78) -- cycle ;
%Straight Lines [id:da550732043590615] 
\draw [color={rgb, 255:red, 139; green, 6; blue, 24 }  ,draw opacity=1 ] [dash pattern={on 4.5pt off 4.5pt}]  (299.55,177.34) -- (353.81,176.34) ;
%Straight Lines [id:da3364633224260798] 
\draw [color={rgb, 255:red, 139; green, 6; blue, 24 }  ,draw opacity=1 ]   (202.45,177.62) -- (249.65,178.1) -- (297.94,177.34) ;
\draw  [color={rgb, 255:red, 139; green, 6; blue, 24 }  ,draw opacity=1 ][fill={rgb, 255:red, 139; green, 6; blue, 24 }  ,fill opacity=1 ] (242.14,180.57) -- (238.37,178.08) -- (242.22,175.72) -- (240.28,178.11) -- cycle ;
%Shape: Rectangle [id:dp8161010067374325] 
\draw  [color={rgb, 255:red, 109; green, 107; blue, 107 }  ,draw opacity=1 ][fill={rgb, 255:red, 192; green, 188; blue, 188 }  ,fill opacity=1 ] (356.69,85.64) -- (499.05,85.64) -- (499.05,176.34) -- (356.69,176.34) -- cycle ;
%Straight Lines [id:da007882411965093872] 
\draw [color={rgb, 255:red, 139; green, 6; blue, 24 }  ,draw opacity=1 ]   (357.81,176.34) -- (499.05,176.44) ;
\draw  [color={rgb, 255:red, 139; green, 6; blue, 24 }  ,draw opacity=1 ][fill={rgb, 255:red, 139; green, 6; blue, 24 }  ,fill opacity=1 ] (459.34,178.61) -- (455.58,176.11) -- (459.44,173.76) -- (457.48,176.15) -- cycle ;

% Text Node
\draw (398.67,63.95) node [anchor=north west][inner sep=0.75pt]    {$S$};
% Text Node
\draw (500.96,169.02) node [anchor=north west][inner sep=0.75pt]    {$a^{( S_{1})}$};
% Text Node
\draw (478.88,91.24) node [anchor=north west][inner sep=0.75pt]    {$S_{1}$};
% Text Node
\draw (205.87,89.24) node [anchor=north west][inner sep=0.75pt]    {$S_{2}$};
% Text Node
\draw (165.1,170.2) node [anchor=north west][inner sep=0.75pt]    {$b^{( S_{2})}$};

\end{tikzpicture}
    \caption{Action of a surface operator on a non-genuine line.}
    \label{fig:Sacttwisted}
\end{figure}
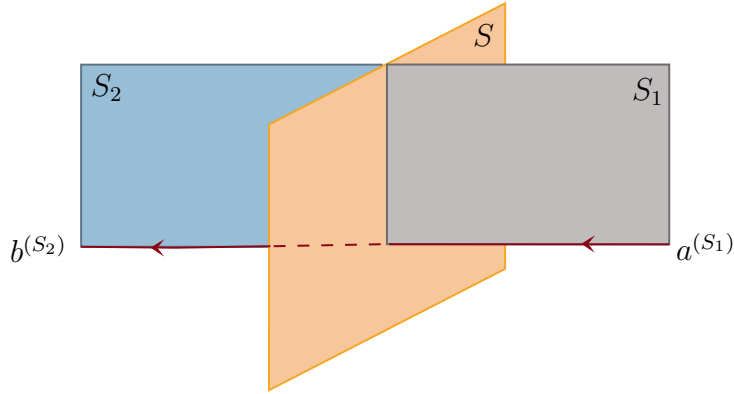

\begin{figure}[h!]
    \centering

\tikzset{every picture/.style={line width=0.75pt}} %set default line width to 0.75pt        

\begin{tikzpicture}[x=0.75pt,y=0.75pt,yscale=-1,xscale=1]
%uncomment if require: \path (0,496); %set diagram left start at 0, and has height of 496

%Shape: Rectangle [id:dp6960540634213839] 
\draw  [color={rgb, 255:red, 96; green, 116; blue, 133 }  ,draw opacity=1 ][fill={rgb, 255:red, 152; green, 189; blue, 214 }  ,fill opacity=1 ] (232.78,66.99) -- (362.06,66.99) -- (362.06,143.86) -- (232.78,143.86) -- cycle ;
%Shape: Rectangle [id:dp3772021515349301] 
\draw  [color={rgb, 255:red, 96; green, 116; blue, 133 }  ,draw opacity=1 ][fill={rgb, 255:red, 152; green, 189; blue, 214 }  ,fill opacity=1 ] (186.74,93.94) -- (316.03,93.94) -- (316.03,170.82) -- (186.74,170.82) -- cycle ;
%Shape: Parallelogram [id:dp6562840942953989] 
\draw  [color={rgb, 255:red, 245; green, 166; blue, 35 }  ,draw opacity=1 ][fill={rgb, 255:red, 248; green, 198; blue, 155 }  ,fill opacity=1 ] (275.25,237.46) -- (275.27,121.14) -- (416.18,32.61) -- (416.16,148.93) -- cycle ;
%Straight Lines [id:da550732043590615] 
\draw [color={rgb, 255:red, 139; green, 6; blue, 24 }  ,draw opacity=1 ] [dash pattern={on 4.5pt off 4.5pt}]  (278.63,144.33) -- (362.06,143.86) ;
%Straight Lines [id:da3364633224260798] 
\draw [color={rgb, 255:red, 139; green, 6; blue, 24 }  ,draw opacity=1 ] [dash pattern={on 4.5pt off 4.5pt}]  (232.78,143.86) -- (274.63,144.33) ;
\draw  [color={rgb, 255:red, 139; green, 6; blue, 24 }  ,draw opacity=1 ][fill={rgb, 255:red, 139; green, 6; blue, 24 }  ,fill opacity=1 ] (267.72,146.37) -- (264.5,144.29) -- (267.78,142.32) -- (266.13,144.32) -- cycle ;
%Shape: Rectangle [id:dp8161010067374325] 
\draw  [color={rgb, 255:red, 109; green, 107; blue, 107 }  ,draw opacity=1 ][fill={rgb, 255:red, 192; green, 188; blue, 188 }  ,fill opacity=1 ] (362.06,66.99) -- (483.42,66.99) -- (483.42,142.83) -- (362.06,142.83) -- cycle ;
%Straight Lines [id:da007882411965093872] 
\draw [color={rgb, 255:red, 139; green, 6; blue, 24 }  ,draw opacity=1 ]   (362.91,143.86) -- (483.31,143.95) ;
\draw  [color={rgb, 255:red, 139; green, 6; blue, 24 }  ,draw opacity=1 ][fill={rgb, 255:red, 139; green, 6; blue, 24 }  ,fill opacity=1 ] (452.02,145.57) -- (448.81,143.48) -- (452.1,141.52) -- (450.44,143.51) -- cycle ;
%Shape: Rectangle [id:dp7519728667469658] 
\draw  [color={rgb, 255:red, 109; green, 107; blue, 107 }  ,draw opacity=1 ][fill={rgb, 255:red, 192; green, 188; blue, 188 }  ,fill opacity=1 ] (316.77,93.9) -- (438.13,93.9) -- (438.13,169.74) -- (316.77,169.74) -- cycle ;
%Straight Lines [id:da8342289947429914] 
\draw [color={rgb, 255:red, 139; green, 6; blue, 24 }  ,draw opacity=1 ]   (316.03,169.82) -- (438.13,169.74) ;
\draw  [color={rgb, 255:red, 139; green, 6; blue, 24 }  ,draw opacity=1 ][fill={rgb, 255:red, 139; green, 6; blue, 24 }  ,fill opacity=1 ] (381.26,172.33) -- (378.06,170.24) -- (381.35,168.28) -- (379.68,170.27) -- cycle ;
%Straight Lines [id:da8088359786823808] 
\draw [color={rgb, 255:red, 139; green, 6; blue, 24 }  ,draw opacity=1 ] [dash pattern={on 4.5pt off 4.5pt}]  (272.31,170.05) -- (316.03,169.82) ;
%Straight Lines [id:da8069500827877811] 
\draw [color={rgb, 255:red, 139; green, 6; blue, 24 }  ,draw opacity=1 ]   (186.74,170.82) -- (272.31,170.05) ;
\draw  [color={rgb, 255:red, 139; green, 6; blue, 24 }  ,draw opacity=1 ][fill={rgb, 255:red, 139; green, 6; blue, 24 }  ,fill opacity=1 ] (237.03,172.13) -- (233.82,170.05) -- (237.09,168.07) -- (235.44,170.08) -- cycle ;
%Straight Lines [id:da5751545262143923] 
\draw [color={rgb, 255:red, 96; green, 116; blue, 133 }  ,draw opacity=1 ] [dash pattern={on 4.5pt off 4.5pt}]  (232.65,96.47) -- (232.78,143.86) ;
%Shape: Rectangle [id:dp5517448955433528] 
\draw  [color={rgb, 255:red, 96; green, 116; blue, 133 }  ,draw opacity=1 ][fill={rgb, 255:red, 152; green, 189; blue, 214 }  ,fill opacity=1 ] (352.38,336.2) -- (464.61,336.2) -- (464.61,417.78) -- (352.38,417.78) -- cycle ;
%Shape: Parallelogram [id:dp020339625721702892] 
\draw  [color={rgb, 255:red, 245; green, 166; blue, 35 }  ,draw opacity=1 ][fill={rgb, 255:red, 248; green, 198; blue, 155 }  ,fill opacity=1 ] (422.49,482) -- (422.51,363.04) -- (510.63,308.7) -- (510.61,427.67) -- cycle ;
%Straight Lines [id:da8599958019584925] 
\draw [color={rgb, 255:red, 139; green, 6; blue, 24 }  ,draw opacity=1 ] [dash pattern={on 4.5pt off 4.5pt}]  (424.24,417.52) -- (464.39,416.64) ;
%Straight Lines [id:da7300097754180602] 
\draw [color={rgb, 255:red, 139; green, 6; blue, 24 }  ,draw opacity=1 ]   (352.38,417.78) -- (387.31,418.2) -- (423.04,417.52) ;
\draw  [color={rgb, 255:red, 139; green, 6; blue, 24 }  ,draw opacity=1 ][fill={rgb, 255:red, 139; green, 6; blue, 24 }  ,fill opacity=1 ] (381.75,420.39) -- (378.96,418.18) -- (381.81,416.09) -- (380.37,418.21) -- cycle ;
%Shape: Rectangle [id:dp9850677574445172] 
\draw  [color={rgb, 255:red, 109; green, 107; blue, 107 }  ,draw opacity=1 ][fill={rgb, 255:red, 192; green, 188; blue, 188 }  ,fill opacity=1 ] (466.52,336.16) -- (571.87,336.16) -- (571.87,416.64) -- (466.52,416.64) -- cycle ;
%Straight Lines [id:da8034677991734616] 
\draw [color={rgb, 255:red, 139; green, 6; blue, 24 }  ,draw opacity=1 ]   (467.35,416.64) -- (571.87,416.73) ;
\draw  [color={rgb, 255:red, 139; green, 6; blue, 24 }  ,draw opacity=1 ][fill={rgb, 255:red, 139; green, 6; blue, 24 }  ,fill opacity=1 ] (542.48,418.66) -- (539.7,416.43) -- (542.55,414.35) -- (541.11,416.47) -- cycle ;

% Text Node
\draw (279.73,209.95) node [anchor=north west][inner sep=0.75pt]    {$S$};
% Text Node
\draw (484.43,138.37) node [anchor=north west][inner sep=0.75pt]    {$a^{( S_{1})}$};
% Text Node
\draw (466.42,67.75) node [anchor=north west][inner sep=0.75pt]    {$S_{1}$};
% Text Node
\draw (235.39,68.59) node [anchor=north west][inner sep=0.75pt]    {$S_{2}$};
% Text Node
\draw (198.39,133.46) node [anchor=north west][inner sep=0.75pt]    {$b^{( S_{2})}$};
% Text Node
\draw (417.83,97.02) node [anchor=north west][inner sep=0.75pt]    {$S_{3}$};
% Text Node
\draw (187.05,95.38) node [anchor=north west][inner sep=0.75pt]    {$S_{4}$};
% Text Node
\draw (436.8,172.57) node [anchor=north west][inner sep=0.75pt]    {$a^{( S_{2})}$};
% Text Node
\draw (165.54,174.33) node [anchor=north west][inner sep=0.75pt]    {$b^{( S_{4})}$};
% Text Node
\draw (340.6,258) node [anchor=north west][inner sep=0.75pt]  [rotate=-90]  {$\Longrightarrow $};
% Text Node
\draw (427.03,454.05) node [anchor=north west][inner sep=0.75pt]    {$S$};
% Text Node
\draw (576.25,409.18) node [anchor=north west][inner sep=0.75pt]    {$a^{( S_{i})}$};
% Text Node
\draw (552.47,340.15) node [anchor=north west][inner sep=0.75pt]    {$S_{i}$};
% Text Node
\draw (352.44,338.38) node [anchor=north west][inner sep=0.75pt]    {$S_{j}$};
% Text Node
\draw (318.58,410.22) node [anchor=north west][inner sep=0.75pt]    {$b^{( S_{j})}$};
% Text Node
\draw (0,362.4) node [anchor=north west][inner sep=0.75pt]    {$\sum\limits_{S_{i} \in S_{1} \times S_{3} ,\ S_{j} \in S_{2} \times S_{4}} \ \sum\limits_{a^{( S_{i})} ,a^{( S_{j})}} \ N_{a^{( S_{1})} a^{( S_{2})}}^{a^{( S_{i})}} N_{b^{( S_{1})} b^{( S_{2})}}^{b^{( S_{i})}}$};

\end{tikzpicture}
    \caption{Condition on $S$-action from fusion of non-genuine lines}
    \label{fig:Sacttwistconst}
\end{figure}
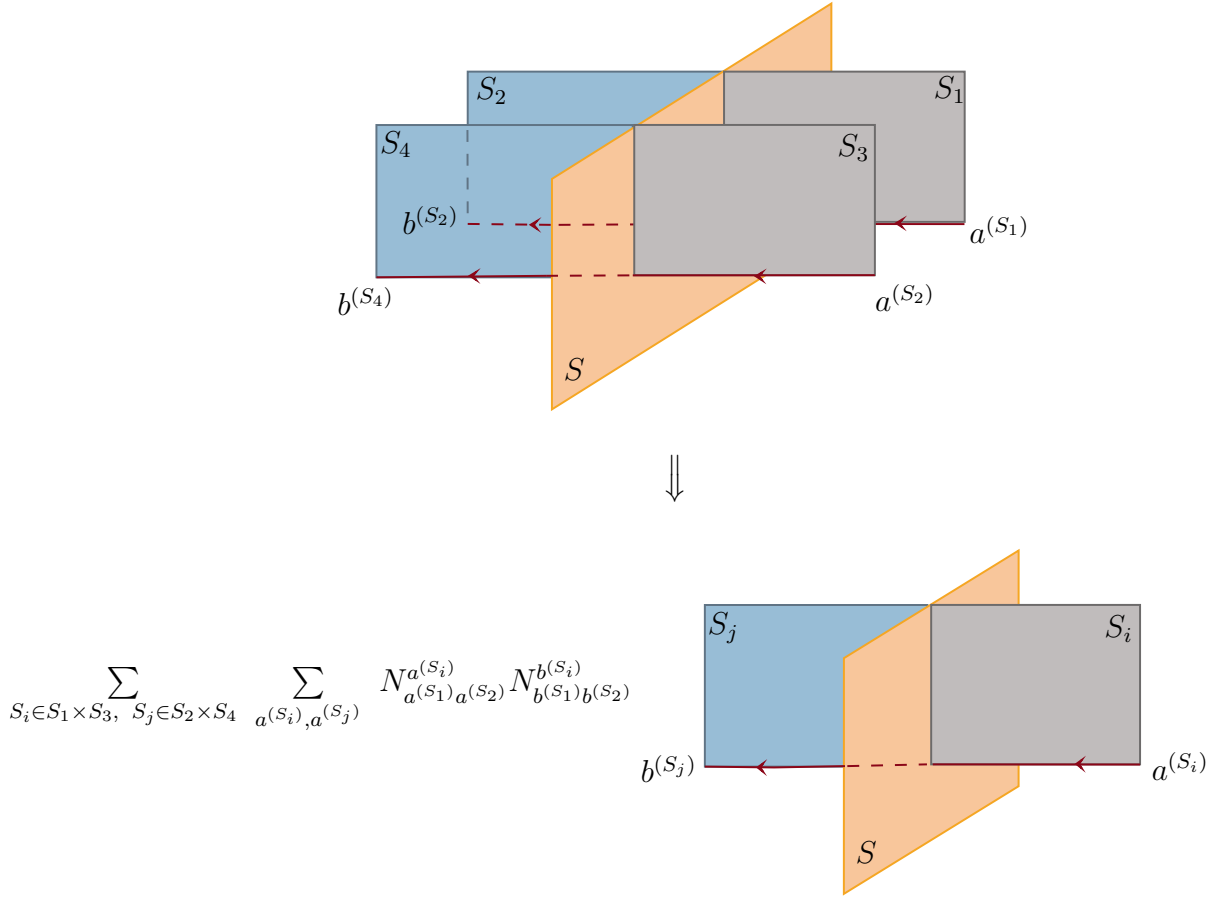

Let $a^{(S_1)}$ be a non-genuine line at the end of surface operators $S_1$. Then the action of a surface operator, $S$, on $a^{(S_1)}$ is given by 
\be
S \cdot a^{(S_1)}= \sum_{S_2} \sum_{b^{(S_2)}} n_{S \cdot a^{(S_1)}}^{b^{(S_2)}} ~ b^{(S_2)}~,
\ee
where the fusion coefficient $n_{S \cdot a^{(S_1)}}^{b^{(S_2)}}\ne0$ if Fig. \ref{fig:Sacttwisted} is consistent. For $n_{S \cdot a^{(S_1)}}^{b^{(S_2)}}$ to be non-zero, there are various constraints on $S_2$ and $b^{(S_2)}$ that must be satisfied. First of all, there must a non-trivial topological junction between the surface operators
\be
S_1 \times S \text{ and } S \times S_2~.
\ee
In other words,
\be
\label{eq:twistactcond1}
\text{Hom}(S_1 \times S,S \times S_2)\neq 0~.
\ee
In particular, if the surface operators are all invertible, then we get the constraint 
\be
S_2 = S^{-1} \times S_1 \times S~.
\ee
That is, $S_2$ and $S_1$ should be related by conjugation. Also, we can generalize Fig. \ref{fig:nonabnoninvcond} to get Fig. \ref{fig:Sacttwistconst}.
From this diagram, we get the constraint
\be
\label{eq:twistactcond2}
n_{S a^{(S_1)}}^{b^{(S_2)}}n_{S a^{(S_3)}}^{b^{(S_4)}} ~ \leq \sum\limits_{S_{i} \in S_{1}  \times S_{3} ,\ S_{j} \in S_{2} \times S_{4}} \ \sum\limits_{a^{( S_{i})} ,a^{( S_{j})}} \ N_{a^{( S_{1})} a^{( S_{2})}}^{a^{( S_{i})}} N_{b^{( S_{1})} b^{( S_{2})}}^{b^{( S_{i})}} ~~ n_{S a^{(S_i)}}^{b^{(S_j)}}~.
\ee
In the rest of this appendix, we will focus on some examples of abelian TQFTs and study the action of invertible line operators on the non-genuine line operators. 

\subsection{Semion $\times$ Semion TQFT}

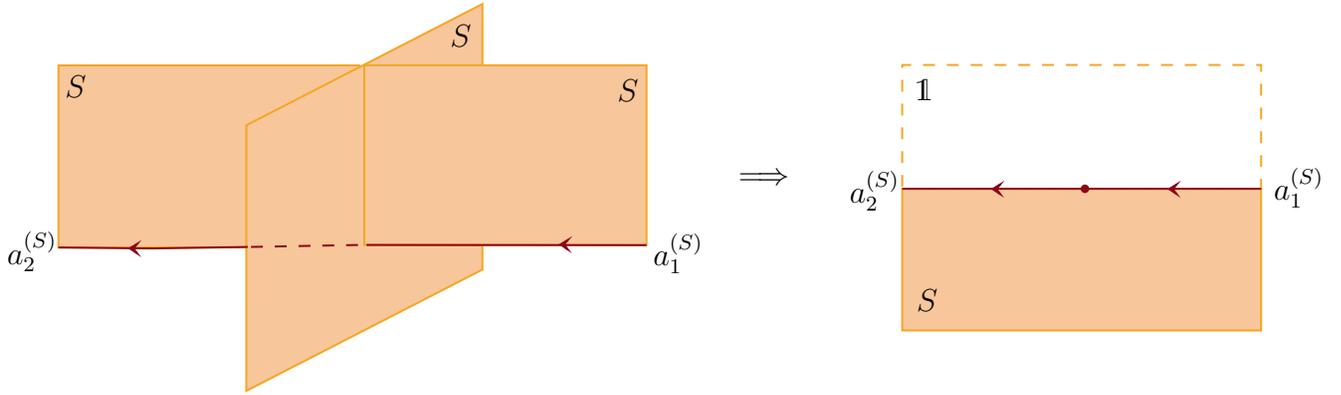
\begin{figure}[h!]
    \centering

\tikzset{every picture/.style={line width=0.75pt}} %set default line width to 0.75pt        

\begin{tikzpicture}[x=0.75pt,y=0.75pt,yscale=-1,xscale=1]

%Shape: Rectangle [id:dp7703513000869485] 
\draw  [color={rgb, 255:red, 245; green, 166; blue, 35 }  ,draw opacity=1 ][dash pattern={on 4.5pt off 4.5pt}] (469,88.5) -- (650,88.5) -- (650,151) -- (469,151) -- cycle ;
%Shape: Rectangle [id:dp6960540634213839] 
\draw  [color={rgb, 255:red, 245; green, 166; blue, 35 }  ,draw opacity=1 ][fill={rgb, 255:red, 248; green, 198; blue, 155 }  ,fill opacity=1 ] (43.45,88.7) -- (195.1,88.7) -- (195.1,180.62) -- (43.45,180.62) -- cycle ;
%Shape: Parallelogram [id:dp6562840942953989] 
\draw  [color={rgb, 255:red, 245; green, 166; blue, 35 }  ,draw opacity=1 ][fill={rgb, 255:red, 248; green, 198; blue, 155 }  ,fill opacity=1 ] (138.19,253) -- (138.22,118.92) -- (257.3,57.7) -- (257.28,191.78) -- cycle ;
%Straight Lines [id:da550732043590615] 
\draw [color={rgb, 255:red, 139; green, 6; blue, 24 }  ,draw opacity=1 ] [dash pattern={on 4.5pt off 4.5pt}]  (140.55,180.34) -- (194.81,179.34) ;
%Straight Lines [id:da3364633224260798] 
\draw [color={rgb, 255:red, 139; green, 6; blue, 24 }  ,draw opacity=1 ]   (43.45,180.62) -- (90.65,181.1) -- (138.94,180.34) ;
\draw  [color={rgb, 255:red, 139; green, 6; blue, 24 }  ,draw opacity=1 ][fill={rgb, 255:red, 139; green, 6; blue, 24 }  ,fill opacity=1 ] (83.14,183.57) -- (79.37,181.08) -- (83.22,178.72) -- (81.28,181.11) -- cycle ;
%Shape: Rectangle [id:dp8161010067374325] 
\draw  [color={rgb, 255:red, 245; green, 166; blue, 35 }  ,draw opacity=1 ][fill={rgb, 255:red, 248; green, 198; blue, 155 }  ,fill opacity=1 ] (197.69,88.64) -- (340.05,88.64) -- (340.05,179.34) -- (197.69,179.34) -- cycle ;
%Straight Lines [id:da007882411965093872] 
\draw [color={rgb, 255:red, 139; green, 6; blue, 24 }  ,draw opacity=1 ]   (198.81,179.34) -- (340.05,179.44) ;
\draw  [color={rgb, 255:red, 139; green, 6; blue, 24 }  ,draw opacity=1 ][fill={rgb, 255:red, 139; green, 6; blue, 24 }  ,fill opacity=1 ] (300.34,181.61) -- (296.58,179.11) -- (300.44,176.76) -- (298.48,179.15) -- cycle ;
%Shape: Rectangle [id:dp7873746155449598] 
\draw  [color={rgb, 255:red, 245; green, 166; blue, 35 }  ,draw opacity=1 ][fill={rgb, 255:red, 248; green, 198; blue, 155 }  ,fill opacity=1 ] (469,151) -- (650,151) -- (650,222.5) -- (469,222.5) -- cycle ;
%Straight Lines [id:da7686150260254263] 
\draw [color={rgb, 255:red, 139; green, 6; blue, 24 }  ,draw opacity=1 ]   (469,151) -- (650,151) ;
%Shape: Circle [id:dp9274683732903845] 
\draw  [color={rgb, 255:red, 139; green, 6; blue, 24 }  ,draw opacity=1 ][fill={rgb, 255:red, 139; green, 6; blue, 24 }  ,fill opacity=1 ] (559.5,151) .. controls (559.5,150.09) and (560.24,149.35) .. (561.15,149.35) .. controls (562.06,149.35) and (562.8,150.09) .. (562.8,151) .. controls (562.8,151.91) and (562.06,152.65) .. (561.15,152.65) .. controls (560.24,152.65) and (559.5,151.91) .. (559.5,151) -- cycle ;
\draw  [color={rgb, 255:red, 139; green, 6; blue, 24 }  ,draw opacity=1 ][fill={rgb, 255:red, 139; green, 6; blue, 24 }  ,fill opacity=1 ] (607.34,153.61) -- (603.58,151.11) -- (607.44,148.76) -- (605.48,151.15) -- cycle ;
\draw  [color={rgb, 255:red, 139; green, 6; blue, 24 }  ,draw opacity=1 ][fill={rgb, 255:red, 139; green, 6; blue, 24 }  ,fill opacity=1 ] (518.34,153.61) -- (514.58,151.11) -- (518.44,148.76) -- (516.48,151.15) -- cycle ;

% Text Node
\draw (239.67,66.95) node [anchor=north west][inner sep=0.75pt]    {$S$};
% Text Node
\draw (341.96,172.02) node [anchor=north west][inner sep=0.75pt]    {$a_{1}^{( S)}$};
% Text Node
\draw (323.88,94.24) node [anchor=north west][inner sep=0.75pt]    {$S$};
% Text Node
\draw (45.45,92.1) node [anchor=north west][inner sep=0.75pt]    {$S$};
% Text Node
\draw (16.1,171.2) node [anchor=north west][inner sep=0.75pt]    {$a_{2}^{( S)}$};
% Text Node
\draw (385,141.4) node [anchor=north west][inner sep=0.75pt]    {$\Longrightarrow $};
% Text Node
\draw (441.1,140.2) node [anchor=north west][inner sep=0.75pt]    {$a_{2}^{( S)}$};
% Text Node
\draw (654.96,139.02) node [anchor=north west][inner sep=0.75pt]    {$a_{1}^{( S)}$};
% Text Node
\draw (474.88,200.24) node [anchor=north west][inner sep=0.75pt]    {$S$};
% Text Node
\draw (474,94.4) node [anchor=north west][inner sep=0.75pt]    {$\mathds{1}$};
\end{tikzpicture}
\caption{Action of $S$ on $a_{1,2}^{(S)}$}
\label{fig:semsemact}
\end{figure}

Consider the Semion $\times$ Semion TQFT with line operators $1,s,\tilde s, s\tilde s$ with topological twists $1,i,i,-1$, respectively. This theory has a $\DZ_2$ symmetry implemented by a surface operator, $S$, which acts as
\be
S\cdot s=\tilde s~.
\ee
Note that $S$ has two topological boundary conditions, and let $a_1^{(S)}$ and $a_2^{(S)}$ be the non-genuine line operators labelling these boundaries. We have the fusion rules
\be
a_{2,1}^{(S)}= s \times a_{1,2}^{(S)} = \tilde s \times a_{1,2}^{(S)}~,
\ee
and 
\be
a_{1,2}^{(S)} \times a_{1,2}^{(S)}= 1 + s \tilde s,~~ a_{1,2}^{(S)} \times a_{2,1}^{(S)}= s + \tilde s~.
\ee
The action of the surface operator $S$ on $a_{1,2}^{(S)}$ must fix them. To understand this statement, suppose we have 
\be
S \cdot a_{1}^{(S)}=a_2^{(S)}~.
\ee
Then we get Fig. \ref{fig:semsemact}
whose R.H.S. is obtained from ``flattening" the L.H.S. In other words, by fusing two patches of the $S$ surface on the left diagram on either side of the surface $S$ in the middle we get the diagram on the right. In the right diagram we have the non-genuine lines, $a_{1,2}^{(S)}$, forming a junction at the end of $S$. This situation is not consistent since $a_{1}^{(S)}$ and $a_{2}^{(S)}$ are distinct simple boundary conditions of $S$. Therefore, $S$ must act trivially on the two non-genuine lines, $a_{1,2}^{(S)}$. 

More generally, we can consider the action of $S$ on non-genuine lines $a^{(\bar S)} $ bounding the inverse surface $\bar S$, and use the argument in Fig. \ref{fig:semsemact} to show that the action of $S$ on any non-genuine line $a^{(S)}$ is trivial. 

\subsection{3-fermion Model}

Consider the 3-fermion model with line operators $1,\psi_1,\psi_2,\psi_3$ and topological twists $\theta_{1}=1, \theta_{\psi_i}=-1$. This TQFT has an $S_3$ 0-form symmetry which permutes the three fermions. Consider the surface operator $S_{(12)}$ which permutes $\psi_1$ and $\psi_2$. We have two non-genuine line operators $a_{1,2}^{(S_{(12)})}$ with fusion rules
\be
a_{2,1}^{(S_{(12)})}= \psi_1 \times a_{1,2}^{(S_{(12)})} = \psi_2 \times a_{1,2}^{(S_{(12)})} ~,
\ee
and 
\be
a_{1,2}^{(S_{(12)})} \times a_{1,2}^{(S_{(12)})}= 1 + \psi_3,~~ a_{1,2}^{(S_{(12)})} \times a_{2,1}^{(S_{(12)})}= \psi_1 + \psi_3~.
\ee
Similarly, we have the surface operators $S_{(13)},S_{(23)}$ and corresponding non-genuine lines. Now, let us consider the surface operators $S_{(123)},S_{(132)}$ which permute all three fermionic lines. These surface operators have a single non-genuine line at their boundary $b^{(S_{(123)})}$, $c^{(S_{(132)})}$. 

The action of the surface operators on these non-genuine lines is mostly fixed by the fact the output and input surface operators must be conjugated by the surface operator being acted with. Moreover, $S$ acts trivially on a non-genuine line at its own boundary.  

Let us consider the action of the surface operator $S_{(123)}$ on $a_{1,2}^{(S_{(12)})}$. We know that 
\be
\label{eq:Sact3ferm}
S_{(123)} \cdot a_{1}^{(S_{(12)})}= n_{S \cdot a_{1}^{(S_{(12)})}}^{a_{1}^{S_{(23)}}} ~ a_{1}^{(S_{(23)})} + n_{S \cdot a_{1}^{(S_{(12)})}}^{a_{2}^{S_{(23)}}} ~ a_{2}^{(S_{(23)})}~.
\ee
Since the surface operator $S_{(123)}$ is invertible, we want the action of $S_{(123)}$ to preserve all gauge-invariant topological data associated with the non-genuine lines. From the fusion rules of non-genuine lines described above we know that both $a_{1,2}^{(S_{(12)})}$ and $a_{1,2}^{(S_{(13)})}$ have quantum dimension $\sqrt{2}$. Therefore, to preserve quantum dimensions only one among $n_{S \cdot a_{1}^{(S_{(12)})}}^{a_{1}^{S_{(23)}}}$ and $n_{S \cdot a_{1}^{(S_{(12)})}}^{a_{2}^{S_{(23)}}}$ must be non-zero. To exactly determine the outcome, we must compare the topological data of $a_{a_1}^{(S_{(12)})}$ with $a_{a_1}^{(S_{(23)})}$ and $a_{a_2}^{(S_{(23)})}$. For a non-genuine line operator $a$, the following combination of the $G$-crossed braided $R$-matrix is gauge invariant
\be
\frac{\sum_{\mu} [R_{aa}^{c}]_{\mu\mu}}{\sum_{\nu} [R_{aa}^{c'}]_{\nu\nu}}~,
\ee
where $\mu,\nu$ run over the basis of the relevant fusion spaces \cite{barkeshli2019symmetry}. Using \cite[Section XI M]{barkeshli2019symmetry}, we find
\be
\frac{R_{{a_1}^{(S_{(12)})}{a_1}^{(S_{(12)})}}^{1}}{R_{{a_1}^{(S_{(12)})}{a_1}^{(S_{(12)})}}^{\psi_3}}= e^{\frac{i\pi}{2}}, ~ \frac{R_{{a_1}^{(S_{(23)})}{a_1}^{(S_{(23)})}}^{1}}{R_{{a_1}^{(S_{(23)})}{a_1}^{(S_{(23)})}}^{\psi_1}}= e^{\frac{i\pi}{2}}, ~ \frac{R_{{a_2}^{(S_{(23)})}{a_2}^{(S_{(23)})}}^{1}}{R_{{a_2}^{(S_{(23)})}{a_2}^{(S_{(23)})}}^{\psi_1}}= e^{\frac{-i\pi}{2}}~.
\ee
Since the action of $S_{(123)}$ must preserve gauge-invariant topological data, we find 
\be
S_{(123)} \cdot a_{1}^{(S_{(12)})}=a_{1}^{(S_{(23)})}~.
\ee

More generally, the action of an invertible surface operator on non-genuine lines must be such that 
\be
\label{eq:twistactcond3}
\frac{\sum_{\mu} [R_{S\cdot a, S\cdot a}^{S\cdot c}]_{\mu\mu}}{\sum_{\nu} [R_{S\cdot a, S\cdot a}^{S\cdot c'}]_{\nu\nu}}=\frac{\sum_{\mu} [R_{a a}^{c}]_{\mu\mu}}{\sum_{\nu} [R_{aa}^{c'}]_{\nu\nu}}~,
\ee
for all $c,c'$. The equations \eqref{eq:twistactcond1},\eqref{eq:twistactcond2} and \eqref{eq:twistactcond3} are necessary conditions to be satisfied for consistent action of invertible surface operators on the genuine and non-genuine line operators. It will be interesting to find a necessary and sufficient set of conditions that determine the action of surface operators on non-genuine lines. Note that the conditions  \eqref{eq:twistactcond1},\eqref{eq:twistactcond2} and \eqref{eq:twistactcond3} are sufficient to determine the action of invertible symmetries on the twisted sectors in the 3-fermion model. For invertible surfaces, this is related to determining which higher representations of the symmetry group are realized in the TQFT \cite{Bhardwaj:2023wzd,Bartsch:2023pzl}. For non-invertible surface operators, this would amount to finding which representations of higher Tube algebra are realized in these TQFTs \cite{Bhardwaj:2023ayw,Bartsch:2023wvv}. We leave this study for future work.

\end{appendices}

\newpage
\bibliography{chetdocbib3}

%bibliography generated by nb.bst v1.01 (C) 2003-2010 Niklas Beisert
\begin{thebibliography}{10}
\ifx\href\asklfhas\newcommand{\href}[2]{#2}\fi
\ifx\arxivref\asklfhas\newcommand{\arxivref}[2]{\href{http://arxiv.org/abs/#1}{#2}}\fi
\ifx\doiref\asklfhas\newcommand{\doiref}[2]{\href{http://dx.doi.org/#1}{#2}}\fi
\parskip 0pt
\normalsize

\bibitem{gaiotto2015generalized}
D.~Gaiotto, A.~Kapustin, N.~Seiberg \& B.~Willett,
\textit{``{Generalized Global Symmetries}''},
\doiref{10.1007/JHEP02(2015)172}{JHEP \textbf{1502}, 172
  (2015)\ignorespaces}\ignorespaces,
\normalsize{\texttt{\arxivref{1412.5148}{arXiv:1412.5148
  \![hep-th]}}}\ignorespaces
\bibitem{cordova2022snowmass}
C.~Cordova, T.~T. Dumitrescu, K.~Intriligator \& S.-H. Shao,
\textit{``{Snowmass White Paper: Generalized Symmetries in Quantum Field Theory
  and Beyond}''},
\normalsize{\texttt{\arxivref{2205.09545}{arXiv:2205.09545
  \![hep-th]}}}\ignorespaces,
in \textit{``{2022 Snowmass Summer Study}''}
\bibitem{Bhardwaj:2022yxj}
L.~Bhardwaj, L.~E. Bottini, S.~Schafer-Nameki \& A.~Tiwari,
\textit{``{Non-invertible higher-categorical symmetries}''},
\doiref{10.21468/SciPostPhys.14.1.007}{SciPost~Phys. \textbf{14}, 007
  (2023)\ignorespaces}\ignorespaces,
\normalsize{\texttt{\arxivref{2204.06564}{arXiv:2204.06564
  \![hep-th]}}}\ignorespaces
\bibitem{Bhardwaj:2022kot}
L.~Bhardwaj, S.~Schafer-Nameki \& A.~Tiwari,
\textit{``{Unifying Constructions of Non-Invertible Symmetries}''},
\normalsize{\texttt{\arxivref{2212.06159}{arXiv:2212.06159
  \![hep-th]}}}\ignorespaces
\bibitem{Bhardwaj:2022maz}
L.~Bhardwaj, L.~E. Bottini, S.~Schafer-Nameki \& A.~Tiwari,
\textit{``{Non-Invertible Symmetry Webs}''},
\normalsize{\texttt{\arxivref{2212.06842}{arXiv:2212.06842
  \![hep-th]}}}\ignorespaces
\bibitem{Copetti:2023mcq}
C.~Copetti, M.~Del~Zotto, K.~Ohmori \& Y.~Wang,
\textit{``{Higher Structure of Chiral Symmetry}''},
\normalsize{\texttt{\arxivref{2305.18282}{arXiv:2305.18282
  \![hep-th]}}}\ignorespaces
\bibitem{kapustin2010surface}
A.~Kapustin \& N.~Saulina,
\textit{``Surface operators in 3d Topological Field Theory and 2d Rational
  Conformal Field Theory''},
\normalsize{\texttt{\arxivref{1012.0911}{arXiv:1012.0911
  \![hep-th]}}}\ignorespaces
\bibitem{Fuchs_2013}
J.~Fuchs, C.~Schweigert \& A.~Valentino,
\textit{``Bicategories for Boundary Conditions and for Surface Defects in 3-d
  {TFT}''},
\doiref{10.1007/s00220-013-1723-0}{Communications~in~Mathematical~Physics
  \textbf{321}, 543 (2013)\ignorespaces}\ignorespaces
\bibitem{Johnson_Freyd_2022}
T.~Johnson-Freyd,
\textit{``On the Classification of Topological Orders''},
\doiref{10.1007/s00220-022-04380-3}{Communications~in~Mathematical~Physics
  \textbf{393}, 989 (2022)\ignorespaces}\ignorespaces
\bibitem{Buican:2021xhs}
M.~Buican \& H.~Jiang,
\textit{``{1-form symmetry, isolated $ \mathcal{N} $ = 2 SCFTs, and Calabi-Yau
  threefolds}''},
\doiref{10.1007/JHEP12(2021)024}{JHEP \textbf{2112}, 024
  (2021)\ignorespaces}\ignorespaces,
\normalsize{\texttt{\arxivref{2106.09807}{arXiv:2106.09807
  \![hep-th]}}}\ignorespaces
\bibitem{JOHNSON_FREYD_2021}
T.~JOHNSON-FREYD \& M.~YU,
\textit{``{FUSION} 2-{CATEGORIES} {WITH} {NO} {LINE} {OPERATORS} {ARE}
  {GROUPLIKE}''},
\doiref{10.1017/s0004972721000095}{Bulletin~of~the~Australian~Mathematical~Society
  \textbf{104}, 434 (2021)\ignorespaces}\ignorespaces
\bibitem{Roumpedakis:2022aik}
K.~Roumpedakis, S.~Seifnashri \& S.-H. Shao,
\textit{``{Higher Gauging and Non-invertible Condensation Defects}''},
\normalsize{\texttt{\arxivref{2204.02407}{arXiv:2204.02407
  \![hep-th]}}}\ignorespaces
\bibitem{Brunner:2007qu}
I.~Brunner \& D.~Roggenkamp,
\textit{``{B-type defects in Landau-Ginzburg models}''},
\doiref{10.1088/1126-6708/2007/08/093}{JHEP \textbf{0708}, 093
  (2007)\ignorespaces}\ignorespaces,
\normalsize{\texttt{\arxivref{0707.0922}{arXiv:0707.0922
  \![hep-th]}}}\ignorespaces
\bibitem{Gaiotto:2012np}
D.~Gaiotto,
\textit{``{Domain Walls for Two-Dimensional Renormalization Group Flows}''},
\doiref{10.1007/JHEP12(2012)103}{JHEP \textbf{1212}, 103
  (2012)\ignorespaces}\ignorespaces,
\normalsize{\texttt{\arxivref{1201.0767}{arXiv:1201.0767
  \![hep-th]}}}\ignorespaces
\bibitem{Konechny:2023xvo}
A.~Konechny,
\textit{``{RG boundaries and Cardy's variational ansatz for multiple
  perturbations}''},
\normalsize{\texttt{\arxivref{2306.13719}{arXiv:2306.13719
  \![hep-th]}}}\ignorespaces
\bibitem{Bachas:2001vj}
C.~Bachas, J.~de~Boer, R.~Dijkgraaf \& H.~Ooguri,
\textit{``{Permeable conformal walls and holography}''},
\doiref{10.1088/1126-6708/2002/06/027}{JHEP \textbf{0206}, 027
  (2002)\ignorespaces}\ignorespaces,
\normalsize{\texttt{\arxivref{hep-th/0111210}{hep-th/0111210}}}\ignorespaces
\bibitem{Bachas:2013nxa}
C.~P. Bachas, I.~Brunner, M.~R. Douglas \& L.~Rastelli,
\textit{``{Calabi\textquoteright{}s diastasis as interface entropy}''},
\doiref{10.1103/PhysRevD.90.045004}{Phys.~Rev.~D \textbf{90}, 045004
  (2014)\ignorespaces}\ignorespaces,
\normalsize{\texttt{\arxivref{1311.2202}{arXiv:1311.2202
  \![hep-th]}}}\ignorespaces
\bibitem{thibault2022drinfeld}
D.~D. Thibault,
\textit{``Drinfeld centers and Morita equivalence classes of fusion
  2-categories''},
arXiv~preprint~arXiv:2211.04917 \textbf{90}, D.~D. Thibault
  (2022)\ignorespaces\ignorespaces
\bibitem{Moore:1988qv}
G.~W. Moore \& N.~Seiberg,
\textit{``{Classical and Quantum Conformal Field Theory}''},
\doiref{10.1007/BF01238857}{Commun.~Math.~Phys. \textbf{123}, 177
  (1989)\ignorespaces}\ignorespaces
\bibitem{etingof2016tensor}
P.~Etingof, S.~Gelaki, D.~Nikshych \& V.~Ostrik,
\textit{``Tensor categories''},
American Mathematical Soc. (2016)\ignorespaces
\bibitem{Kapustin:2010if}
A.~Kapustin \& N.~Saulina,
\textit{``{Surface operators in 3d Topological Field Theory and 2d Rational
  Conformal Field Theory}''},
\normalsize{\texttt{\arxivref{1012.0911}{arXiv:1012.0911
  \![hep-th]}}}\ignorespaces
\bibitem{davydov2011witt}
A.~Davydov, M.~Mueger, D.~Nikshych \& V.~Ostrik,
\textit{``The Witt group of non-degenerate braided fusion categories''},
\normalsize{\texttt{\arxivref{1009.2117}{arXiv:1009.2117
  \![math.QA]}}}\ignorespaces
\bibitem{gaiotto2019condensations}
D.~Gaiotto \& T.~Johnson-Freyd,
\textit{``Condensations in higher categories''},
\normalsize{\texttt{\arxivref{1905.09566}{arXiv:1905.09566
  \![math.CT]}}}\ignorespaces
\bibitem{Choi:2022zal}
Y.~Choi, C.~Cordova, P.-S. Hsin, H.~T. Lam \& S.-H. Shao,
\textit{``{Non-invertible Condensation, Duality, and Triality Defects in 3+1
  Dimensions}''},
\doiref{10.1007/s00220-023-04727-4}{Commun.~Math.~Phys. \textbf{402}, 489
  (2023)\ignorespaces}\ignorespaces,
\normalsize{\texttt{\arxivref{2204.09025}{arXiv:2204.09025
  \![hep-th]}}}\ignorespaces
\bibitem{deligne2002categories}
P.~Deligne,
\textit{``Cat{\'e}gories tensorielles''},
Mosc.~Math.~J \textbf{2}, 227 (2002)\ignorespaces\ignorespaces
\bibitem{Bhardwaj:2022lsg}
L.~Bhardwaj, S.~Schafer-Nameki \& J.~Wu,
\textit{``{Universal Non-Invertible Symmetries}''},
\doiref{10.1002/prop.202200143}{Fortsch.~Phys. \textbf{70}, 2200143
  (2022)\ignorespaces}\ignorespaces,
\normalsize{\texttt{\arxivref{2208.05973}{arXiv:2208.05973
  \![hep-th]}}}\ignorespaces
\bibitem{bartsch2023noninvertible}
T.~Bartsch, M.~Bullimore, A.~E.~V. Ferrari \& J.~Pearson,
\textit{``Non-invertible Symmetries and Higher Representation Theory I''},
\normalsize{\texttt{\arxivref{2208.05993}{arXiv:2208.05993
  \![hep-th]}}}\ignorespaces
\bibitem{greenough2010monoidal}
J.~Greenough,
\textit{``Monoidal 2-structure of bimodule categories''},
Journal~of~Algebra \textbf{324}, 1818 (2010)\ignorespaces\ignorespaces
\bibitem{Radhakrishnan:2023zcq}
R.~Radhakrishnan,
\textit{``{On Reconstructing Finite Gauge Group from Fusion Rules}''},
\normalsize{\texttt{\arxivref{2302.08419}{arXiv:2302.08419
  \![hep-th]}}}\ignorespaces
\bibitem{Kaidi:2022cpf}
J.~Kaidi, K.~Ohmori \& Y.~Zheng,
\textit{``{Symmetry TFTs for Non-Invertible Defects}''},
\normalsize{\texttt{\arxivref{2209.11062}{arXiv:2209.11062
  \![hep-th]}}}\ignorespaces
\bibitem{Kaidi:2023maf}
J.~Kaidi, E.~Nardoni, G.~Zafrir \& Y.~Zheng,
\textit{``{Symmetry TFTs and Anomalies of Non-Invertible Symmetries}''},
\normalsize{\texttt{\arxivref{2301.07112}{arXiv:2301.07112
  \![hep-th]}}}\ignorespaces
\bibitem{Choi:2022rfe}
Y.~Choi, H.~T. Lam \& S.-H. Shao,
\textit{``{Noninvertible Time-Reversal Symmetry}''},
\doiref{10.1103/PhysRevLett.130.131602}{Phys.~Rev.~Lett. \textbf{130}, 131602
  (2023)\ignorespaces}\ignorespaces,
\normalsize{\texttt{\arxivref{2208.04331}{arXiv:2208.04331
  \![hep-th]}}}\ignorespaces
\bibitem{Belov:2005ze}
D.~Belov \& G.~W. Moore,
\textit{``{Classification of Abelian spin Chern-Simons theories}''},
\normalsize{\texttt{\arxivref{hep-th/0505235}{hep-th/0505235}}}\ignorespaces
\bibitem{stirling2008abelian}
S.~D. Stirling,
\textit{``Abelian Chern-Simons theory with toral gauge group, modular tensor
  categories, and group categories''},
\normalsize{\texttt{\arxivref{0807.2857}{arXiv:0807.2857
  \![hep-th]}}}\ignorespaces
\bibitem{Kapustin_2011}
A.~Kapustin \& N.~Saulina,
\textit{``Topological boundary conditions in abelian Chern{\textendash}Simons
  theory''},
\doiref{10.1016/j.nuclphysb.2010.12.017}{Nuclear~Physics~B \textbf{845}, 393
  (2011)\ignorespaces}\ignorespaces
\bibitem{Delmastro:2019vnj}
D.~Delmastro \& J.~Gomis,
\textit{``{Symmetries of Abelian Chern-Simons Theories and Arithmetic}''},
\doiref{10.1007/JHEP03(2021)006}{JHEP \textbf{2103}, 006
  (2021)\ignorespaces}\ignorespaces,
\normalsize{\texttt{\arxivref{1904.12884}{arXiv:1904.12884
  \![hep-th]}}}\ignorespaces
\bibitem{Wang_2020}
L.~Wang \& Z.~Wang,
\textit{``In and around abelian anyon models''},
\doiref{10.1088/1751-8121/abc6c0}{Journal~of~Physics~A:~Mathematical~and~Theoretical
  \textbf{53}, 505203 (2020)\ignorespaces}\ignorespaces
\bibitem{Kong_2009}
L.~Kong \& I.~Runkel,
\textit{``Cardy Algebras and Sewing Constraints, I''},
\doiref{10.1007/s00220-009-0901-6}{Communications~in~Mathematical~Physics
  \textbf{292}, I.~Runkel (2009)\ignorespaces}\ignorespaces
\bibitem{gelaki2009centers}
S.~Gelaki, D.~Naidu \& D.~Nikshych,
\textit{``Centers of graded fusion categories''},
\normalsize{\texttt{\arxivref{0905.3117}{arXiv:0905.3117
  \![math.QA]}}}\ignorespaces
\bibitem{muger2003structure}
M.~M{\"u}ger,
\textit{``On the structure of modular categories''},
Proceedings~of~the~London~Mathematical~Society \textbf{87}, 291
  (2003)\ignorespaces\ignorespaces
\bibitem{braidedcat1}
V.~Drinfeld, S.~Gelaki, D.~Nikshych \& V.~Ostrik,
\textit{``On braided fusion categories I''},
\href{https://arxiv.org/abs/0906.0620}{\texttt{https://arxiv.org/abs/0906.0620}}
\bibitem{Buican:2021axn}
M.~Buican \& R.~Radhakrishnan,
\textit{``{Galois orbits of TQFTs: symmetries and unitarity}''},
\doiref{10.1007/JHEP01(2022)004}{JHEP \textbf{2201}, 004
  (2022)\ignorespaces}\ignorespaces,
\normalsize{\texttt{\arxivref{2109.02766}{arXiv:2109.02766
  \![hep-th]}}}\ignorespaces
\bibitem{Buican:2021uyp}
M.~Buican, A.~Dymarsky \& R.~Radhakrishnan,
\textit{``{Quantum codes, CFTs, and defects}''},
\doiref{10.1007/JHEP03(2023)017}{JHEP \textbf{2303}, 017
  (2023)\ignorespaces}\ignorespaces,
\normalsize{\texttt{\arxivref{2112.12162}{arXiv:2112.12162
  \![hep-th]}}}\ignorespaces
\bibitem{Kawabata:2023rlt}
K.~Kawabata \& S.~Yahagi,
\textit{``{Elliptic genera from classical error-correcting codes}''},
\normalsize{\texttt{\arxivref{2308.12592}{arXiv:2308.12592
  \![hep-th]}}}\ignorespaces
\bibitem{Kawabata:2023iss}
K.~Kawabata, T.~Nishioka \& T.~Okuda,
\textit{``{Narain CFTs from quantum codes and their $\mathbb{Z}_2$ gauging}''},
\normalsize{\texttt{\arxivref{2308.01579}{arXiv:2308.01579
  \![hep-th]}}}\ignorespaces
\bibitem{Kawabata:2023usr}
K.~Kawabata, T.~Nishioka \& T.~Okuda,
\textit{``{Supersymmetric conformal field theories from quantum stabilizer
  codes}''},
\normalsize{\texttt{\arxivref{2307.14602}{arXiv:2307.14602
  \![hep-th]}}}\ignorespaces
\bibitem{Alam:2023qac}
Y.~F. Alam, K.~Kawabata, T.~Nishioka, T.~Okuda \& S.~Yahagi,
\textit{``{Narain CFTs from nonbinary stabilizer codes}''},
\normalsize{\texttt{\arxivref{2307.10581}{arXiv:2307.10581
  \![hep-th]}}}\ignorespaces
\bibitem{Furuta:2023xwl}
Y.~Furuta,
\textit{``{On the Rationality and the Code Structure of a Narain CFT, and the
  Simple Current Orbifold}''},
\normalsize{\texttt{\arxivref{2307.04190}{arXiv:2307.04190
  \![hep-th]}}}\ignorespaces
\bibitem{Kawabata:2023nlt}
K.~Kawabata \& S.~Yahagi,
\textit{``{Fermionic CFTs from classical codes over finite fields}''},
\doiref{10.1007/JHEP05(2023)096}{JHEP \textbf{2305}, 096
  (2023)\ignorespaces}\ignorespaces,
\normalsize{\texttt{\arxivref{2303.11613}{arXiv:2303.11613
  \![hep-th]}}}\ignorespaces
\bibitem{Dymarsky:2020qom}
A.~Dymarsky \& A.~Shapere,
\textit{``{Quantum stabilizer codes, lattices, and CFTs}''},
\doiref{10.1007/JHEP03(2021)160}{JHEP \textbf{2021}, 160
  (2020)\ignorespaces}\ignorespaces,
\normalsize{\texttt{\arxivref{2009.01244}{arXiv:2009.01244
  \![hep-th]}}}\ignorespaces
\bibitem{Kibe:2023nnj}
T.~Kibe, A.~Mukhopadhyay \& P.~Padmanabhan,
\textit{``{A stabilizer code model with non-invertible symmetries: Strange
  fractons, confinement, and non-commutative and non-Abelian fusion rules}''},
\normalsize{\texttt{\arxivref{2309.10037}{arXiv:2309.10037
  \![hep-th]}}}\ignorespaces
\bibitem{ostrik2001module}
V.~Ostrik,
\textit{``Module categories, weak Hopf algebras and modular invariants''},
\normalsize{\texttt{\arxivref{math/0111139}{math/0111139
  \![math.QA]}}}\ignorespaces
\bibitem{Fuchs:2002cm}
J.~Fuchs, I.~Runkel \& C.~Schweigert,
\textit{``{TFT construction of RCFT correlators 1. Partition functions}''},
\doiref{10.1016/S0550-3213(02)00744-7}{Nucl.~Phys.~B \textbf{646}, 353
  (2002)\ignorespaces}\ignorespaces,
\normalsize{\texttt{\arxivref{hep-th/0204148}{hep-th/0204148}}}\ignorespaces
\bibitem{davydov2015unphysical}
A.~Davydov,
\textit{``Unphysical diagonal modular invariants''},
\normalsize{\texttt{\arxivref{1412.8505}{arXiv:1412.8505
  \![math.QA]}}}\ignorespaces
\bibitem{Kawahigashi_2015}
Y.~Kawahigashi,
\textit{``A Remark on Gapped Domain Walls Between Topological Phases''},
\doiref{10.1007/s11005-015-0766-x}{Letters~in~Mathematical~Physics
  \textbf{105}, 893 (2015)\ignorespaces}\ignorespaces
\bibitem{Buican:2020who}
M.~Buican, L.~Li \& R.~Radhakrishnan,
\textit{``{$a\times b=c$ in $2+1$D TQFT}''},
\doiref{10.22331/q-2021-06-04-468}{Quantum \textbf{5}, 468
  (2021)\ignorespaces}\ignorespaces,
\normalsize{\texttt{\arxivref{2012.14689}{arXiv:2012.14689
  \![hep-th]}}}\ignorespaces
\bibitem{Davydov_2014}
A.~Davydov,
\textit{``Bogomolov multiplier, double class-preserving automorphisms, and
  modular invariants for orbifolds''},
\doiref{10.1063/1.4895764}{Journal~of~Mathematical~Physics \textbf{55},
  A.~Davydov (2014)\ignorespaces}\ignorespaces
\bibitem{Cong_2017}
I.~Cong, M.~Cheng \& Z.~Wang,
\textit{``Hamiltonian and Algebraic Theories of Gapped Boundaries in
  Topological Phases of Matter''},
\doiref{10.1007/s00220-017-2960-4}{Communications~in~Mathematical~Physics
  \textbf{355}, 645 (2017)\ignorespaces}\ignorespaces
\bibitem{barkeshli2019symmetry}
M.~Barkeshli, P.~Bonderson, M.~Cheng \& Z.~Wang,
\textit{``Symmetry fractionalization, defects, and gauging of topological
  phases''},
Physical~Review~B \textbf{100}, 115147 (2019)\ignorespaces\ignorespaces
\bibitem{Moore:1988ss}
G.~W. Moore \& N.~Seiberg,
\textit{``{Naturality in Conformal Field Theory}''},
\doiref{10.1016/0550-3213(89)90511-7}{Nucl.~Phys.~B \textbf{313}, 16
  (1989)\ignorespaces}\ignorespaces
\bibitem{Dijkgraaf:1988tf}
R.~Dijkgraaf \& E.~P. Verlinde,
\textit{``{Modular Invariance and the Fusion Algebra}''},
\doiref{10.1016/0920-5632(88)90371-4}{Nucl.~Phys.~B~Proc.~Suppl. \textbf{5},
  87}\ignorespaces
\bibitem{davydov2011commutative}
A.~Davydov \& T.~Booker,
\textit{``Commutative Algebras in Fibonacci Categories''},
\normalsize{\texttt{\arxivref{1103.3537}{arXiv:1103.3537
  \![math.CT]}}}\ignorespaces
\bibitem{Neupert_2016}
T.~Neupert, H.~He, C.~von~Keyserlingk, G.~Sierra \& B.~A. Bernevig,
\textit{``Boson condensation in topologically ordered quantum liquids''},
\doiref{10.1103/physrevb.93.115103}{Physical~Review~B \textbf{93}, B.~A.
  Bernevig (2016)\ignorespaces}\ignorespaces
\bibitem{neupert2016no}
T.~Neupert, H.~He, C.~Von~Keyserlingk, G.~Sierra \& B.~A. Bernevig,
\textit{``No-go theorem for boson condensation in topologically ordered quantum
  liquids''},
New~Journal~of~Physics \textbf{18}, 123009 (2016)\ignorespaces\ignorespaces
\bibitem{Fuchs_2004}
J.~Fuchs, I.~Runkel \& C.~Schweigert,
\textit{``{TFT} construction of {RCFT} correlators III:Simple currents''},
\doiref{10.1016/j.nuclphysb.2004.05.014}{Nuclear~Physics~B \textbf{694}, 277
  (2004)\ignorespaces}\ignorespaces
\bibitem{Komargodski:2020mxz}
Z.~Komargodski, K.~Ohmori, K.~Roumpedakis \& S.~Seifnashri,
\textit{``{Symmetries and strings of adjoint QCD$_{2}$}''},
\doiref{10.1007/JHEP03(2021)103}{JHEP \textbf{2103}, 103
  (2021)\ignorespaces}\ignorespaces,
\normalsize{\texttt{\arxivref{2008.07567}{arXiv:2008.07567
  \![hep-th]}}}\ignorespaces
\bibitem{rowell2009classification}
E.~Rowell, R.~Stong \& Z.~Wang,
\textit{``On classification of modular tensor categories''},
\normalsize{\texttt{\arxivref{0712.1377}{arXiv:0712.1377
  \![math.QA]}}}\ignorespaces
\bibitem{ostrik2006module}
V.~Ostrik,
\textit{``Module categories over the Drinfeld double of a finite group''},
\normalsize{\texttt{\arxivref{math/0202130}{math/0202130
  \![math.QA]}}}\ignorespaces
\bibitem{Kong_2008}
L.~Kong \& I.~Runkel,
\textit{``Morita classes of algebras in modular tensor categories''},
\doiref{10.1016/j.aim.2008.07.004}{Advances~in~Mathematics \textbf{219}, 1548
  (2008)\ignorespaces}\ignorespaces
\bibitem{Sawin:1999xt}
S.~F. Sawin,
\textit{``{Invariants of spin three manifolds from Chern-Simons theory and
  finite dimensional Hopf algebras}''},
\normalsize{\texttt{\arxivref{math/9910106}{math/9910106}}}\ignorespaces
\bibitem{davydov2012picard}
A.~Davydov \& D.~Nikshych,
\textit{``The Picard crossed module of a braided tensor category''},
\normalsize{\texttt{\arxivref{1202.0061}{arXiv:1202.0061
  \![math.QA]}}}\ignorespaces
\bibitem{ng2023classification}
S.-H. Ng, E.~C. Rowell \& X.-G. Wen,
\textit{``Classification of modular data up to rank 11''},
arXiv~preprint~arXiv:2308.09670 \textbf{219}, X.
  (2023)\ignorespaces\ignorespaces
\bibitem{Geiko:2022qjy}
R.~Geiko \& G.~W. Moore,
\textit{``{When Does A Three-Dimensional Chern-Simons-Witten Theory Have A Time
  Reversal Symmetry?}''},
\normalsize{\texttt{\arxivref{2209.04519}{arXiv:2209.04519
  \![hep-th]}}}\ignorespaces
\bibitem{Kaidi:2021gbs}
J.~Kaidi, Z.~Komargodski, K.~Ohmori, S.~Seifnashri \& S.-H. Shao,
\textit{``{Higher central charges and topological boundaries in 2+1-dimensional
  TQFTs}''},
\doiref{10.21468/SciPostPhys.13.3.067}{SciPost~Phys. \textbf{13}, 067
  (2022)\ignorespaces}\ignorespaces,
\normalsize{\texttt{\arxivref{2107.13091}{arXiv:2107.13091
  \![hep-th]}}}\ignorespaces
\bibitem{Lan:2014uaa}
T.~Lan, J.~C. Wang \& X.-G. Wen,
\textit{``{Gapped Domain Walls, Gapped Boundaries and Topological
  Degeneracy}''},
\doiref{10.1103/PhysRevLett.114.076402}{Phys.~Rev.~Lett. \textbf{114}, 076402
  (2015)\ignorespaces}\ignorespaces,
\normalsize{\texttt{\arxivref{1408.6514}{arXiv:1408.6514
  \![cond-mat.str-el]}}}\ignorespaces
\bibitem{naidu2006categorical}
D.~Naidu,
\textit{``Categorical Morita equivalence for group-theoretical categories''},
\normalsize{\texttt{\arxivref{math/0605530}{math/0605530
  \![math.QA]}}}\ignorespaces
\bibitem{Freed:2012bs}
D.~S. Freed \& C.~Teleman,
\textit{``{Relative quantum field theory}''},
\doiref{10.1007/s00220-013-1880-1}{Commun.~Math.~Phys. \textbf{326}, 459
  (2014)\ignorespaces}\ignorespaces,
\normalsize{\texttt{\arxivref{1212.1692}{arXiv:1212.1692
  \![hep-th]}}}\ignorespaces
\bibitem{Freed:2022qnc}
D.~S. Freed, G.~W. Moore \& C.~Teleman,
\textit{``{Topological symmetry in quantum field theory}''},
\normalsize{\texttt{\arxivref{2209.07471}{arXiv:2209.07471
  \![hep-th]}}}\ignorespaces
\bibitem{Cheng_2020}
M.~Cheng \& D.~J. Williamson,
\textit{``Relative anomaly in $(1+1)d$ rational conformal field theory''},
\doiref{10.1103/physrevresearch.2.043044}{Physical~Review~Research \textbf{2},
  D.~J. Williamson (2020)\ignorespaces}\ignorespaces
\bibitem{choi2023comments}
Y.~Choi, B.~C. Rayhaun, Y.~Sanghavi \& S.-H. Shao,
\textit{``Comments on Boundaries, Anomalies, and Non-Invertible Symmetries''},
\normalsize{\texttt{\arxivref{2305.09713}{arXiv:2305.09713
  \![hep-th]}}}\ignorespaces
\bibitem{Bhardwaj:2023wzd}
L.~Bhardwaj \& S.~Schafer-Nameki,
\textit{``{Generalized Charges, Part I: Invertible Symmetries and Higher
  Representations}''},
\normalsize{\texttt{\arxivref{2304.02660}{arXiv:2304.02660
  \![hep-th]}}}\ignorespaces
\bibitem{Bartsch:2023pzl}
T.~Bartsch, M.~Bullimore \& A.~Grigoletto,
\textit{``{Higher representations for extended operators}''},
\normalsize{\texttt{\arxivref{2304.03789}{arXiv:2304.03789
  \![hep-th]}}}\ignorespaces
\bibitem{Bhardwaj:2023ayw}
L.~Bhardwaj \& S.~Schafer-Nameki,
\textit{``{Generalized Charges, Part II: Non-Invertible Symmetries and the
  Symmetry TFT}''},
\normalsize{\texttt{\arxivref{2305.17159}{arXiv:2305.17159
  \![hep-th]}}}\ignorespaces
\bibitem{Bartsch:2023wvv}
T.~Bartsch, M.~Bullimore \& A.~Grigoletto,
\textit{``{Representation theory for categorical symmetries}''},
\normalsize{\texttt{\arxivref{2305.17165}{arXiv:2305.17165
  \![hep-th]}}}\ignorespaces
\end{thebibliography}

\end{document}